%% file: xrb8v8.tex
\documentclass[12pt,preprint]{aastex}

\shorttitle{Hydrodynamic Models of Type I XRBs: Metallicity Effects}
\shortauthors{Jos\'e et al.}

\newcommand{\zapj}{ApJ~}

\newcommand{\gap}{\mathrel{ \rlap{\raise.5ex\hbox{$>$}}
      {\lower.5ex\hbox{$\sim$}} } }
\newcommand{\lap}{\mathrel{ \rlap{\raise.5ex\hbox{$<$}}
		 {\lower.5ex\hbox{$\sim$}} } }

\begin{document}

\title{Hydrodynamic Models of Type I X-Ray Bursts: Metallicity Effects}

\author{Jordi Jos\'e}
\affil{Departament de F\'\i sica i Enginyeria Nuclear, EUETIB,
      Universitat Polit\`ecnica de Catalunya, C./ Comte d'Urgell 187,
      E-08036 Barcelona,  and Institut d'Estudis Espacials de Catalunya (IEEC), 
      Ed. Nexus-201, C/ Gran Capit\`a 2-4, E-08034, Barcelona, Spain; jordi.jose@upc.edu}
\author{Ferm\'\i n Moreno}
\affil{Departament de F\'\i sica i Enginyeria Nuclear, EUETIB,
      Universitat Polit\`ecnica de Catalunya, C./ Comte d'Urgell 187,
      E-08036 Barcelona, Spain; moreno@ieec.fcr.es}
\author{Anuj Parikh}
\affil{Physik Department E12,
       Technische Universit\"at M\"unchen, James-Franck-Strasse,
              D-85748 Garching, Germany; anuj.parikh@ph.tum.de}
\and
\author{Christian Iliadis}
\affil{Department of Physics and Astronomy, University of North Carolina,
       Chapel Hill, NC 27599-3255; and Triangle Universities Nuclear Laboratory,
       Durham, NC 27708-0308; iliadis@unc.edu}

\received{}
\accepted{}

\slugcomment{\underline{Submitted to}: \zapj \underline{Version}:
\today}

\clearpage

\begin{abstract}

Type I X-ray bursts are thermonuclear stellar explosions driven 
by charged-particle reactions. In the regime for combined H/He-ignition, 
the main nuclear flow is dominated by the {\it rp-process} (rapid
proton-captures and $\beta^+$-decays), the $3\alpha$-reaction,
and the {\it $\alpha$p-process} (a suite of ($\alpha$,p) and (p,$\gamma$) 
reactions). The main flow is expected to proceed away from the valley of
stability, eventually reaching the proton drip-line beyond A = 38.
Detailed analysis of the relevant reactions along the main path has only been scarcely addressed,
mainly in the context of parameterized one-zone models. In this paper,
we present a detailed study of the nucleosynthesis and nuclear processes
powering type I X-ray bursts. 
The reported 11 bursts have been computed by means of a spherically symmetric
(1D), Lagrangian, hydrodynamic code, linked to a nuclear reaction
network that contains 325 isotopes 
(from $^1$H to $^{107}$Te), and 1392 nuclear processes. These 
evolutionary sequences, followed from the onset of accretion up to
the explosion and expansion stages, have been performed for 2 different
metallicities to explore the dependence between the extension
of the main nuclear flow and the initial metal content. 
We carefully analyze the dominant reactions and the products of
nucleosynthesis,
together with the the physical parameters that determine the light curve (including recurrence times,
ratios between persistent and burst luminosities, or the extent of the envelope expansion).
Results are in qualitative agreement with the observed properties of 
some well-studied bursting sources. 
Leakage from the predicted SbSnTe-cycle cannot be discarded in some of
our models.
Production of $^{12}$C (and implications
for the mechanism that powers superbursts), light p-nuclei, and the amount
of H left over after the bursting episodes will also be discussed.  
 \end{abstract}

\clearpage

\section{Introduction}

Type I X-ray bursts (hereafter, XRBs) are cataclysmic stellar events. They are powered by
thermonuclear runaways (TNR) in the H/He-rich envelopes
accreted onto neutron stars in close binary systems (see reviews by
Bildsten 1998; Lewin et al.  1993, 1995; Psaltis 2006; Schatz \& Rehm 2006;
Strohmayer \& Bildsten 2006).  These events constitute the most frequent
type of thermonuclear stellar explosion in the Galaxy (the third, in terms of 
total energy output after supernovae and classical novae), in part 
because of their
short recurrence period (hours to days). About $90$ Galactic low-mass X-ray
binaries exhibiting such bursting behavior (with burst durations of 
$\tau_{burst} \sim$ 10 - 100 s)
have been found since the discovery of XRBs by Grindlay et al. (1976),
and independently, by Belian et al. (1976).
Type I X-ray bursts and their associated nucleosynthesis 
have been extensively modeled by different groups
(see pioneering work by Woosley \& Taam 1976, Maraschi \& Cavaliere 1977, and Joss 1977), 
reflecting the astrophysical interest in
determining the nuclear processes that power the explosion, the light curve, as well as
in providing reliable estimates for the chemical composition of the neutron star surface
(see Schatz et al. 1999, Parikh et al. 2008, and references therein).

With a neutron star hosting the explosion, temperatures and densities in the accreted envelope 
reach high values: T$_{peak} > 10^9$ K, and $\rho \sim 10^6$ g cm$^{-3}$.  
As a result, detailed
nucleosynthesis studies require the use of hundreds of isotopes,
linked by thousands of nuclear
interactions, extending all the way up to the SnSbTe-mass region
(Schatz et al. 2001) or beyond 
(the extent of the nuclear activity\footnote{The nuclear 
activity reflects the changes in composition driven by different nuclear
processes (i.e., p- and $\alpha$-capture reactions, $\beta$-decays, ...)
that take place in the envelope at different stages of the burst.
In this work, 
the extent of the nuclear activity is arbitrarily defined by the heaviest nucleus that achieves a mass fraction $>10^{-9}$.}  
in the XRB nucleosynthesis study of Koike et al. 2004 reaches $^{126}$Xe). 
Indeed, the extent of the rp-process in XRBs is still
not clear: recent experimental work now shows that it will be more
difficult to reach the SnSbTe-mass region (Elomaa et al. 2009).
 Because of computational constraints, XRB nucleosynthesis studies
have been traditionally performed using limited nuclear reaction networks, truncated 
near Ni (Woosley \& Weaver 1984; Taam et al. 1993; Taam, Woosley, \& Lamb 1996 --all using a 
19-isotope network), Kr (Hanawa, Sugimoto, \& Hashimoto 1983 --274-isotope network; 
Koike et al. 1999 --463 nuclides), Cd (Wallace \& Woosley 1984 --16-isotope network), 
or Y (Wallace \& Woosley 1981 --250-isotope network).  
On the other hand, Schatz et al. (1999, 2001)
have carried out very detailed nucleosynthesis calculations with a network containing 
more than 600 isotopes (up to Xe, in Schatz et al. 2001), 
but using a one-zone approach. Koike et al. (2004) have also performed detailed one-zone nucleosynthesis 
calculations, with temperature and density profiles obtained from a spherically symmetric evolutionary code,
linked to a 1270-isotope network extending up to $^{198}$Bi.

Until recently, it has not been possible to couple hydrodynamic stellar calculations (in 1-D)
and detailed networks. Recent efforts include Fisker et al. (2004, 2006, 2007, 2008), and Tan et al. (2007) 
($\sim$ 300 isotopes, up to $^{107}$Te), Jos\'e \& Moreno (2006) (2640 reactions 
and 478 isotopes, up to Te), or Woosley et al. (2004) and Heger et al. (2007) 
(up to 1300 isotopes with an adaptive network).
This has prompted a detailed analysis of the nuclear activity powering the bursts.
The most detailed work to date is that of Fisker et al. (2008), 
in the context of the 1-D general relativistic hydrodynamic code AGILE (Liebend\"orfer et al. 2002),
linked to a nuclear reaction network containing 304 isotopes: 
a thorough analysis of the main nuclear activity in one characteristic burst 
is reported (although details for a sequence of 5 consecutive, 'representative' bursts are also outlined). 
However, because of the specific choice of metallicity (Z = $10^{-3}$, for the 
accreted matter) and mass-accretion rate ($\dot M \sim 10^{17}$ g s$^{-1}$)  
adopted, the nuclear activity does not extend much beyond mass $A \sim 65$, 
as a result of compositional inertia effects, that quench further extension of the nuclear path. Hence, the flow 
does not reach the {\it SnSbTe}-mass region, which was suggested as a natural 
endpoint in XRB nucleosynthesis studies (see Schatz et al. 1999,2001). 

Clearly, the identification of the most relevant reactions in the $A \sim 65 - 100$ mass region remains to
be addressed in detail in the framework of hydrodynamic simulations. This is particularly relevant since,
as first pointed out by Hanawa et al. (1983), proton captures on heavy nuclei (i.e., the {\it rp}-process) have
a dramatic effect on the shape of XRB light curves.
To this end, a new set
of type I X-ray bursts have been computed with SHIVA,
a 1-D, spherically symmetric, hydrodynamic, implicit, Lagrangian code, used extensively in the modeling of classical
nova outbursts (see Jos\'e \& Hernanz 1998). The code has been linked
 to a fully updated nuclear reaction network containing 324 nuclides and 1392 nuclear processes,
 a subset of that used in Parikh et al. (2008), and includes the most relevant charged-particle
induced reactions occurring between $^1$H and $^{107}$Te, as well as their corresponding reverse processes. 
It is worth noting that the size of this network is similar (though slightly larger) to that adopted by Fisker et al. (2008).
In order to set up the reaction rate library for our study, we started by adopting the proton drip line predicted by Audi et al. (2003a, 2003b).
Experimental rates are available for a small subset of reactions (adopted from Angulo et al. 1999, 
Iliadis et al. 2001, and some recent updates for selected reactions). 
For all other reactions for which experimental rates are not available, we used the rates from the Hauser-Feshbach codes 
MOST (Goriely 1998; Arnould \& Goriely 2006) and NON-SMOKER (Rauscher \& Thielemann 2000; for details see Parikh et al. 2008).
Neutron captures are disregarded
since our early test calculations revealed that they play a minor role in XRB nucleosynthesis. All reaction rates incorporate
the effects of thermal excitations in the target nuclei (Rauscher \& Thielemann 2000). 
Screening factors are taken from Graboske et al. (1973) and DeWitt, Graboske, \& Cooper (1973).
For the weak interactions, $\beta$-delayed nucleon emission and laboratory decay rates (Audi et al. 2003a) have been adopted.
For a discussion of employing stellar versus laboratory decay rates, see Woosley et al. (2004). It is worth noting, however, that many computed stellar decay rates ( Fuller et al. 1982a, 1982b; Langanke \& Martinez-Pinedo 2000)
do not converge to their laboratory values at lower temperatures and densities, calling into question the model used for these calculations.
Studies employing properly converging stellar decay rates for all
isotopes relevant to XRB nucleosynthesis have not been performed by
any group yet, and would certainly be interesting,
although the results presented in this work would not be dramatically 
affected by their inclusion. 

SHIVA uses a time-dependent formalism for convective transport whenever the characteristic convective timescale becomes larger
that the integration time step. Partial mixing between adjacent convective shells is treated by means of a diffusion equation (Prialnik, Shara, \& Shaviv 1979).  
No additional semiconvection or thermohaline mixing is considered.
Models make use of Iben's (1975) opacity fits, better suited than the OPAL 
opacities for astrophysical environments that exhibit strong variations in metallicity, as in XRB nucleosynthesis.
However, plans to incorporate these more realistic opacities are currently underway. 
The adopted equation of state includes contributions from the electron
gas (with different degrees of degeneracy; Blinnikov et al. 1996), a multicomponent ion plasma,
and radiation; Coulomb corrections to the electronic pressure are also taken into account.

Accretion is computed by redistributing material through a constant 
number of envelope shells (see Kutter \& Sparks 1980, for details). To handle this, a tiny envelope, 
containing $1.1 \times 10^{18}$ g of material (less than 1 permil of the total envelope mass accreted during the first bursting
episode), distributed through all the envelope shells, is put initially in place(the influence of the number of envelope shells on burst
properties will be discussed in Section 3).
The model is then relaxed using a few, very large timesteps, to guarantee  
hydrostatic equilibrium. 
The temperature at the bottom of the envelope barely reaches $2.7 \times 10^7$ K, whereas the density is just 
$1.4 \times 10^3$ g cm$^{-3}$ (corresponding to a pressure of $5.7 \times 10^{18}$ dyn cm$^{-2}$).  Mass accretion 
and nuclear reactions are then initiated.

Special emphasis is placed on the effect of the initial metallicity of the accreted
matter on the main nuclear path, which in turn, 
will affect the final post-burst envelope composition 
and the shape of the light curves. 

The structure of the manuscript is as follows: in Section 2, we analyze the main features (nuclear
path, nucleosynthesis, light curves, etc) of a series of 4 bursts computed in a model with solar-like accreted material.
The effect of the resolution adopted in this model is discussed in Section 3. A detailed analysis of the impact
of the metallicity of the accreted material on burst properties is given in Section 4.
Finally, a comparison with previous work, together with a thorough analysis of the corrections posed by
general relativity, are discussed in Section 5.

\section{Model 1\label{sec:Model 1}}
We summarize the gross properties of a series of thermonuclear bursts 
driven by mass accretion onto a 1.4 M$_\odot$ neutron star 
(L$_{ini}$ = $1.6 \times 10^{34}$ erg s$^{-1}$ = 4.14 L$_\odot$),
at a rate $\rm{\dot M}$$_{acc}$ = $1.75 \times 10^{-9}$ M$_\odot$ yr$^{-1}$  
(corresponding to 0.08 $\rm{\dot M}$$_{Edd}$). The composition
of the accreted material (see Table 1) is assumed to be solar-like (X=0.7048, Y=0.2752, Z=0.02).
All metals are initially assumed to be in the form of $^{14}$N, 
following the rapid rearrangement of CNO isotopes that naturally occurs early in the burst (see Woosley et al. 2004).
This model is qualitatively similar to model ZM, computed by Woosley
et al. (2004) in the framework of the 1-D, hydrodynamic, implicit code 
{\it KEPLER}. This choice is made intentionally to compare
with previous hydrodynamic studies.
Notice, however, 
that Woosley et al. assume a value of 10 km for the neutron star radius. In contrast, our model yields a value of 13.1 km, 
following the integration of the neutron star structure\footnote{The Harrison-Wheeler equation of state
(see Shapiro \& Teukolsky 1983) has been adopted to build up the initial neutron star configuration in hydrostatic equilibrium conditions. Although this
equation of state is a rather crude approximation to the physical conditions in the very deep interior of a neutron star, the radius obtained is in good
agreement with the values derived with more accurate equations of state (Lattimer 2009).}
from the core to its surface, in hydrostatic
equilibrium. Differences in the neutron star size
(and in turn, in surface gravity) may effect the strength of the
explosion (mass accreted, peak temperature, nucleosynthesis, etc).
 
\subsection{First burst}
The piling up of solar-like material on top of the neutron star during the accretion stage progressively compresses and heats the envelope (consisting of
60 shells).  Indeed, only 145 seconds since the 
beginning of accretion, the temperature at the base of the envelope reaches 
T$_{base} = 5 \times 10^7$ K (with $\rho_{base}$ exceeding $10^4$ g cm$^{-3}$). 

\begin{figure*}
 \centering
 \includegraphics[width=0.85\textwidth]{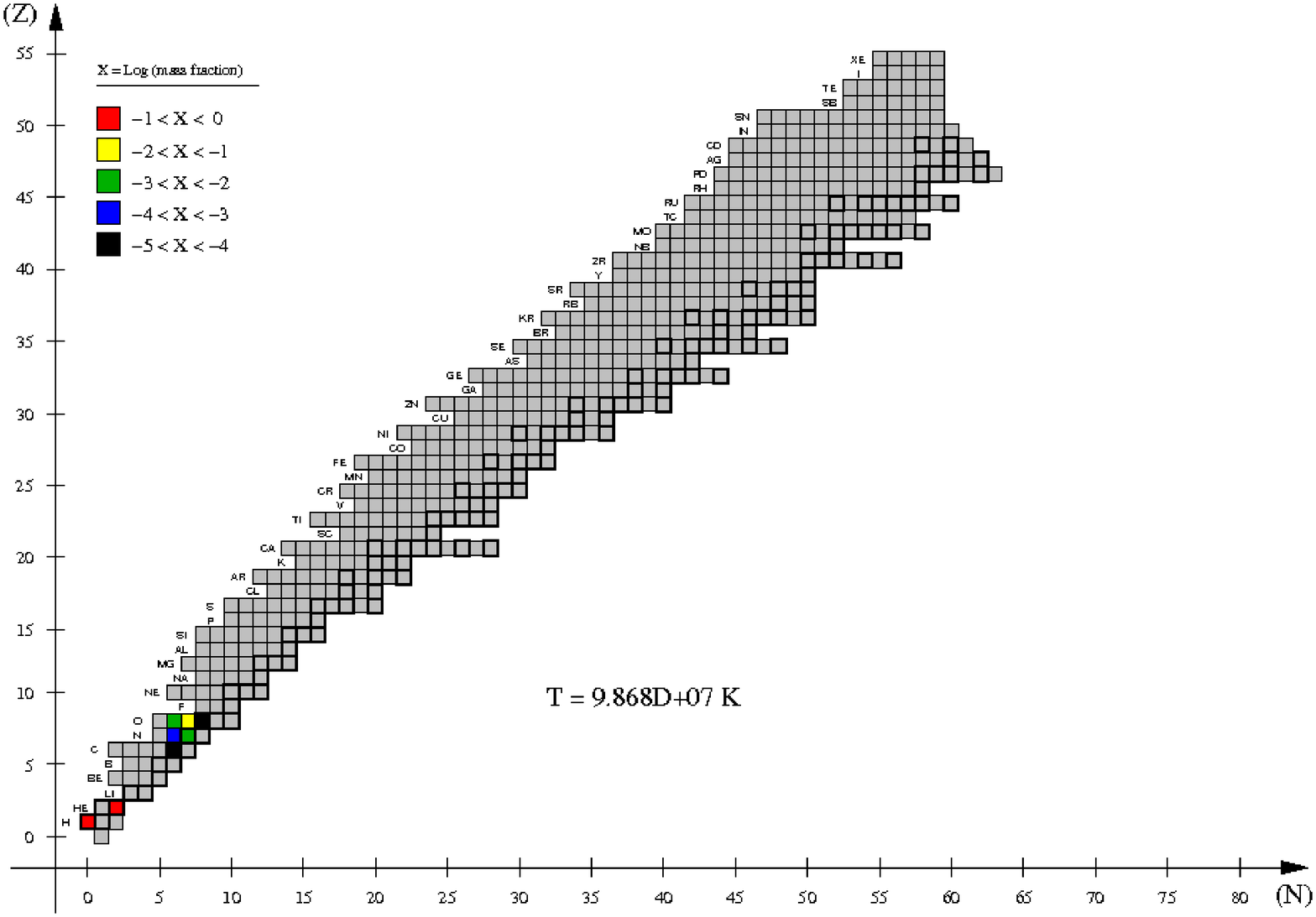}
 \includegraphics[width=0.85\textwidth]{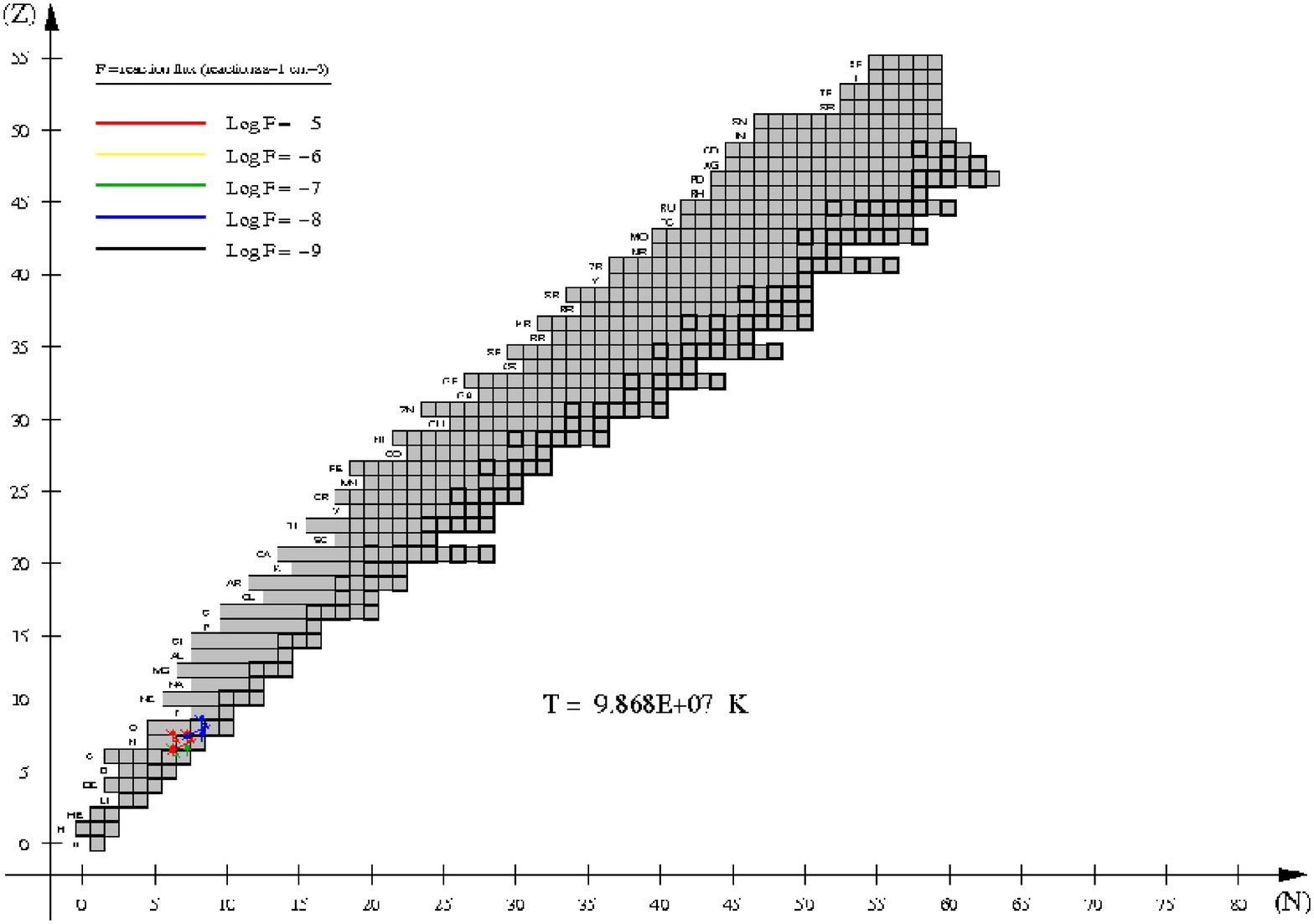}
\caption{Main nuclear activity at the innermost envelope shell for model 1
($\rm{M}_{NS}$ = 1.4 M$_\odot$, $\rm{\dot M}$$_{acc}$ = $1.75 \times 10^{-9}$ M$_\odot$ yr$^{-1}$, 
 Z = 0.02), at the early stages of accretion ( T$_{base} = 9.9 \times 10^{7}$ K). Upper panel:
 mass fractions of the most abundant species (X $> 10^{-5}$); 
 Lower panel: main reaction fluxes (F $\geq 10^{-9}$ reactions s$^{-1}$ cm$^{-3}$).  }
\label{fig:M1000}
\end{figure*}

The early nuclear activity is fully dominated by H-burning through hot 
CNO-cycle reactions,
initiated by proton-captures on $^{14}$N nuclei. At this stage
(t=2327 s), the envelope achieves T$_{base} \sim 10^8$ K ($\rho_{base} \sim 6.5 
\times 10^4$ g cm$^{-3}$), with an energy generation rate of $\epsilon_{nuc} \sim 1.2 \times 10^{14}$ erg g$^{-1}$ s$^{-1}$.
The main reaction path (see Fig. \ref{fig:M1000}) 
is led by $^{15}$N(p, $\alpha$)$^{12}$C, which powers $^{12}$C(p, $\gamma$)$^{13}$N(p, 
$\gamma$)$^{14}$O($\beta^+$)$^{14}$N(p, $\gamma$)$^{15}$O($\beta^+$)$^{15}$N. This suite of 
nuclear processes competes with $^{13}$N($\beta^+$)$^{13}$C(p, 
$\gamma$)$^{14}$N, and to a lesser extent, with $^{15}$N(p, $\gamma$)$^{16}$O(p, 
$\gamma$)$^{17}$F($\beta^+$)$^{17}$O(p, $\alpha$)$^{14}$N. 
Besides H (X=0.689) and $^4$He (Y=0.290), the next most abundant species in the envelope
is now $^{15}$O ($1.2 \times 10^{-2}$). The amount of unburned $^{14}$N has dropped
to $2.4 \times 10^{-3}$ (in the following, when discussing
the nucleosynthesis, we will refer to abundances by mass, i.e., mass 
fractions). 
Other CNO-group nuclei, such as  $^{12}$C ($5.4 \times 10^{-5}$), 
$^{13}$N ($4.2 \times 10^{-4}$), $^{14}$O ($6.1 \times 10^{-3}$), or $^{16}$O ($1.1 \times 10^{-5}$), 
have already achieved an abundance $\geq 10^{-5}$.  

\begin{figure}
 \centering
   \includegraphics[width=0.90\textwidth]{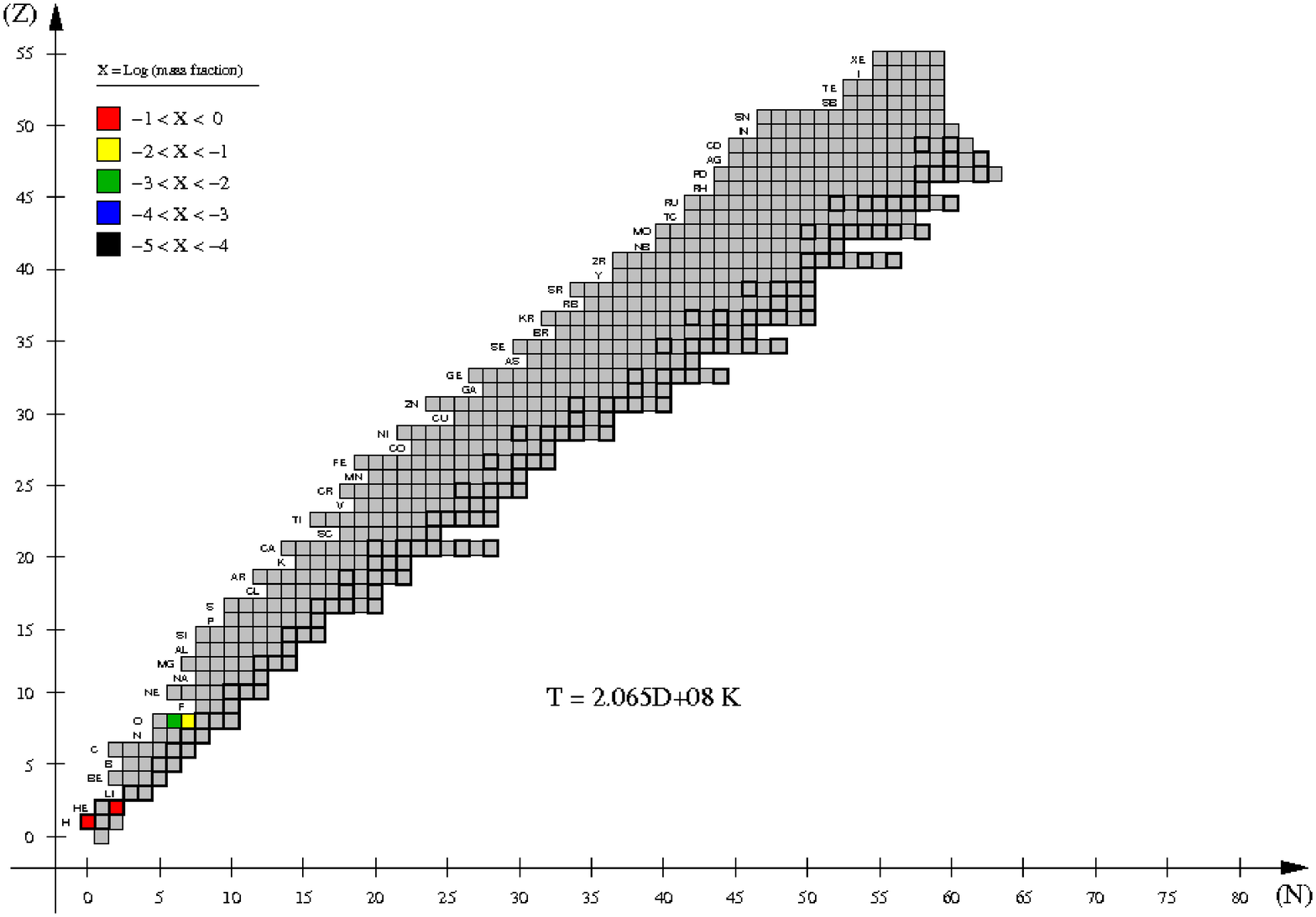}
   \includegraphics[width=0.90\textwidth]{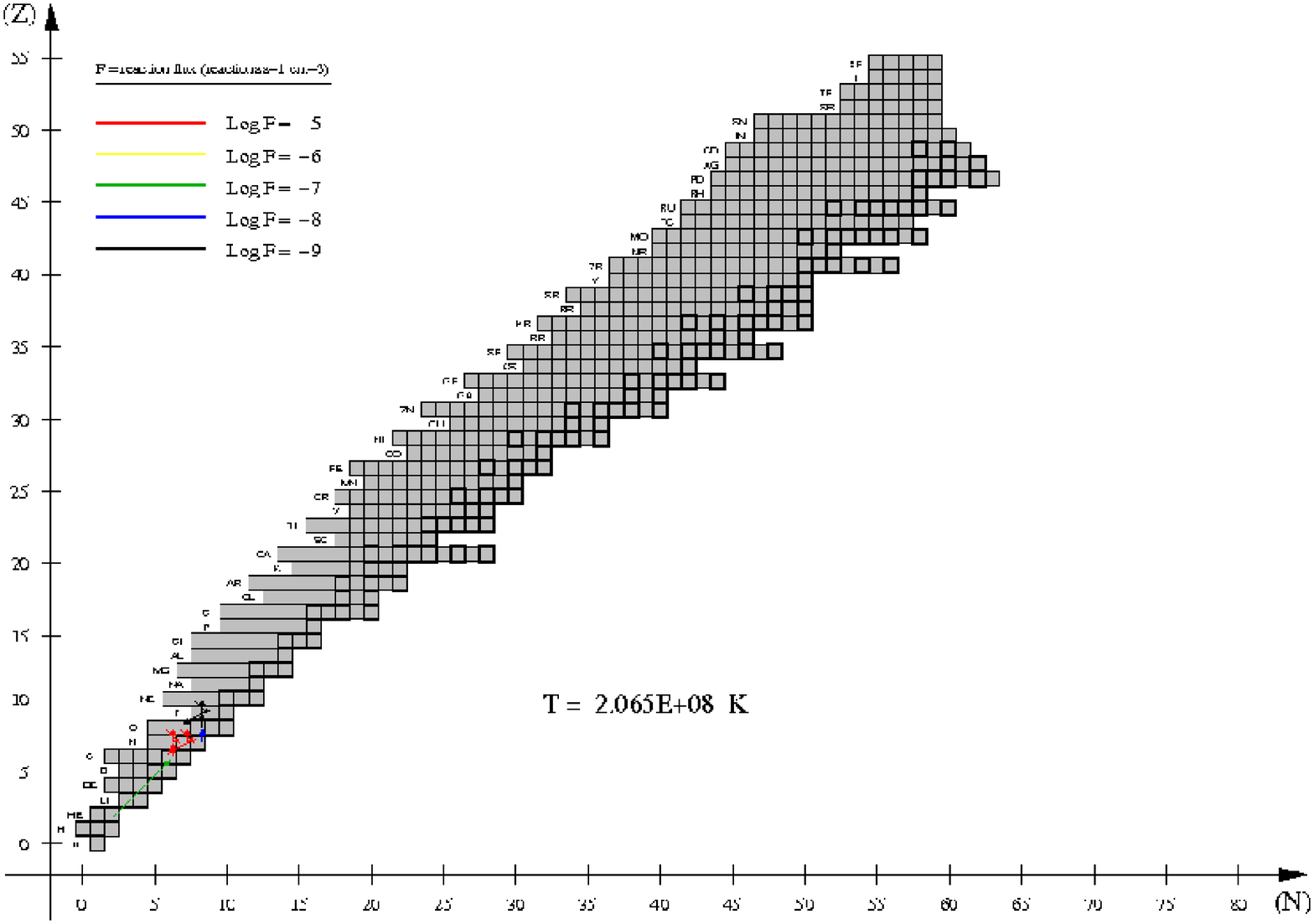}
\caption{Same as Fig. \ref{fig:M1000},  but for T$_{base} = 2.1 \times 10^{8}$ K.}
\label{fig:M3000}
\end{figure}

4.49 hours (16,163 s) after the beginning of accretion, T$_{base}$  
reaches $2.1 \times 10^8$ K. Mass accretion in highly degenerate conditions has compressed the envelope base to 
a density of $\rho_{base} = 2.7 \times 10^5$ g cm$^{-3}$ (P$_{base} = 9.1 \times 10^{21}$ dyn cm$^{-2}$).  The total luminosity of the star
has now increased to a value of $2.5 \times 10^{35}$ erg s$^{-1}$.
The main nuclear activity\footnote{Equilibrium (p,$\gamma$)-($\gamma$),p) 
pairs have been removed from the accompanying plots to
highlight those reactions of lower flux that directly lead to the
production of heavier species during the burst.}
(Fig. \ref{fig:M3000}) is still dominated by proton-captures and $\beta^+$-decays, characteristics of the hot CNO-cycle,
now supplemented by 
$^{15}$N(p, $\gamma$)$^{16}$O(p, $\gamma$)$^{17}$F(p, $\gamma$)$^{18}$Ne($\beta^+$)$^{18}$F(p, $\alpha$)$^{15}$O, 
and by the 3$\alpha$ reaction. 
The numerous p-captures have reduced the 
hydrogen content to 0.408. In turn, $^4$He has increased to 0.570, 
becoming now the most abundant species at the base of the envelope (followed by
the short-lived species $^{14}$O 
[$7.7 \times 10^{-3}$] and $^{15}$O [$1.4 \times 10^{-2}$]), while most of the CNO nuclei 
have been reduced to $10^{-7} - 10^{-9}$, by mass). The extension of the main nuclear activity reaches $^{40}$Ca. Indeed, $^{32}$S 
and $^{40}$Ca are the only species in the 
Ne-Ca mass region with abundances exceeding $10^{-9}$.

\begin{figure}
 \centering
   \includegraphics[width=0.45\textwidth]{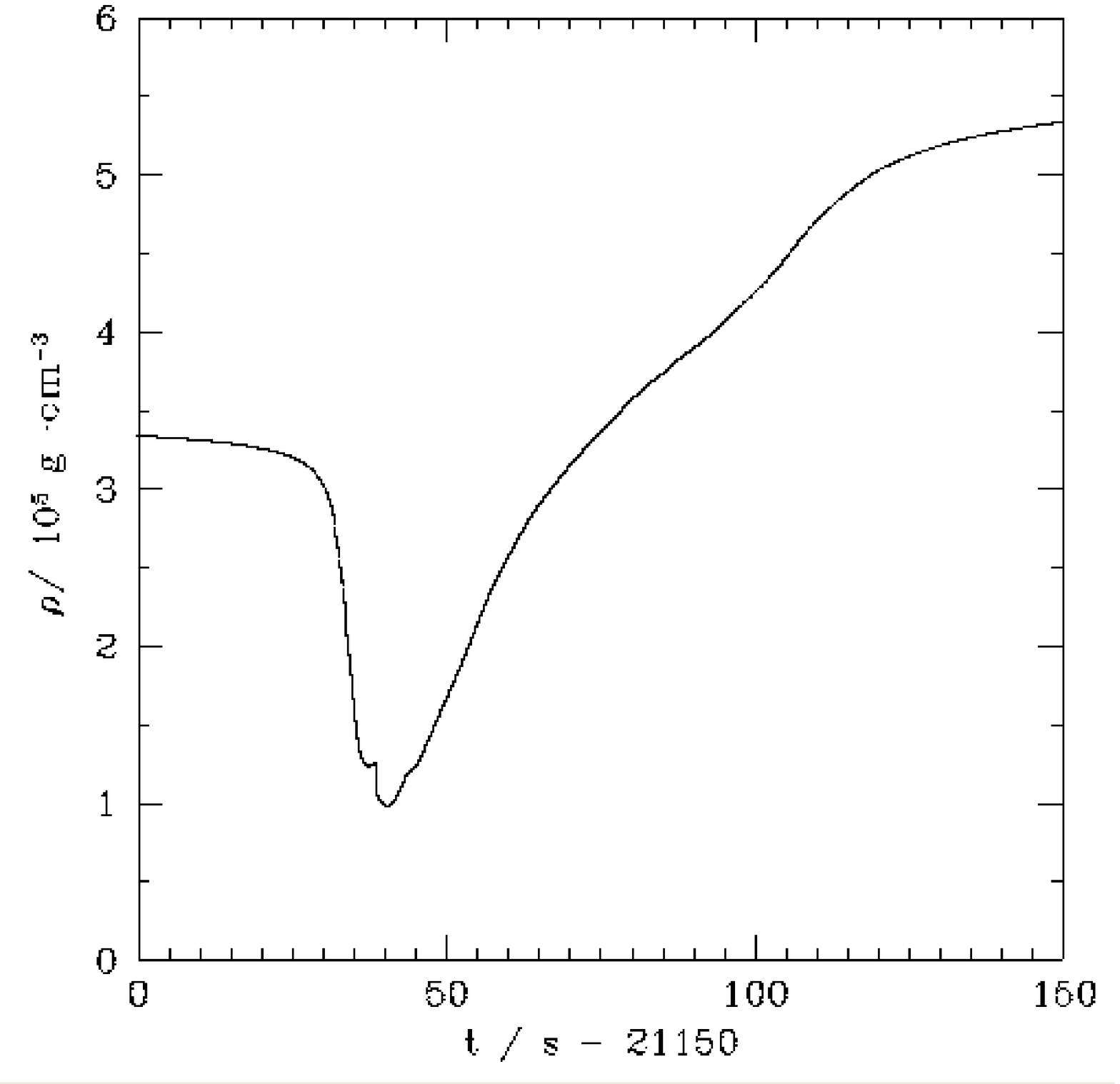}
   \includegraphics[width=0.45\textwidth]{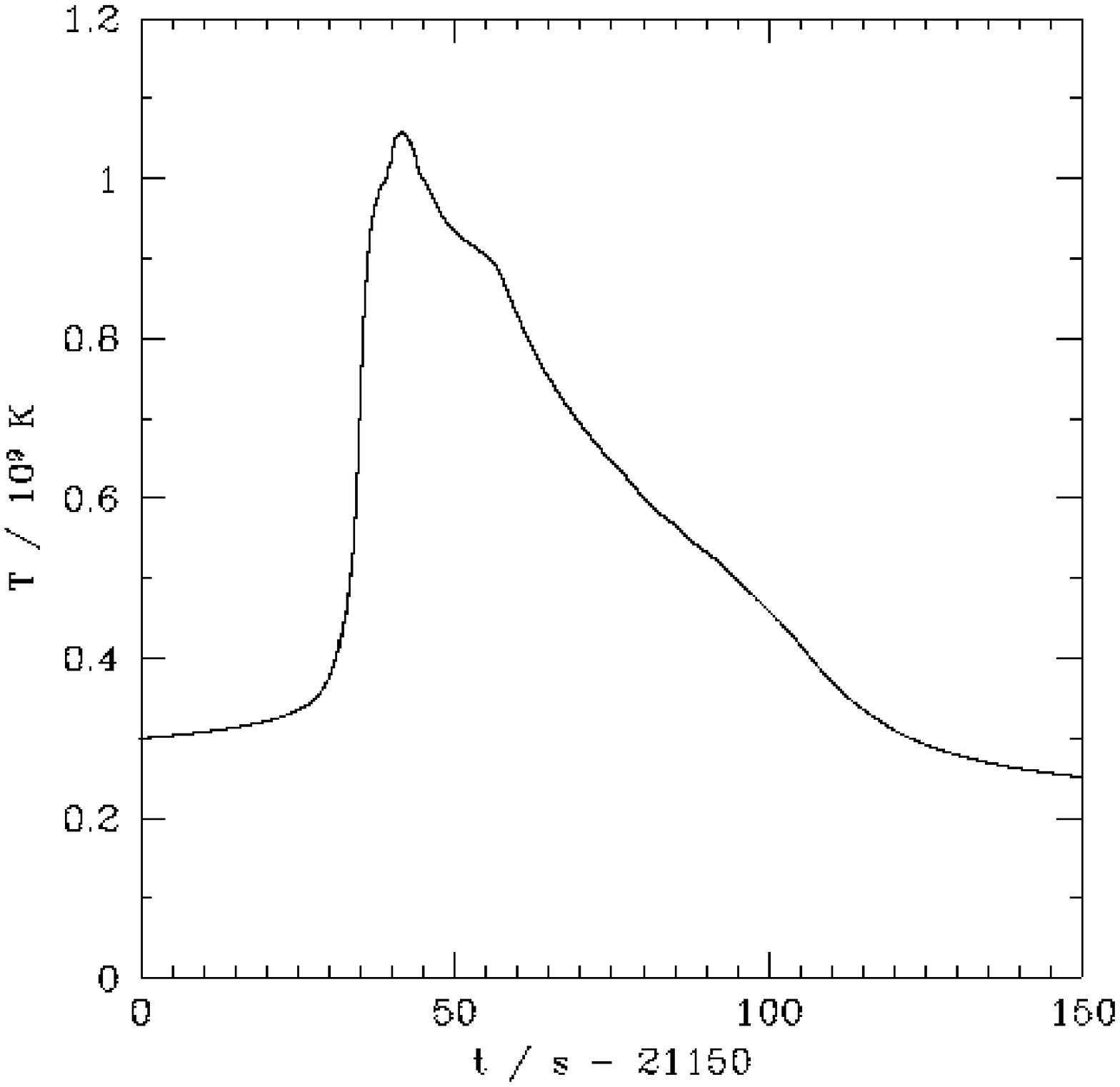}
   \includegraphics[width=0.45\textwidth]{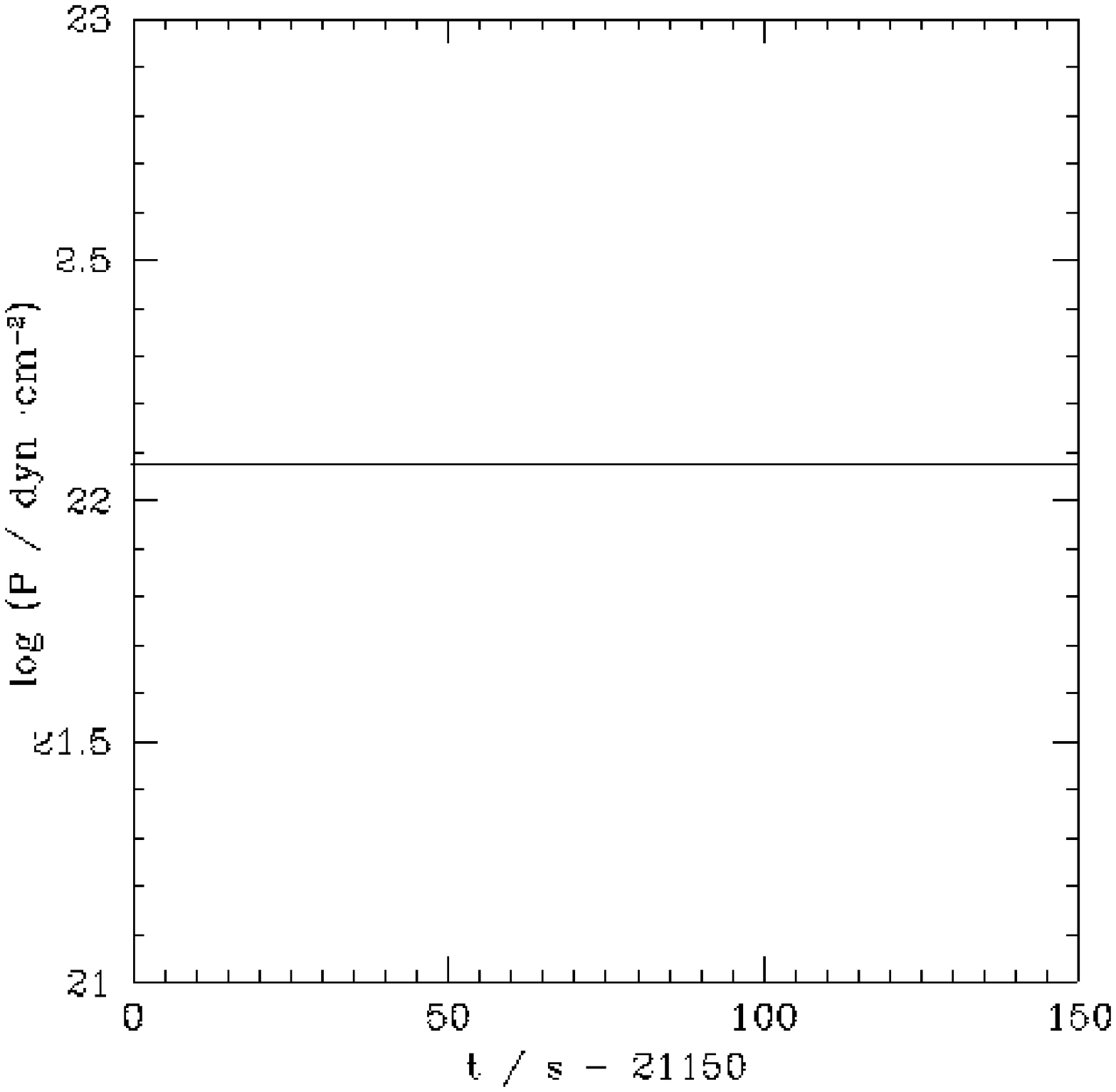}
   \includegraphics[width=0.45\textwidth]{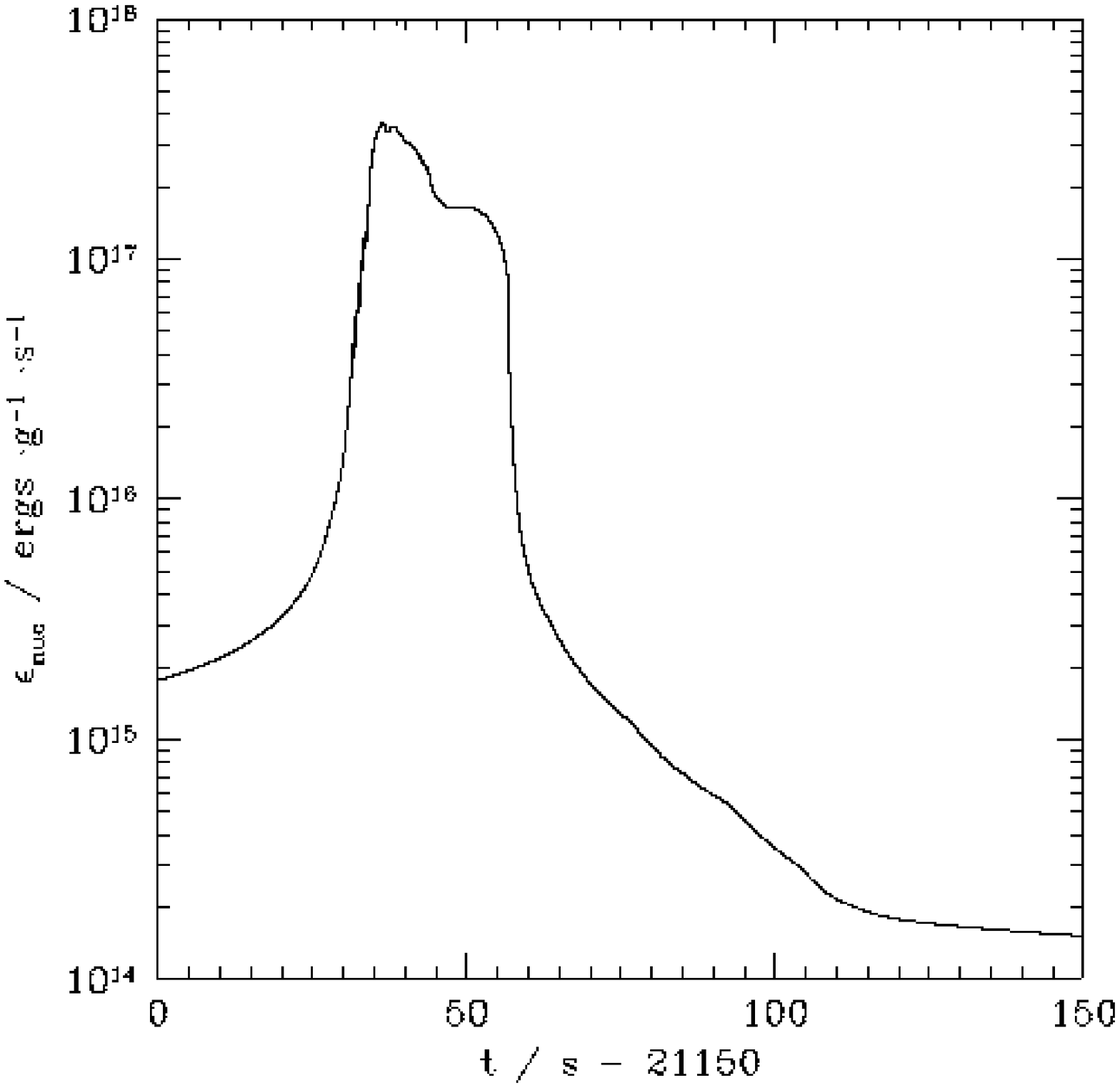}
\caption{Time evolution of density (upper left panel), temperature (upper right), pressure (lower left), and
          nuclear energy generation rate (lower right), at the innermost envelope shell for model 1
	  ($\rm{M}_{NS}$ = 1.4 M$_\odot$, $\rm{\dot M}$$_{acc}$ = $1.75 \times 10^{-9}$ M$_\odot$ yr$^{-1}$,
	   Z = 0.02), along the first bursting episode.  The origin of the time coordinate is arbitrarily
	   chosen as 21,150 s, for which T$_{base} \sim 3 \times 10^8$ K.}
\label{fig:rhotpenuc_t1}
\end{figure}

\begin{figure}
 \centering
   \includegraphics[width=0.45\textwidth]{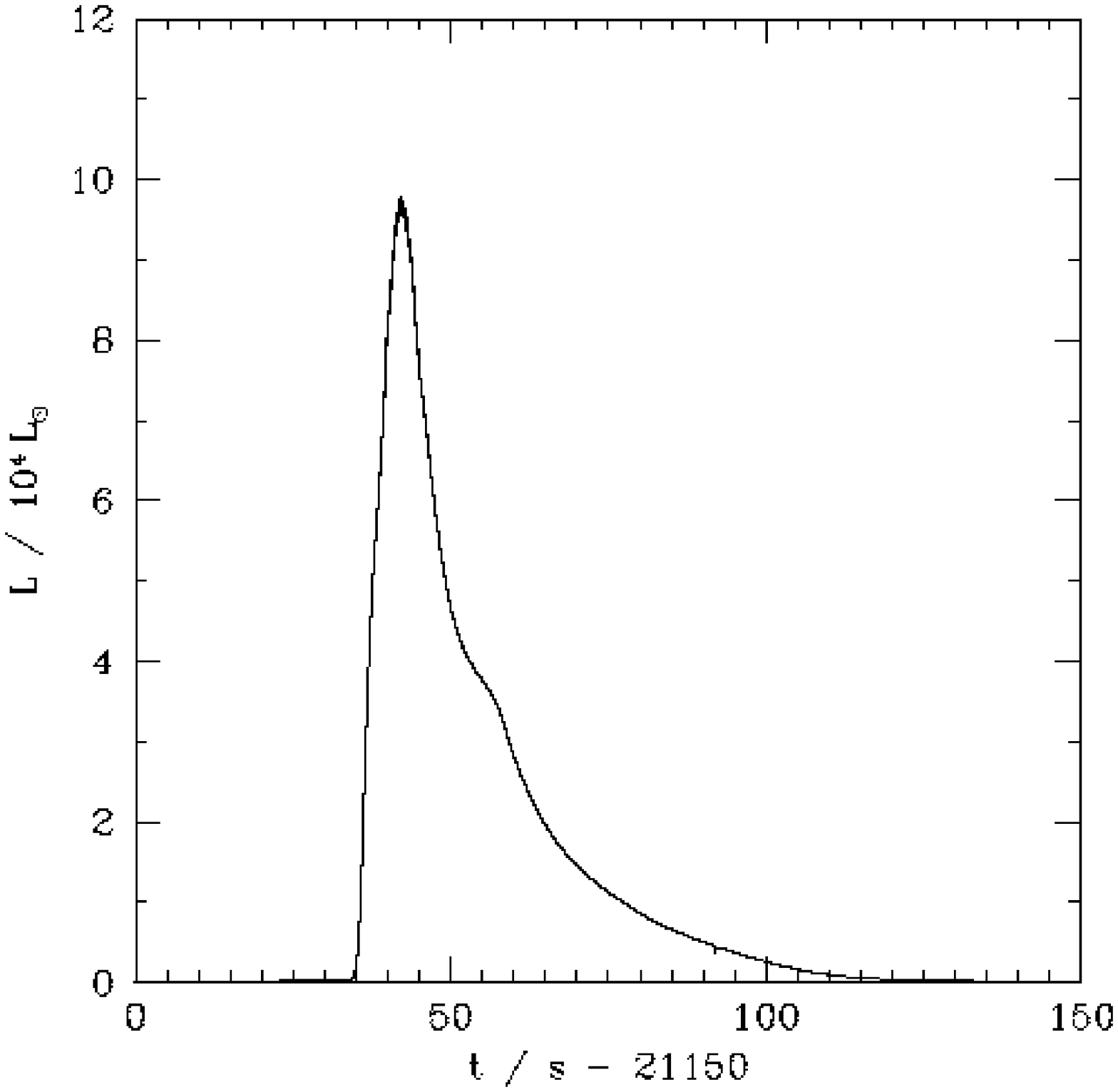}
   \includegraphics[width=0.45\textwidth]{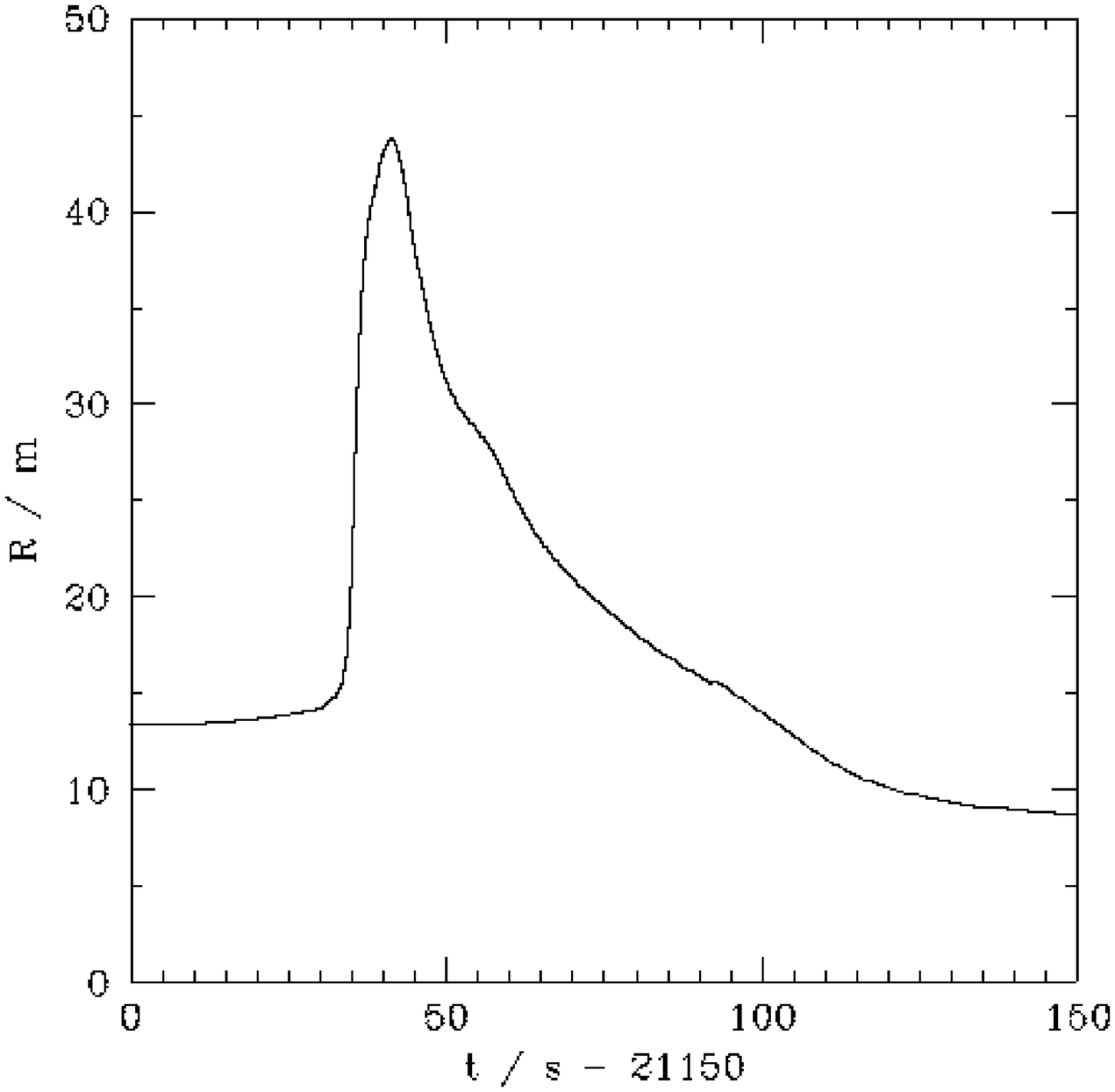}
\caption{Same as Fig.  \ref{fig:rhotpenuc_t1}, but for the overall neutron star luminosity (left panel), and envelope size (right panel),
         as measured from the core-envelope interface.}
\label{fig:lumrad_t1}
\end{figure}

Convection sets in erratically, at $\sim 1$ m above the core-envelope interface (the overall envelope size, $\Delta z$, is $\sim$ 14 m),
when T$_{base}$ reaches $3.9 \times 10^{8}$ K, 
 and progressively extends throughout the whole envelope.
Time evolution of density, temperature, pressure, and rate of nuclear energy generation, at the innermost envelope shell,
as well as of the overall neutron star luminosity and envelope size, are shown in Figs.  \ref{fig:rhotpenuc_t1} and \ref{fig:lumrad_t1}.

\begin{figure}
 \centering
   \includegraphics[width=0.90\textwidth]{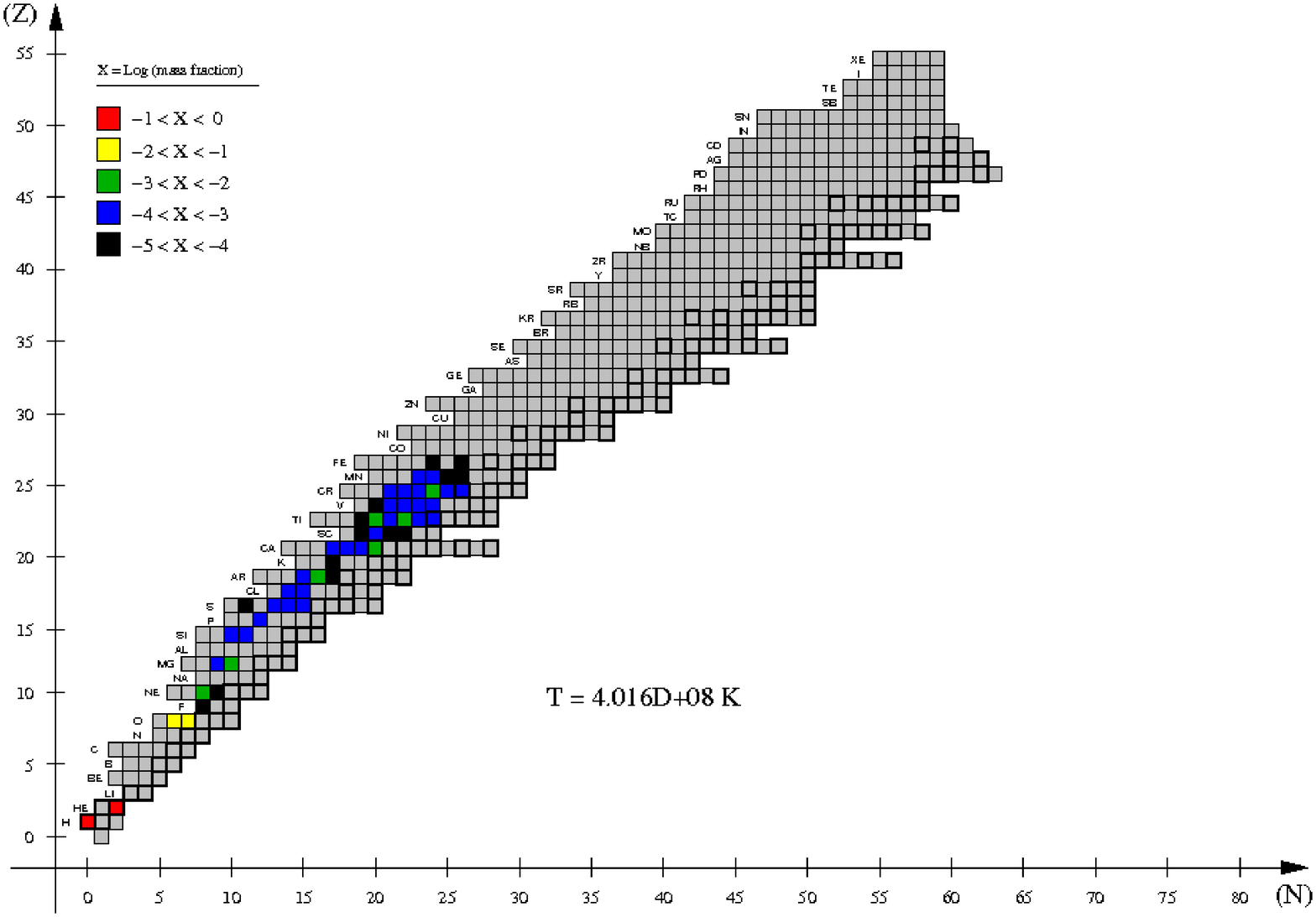}
   \includegraphics[width=0.90\textwidth]{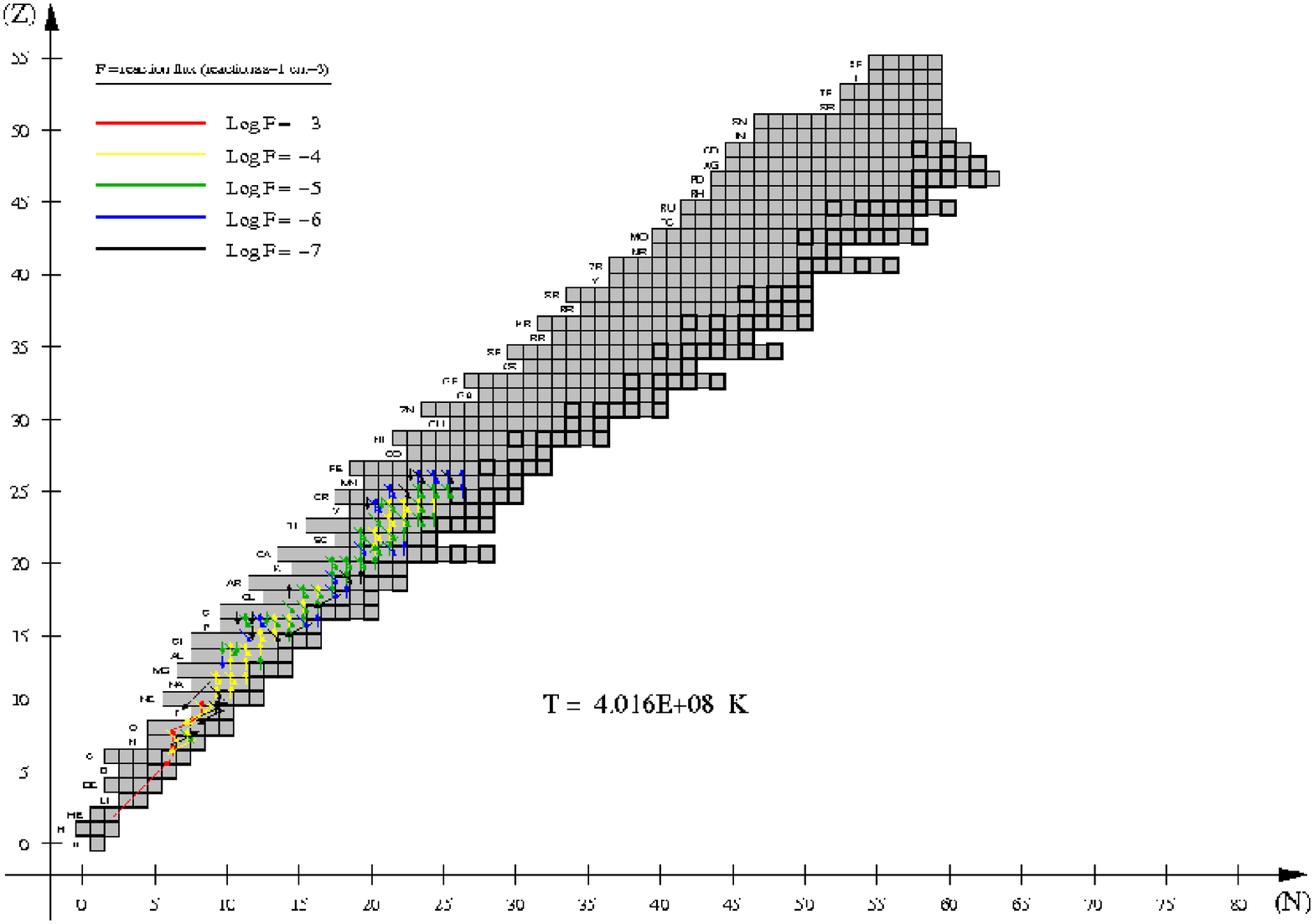}
\caption{Same as Fig. \ref{fig:M1000}, but for T$_{base} = 4 \times 10^{8}$ K.}
\label{fig:M14000}
\end{figure}

 Shortly after, at t=5.88 hours (21,181 s), 
T$_{base}$ reaches $4 \times 10^8$ K (with $\rho_{base} = 
2.9 \times 10^5$ g cm$^{-3}$, and P$_{base}=1.2 \times 10^{22}$ dyn cm$^{-2}$).
The hydrogen content has dropped to 0.209, whereas $^4$He achieves 0.625. In turn, the rate of nuclear energy generation
has increased to a value of $2.8 \times 10^{16}$ erg g$^{-1}$ s$^{-1}$.
The metallicity of this innermost envelope shell has 
increased from an initial value of 0.02 to 0.17, due to leakage from CNO cycle 
(mainly powered by $^{15}$O($\alpha$, $\gamma$)). 
As before, the next most abundant species are $^{14}$O ($6.9 \times 10^{-2}$), and 
$^{15}$O ($6.5 \times 10^{-2}$), but the number of isotopes with moderately large abundances
has now increased. Indeed, 
$^{40}$Ca, $^{22}$Mg, $^{18}$Ne, $^{34}$Ar, $^{48}$Cr, and $^{42,44}$Ti, have achieved mass fractions
of the order of $10^{-3}$. 
The nuclear activity extends as far as $^{53}$Co. The largest reaction fluxes 
(number of reactions per unit time and volume)
correspond to the equilibrium processes $^{21}$Mg(p, $\gamma$)$^{22}$Al($\gamma$, p)$^{21}$Mg, 
$^{30}$S(p, $\gamma$)$^{31}$Cl($\gamma$, p)$^{30}$S, 
and $^{25}$Si(p, $\gamma$)$^{26}$P($\gamma$, p)$^{25}$Si.
Additional activity is powered by 3$\alpha \rightarrow ^{12}$C(p, $\gamma$)$^{13}$N(p, $\gamma$)$^{14}$O, 
followed by $^{14}$O($\alpha$, p)$^{17}$F(p, $\gamma$)$^{18}$Ne.
    The suite of secondary nuclear paths is rich and complex (see Fig. \ref{fig:M14000}), and is mainly dominated by p-capture reactions 
    and $\beta^+$-decays, as well as by the CNO-breakout reaction 
    $^{15}$O($\alpha$, $\gamma$)$^{19}$Ne.
It is worth noting that the main nuclear path above Ca begins to move away from the valley of stability, towards the proton-drip line 
(see Fig. \ref{fig:M14000}, lower panel).

 Just 2.3 seconds later (t = 21,183 s), T$_{base}$ achieves $5 \times 10^8$ K. 
$\rho_{base}$ has slightly decreased to $2.3 \times 10^5$ g cm$^{-3}$ because of a 
mild envelope expansion ($\Delta z \sim 15.5$ m). Notice, however, that P$_{base} = 1.2 \times 10^{22}$ dyn cm$^{-2}$. 
Hence, the TNR is taking place nearly at constant pressure.  
A time-dependent, convective mixing with adjacent shells, with a characteristic timescale of $\tau_{conv} \sim 10^{-4}$ s 
($v_{conv} \sim 10^3-10^5$ cm s$^{-1}$), 
causes a slight increase in the H abundance at the base of the envelope. Indeed, the
H abundance is now 0.288, by mass, whereas the $^4$He content has slightly decreased to 0.563 (due to the
high temperatures, favoring $\alpha$-captures).  
The next most abundant species is now $^{18}$Ne ($4.4 \times 10^{-2}$), together 
with $^{14,15}$O ($4.2 \times 10^{-2}$, and $1.8 \times 10^{-2}$, respectively).
Several isotopes, such as $^{21,22}$Mg, 
$^{29,30}$S, $^{50,52}$Fe, $^{27}$P, $^{24,25}$Si, $^{49,50,51}$Mn, and $^{34}$Ar,
have achieved abundances an order of magnitude lower ($\sim 10^{-3}$). 
The nuclear activity extends up to $^{57}$Cu now, powering an energy generation rate of
 $1.2 \times 10^{17}$ erg g$^{-1}$ s$^{-1}$, and an overall luminosity of $1.3 \times 10^{36}$ erg s$^{-1}$.
The largest fluxes still correspond to the forward 
and reverse reactions
$^{30}$S(p, $\gamma$)$^{31}$Cl($\gamma$, p)$^{30}$S,
$^{21}$Mg(p, $\gamma$)$^{22}$Al($\gamma$, p)$^{21}$Mg,
and $^{25}$Si(p, $\gamma$)$^{26}$P($\gamma$, p)$^{25}$Si,
together with
$^{14}$O($\alpha$, p)$^{17}$F(p, $\gamma$)$^{18}$Ne.
$^{14}$O+$\alpha$ becomes the most important $\alpha$-capture reaction,
overcoming $^{15}$O($\alpha$, $\gamma$), or the 3$\alpha$.
Additional activity is driven by $^{18}$Ne($\beta^+$)$^{18}$F(p, $\alpha$)$^{15}$O, 
 $^{19}$Ne(p, $\gamma$)$^{20}$Na(p, $\gamma$)$^{21}$Mg, $^{21}$Na(p, $\gamma$)$^{22}$Mg(p, $\gamma$)$^{23}$Al,
 $^{12}$C(p, $\gamma$)$^{13}$N(p, $\gamma$)$^{14}$O, 
      and $^{26}$Si(p, $\gamma$)$^{27}$P.
      
\begin{figure}
 \centering
   \includegraphics[width=0.90\textwidth]{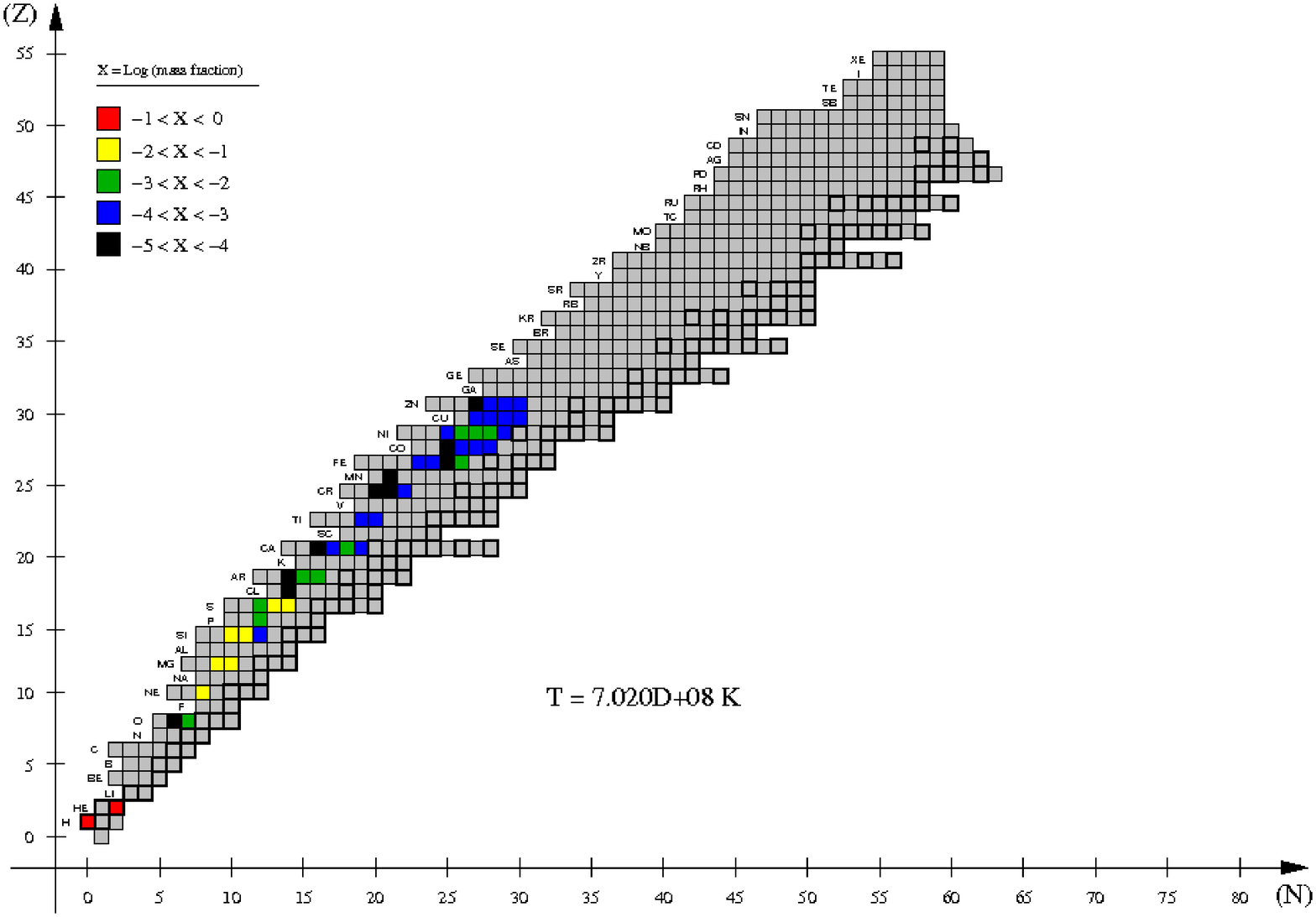}
   \includegraphics[width=0.90\textwidth]{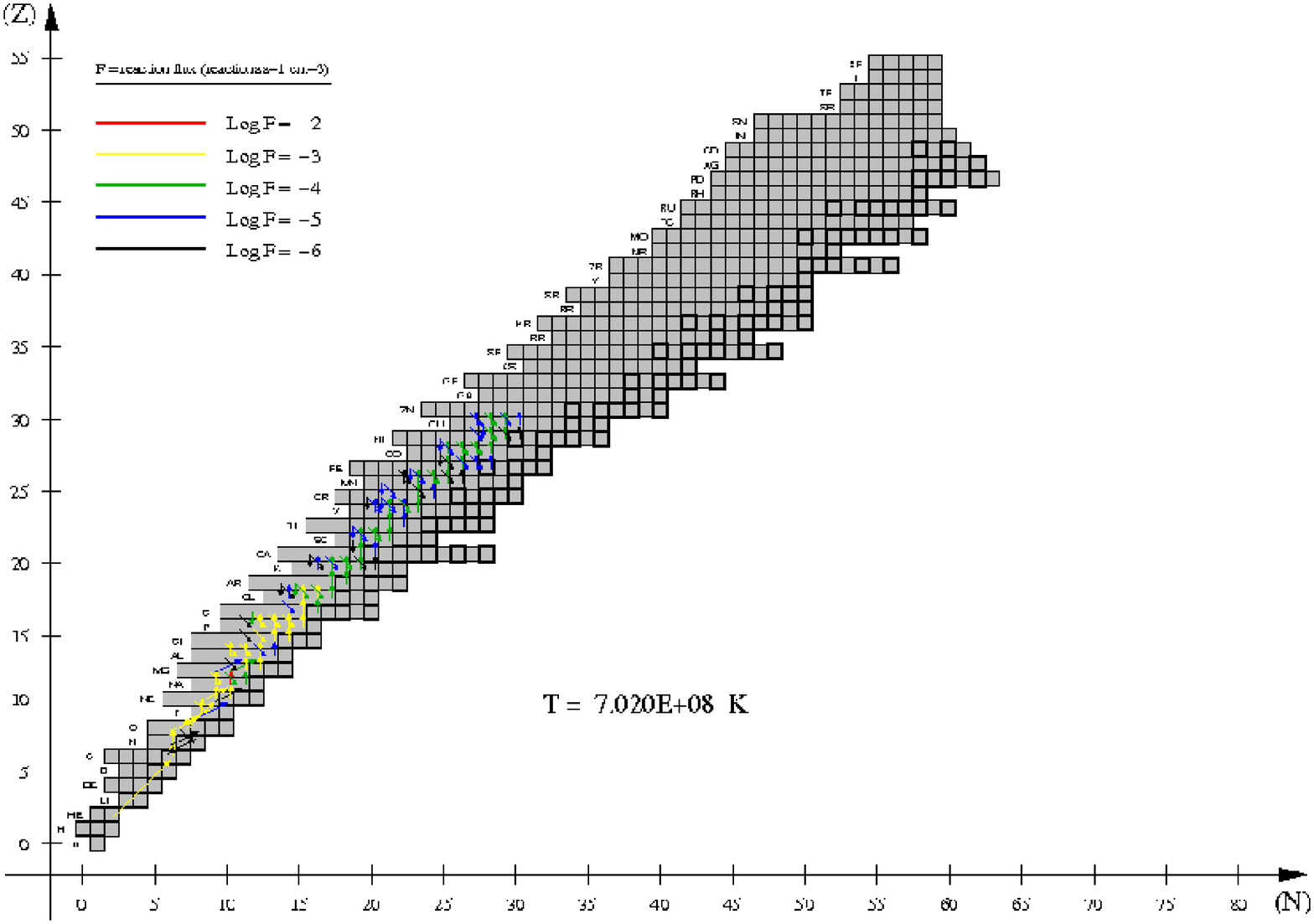}
\caption{Same as Fig. \ref{fig:M1000}, but for T$_{base} = 7 \times 10^{8}$ K.}
\label{fig:M284998}
\end{figure}

A qualitatively similar picture is found when T$_{base}$ achieves $7 \times 10^8$ K
(t = 21,185 s), with the most abundant species at the envelope base being 
 H (0.308), $^4$He (0.507), $^{18}$Ne ($4.8 \times 10^{-2}$), $^{22}$Mg ($2.5 \times 10^{-2}$), $^{29,30}$S ($1.1 \times 10^{-2}$, 
 and $2.2 \times 10^{-2}$, respectively), and
$^{24,25}$Si ($1.1 \times 10^{-2}$ and $1.7 \times 10^{-2}$, respectively).
The number of species with abundances of the order of $10^{-3}$ includes now $^{54,55,56}$Ni, $^{15}$O, 
$^{28}$S, $^{52}$Fe, $^{27}$P, $^{38}$Ca, and $^{33,34}$Ar,
with the main nuclear activity (see Fig. \ref{fig:M284998}) extending all the way up to $^{60}$Zn. 
The largest reaction fluxes are achieved by the equilibrium processes described before,
supplemented now by
$^{26}$Si(p, $\gamma$)$^{27}$P($\gamma$, p)$^{26}$Si,
$^{22}$Mg(p, $\gamma$)$^{23}$Al($\gamma$, p)$^{22}$Mg,
$^{29}$S(p, $\gamma$)$^{30}$Cl($\gamma$, p)$^{29}$S, and
$^{16}$O(p, $\gamma$)$^{17}$F($\gamma$, p)$^{16}$O, followed by
p-capture reactions (and $\beta^+$-decays) on Ne-Mg nuclei, such
as $^{19}$Ne(p, $\gamma$)$^{20}$Na(p, $\gamma$)$^{21}$Mg($\beta^+$)$^{21}$Na(p, $\gamma$)$^{22}$Mg,
or $^{17}$F(p, $\gamma$)$^{18}$Ne($\beta^+$)$^{18}$F(p, $\alpha$)$^{15}$O,
and by the $\alpha$-capture reactions 
$^{15}$O($\alpha$, $\gamma$)$^{19}$Ne, $^{14}$O($\alpha$, p)$^{17}$F, and the 3$\alpha$.
	 
\begin{figure}
 \centering
   \includegraphics[width=0.90\textwidth]{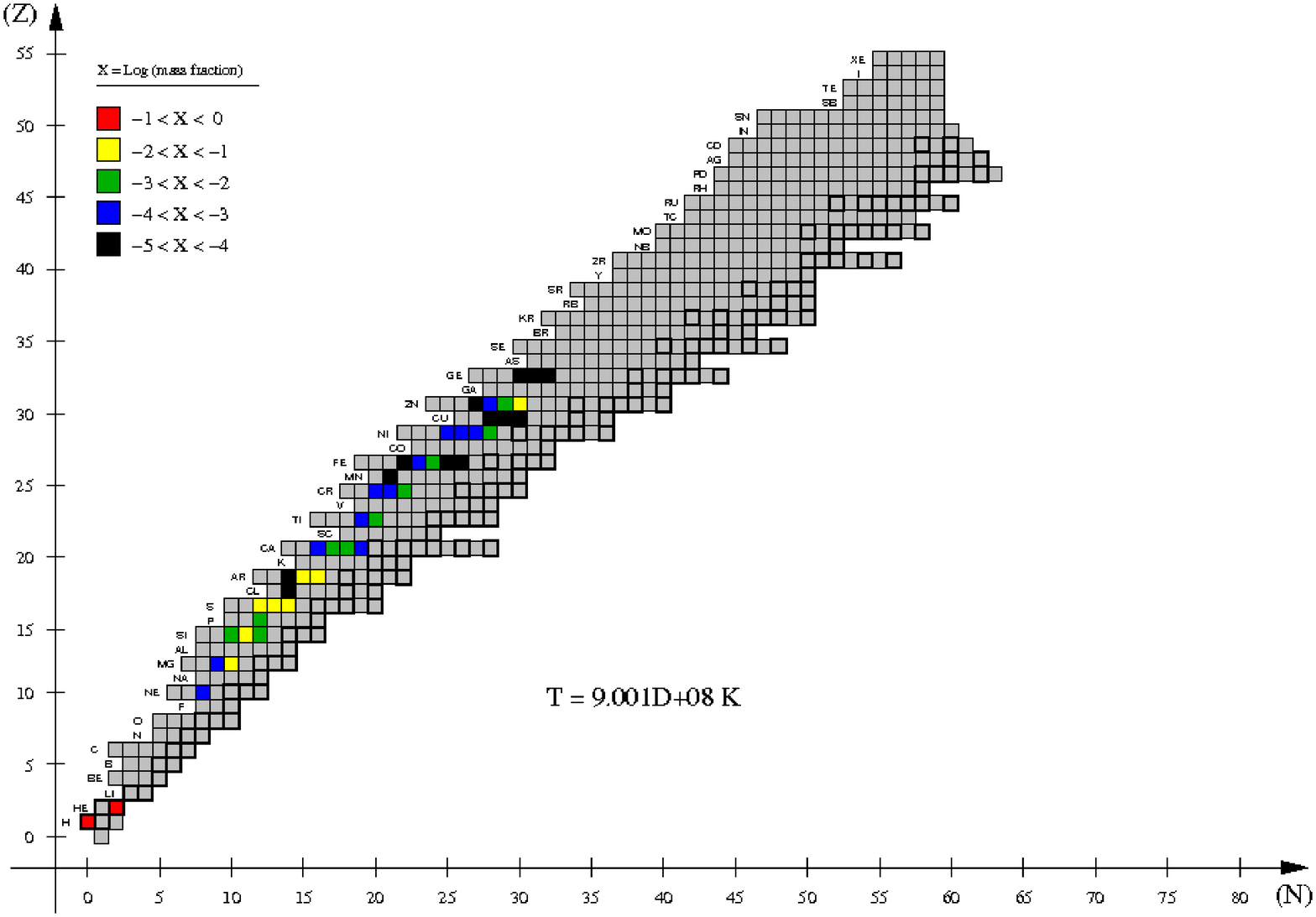}
   \includegraphics[width=0.90\textwidth]{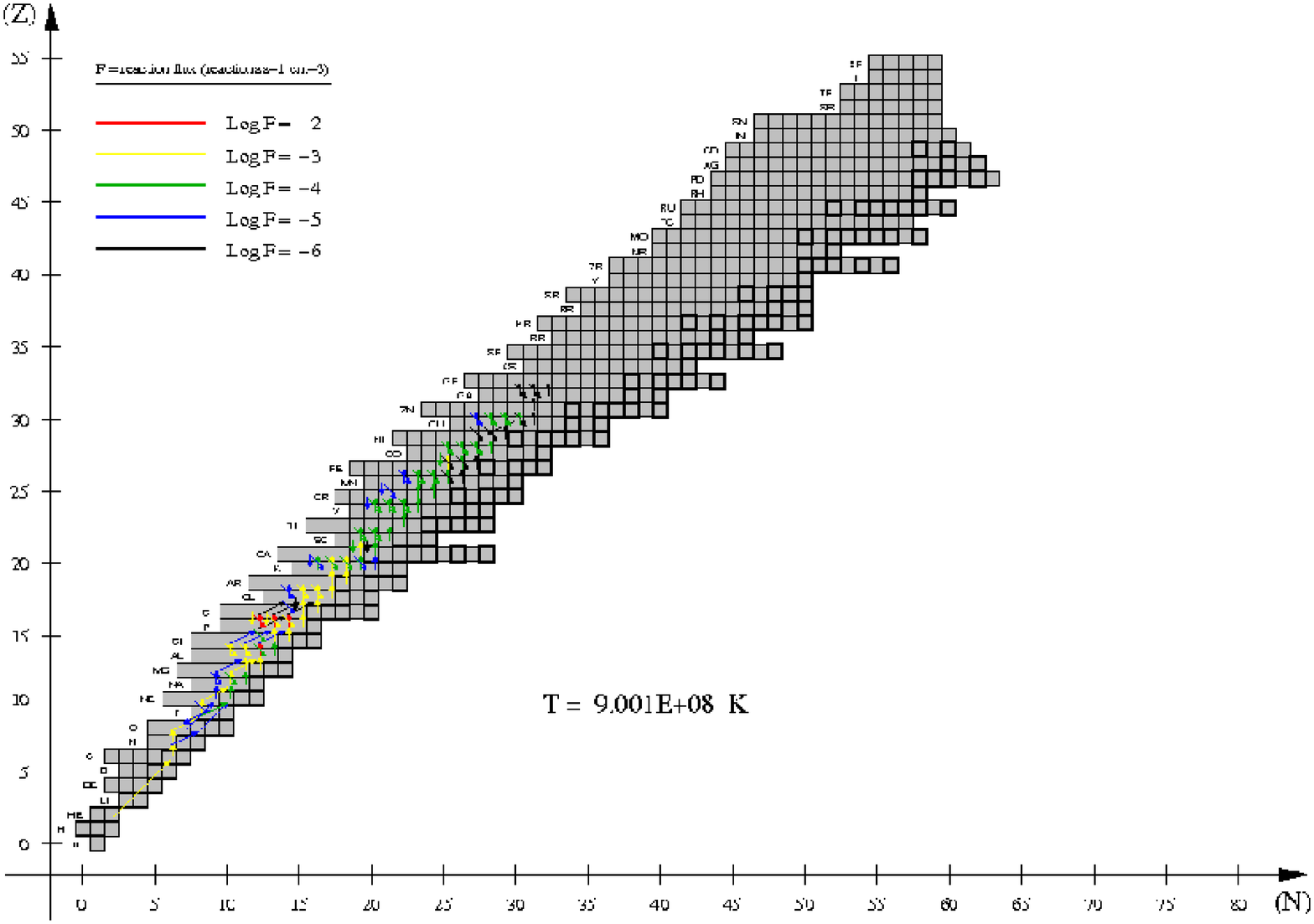}
\caption{Same as Fig. \ref{fig:M1000}, but for T$_{base} = 9 \times 10^{8}$ K.}
\label{fig:M430998}
\end{figure}

One second later (t = 21,186 s), T$_{base}$ achieves $9 \times 10^8$ K. The hectic nuclear
activity, which at this stage releases $\epsilon_{nuc} \sim 3.7 \times 10^{17}$ erg g$^{-1}$ s$^{-1}$, 
has reduced the H and $^4$He abundances down to 0.262 and 0.457, respectively. 
The next most abundant species are now $^{22}$Mg, $^{25}$Si, $^{28,29,30}$S, $^{33,34}$Ar, and $^{60}$Zn,
all with mass fractions $\sim 10^{-2}$. The main nuclear activity has extended up to $^{68}$Se. Aside from  
equilibrium (p, $\gamma$) and ($\gamma$, p) pairs
(that involve $^{16}$O-$^{17}$F and a handful of species in the mass range Mg-Zn), the largest reaction fluxes correspond to a suite of p-captures and $\beta^+$-decays (see Fig. \ref{fig:M430998}, lower panel),
 mainly $^{25}$Al(p, $\gamma$)$^{26}$Si,
 and $^{27,28,29}$P(p, $\gamma$)$^{28,29,30}$S. Moreover, the most important $\alpha$-capture
reactions, $^{22}$Mg($\alpha$, p)$^{25}$Al, the 3$\alpha$, $^{18}$Ne($\alpha$, p)$^{21}$Na, and $^{14}$O($\alpha$, p)$^{17}$F, 
have fluxes 
of the order of log F $\sim$ -3 (notice the moderate extension of $\alpha$-captures towards
heavier species as a result of the higher temperatures). 
A very limited nuclear activity in the A=65-100 mass region is, at this stage, driven by  
$^{65}$Ge(p, $\gamma$)$^{66}$As(p, $\gamma$)$^{67}$Se($\beta^+$)$^{67}$As(p, $\gamma$)$^{68}$Se (with log F $\sim$ -8),
and $^{66}$As($\beta^+$)$^{66}$Ge(p, $\gamma$)$^{67}$As (log F $\sim$ -9).

 At t = 21,188 s, when T$_{base}$ achieves $1 \times 10^9$ K, the energy generation rate by nuclear
 reactions reaches its maximum value: 
 $\epsilon_{nuc,max} \sim 4.1 \times 10^{17}$ erg g$^{-1}$ s$^{-1}$. Two seconds later, the envelope will attain maximum
 expansion, with a size $\Delta z_{max} \sim 44$ m.

\begin{figure}
 \centering
   \includegraphics[width=0.90\textwidth]{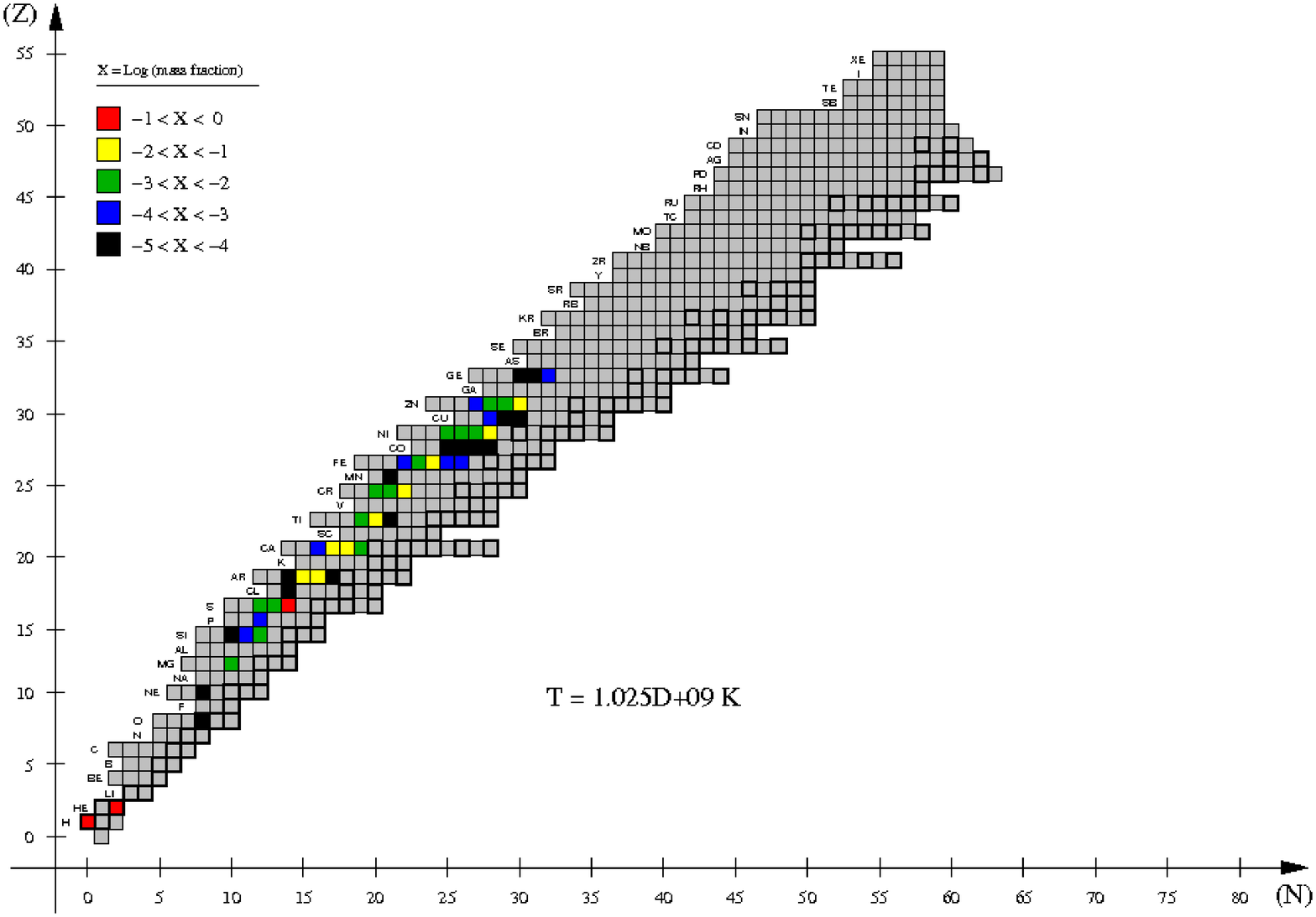}
   \includegraphics[width=0.90\textwidth]{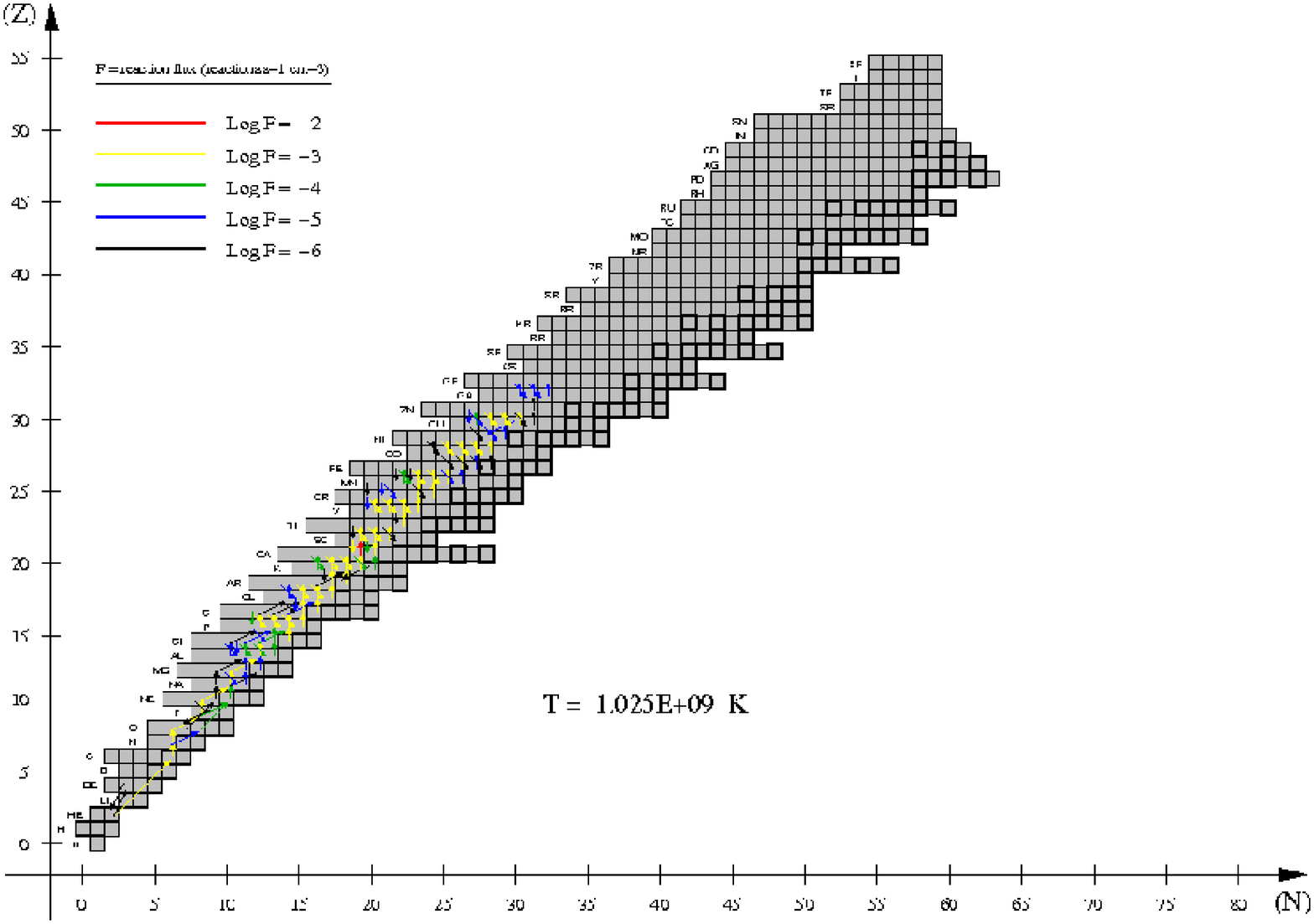}
\caption{Same as Fig. \ref{fig:M1000}, but for T$_{base} = 10^{9}$ K.}
\label{fig:M721998}
\end{figure}

Proton and $\alpha$-captures
continue to reduce the overall H and He abundances at the envelope base (0.191 and 0.400, respectively). 
The next most abundant species is now $^{30}$S (0.103) -a waiting point for the main nuclear path-, followed by
$^{33,34}$Ar, $^{37,38}$Ca, $^{42}$Ti, $^{46}$Cr, $^{50}$Fe, $^{56}$Ni, and $^{60}$Zn (all with X$_i \sim 10^{-2}$).
The nuclear activity has reached $^{76}$Sr (see Fig. \ref{fig:M721998}).
In the A=65-100 mass region, in particular, the nuclear activity is now dominated by the chains 
$^{65}$Ge(p, $\gamma$)$^{66}$As(p, $\gamma$)$^{67}$Se($\beta^+$)$^{67}$As(p, $\gamma$)$^{68}$Se (with log F $\sim$ -7),
$^{66}$As($\beta^+$)$^{66}$Ge(p, $\gamma$)$^{67}$As (log F $\sim$ -8), 
$^{65}$As(p, $\gamma$)$^{66}$Se($\beta^+$)$^{66}$As, and 
$^{68}$Se($\beta^+$)$^{68}$As(p, $\gamma$)$^{69}$Se(p, 
$\gamma$)$^{70}$Br(p, $\gamma$)$^{71}$Kr($\beta^+$)$^{71}$Br(p, $\gamma$)$^{72}$Kr  (log F $\sim$ -9).  In terms of energy production, the most important 
contributions come from $^{39}$Ca(p, $\gamma$)$^{40}$Sc (log F $\sim$ -2),
multiple (p, $\gamma$)-reactions involving species in the mass range A $\sim$ 20-60, and a handful of
$\alpha$-capture reactions, such as $^{22}$Mg($\alpha$, p)$^{25}$Al, $^{18}$Ne($\alpha$, p)$^{21}$Na, the 3$\alpha$, and $^{14}$O($\alpha$, p)$^{17}$F
(log F $\sim$ -3).

\begin{figure}
 \centering
   \includegraphics[width=0.90\textwidth]{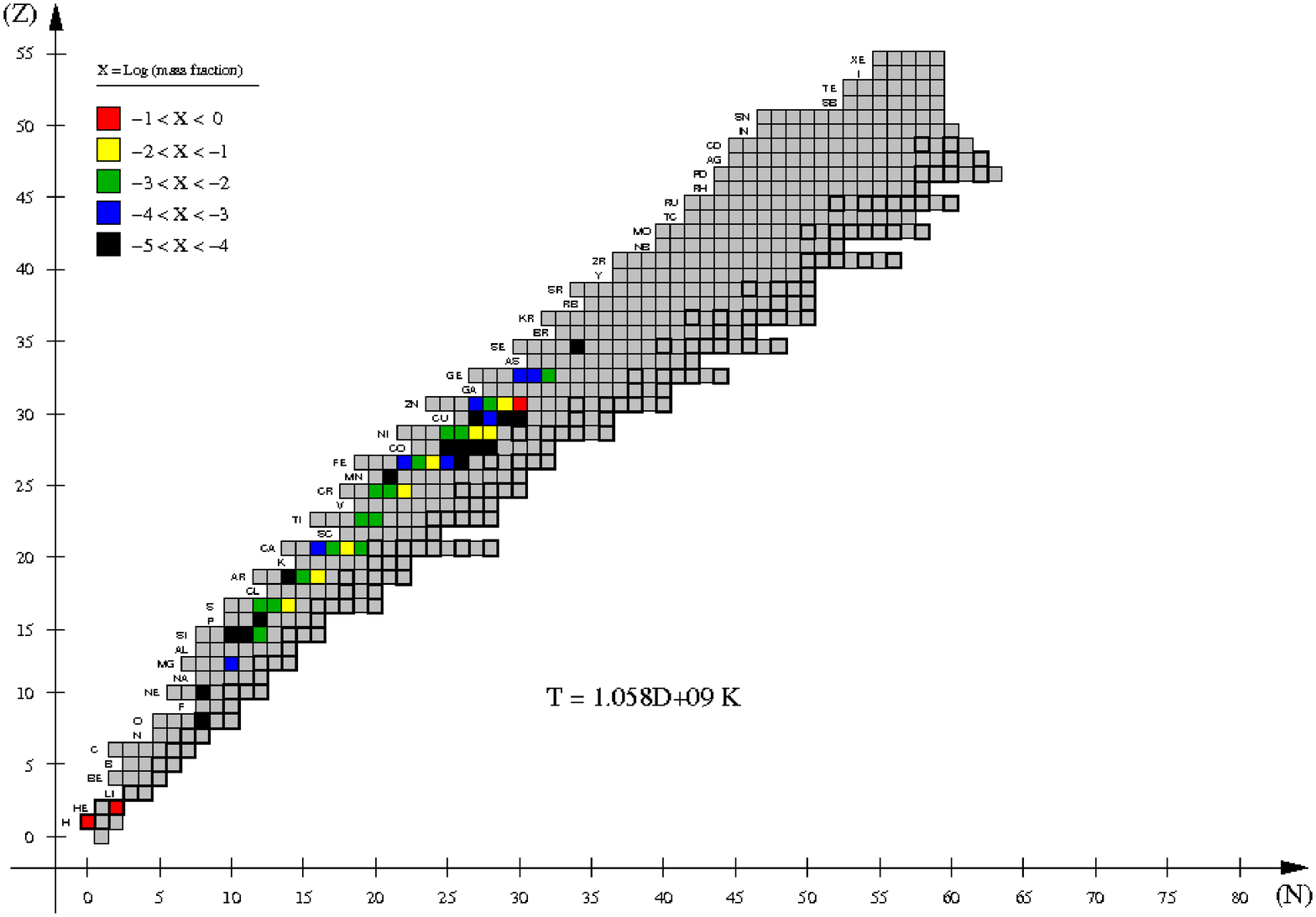}
   \includegraphics[width=0.90\textwidth]{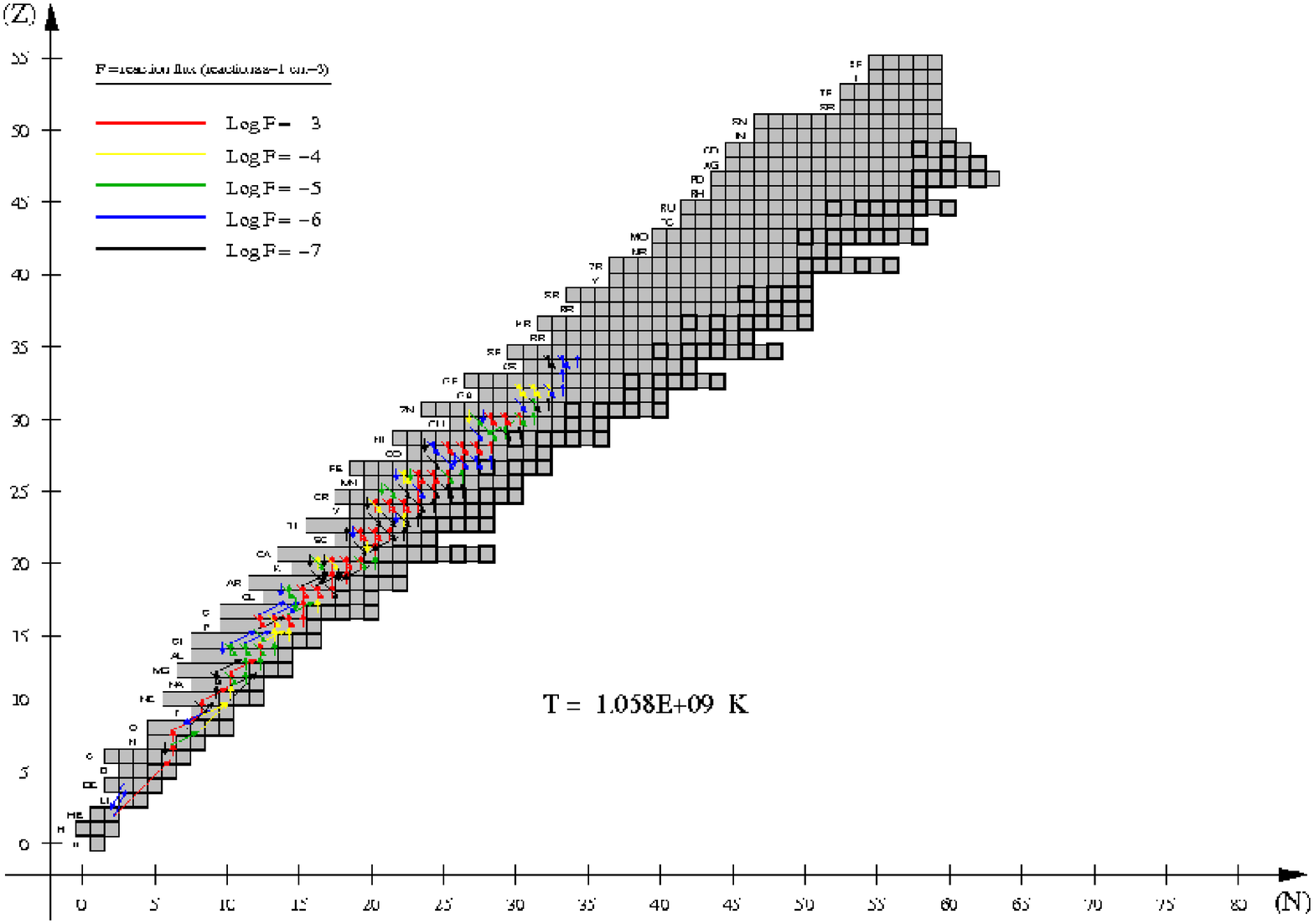}
\caption{Same as Fig. \ref{fig:M1000}, but for the time when temperature at the envelope base reaches a peak value of 
         $T_{peak} = 1.06 \times 10^{9}$ K.}
\label{fig:M999998}
\end{figure}

Four seconds later, the envelope base achieves a maximum temperature of $T_{peak} \sim 1.06 \times 10^9$ K (similar values are reported in the simulations 
by Fisker et al. 2008).
Besides H (0.220) and $^4$He (0.370), the next most abundant isotope is now $^{60}$Zn (0.159) -another waiting
point for the nuclear flow-, followed by
$^{30}$S ($3.3 \times 10^{-2}$), $^{34}$Ar ($3.2 \times 10^{-2}$), $^{38}$Ca ($2.4 \times 10^{-2}$),
$^{46}$Cr ($1.8 \times 10^{-2}$), $^{50}$Fe ($1.3 \times 10^{-2}$), 
$^{55,56}$Ni ($1.8 \times 10^{-2}$, and $2.3 \times 10^{-2}$, respectively), 
and $^{59}$Zn ($1.8 \times 10^{-2}$). As shown in Fig. \ref{fig:M999998}, the extension of the main nuclear path
reaches $^{80}$Zr. 
The nuclear activity in the A=65-100 mass region is now dominated by 
$^{65}$Ge(p, $\gamma$)$^{66}$As(p, $\gamma$)$^{67}$Se($\beta^+$)$^{67}$As(p, $\gamma$)$^{68}$Se (log F $\sim$ -6),
$^{65}$As(p, $\gamma$)$^{66}$Se($\beta^+$)$^{66}$As (log F $\sim$ -7),
and $^{66}$As($\beta^+$)$^{66}$Ge(p, $\gamma$)$^{67}$As, 
$^{68}$Se($\beta^+$)$^{68}$As(p, $\gamma$)$^{69}$Se(p, 
$\gamma$)$^{70}$Br(p, $\gamma$)$^{71}$Kr($\beta^+$)$^{71}$Br(p, $\gamma$)$^{72}$Kr (log F $\sim$ -8). 
Energy production is due to dozens of (p, $\gamma$)-reactions involving species in the mass range A $\sim$ 20-60, plus
some $\alpha$-capture reactions, such as the 3$\alpha$, $^{14}$O($\alpha$, p)$^{17}$F, 
$^{22}$Mg($\alpha$, p)$^{25}$Al, and  $^{18}$Ne($\alpha$, p)$^{21}$Na (log F $\sim$ -3).
Less than a second later (t = 21,192.3 s), the neutron star reaches maximum luminosity,
L$_{max}$ = $3.8 \times 10^{38}$ erg s$^{-1}$ ($9.8 \times 10^4$ L$_\odot$).

The numerous proton-captures on many species during the decline from T$_{peak}$ reduce dramatically
the H content in the innermost shell. Indeed, when T$_{base}$ achieves $9.3 \times 10^8$ K (t = 21,200 s),
the H abundance drops below 0.1, while X($^4$He) = 0.283. Actually, the most abundant species
in this shell is now $^{60}$Zn (0.43 by mass), followed by $^{30}$S ($2.7 \times 10^{-2}$),
$^{34}$Ar ($2.1 \times 10^{-2}$), $^{38}$Ca ($1.3 \times 10^{-2}$), $^{56}$Ni ($1.8 \times 10^{-2}$), 
and $^{64}$Ge ($3.3 \times 10^{-2}$).
The nuclear activity reaches $^{90}$Ru. 
  
\begin{figure}
 \centering
   \includegraphics[width=0.90\textwidth]{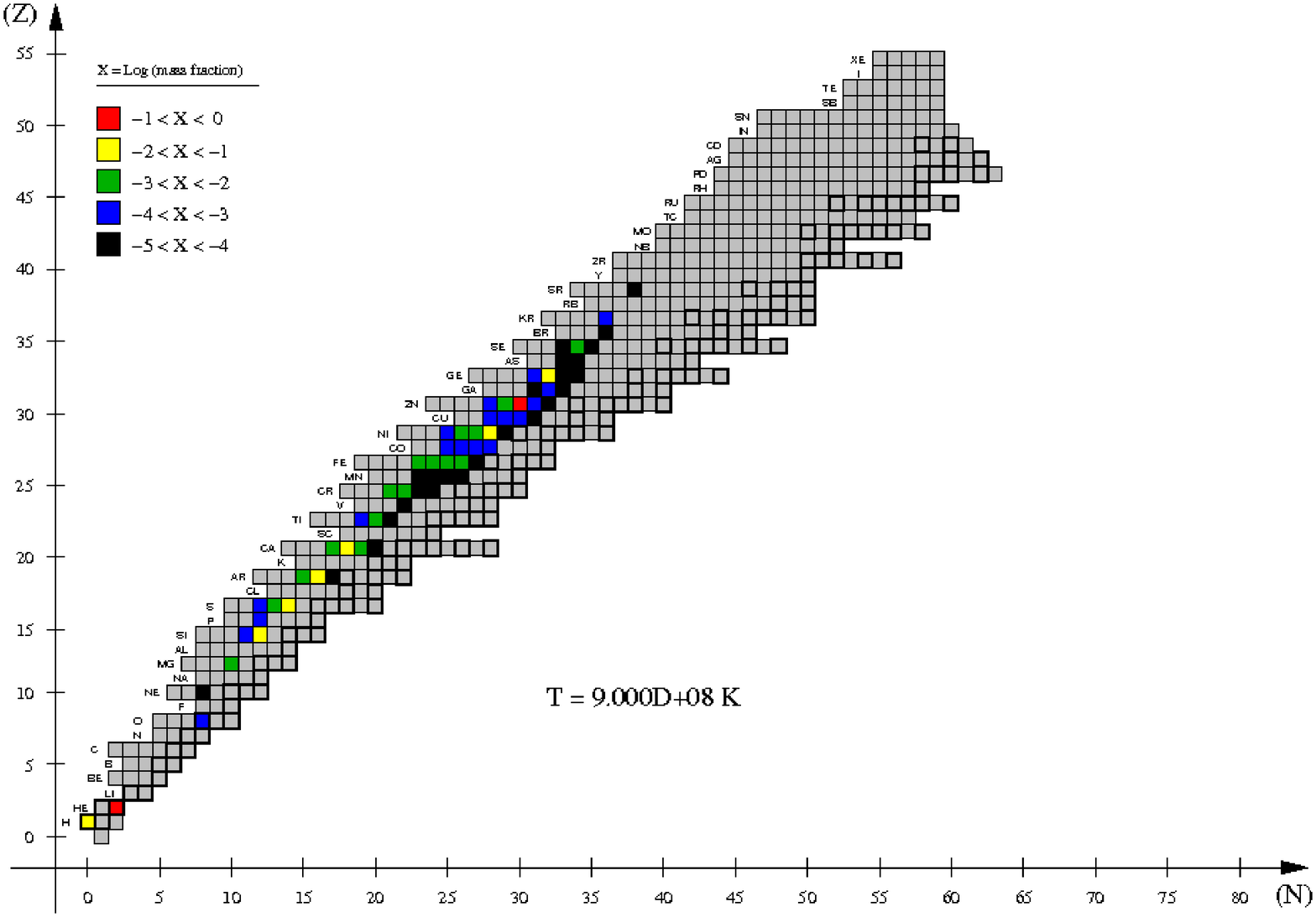}
   \includegraphics[width=0.90\textwidth]{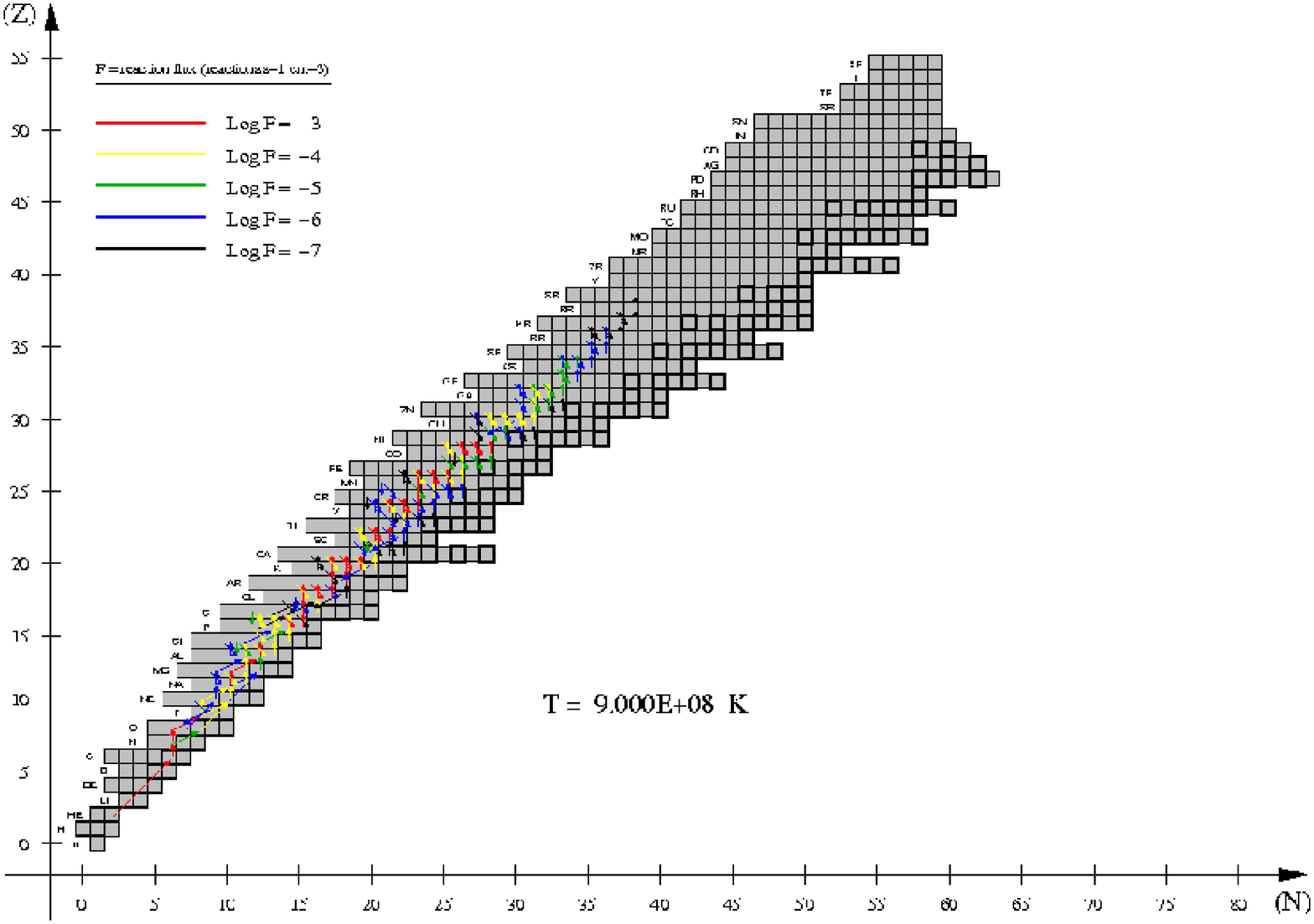}
\caption{Same as Fig. \ref{fig:M1000}, but for $T_{base} = 9 \times 10^{8}$ K.}
\label{fig:M1696998}
\end{figure}

Five seconds later (t = 21,205 s), 
when T$_{base}$ drops to $9.0 \times 10^8$ K, $^{60}$Zn achieves a maximum abundance of
0.519, by mass. H has been reduced to $1.3 \times 10^{-2}$ (X($^4$He)=0.226).
The next most abundant species are now $^{26}$Si, $^{30}$S,
$^{34}$Ar, $^{38}$Ca,
$^{56}$Ni, and $^{64}$Ge (all with mass fractions $\sim 10^{-2}$).
The nuclear activity has not progressed beyond $^{90}$Ru. The largest
reaction fluxes correspond to 
proton-captures and reverse photodisintegration reactions at equilibrium. 
Many other nuclear processes ($\beta^+$-decays and
$\alpha$-induced reactions like
the triple-$\alpha$, $^{14}$O($\alpha$, p)$^{17}$F, or $^{22}$Mg($\alpha$, p))
contribute to the overall nuclear activity (see Fig. \ref{fig:M1696998}).
In the A=65-100 mass region, this is driven by  
$^{65}$Ge(p, $\gamma$)$^{66}$As (followed either by $^{66}$As(p, $\gamma$)$^{67}$Se($\beta^+$)$^{67}$As, or by 
$^{66}$As($\beta^+$)$^{66}$Ge(p, $\gamma$)$^{67}$As), $^{67}$As(p, $\gamma$)$^{68}$Se($\beta^+$)$^{68}$As(p, $\gamma$)$^{69}$Se
(log F $\sim$ -5),
$^{69}$Se(p, $\gamma$)$^{70}$Br($\beta^+$)$^{70}$Se(p, $\gamma$)$^{71}$Br($\beta^+$)$^{71}$Se, 
$^{74}$Rb($\beta^+$)$^{74}$Kr (log F $\sim$ -6),
and $^{70}$Br(p, $\gamma$)$^{71}$Kr($\beta^+$)$^{71}$Br, $^{72}$Kr($\beta^+$)$^{72}$Br(p, $\gamma$)$^{73}$Kr(p, $\gamma$)$^{74}$Rb, 
$^{74}$Kr(p, $\gamma$)$^{75}$Rb(p, $\gamma$)$^{76}$Sr (log F $\sim$ -7).
Energy production is due to (p, $\gamma$)-reactions involving species in the mass range A $\sim$ 20-55, and also to 
 two $\alpha$-capture reactions, $^{14}$O($\alpha$, p)$^{17}$F, and $^{22}$Mg($\alpha$, p)$^{25}$Al (log F $\sim$ -3).

\begin{figure}
 \centering
   \includegraphics[width=0.90\textwidth]{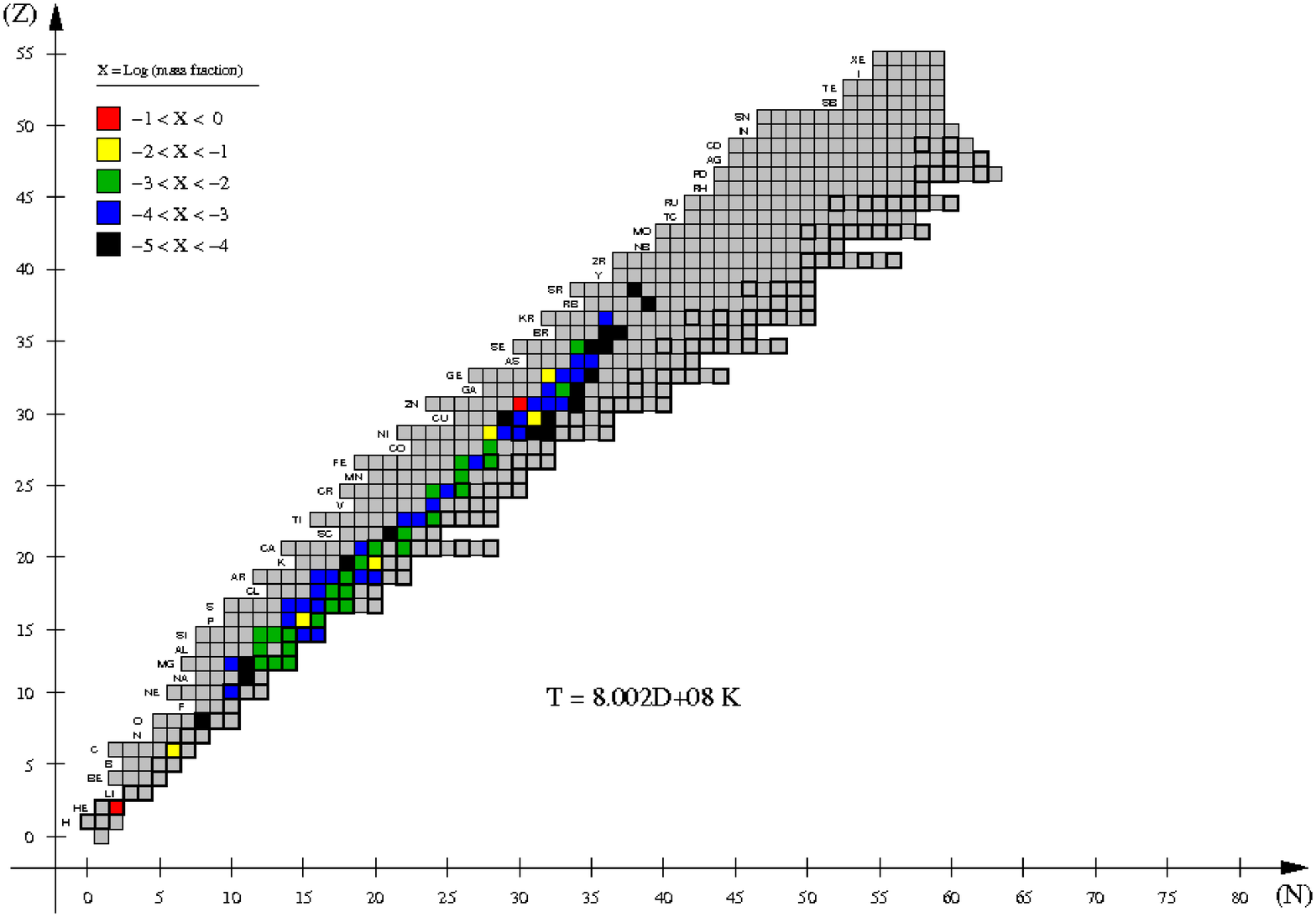}
   \includegraphics[width=0.90\textwidth]{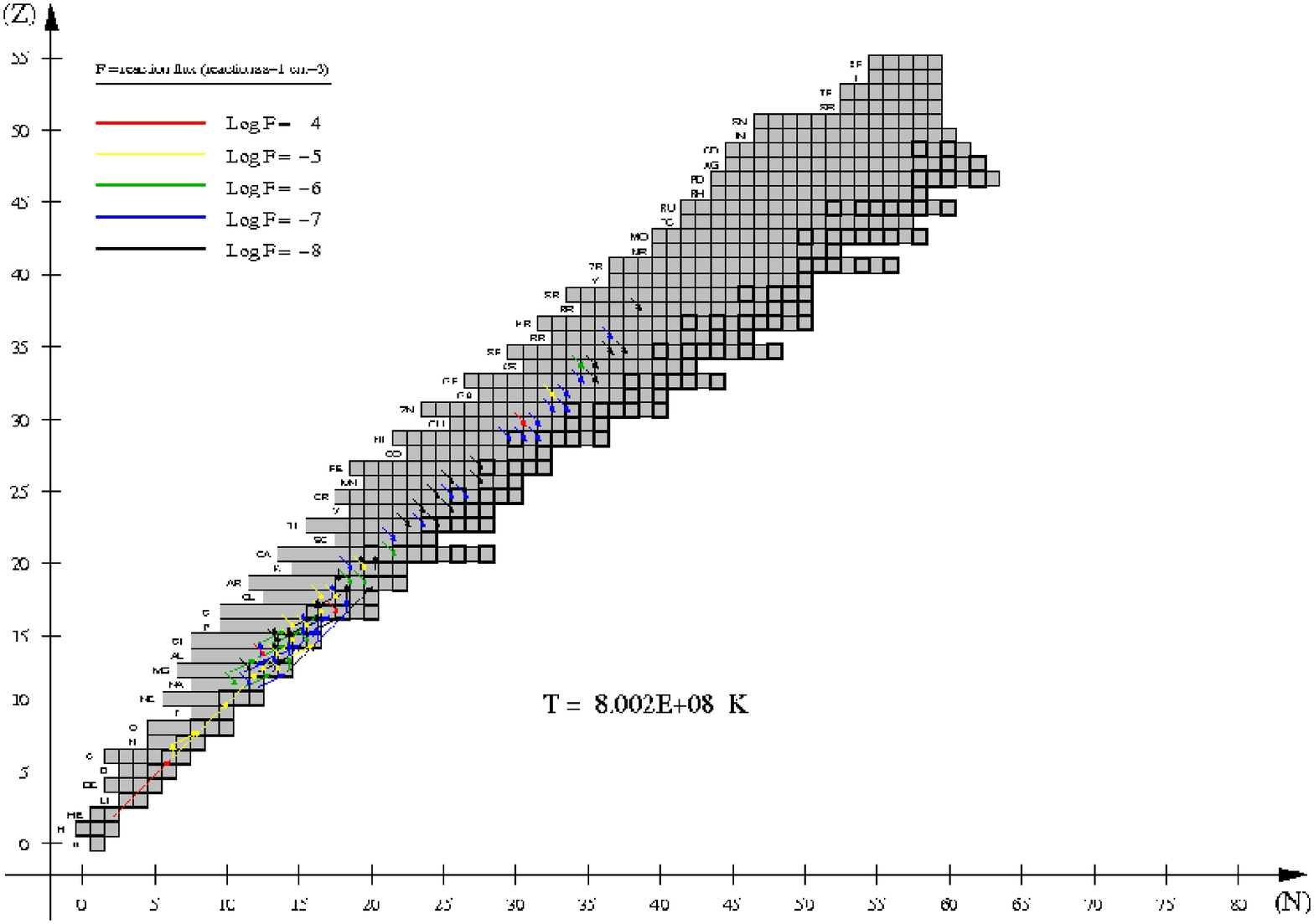}
\caption{Same as Fig. \ref{fig:M1000}, but for $T_{base} = 8 \times 10^{8}$ K.}
\label{fig:M1986998}
\end{figure}

Following the fast decline in temperature, when T$_{base} = 8.0 \times 10^8$ K (t = 21,212 s), the $^{60}$Zn 
abundance has dropped to 0.509, due to $\beta^+$-decays. H has been heavily depleted ($5 \times 10^{-11}$), whereas $^4$He has been slightly reduced to 
an abundance of 0.190.
The next most abundant species are $^{12}$C, $^{30}$P, $^{39}$K, 
$^{56}$Ni, $^{60}$Cu, and $^{64}$Ge.
The extent of the nuclear activity (Fig. \ref{fig:M1986998}) is still 
restricted to $^{90}$Ru; it will
not proceed beyond this endpoint, first because of the heavy H depletion, and second, because the temperature is already too low to allow
proton- or $\alpha$-captures on heavier species due to their large Coulomb barriers.
At this stage, the single, most important reaction, in terms of reaction fluxes, is the triple-$\alpha$,
followed by a suite of $\beta^+$-decay reactions, such as $^{26}$Si($\beta^+$)$^{26m}$Al($\beta^+$)$^{26}$Mg, $^{34}$Cl($\beta^+$)$^{34}$S, 
$^{60}$Zn($\beta^+$)$^{60}$Cu, or $^{27}$Si($\beta^+$)$^{27}$Al.
Several $\alpha$-captures follow the triple-$\alpha$ reaction as a chain: 
$^{12}$C($\alpha$, $\gamma$)$^{16}$O($\alpha$, $\gamma$)$^{20}$Ne($\alpha$, 
$\gamma$)$^{24}$Mg($\alpha$, $\gamma$)$^{28}$Si($\alpha$, $\gamma$)$^{32}$S,
or through alternative paths, proceeding close to the valley of stability, up to $\sim$ Ar, such as
 $^{13}$N($\alpha$, p)$^{16}$O, $^{25,27}$Al($\alpha$, p)$^{28,30}$Si, 
$^{22}$Mg($\alpha$, p)$^{25}$Al,  $^{22}$Na($\alpha$, p)$^{25}$Mg, or
$^{26,27}$Si($\alpha$, p)$^{29,30}$P, to quote some representative cases (see Fig. \ref{fig:M1986998}). 
$\epsilon_{nuc}$ has already declined to a value of $\sim 3.8 \times 10^{15}$ erg g$^{-1}$ s$^{-1}$. 

\begin{figure}
 \centering
   \includegraphics[width=0.90\textwidth]{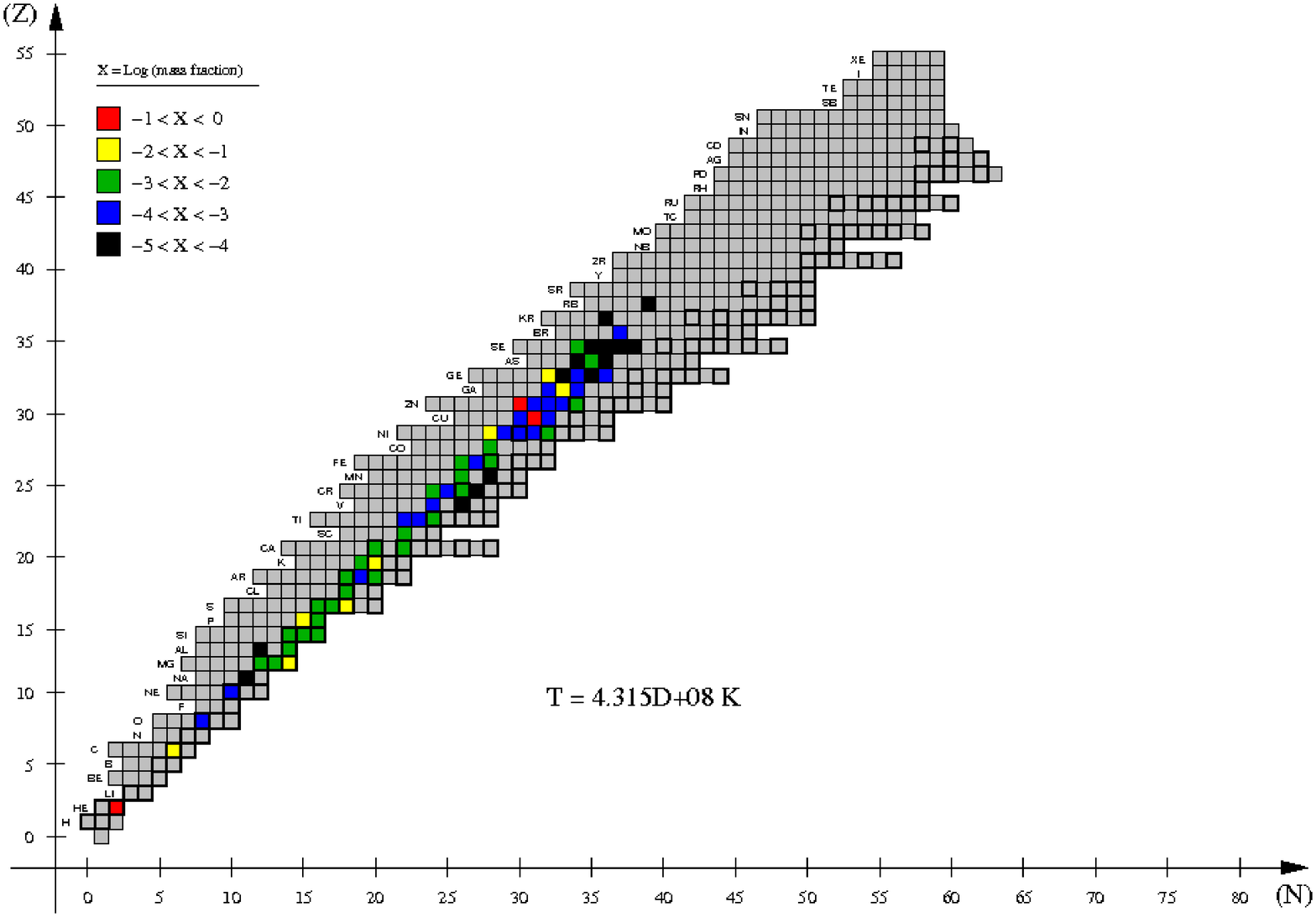}
   \includegraphics[width=0.90\textwidth]{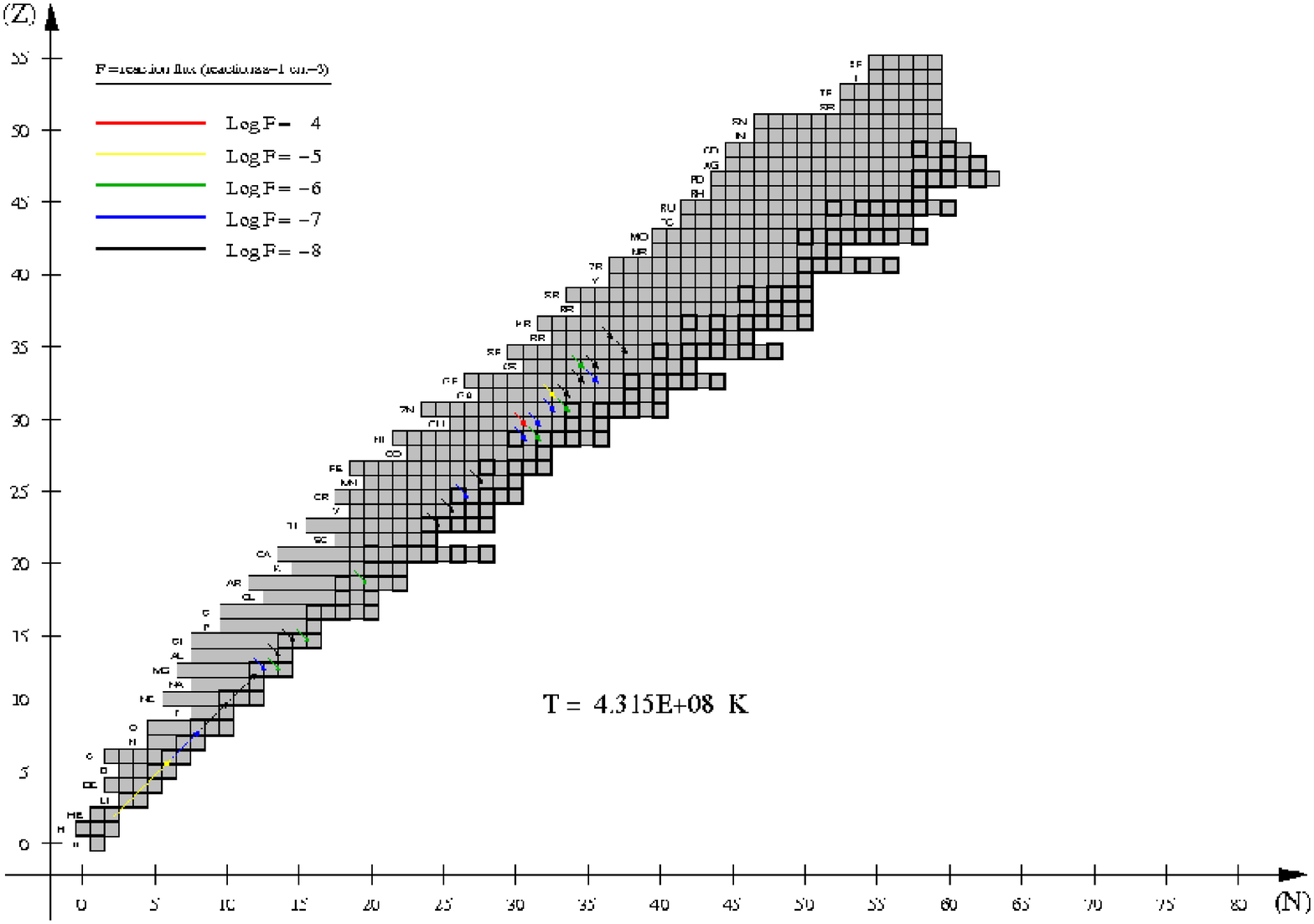}
\caption{Same as Fig. \ref{fig:M1000}, 
but for $T_{base} = 4.3 \times 10^{8}$ K.}
\label{fig:M2675998}
\end{figure}

When T$_{base}$ drops to $4.3 \times 10^8$ K (t = 21,254 s), the nuclear energy generation rate 
has declined to a value of $\epsilon_{nuc} \sim 3 \times 10^{14}$ erg g$^{-1}$ s$^{-1}$.
As shown in Fig. \ref{fig:M2675998},
the nuclear activity is dominated by $^{60}$Zn($\beta^+$)$^{60}$Cu, because of its very large abundance (X($^{60}$Zn)=0.416),
and is followed by the triple-$\alpha$ reaction (X($^{4}$He)=0.137), and by a suite of $\beta^+$-decays of very abundant
isotopes, such as $^{64}$Ge, $^{30}$P,  $^{64}$Ga, or $^{60}$Cu (all with mass fractions $\sim 10^{-2}$, except $^{60}$Cu [0.104]), 
followed by those of $^{38}$K, $^{26m}$Al, $^{68}$Se, $^{25}$Al, 
$^{68}$As, $^{63}$Ga, $^{59}$Cu, and $^{61}$Zn. 
Other species, such as $^{12}$C, $^{26}$Mg,  $^{34}$S, $^{39}$K, 
or $^{56}$Ni, 
have achieved an abundance of $10^{-2}$ by mass at this stage.
The envelope has already shrunk to a size $\Delta z \sim 13$ m, whereas the overall
luminosity of the star has decreased to L$_{NS}$ = $7.7 \times 10^{36}$ erg s$^{-1}$.
 
\begin{figure}[htbp]
 \centering
   \includegraphics[width=0.90\textwidth]{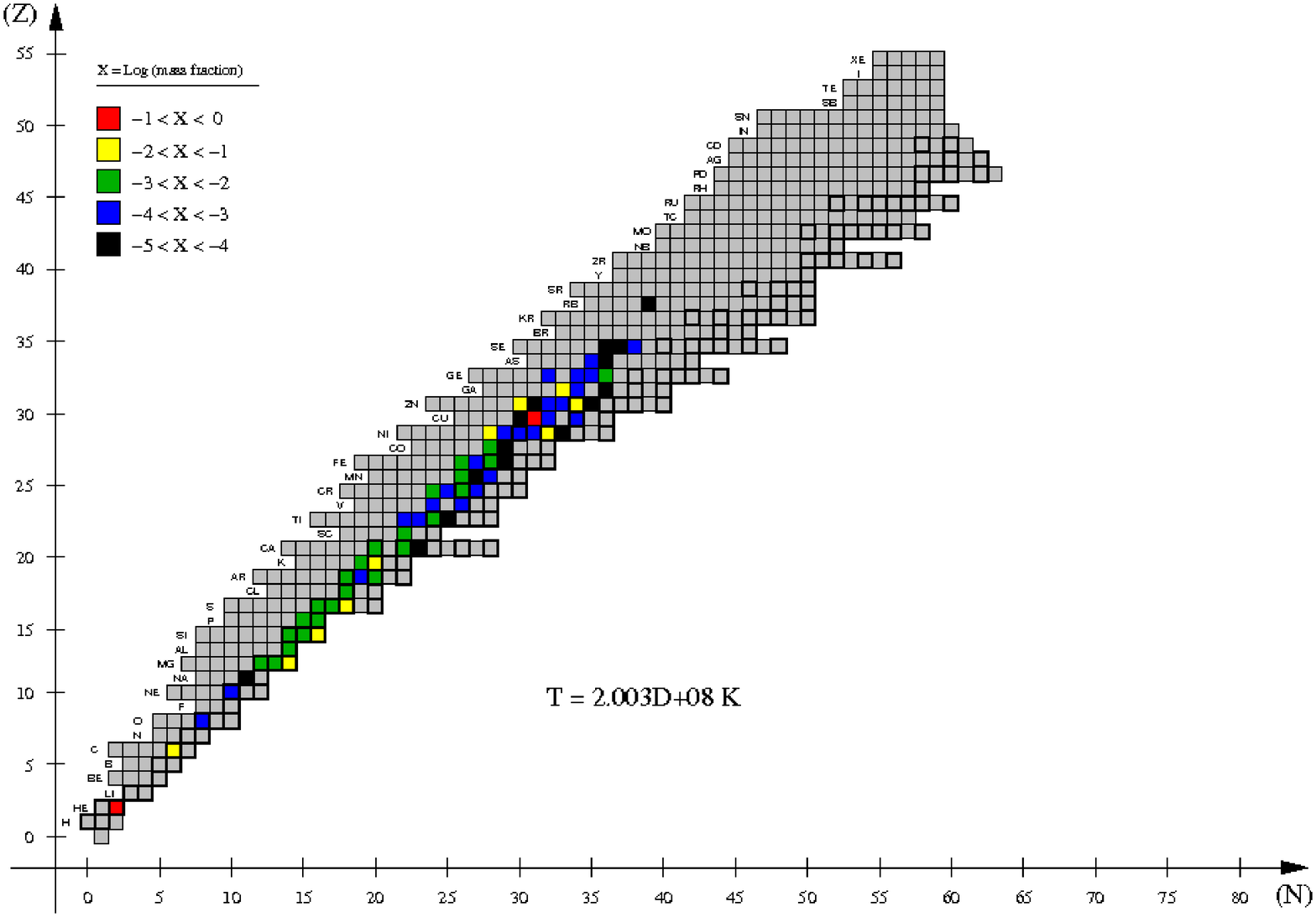}
   \includegraphics[width=0.90\textwidth]{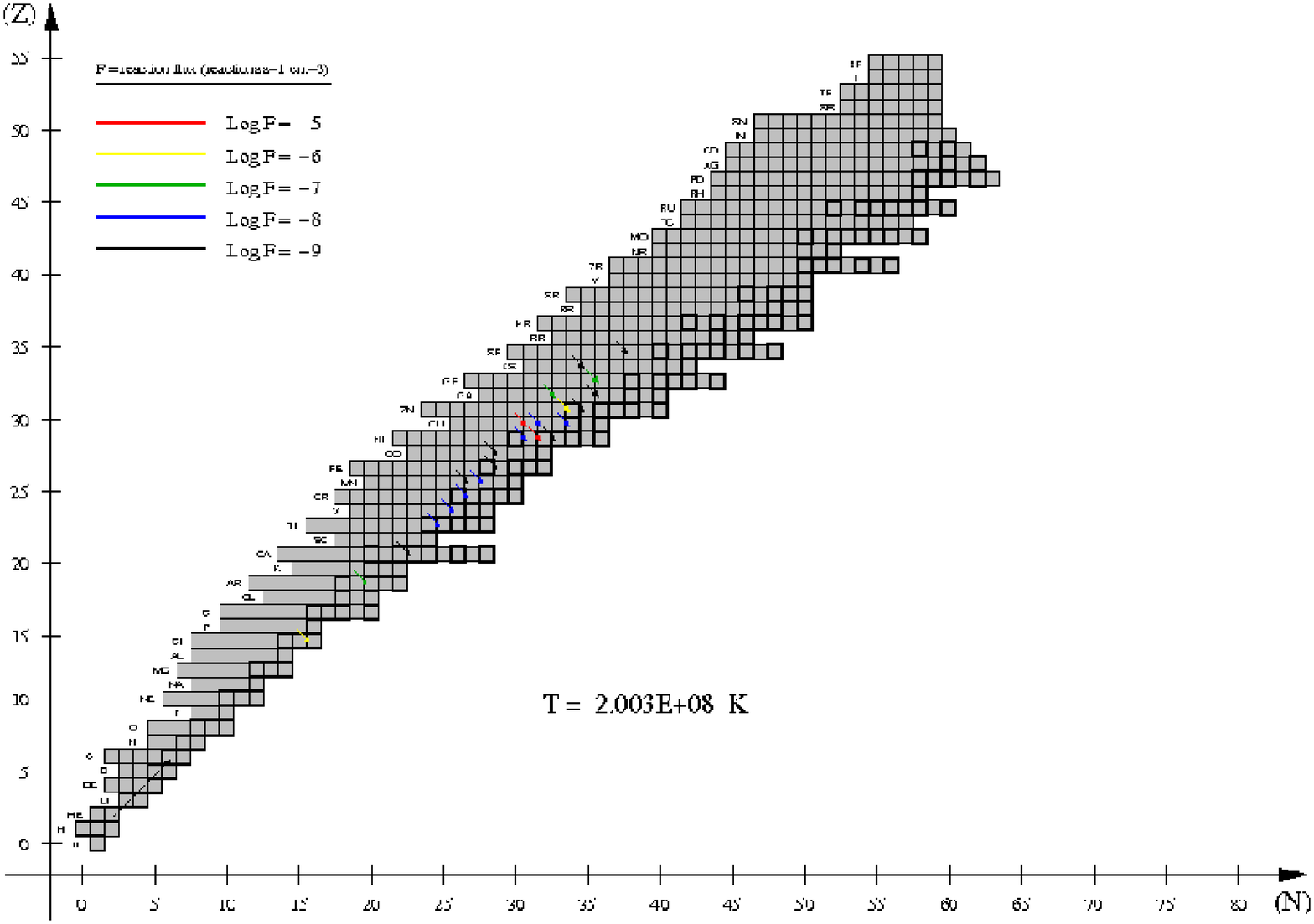}
\caption{Same as fig. \ref{fig:M1000}, but for $T_{base} = 2 \times 10^{8}$ K.}
\label{fig:M2702998}
\end{figure}

When T$_{base}$ reaches $2 \times 10^8$ K (t = 21,618 s), $^{60}$Zn has remarkably decayed into $^{60}$Cu,
which now constitutes the most abundant species (with 0.393) at the base of the envelope.
Because of the relatively low temperatures, the $^4$He abundance is kept constant, at about 0.136.
The next most abundant species are now $^{12}$C, $^{26}$Mg, $^{30}$Si, $^{34}$S,
$^{39}$K, $^{56}$Ni, $^{60}$Ni, $^{64}$Ga, and $^{64}$Zn.
With the exception of the limited contribution of the triple-$\alpha$ reaction,
the main nuclear path (Fig. \ref{fig:M2702998}) is fully dominated by a suite of $\beta^+$-decays on numerous
species, all the way up to $^{72}$Br.

\begin{figure}[htbp]
 \centering
   \includegraphics[width=0.90\textwidth]{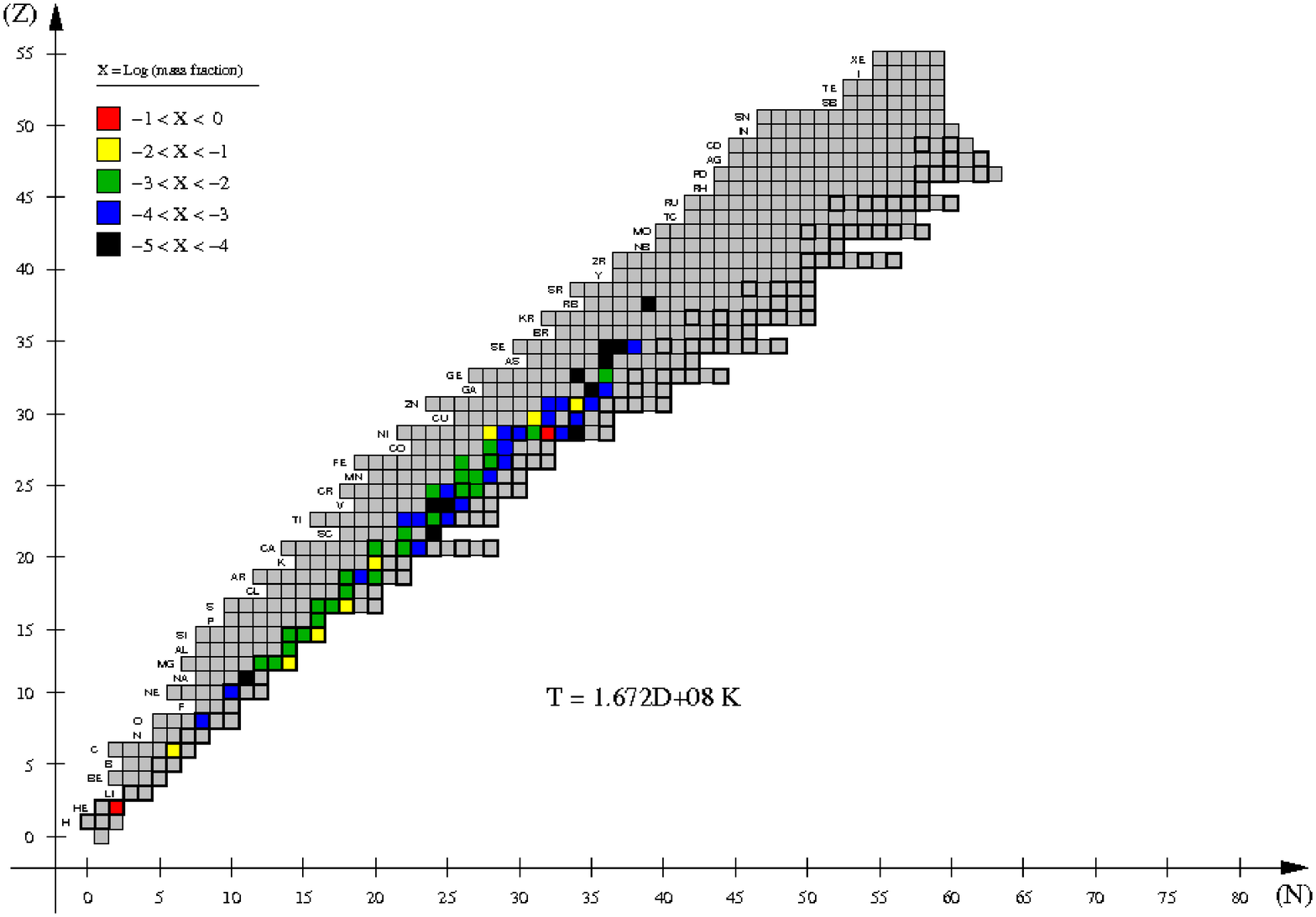}
   \includegraphics[width=0.90\textwidth]{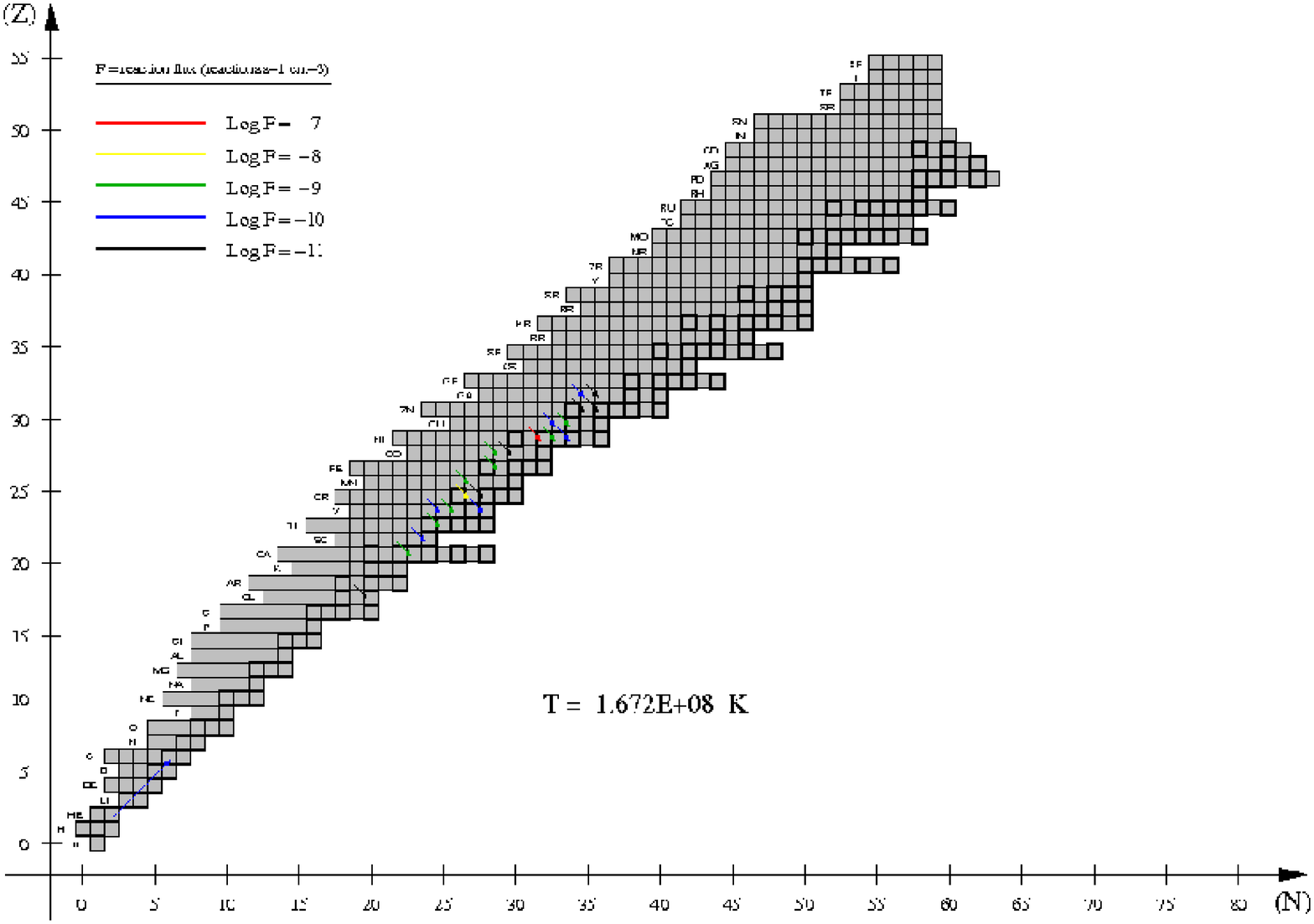}
\caption{Same as Fig. \ref{fig:M1000}, but for the time when temperature 
at the envelope base achieves a minimum value of $T_{min} = 1.67 \times 10^{8}$ K.}
\label{fig:M2917998}
\end{figure}

When t = 28,250 s, T$_{base}$ reaches a minimum value of $1.67 \times 10^8$ K, which we consider to mark the end of the
first burst in our simulations. $^{60}$Cu has decayed already into $^{60}$Ni, now the most
abundant species at the envelope base with a mass fraction of 0.504, followed by $^4$He (0.136, not fully consumed during
the TNR), and by $^{12}$C, $^{26}$Mg, $^{30}$Si, 
$^{34}$S, $^{39}$K, $^{56}$Ni, $^{60}$Cu, and $^{64}$Zn (see Fig. \ref{fig:M2917998}).
The marginal nuclear activity played by a suite of $\beta$-decays powers a rate of nuclear energy generation of $\epsilon_{nuc} 
\sim 8.8 \times 10^{11}$ erg g$^{-1}$ s$^{-1}$. 
At this stage, the size of the envelope has shrunk to $\Delta z \sim 8$ m, whereas the overall
luminosity of the star has decreased to L$_{NS}$ = $8.3 \times 10^{34}$ erg s$^{-1}$.

\begin{figure}[htbp]
 \centering
   \includegraphics[width=0.45\textwidth]{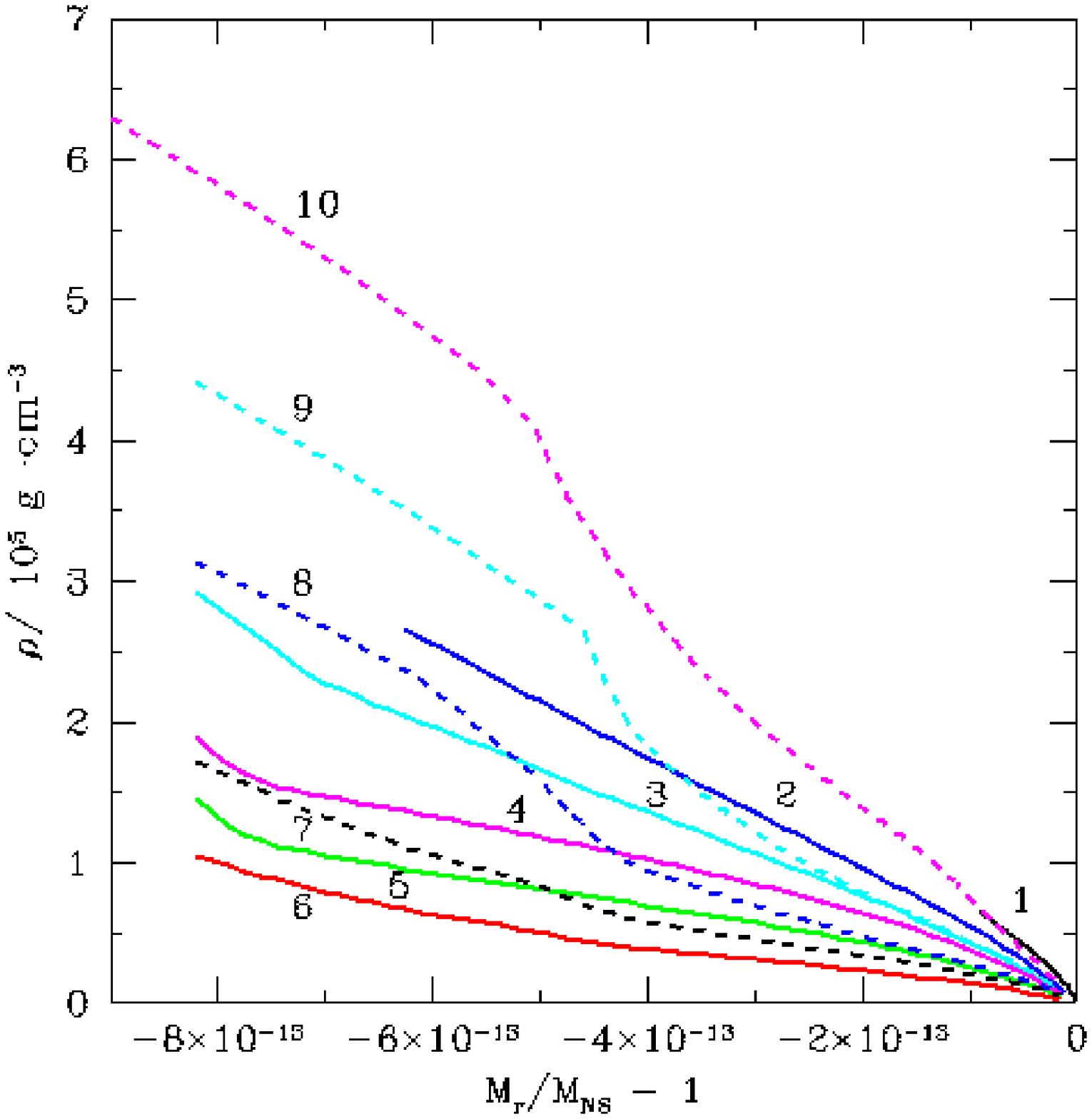}
   \includegraphics[width=0.45\textwidth]{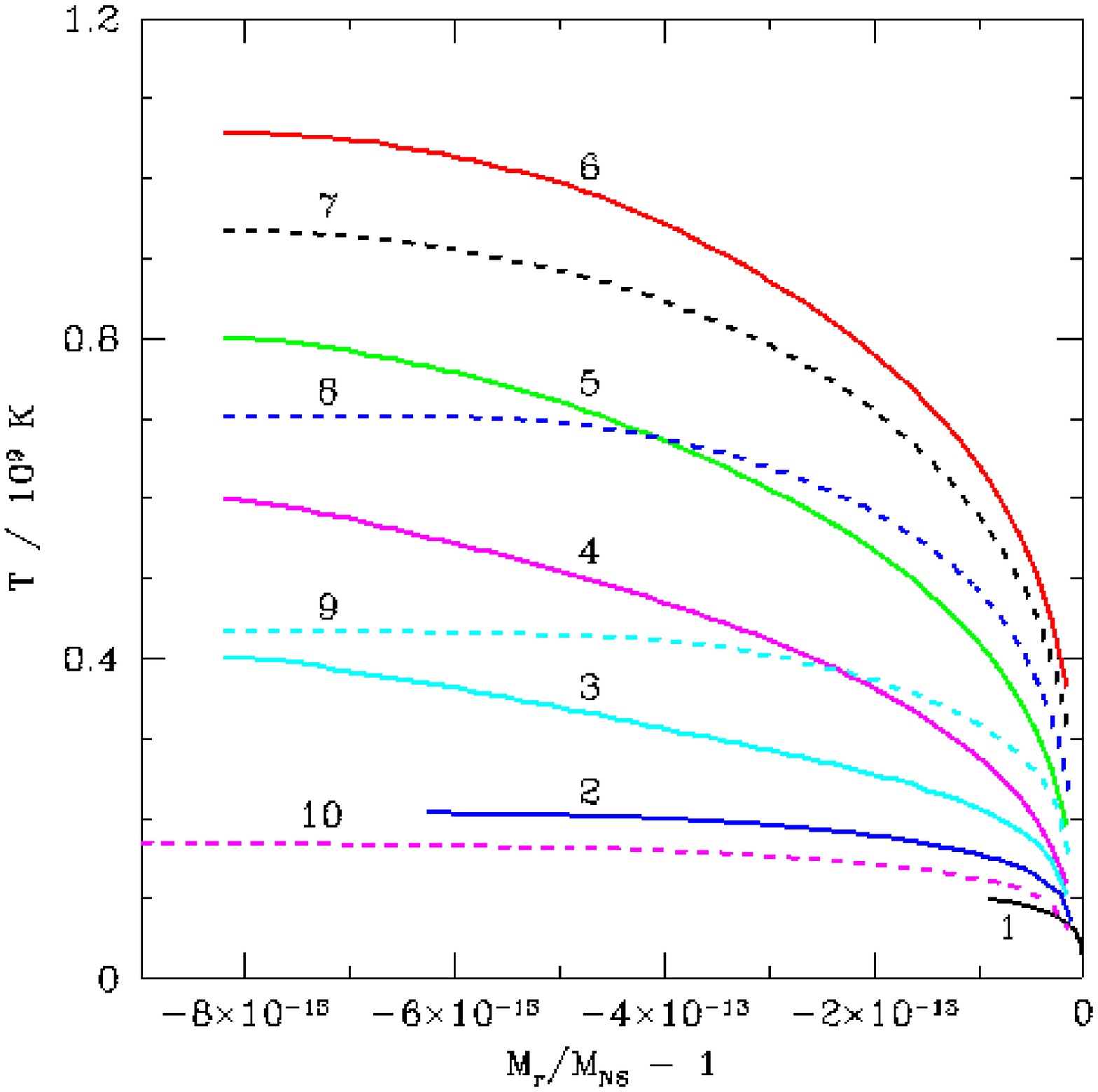}
   \includegraphics[width=0.45\textwidth]{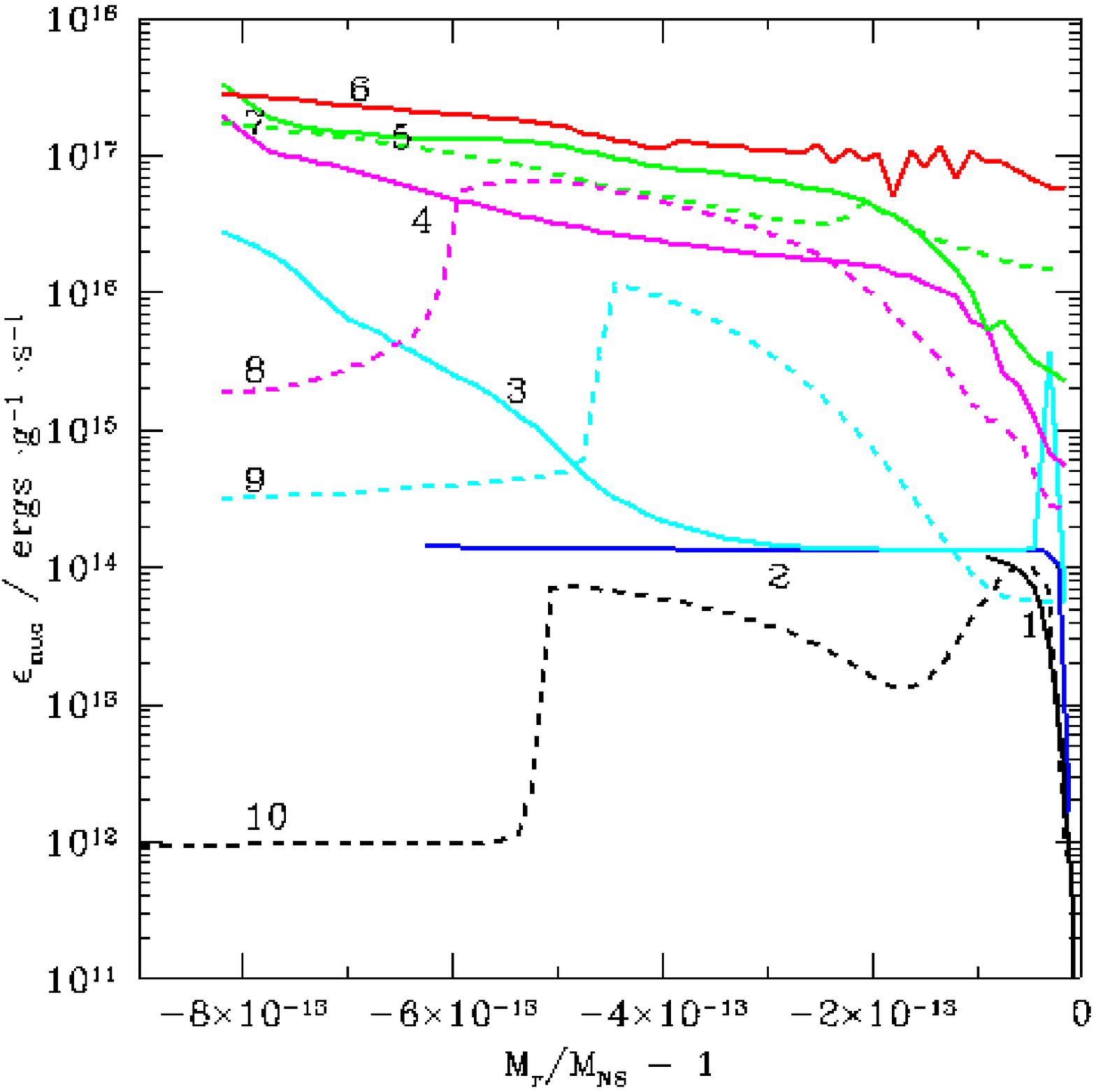}
   \includegraphics[width=0.45\textwidth]{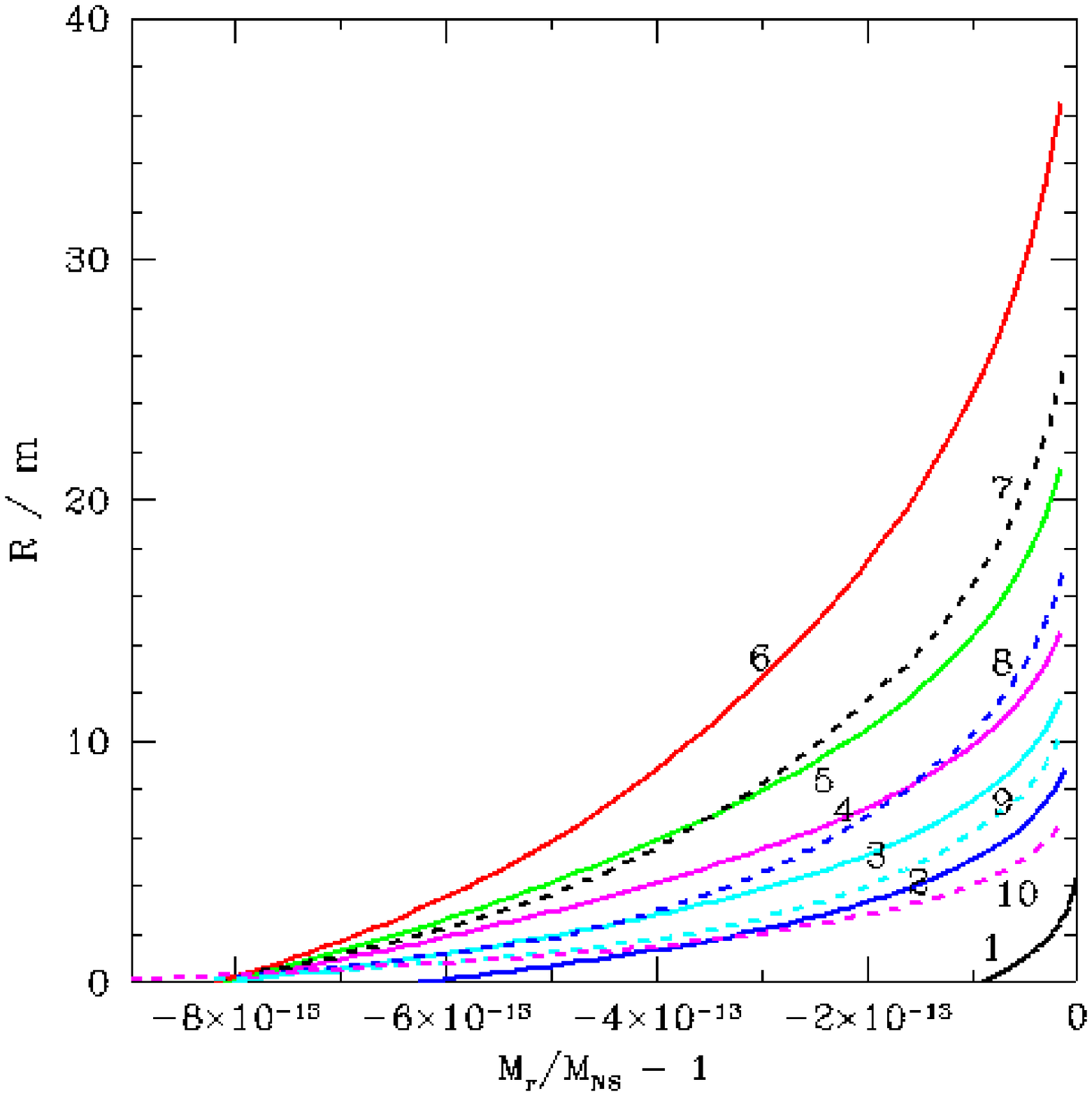}
\caption{Profiles of density (upper left panel), temperature (upper right), 
     nuclear energy generation rate (lower left), and envelope size (measured from the core-envelope
     interface; lower right panel), for model 1
    ($\rm{M}_{NS}$ = 1.4 M$_\odot$, $\rm{\dot M}$$_{acc}$ = $1.75 \times 10^{-9}$ M$_\odot$ yr$^{-1}$,
    Z = 0.02), along the first bursting episode. Labels indicate different moments during the TNR,
    for which T$_{base}$ reaches a value of:
     (1) $9.9 \times 10^7$ K, (2) $2.1 \times 10^8$ K,
     (3) $4 \times 10^8$ K, (4) $6 \times 10^8$ K,
     (5) $8 \times 10^8$ K, (6) $1.06 \times 10^9$ K (T$_{peak}$),
     (7) $9.3 \times 10^8$ K, (8) $7 \times 10^8$ K,
     (9) $4.3 \times 10^8$ K, and (10) $1.7 \times 10^8$ K (T$_{min}$). Notice the dramatic decrease in $\varepsilon_{nuc}$
     in the innermost layers of the envelope (profiles 8 to 10) as H is depleted.}
\label{fig:rhotenuc_m1}
\end{figure}

\begin{figure}[htbp]
 \centering
   \includegraphics[width=0.45\textwidth]{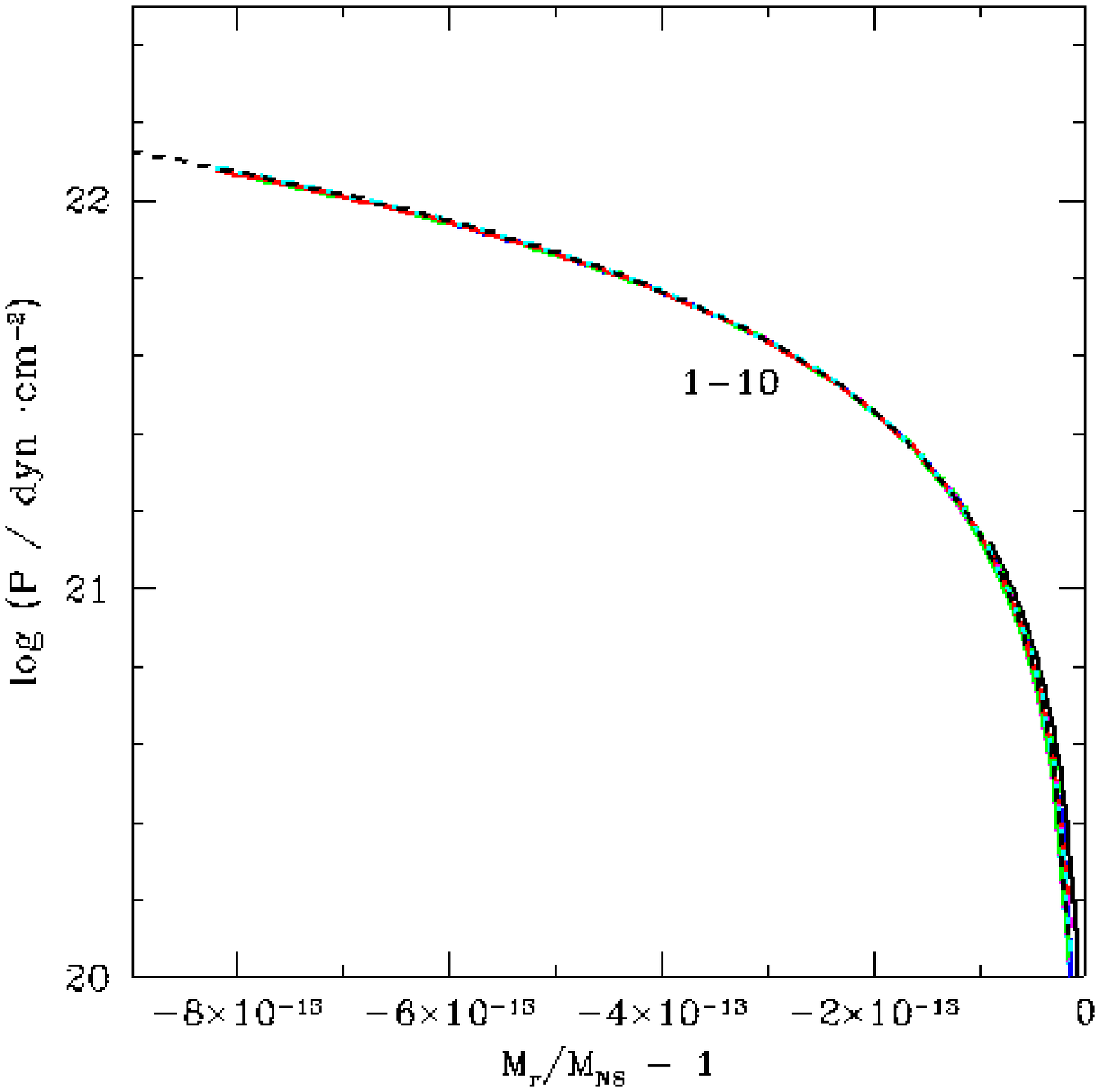}
   \includegraphics[width=0.45\textwidth]{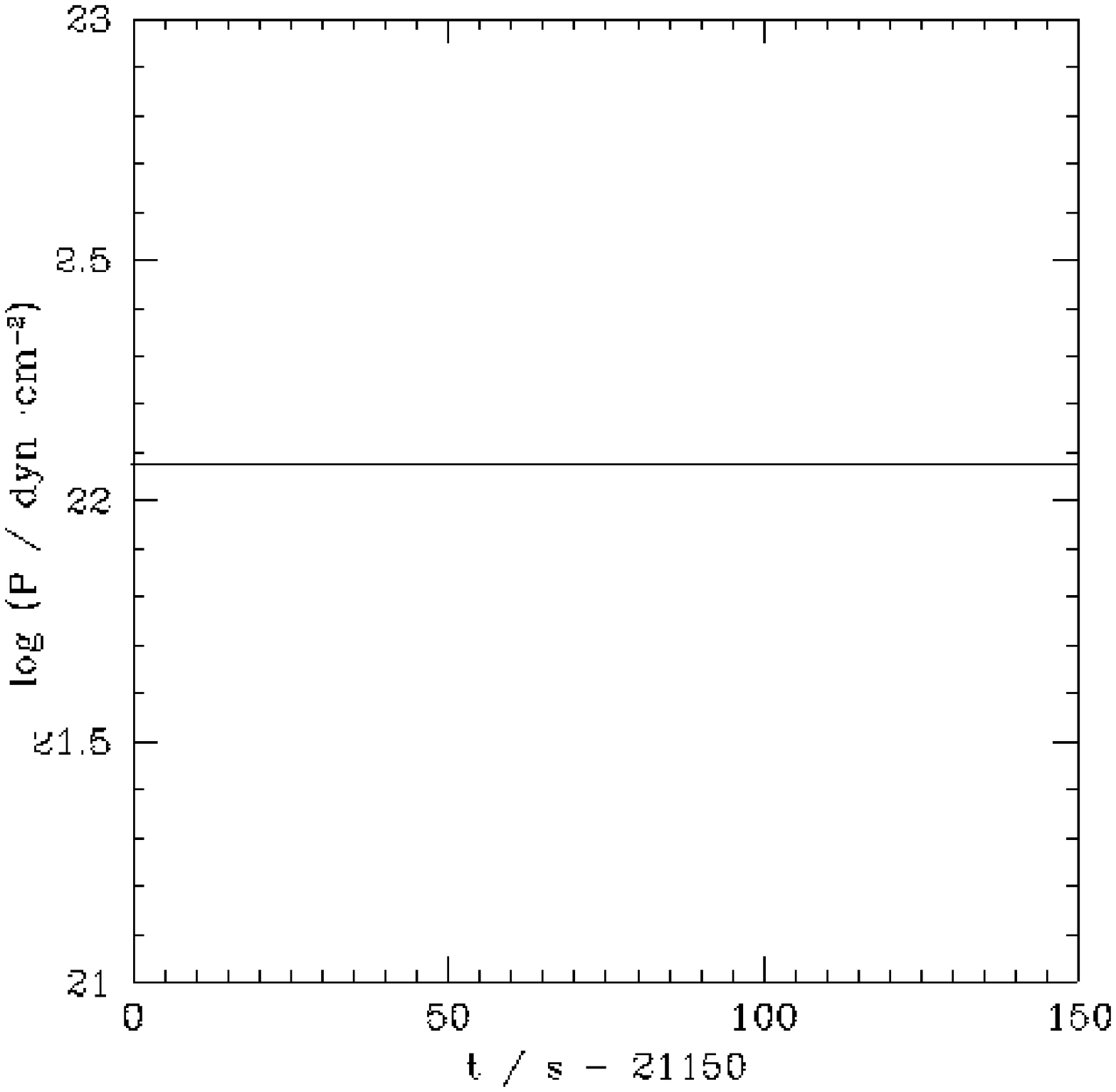}
\caption{Left panel: same as Fig.  \ref{fig:rhotenuc_m1}, but for pressure. Right panel: Time evolution of the
total pressure at the outermost envelope shell.}
\label{fig:prerad_m1}
\end{figure}

Profiles of density, temperature, rate of energy generation, pressure, and size, along the accreted envelope, for different snapshots during
the first bursting episode computed in this model, are shown in Figs. \ref{fig:rhotenuc_m1} and \ref{fig:prerad_m1}.

It is worth noting that the main nucleosynthetic activity takes
place at the innermost, hottest envelope shell.
Because of their lower temperatures and densities, all layers above the
ignition shell show a similar nucleosynthetic pattern but somewhat diluted,
limiting the extent of the nuclear activity to lower masses.
Even though the specific reaction sequences have a clear dependence 
with depth (see Fisker et al. 2008, for the corresponding 
analysis in one bursting episode), it is clear that the identification
of the main nuclear processes responsible for the nucleosynthesis in
XRBs can rely on an accurate account of the activity at the ignition shell,
at different stages of the TNR, as performed in this paper.
However, we would like to outline schematically how depth influences the
extent of the nuclear activity throughout the envelope: while in the innermost
shells of our computational domain (encompassing
$6.8 \times 10^{20}$ g) the nuclear activity reaches $^{90}$Ru,
at $1.7 \times 10^{21}$ g above the core-envelope interface, the activity stops around $^{78}$Sr (the final
mass fraction of $^{90}$Ru barely reaches $10^{-15}$, by mass), while
close to the surface ($2.3 \times 10^{21}$ g), the nuclear activity 
does not extend beyond $^{72}$Se (X($^{90}$Ru) $\sim 10^{-18}$).
Moreover, the purely nucleosynthetic imprint in these shells is difficult
to assess since it is partially poisoned by changes in the chemical 
composition driven by convective mixing. 
 
All in all, the mean, mass-averaged chemical composition of the envelope at the end of this first burst, is mainly dominated by the
presence of intermediate-mass elements (far below the SnSbTe-mass region). 
This includes 
$^{60}$Ni (0.32), $^4$He (0.31), $^1$H (0.17), $^{64}$Zn (0.03), $^{12}$C (0.02), $^{52}$Fe (0.02), or $^{56}$Fe (0.02)
(see Table 2, for the mean composition of all species -stable or with a half-life $> 1$ hr- which achieve X$_i > 10^{-9}$),
with a nucleosynthesis endpoint (defined by the heaviest isotope with X$_i > 10^{-9}$) around $^{89}$Nb (in agreement with the results reported by Fisker
et al. 2008). 
In terms of overproduction factors, $f$ (ratio of the mass-averaged composition of a given isotope over its solar
abundance; see also Figs. \ref{fig:abun_b1} and \ref{fig:abun_b4}), $^{43}$Ca, $^{45}$Sc, $^{49}$Ti, $^{51}$V, $^{60,61}$Ni,  
$^{63,65}$Cu, $^{64,67,68}$Zn, $^{69}$Ga, $^{74}$Se, and $^{78}$Kr achieve a value of $f \sim 10^4$. 

It is important to stress that the presence of unburned H and $^4$He in the envelope, at the end of the first burst, 
will have consequences for the subsequent eruptions (see Section 1.2). Notice, however (Fig. \ref{fig:abun_b1}), that since the innermost envelope
is devoid of H, the next burst will likely initiate well above the core-envelope interface. Moreover, the presence of unburned $^{12}$C, particularly in 
the inner envelope layers, has important implications for studies of the physical mechanism that powers superbursts (see Sect. 2.2).

\subsection{Second, third, and fourth bursts}
For conciseness, we will focus here on the main differences between the first and successive bursts
computed for model 1. A first, remarkable difference is due to the so-called {\it compositional inertia} (Taam 1980; Woosley
et al. 2004), which accounts for differences in the gross properties of the bursts driven by changes in the chemical
content of the envelope. Indeed, after the first burst, the accreted matter will pile up on top of a 
metal-enriched envelope (the initial metallicity, Z$_{ini} \sim 0.02$, 
has risen to a mass-averaged value of Z $\sim$ 0.52, at the end of the first burst) that is devoid of H at its innermost layers.
This will cause a shift in the location of the ignition region, progressively moving away from the core-envelope interface in successive
flashes (see Fig. \ref{fig:conv_tem_b4}, right panel).

This is schematically shown as well in Fig. 
\ref{fig:conv_tem_b4} (left panel), which depicts the mass
above the neutron star core, as well as
the extension of the convective regions
throughout the envelope, for the four bursting episodes computed in this model: 
the peaks of the explosions
($\sim 10-100$ s) correspond to the flat regions of the diagram, whereas the 
stages of steady accretion ($\sim 5$ hr) are indicated
by the steep slopes. Notice that, in agreement with previous work 
(Woosley et al. 2004, Fisker et al. 2008),
convection mainly develops around the peak of the bursts (during most of 
the explosion, energy transport
is carried by radiation only). Notice also that because of fuel consumption (H, 
He), the location of the ignition shell (and the extent of
the convective regions) moves progressively away in mass from the 
neutron star core.

\begin{figure}[bhtp]
 \centering
 \includegraphics[width=0.45\textwidth]{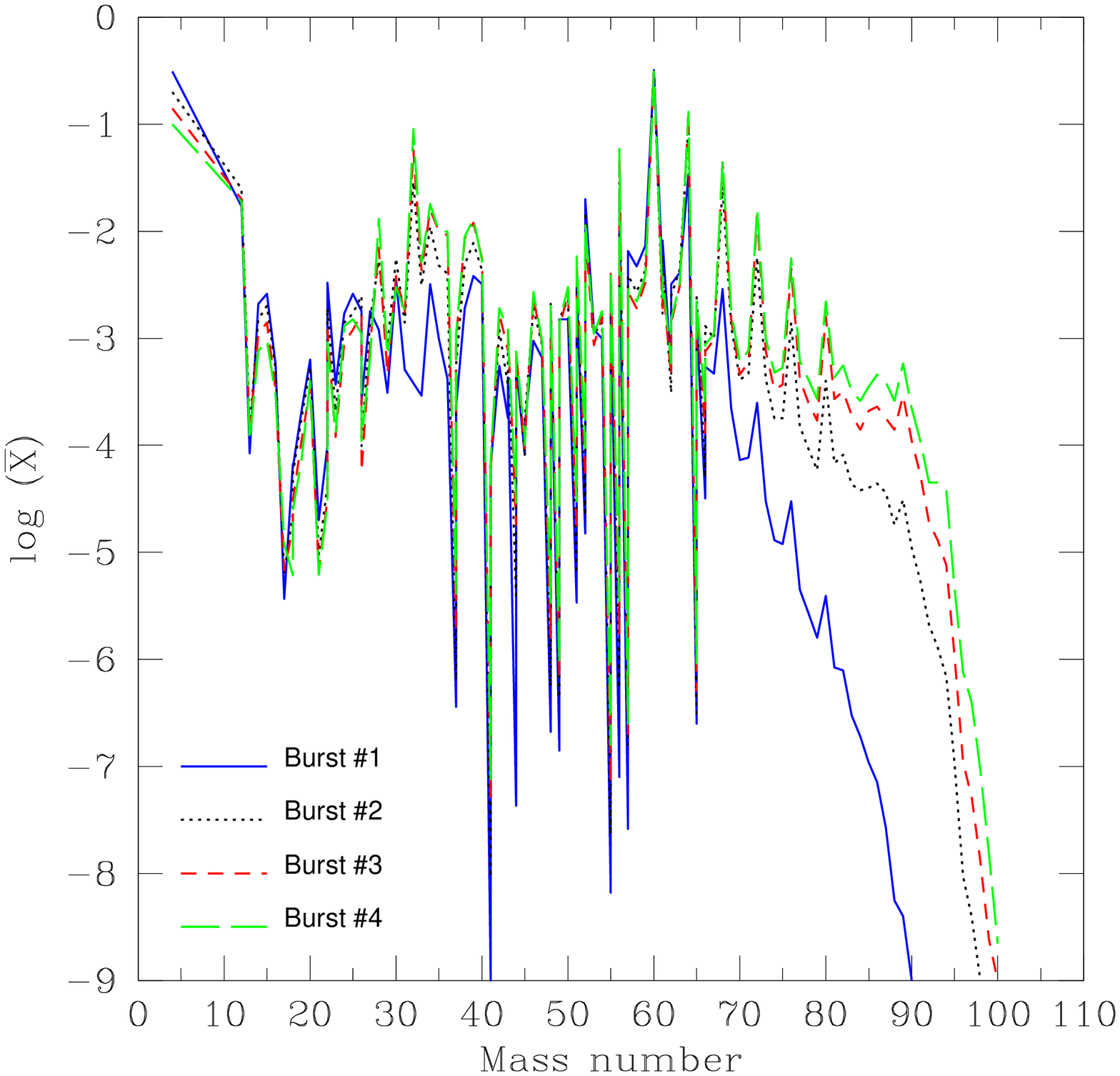}
 \includegraphics[width=0.45\textwidth]{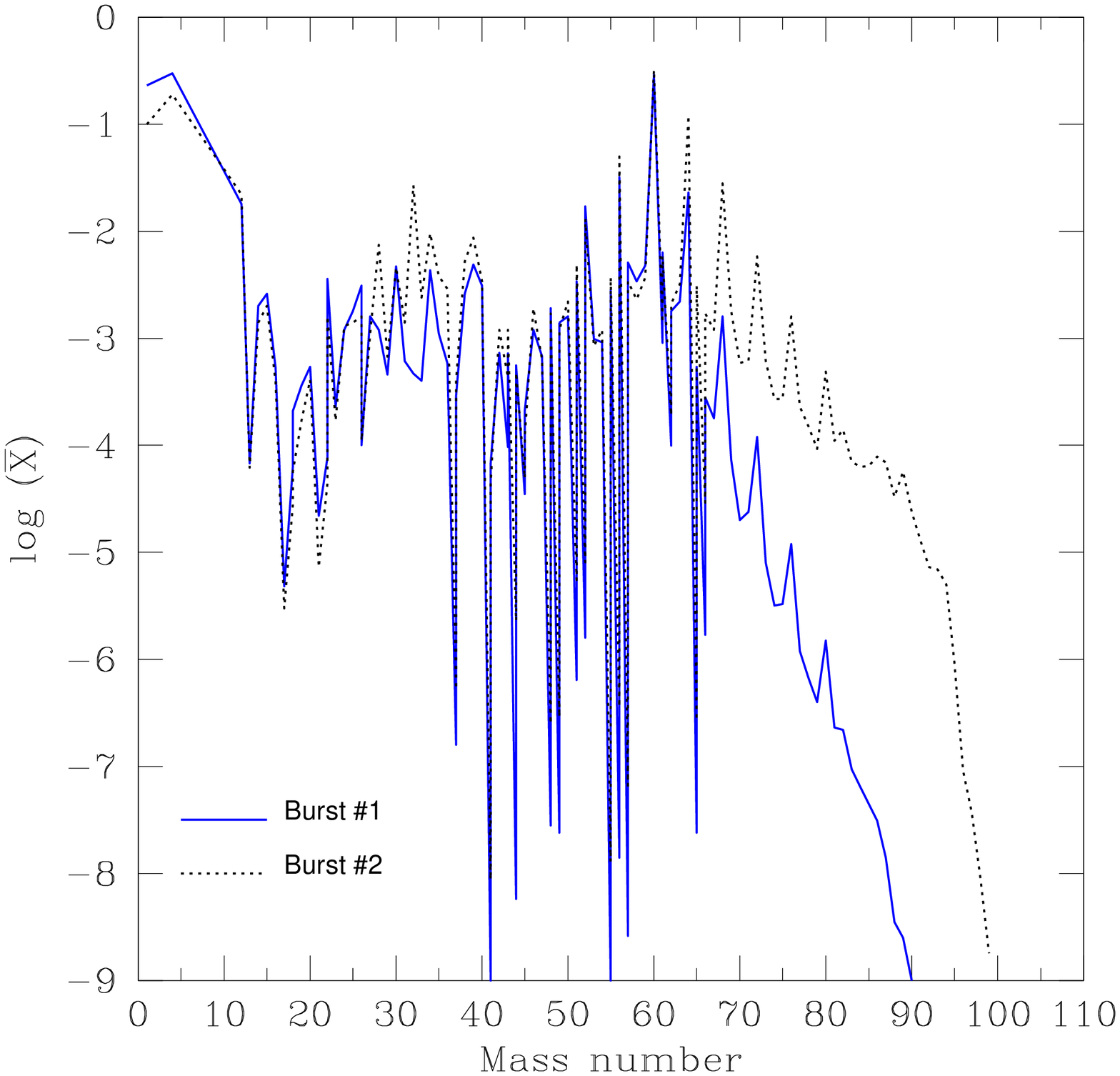}
\caption{Mean post-burst composition in the envelope for each of the bursting episodes computed for model 1 (left) and model 2 (right). 
 }
\label{fig:abun_b4}
\end{figure}

\begin{figure}[htbp]
 \centering
   \includegraphics[width=0.85\textwidth]{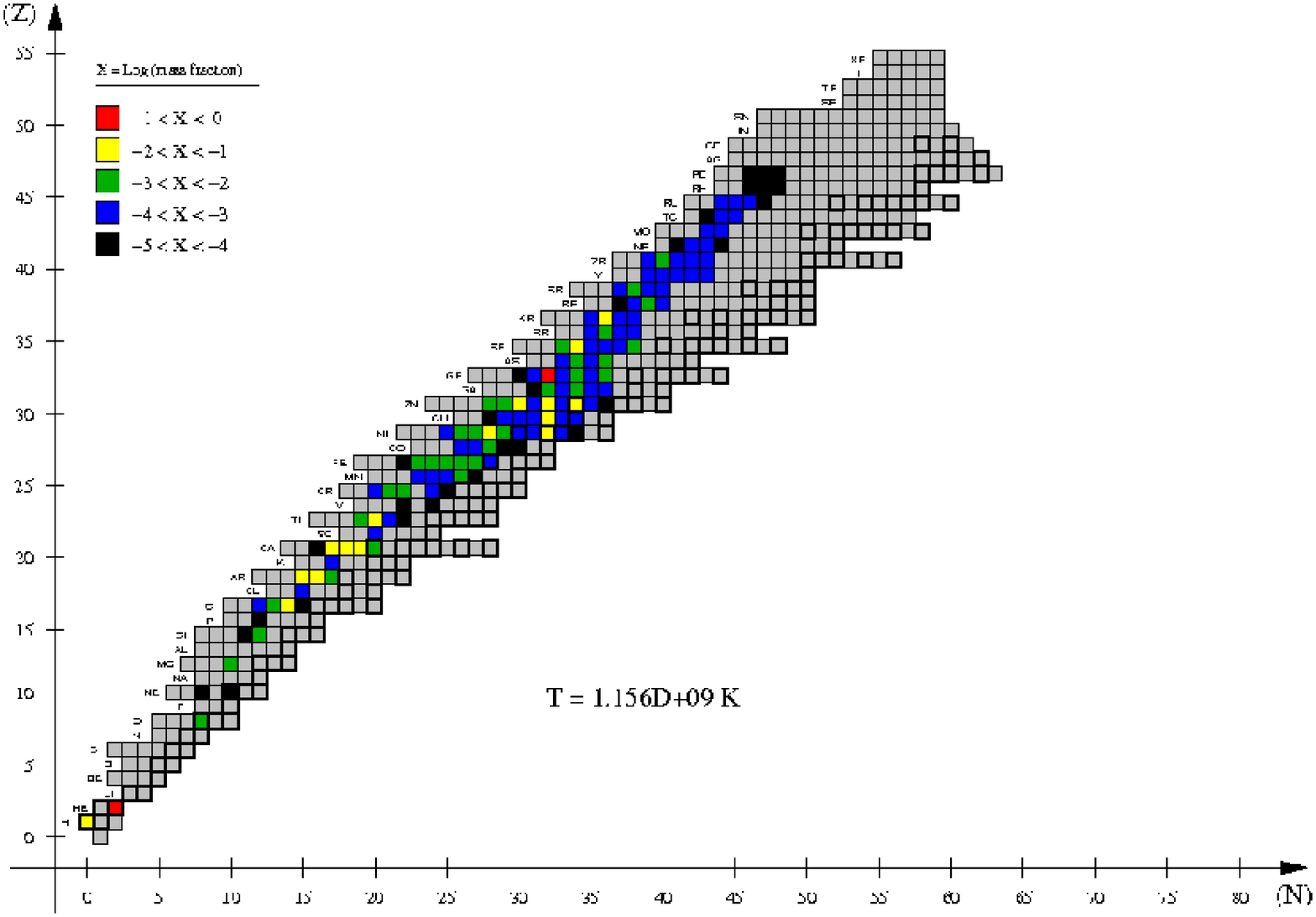}
   \includegraphics[width=0.85\textwidth]{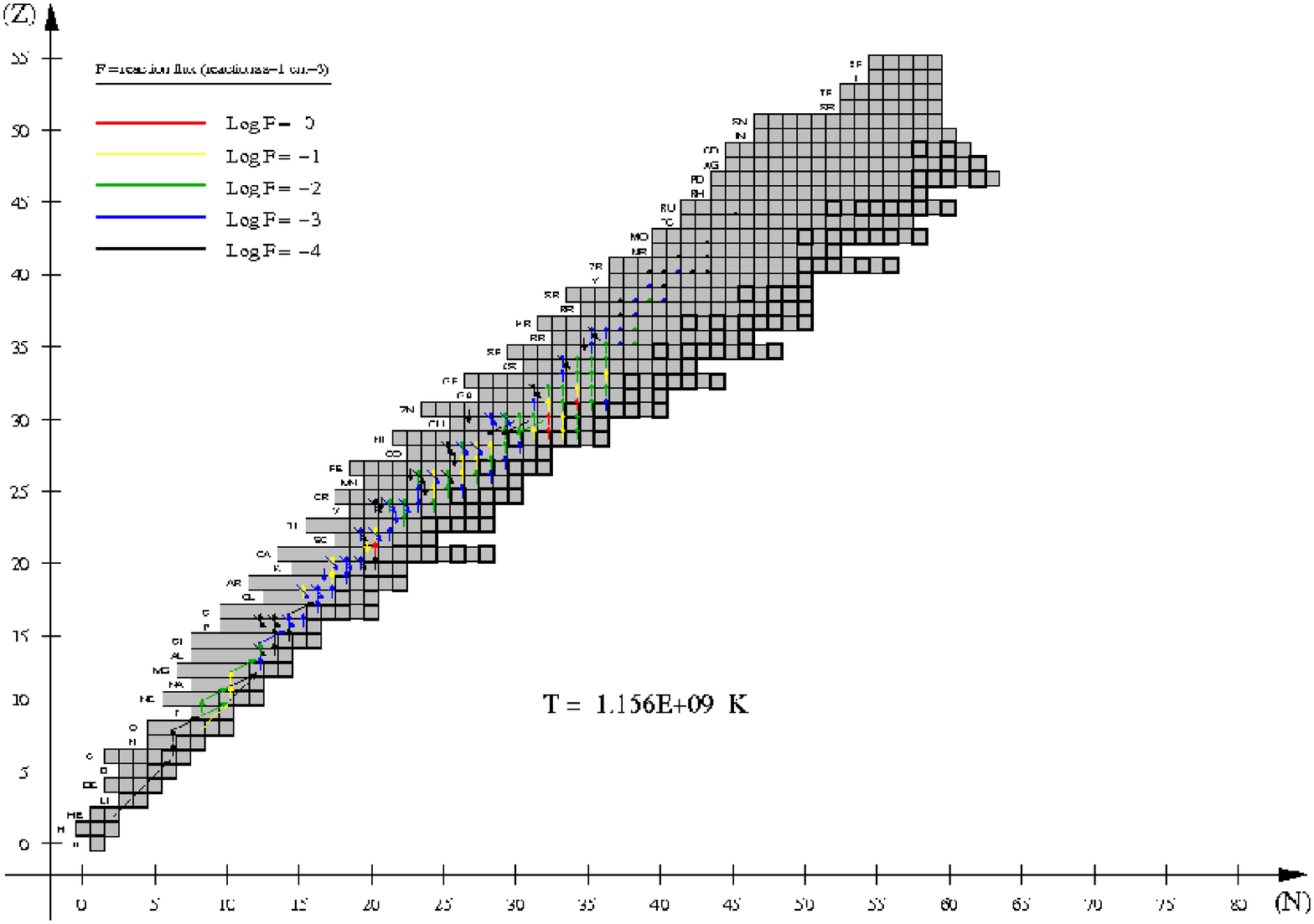}
\caption{Upper panel: main nuclear activity at the ignition shell ($\sim$5.6 m above 
the core-envelope interface),
 when temperature reaches a peak value of $T_{peak} = 1.16 \times 10^{9}$ K, during the 4$^{th}$ burst computed
 for model 1. 
 Lower panel: main reaction fluxes (F $\geq 10^{-9}$ reactions s$^{-1}$ cm$^{-3}$).  }
\label{fig:M1B4TPK}
\end{figure}

From the nucleosynthesis viewpoint (Table 2 and Figs. \ref{fig:abun_b1} and \ref{fig:abun_b4}), 
the nuclear activity extends progressively towards heavier species, 
reaching endpoints (X$_i > 10^{-9}$) around $^{89}$Nb (1$^{st}$ burst), $^{97}$Ru (2$^{nd}$ burst),
$^{99}$Rh (3$^{rd}$ burst), and $^{100}$Pd (4$^{th}$ burst). 
The main nuclear activity and the dominant reaction fluxes at peak temperature,
for the 4$^{th}$ burst computed in this model, are shown in Fig. \ref{fig:M1B4TPK}.
The overall mean metallicity of the envelope at the end of each burst is 0.52 (1$^{st}$ burst), 0.71 (2$^{nd}$ burst), 0.80 (3$^{rd}$ burst), and 0.86 (4$^{th}$ burst). 
This increase in Z reflects both the nuclear activity during each individual burst and the accumulated ashes from previous bursts.
 A similar mass-averaged $^{12}$C yield of $\sim 0.02$ is systematically obtained at the end
of each of the 4 bursts computed. This is not 
 enough to power a superburst, which requires
X($^{12}$C)$_{min} \geq 0.1$, at the envelope base (see Cumming \& Bildsten 2001, Strohmayer \& Brown 2002, Brown 2004,
Cooper \& Narayan 2004, 2005, Cumming 2005, or Cooper et al. 2006). 
With respect to overproduction factors, the increase in nuclear activity reported for
successive bursts translates also into larger values, as high as $f \sim 10^6$, for $^{76}$Se, $^{78,80}$Kr, and
$^{84}$Sr, or $f \sim 10^5$, for
species such as $^{64,68}$Zn, $^{72,73}$Ge, $^{74,77}$Se, $^{82}$Kr, $^{86,87}$Sr, $^{89}$Y, and $^{94}$Mo, 
for the 4$^{th}$ bursting episode (see Figs. \ref{fig:abun_b1} and \ref{fig:abun_b4}).

\begin{figure}[htbp]
 \centering
   \includegraphics[width=0.45\textwidth]{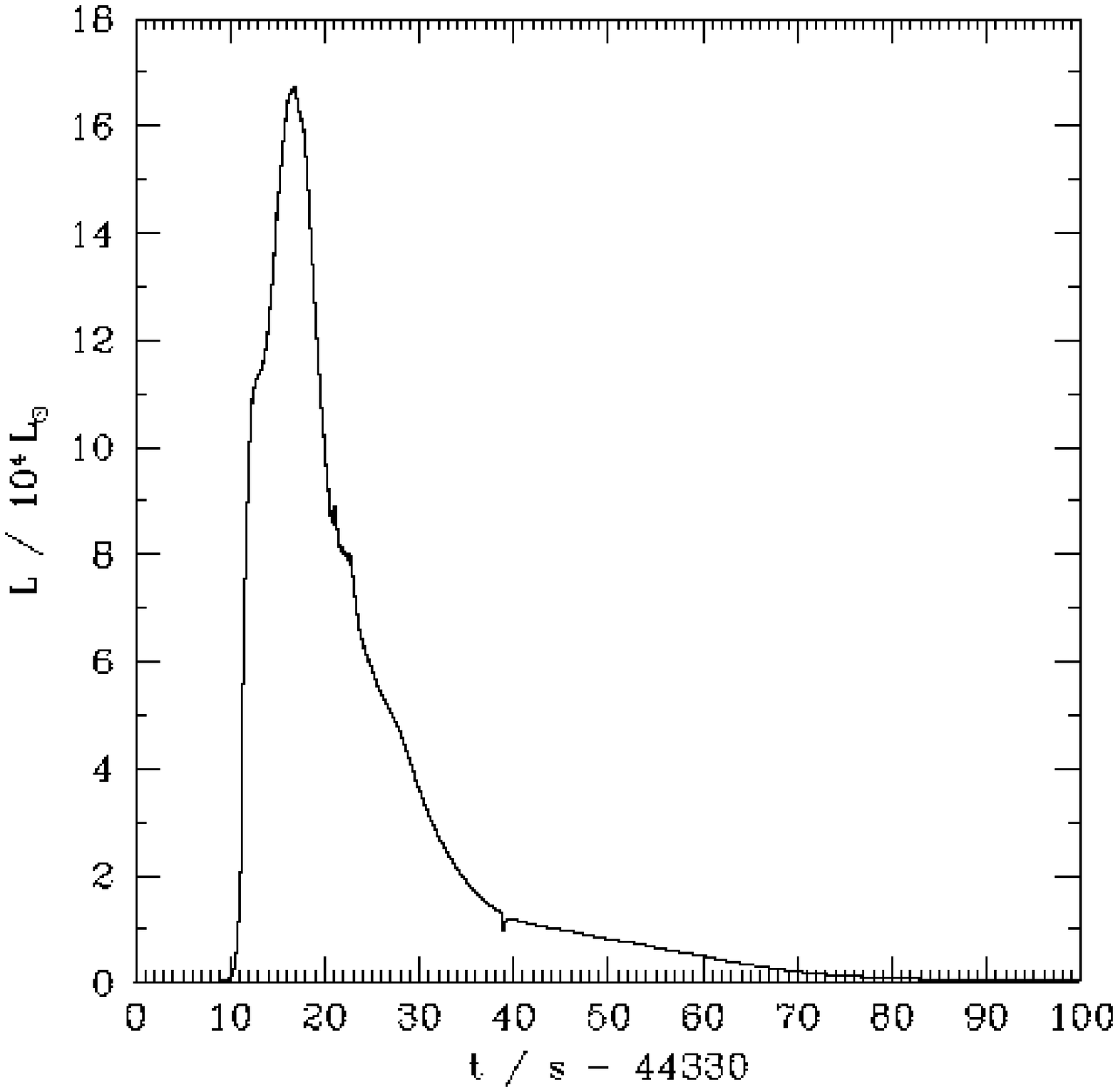}
   \includegraphics[width=0.45\textwidth]{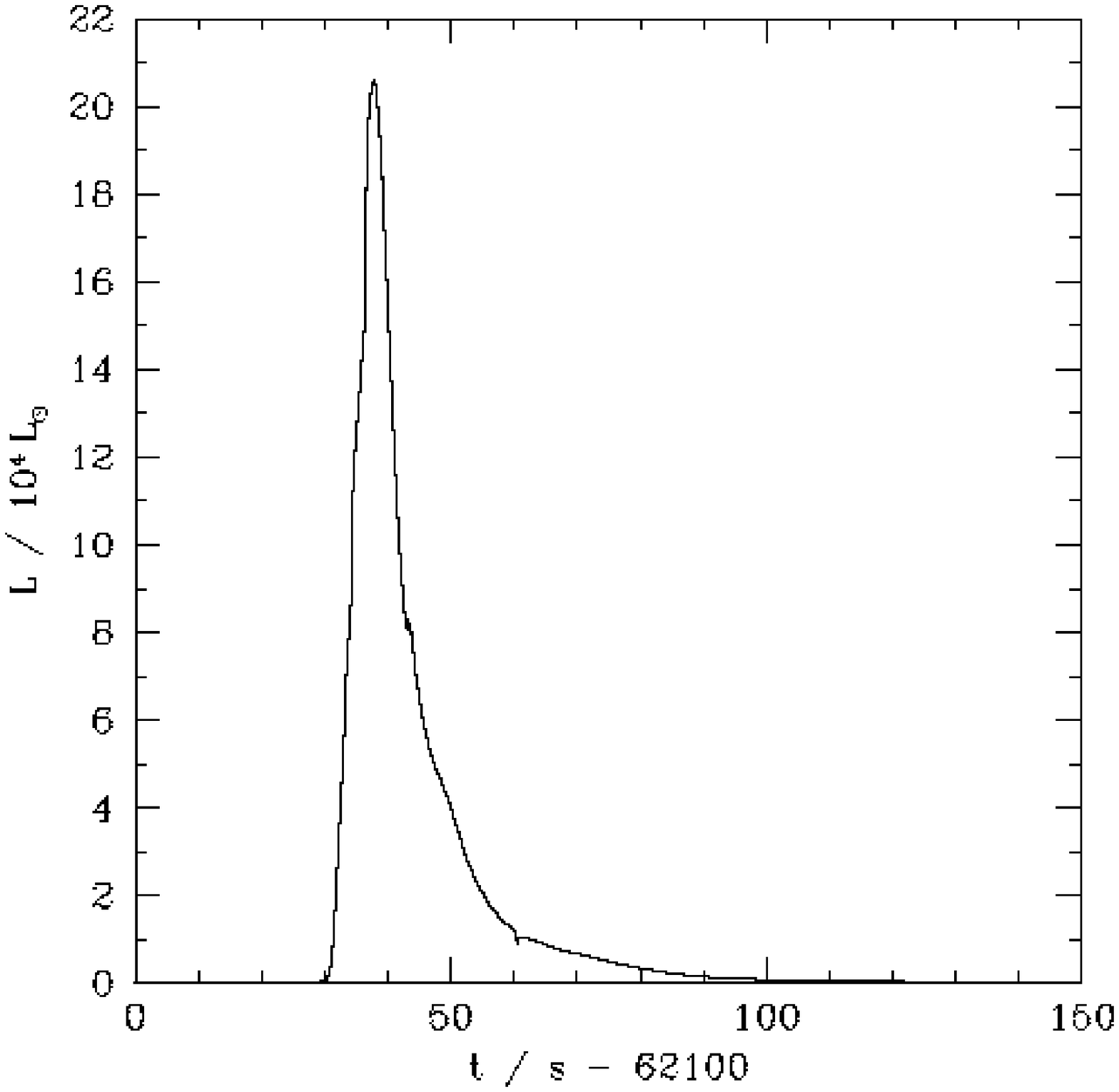}
   \includegraphics[width=0.45\textwidth]{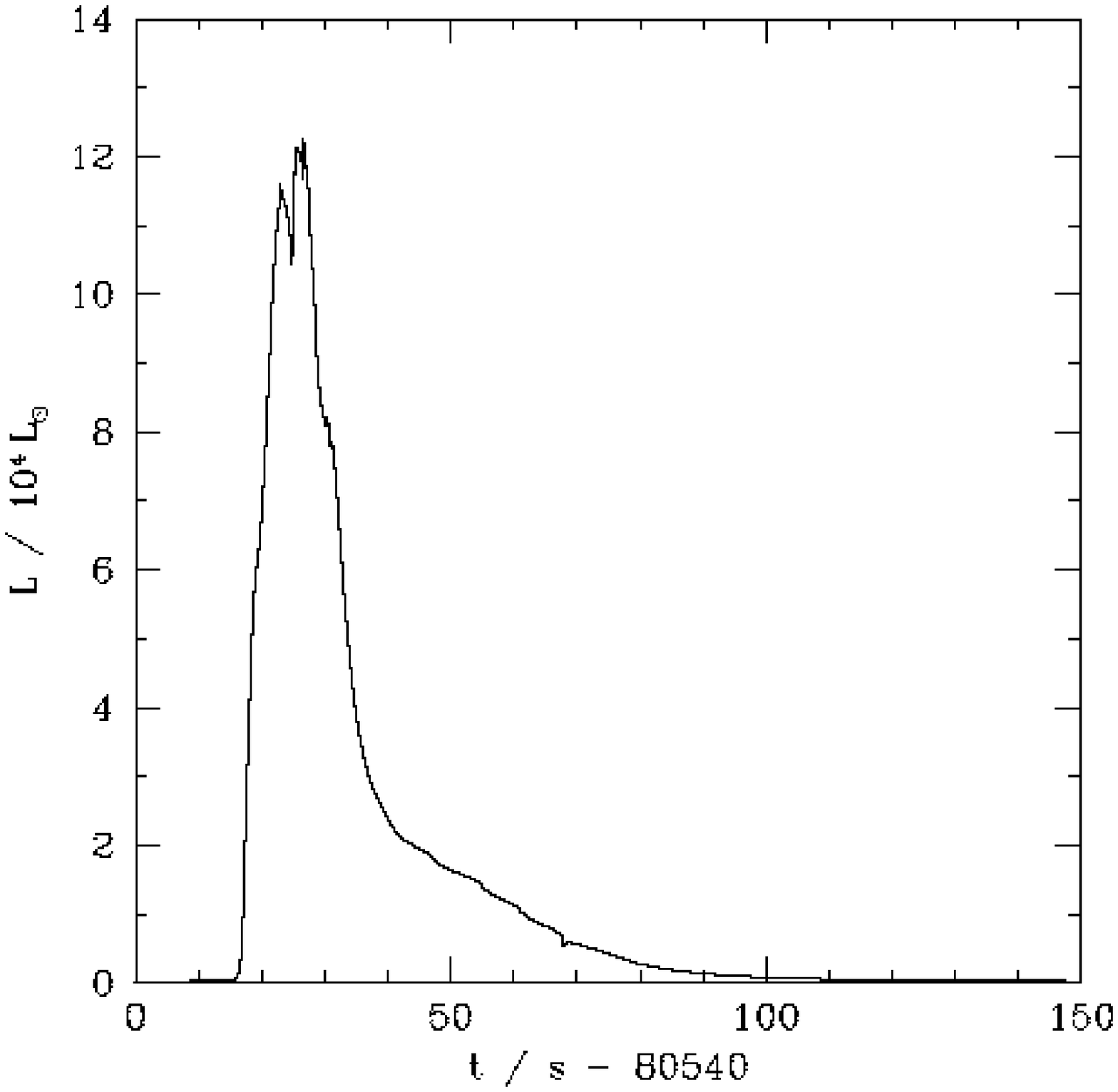}
   \includegraphics[width=0.45\textwidth]{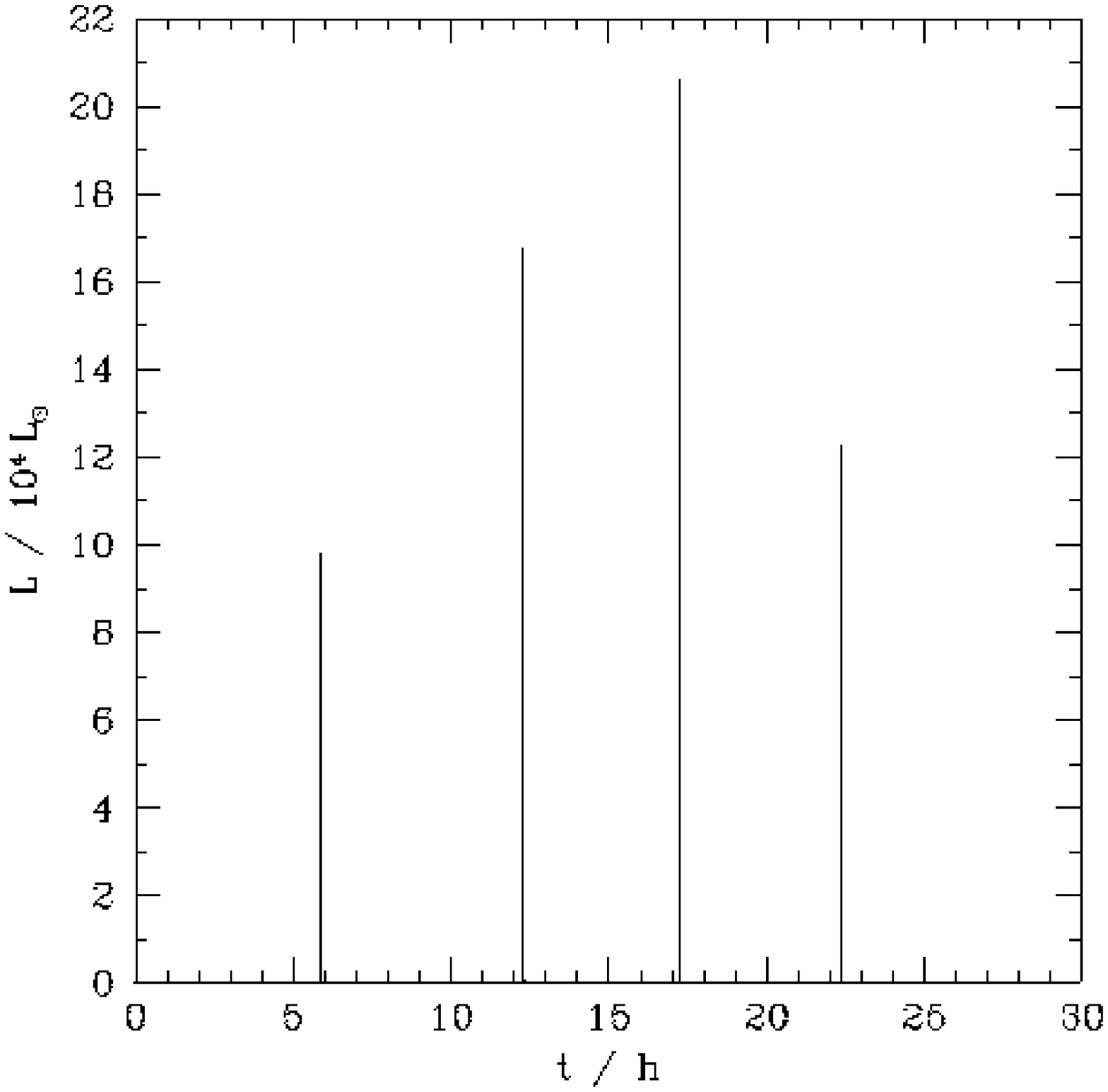}
\caption{Light curves corresponding to the second (upper left panel), third (upper right), and fourth bursts (lower left), 
         and for the overall computed time (lower right), for model 1. }
\label{fig:lum_4b}
\end{figure}

A summary of the gross properties of the four bursts computed in this model 
is given in Table 3. Peak temperatures and luminosities amount to $T_{peak} \sim (1.1 - 1.3) \times 10^9$ K, 
and $L_{peak} \sim (1 - 2) \times 10^5 L_\odot$, respectively.
Recurrence times between bursts of $\tau_{rec} \sim 5 - 6.5$ hr and ratios between 
persistent\footnote{We define $\alpha = \int_t^{t+\tau_{rec}} L(t)\,dt$/$\int_{t'}^{t'+\tau_{0.01}} L(t)\,dt$,
with the latter term integrated over 
the time during which the burst exceeds 1\% of its peak luminosity, $\tau_{0.01}$. 
Notice that during the interburst period, the accretion luminosity,
$L_{acc} = G M \dot{M} / R \sim 1.5 \times 10^{37} erg s^{-1}$, 
will hide the thermal emission from the cooling ashes, as shown in 
Figs. \ref{fig:lumrad_t1} and \ref{fig:lum_4b}.}
and burst luminosities of $\alpha \sim 35 - 40$ (except for the first burst) have been obtained. 
These values are in agreement with those inferred from some observed XRB sources (see Galloway et al. 2008) 
such as the {\it textbook burster} 
GS 1826-24 [$\tau_{rec} = 5.74 \pm 0.13$ hr, $\alpha = 41.7 \pm 1.6$], 
4U 1323-62 [$\tau_{rec} = 5.3$ hr, $\alpha = 38 \pm 4$], 
or 4U 1608-52 [$\tau_{rec} = 4.14 - 7.5$ hr, $\alpha = 41 - 54$]. 
As reported by Woosley et al. (2004), there is also some trend towards stabilization of these values with increasing burst number.

Fig. \ref{fig:lum_4b} depicts the corresponding light curves from the second to the fourth burst. 
A quite interesting feature, observed in some XRBs such as
4U 1608-52  (Penninx et al. 1989), 4U 17+2 (Kuulkers et al. 2002), or 4U 1709-267 (Jonker et al. 2004),
is the appearance of a double-peaked burst in Fig. 
\ref{fig:lum_4b} (lower left panel). Double (or triple) peaked bursts can be classified in two categories (Watts \& Maurer 2007):
the first one corresponds to the so-called {\it photospheric radius expansion} bursts, which exhibit 
multi-peaked bursts in the X-ray band but not in the bolometric luminosity. The second type of multi-peaked
events are also visible in the bolometric light curves, and have been attributed to different causes, such as 
a stepped release of thermonuclear energy caused either by mixing induced by hydrodynamic instabilities
(Fujimoto et al. 1988) or driven by a nuclear waiting-point impedance in the thermonuclear reaction flow
(Fisker et al. 2004). A preliminary analysis of the 4$^{th}$ burst reported for model 1 suggests a likely
nuclear physics origin (waiting-point impedance) for this double-peaked feature (see Jos\'e \& Moreno 2010).

\section{Model 2\label{sec:Model 1}}

In the previous Section, we have reported results from a sequence of type I X-ray bursts computed
with a coarse resolution, in which the accreted envelope was discretized in 60 shells. We have checked
the influence of the adopted number of envelope shells on the gross properties of the bursts by performing
another simulation, identical to model 1, but computed with a finer resolution: 200 shells (hereafter, model 2).

A summary of the main properties of the two bursts computed for model 2 is given in Table 4: the 
recurrence times obtained are in the same range as those reported for model 1 (4 - 6 hr). The same applies
to the ratios between persistent and burst luminosities, as well as to peak temperatures and luminosities.
Similar light curves have also been obtained.

There is also good agreement from the nucleosynthesis viewpoint, with only minor differences in the final,
mass-averaged abundances, as shown in Table 5 (particularly, for the heavier species of the network, since 
their low abundances are very 
sensitive to the specific thermal history of the explosion; see also Fig. \ref{fig:abun_b4}). It is worth noting that both models reach 
almost identical
nucleosynthesis endpoints ($X_i > 10^{-9}$): $^{89}$Nb, for the first burst computed in both models, and $^{97}$Ru
(model 1) and $^{99}$Rh (model 2), for the second burst.
Furthermore, the amounts of unburned H, $^4$He, and $^{12}$C 
are very similar in both models. As expected from the abovementioned similarities, 
there is also good agreement in terms of overproduction factors, dominated by $^{64}$Zn and $^{60}$Ni 
(with $f \sim 10^4$) in the first burst, and by $^{64}$Zn, $^{72}$Ge, $^{74,76}$Se, $^{78,80}$Kr,
and $^{84}$Sr ($f \sim 10^5$), in the second, for both models.

All in all, we conclude that the resolution adopted in model 1 is appropriate for XRB simulations. This is
in agreement with the studies performed by Fisker et al. (2004), who concluded that the minimum 
discretization of the envelope, in 1-D hydrodynamic simulations of X-ray bursts, is about 25 shells.

\section{Model 3\label{sec:Model 3}}

To test the impact of the metallicity of the accreted material (which reflects the 
surface composition of the companion star) on the overall properties of the bursts, we have computed
another series of bursts (hereafter, model 3), driven by accretion of metal-deficient material
(Z $\sim Z_\odot$/20) onto a 1.4 M$_\odot$ neutron star (L$_{ini}$ = $1.6 \times 10^{34}$ erg s$^{-1}$ = 4.14 L$_\odot$),
at a rate $\rm{\dot M}$$_{acc}$ = $1.75 \times 10^{-9}$ M$_\odot$ yr$^{-1}$. 
The composition of the accreted material is assumed to be X=0.759, Y=0.240, and Z=10$^{-3}$,
and as for model 1, all metals are initially added up in the form of $^{14}$N. This model is indeed qualitatively 
similar to model zM, from Woosley et al. (2004) (see also Fisker et al. 2008, and Schatz et al. 2001). Both the envelope zoning and the initial relaxation phase 
are identical to those described for model 1.

\subsection{First burst}
The piling up of matter on top of the neutron star during the accretion stage progressively compresses 
and heats the envelope. 

At t=4337 s, the envelope achieves T$_{base} \sim 10^8$ K ($\rho_{base} \sim 9.8 
\times 10^4$ g cm$^{-3}$). The nuclear activity 
is fully dominated by the CNO-cycle.  In contrast to model 1, the smaller metallicity 
of this model limits substantially the role of proton captures. 
Indeed, at this stage, H has only been reduced to 0.757 at the envelope base.  
The main reaction fluxes are actually an order
of magnitude lower than those reported from model 1, for the same temperature, 
powering an energy generation rate of $\varepsilon_{nuc} \sim 6.2 \times 10^{12}$ erg g$^{-1}$ s$^{-1}$.
Because of the lower metallicity of this model, the time required to
achieve  T$_{base} \sim 10^8$ K is about twice the value reported for model 1,
resulting in a thicker, more massive envelope which will affect
the forthcoming explosion.
Besides H and He (0.242), by far the most abundant nuclei in the envelope, the nuclear activity in the CNO region
increases the chemical abundances of many species in this mass range, with
$^{15}$O ($6.3 \times 10^{-4}$) being the most abundant CNO-group nucleus at the
envelope base. 

16.8 hours (60,347 s) after the beginning of accretion, T$_{base}$ has 
reached $2 \times 10^8$ K (with $\rho_{base}$ = $5.7 \times 10^5$ g cm$^{-3}$, and  
P$_{base} = 3.4 \times 10^{22}$ dyn cm$^{-2}$).  The total luminosity of the star
is only $6.9 \times 10^{34}$ erg s$^{-1}$.
The main nuclear activity is governed by $^{15}$N(p, $\alpha$)$^{12}$C and other
reactions of the CNO cycle. 
$^7$Be(p, $\gamma$)$^8$B is at equilibrium with its reverse photodisintegration reaction $^8$B($\gamma$, p)$^7$Be. 
Because of the limited number of CNO-catalysts in this low-metallicity
     model, some proton-proton chain reactions,
     such as the {\it pep}, $^3$He($\alpha$, $\gamma$)$^7$Be, or $^8$Be $\rightarrow$ 2$^4$He, 
     are relatively frequent. 
In terms of chemical abundances, the now frequent p-captures have reduced the 
hydrogen content down to a value of 0.695 (while X($^4$He) = 0.303). 
The next most abundant isotopes in the network are the short-lived species $^{15}$O 
($1.4 \times 10^{-3}$) and $^{14}$O ($8 \times 10^{-4}$). 
The nuclear activity ($X_i > 10^{-9}$) reaches $^{40}$Ca at this stage.

 18.1 hours (65,081 s) from the onset of accretion, 
T$_{base}$ reaches $4 \times 10^8$ K.
Hydrogen continues to decrease smoothly (X(H) = 0.689), whereas the $^4$He abundance reaches 0.280. 
The next most abundant nuclei are $^{14}$O ($1.4 \times 10^{-2}$), and 
$^{15}$O, now followed by $^{52}$Fe and $^{18}$Ne, with mass fractions of the order of $10^{-3}$. 
The nuclear activity extends up to $^{58}$Cu.  The largest reaction fluxes
correspond to different processes that operate almost at equilibrium with their inverse photodisintegration reactions, such as
$^{21}$Mg(p, $\gamma$)$^{22}$Al($\gamma$, p)$^{21}$Mg, 
$^{30}$S(p, $\gamma$)$^{31}$Cl($\gamma$, p)$^{30}$S, $^{25}$Si(p, $\gamma$)$^{26}$P($\gamma$, p)$^{25}$Si, and
$^{7}$Be(p, $\gamma$)$^{8}$B($\gamma$, p)$^{7}$Be. 

 6 seconds later (t = 65,087 s), T$_{base}$ achieves $5 \times 10^8$ K. 
$\rho_{base}$ has slightly decreased to $4.3 \times 10^5$ g cm$^{-3}$ because of a 
mild envelope expansion ($\Delta z \sim 19.7$ m). 
$^4$He has slightly decreased to 0.267, as  a result of the frequent
 $\alpha$-captures driven by the high temperatures achieved.  
The next most abundant species are now $^{18}$Ne ($10^{-2}$), together 
with $^{14,15}$O ($3.5 \times 10^{-3}$ and $7.5 \times 10^{-3}$, respectively),
$^{52}$Fe ($2.8 \times 10^{-3}$), and $^{34}$Ar ($1.1 \times 10^{-3}$). 
The nuclear activity reaches $^{61}$Ga, powering an energy generation rate of
 $2.8 \times 10^{16}$ erg g$^{-1}$ s$^{-1}$. The overall stellar luminosity is now $2.5 \times 10^{35}$ erg s$^{-1}$.

A qualitatively similar picture is found when T$_{base}$ achieves $7 \times 10^8$ K
(at t = 65,090 s), with 
 the nuclear activity extending all the way up to $^{68}$Se. 

\begin{figure}[htbp]
 \centering
   \includegraphics[width=0.90\textwidth]{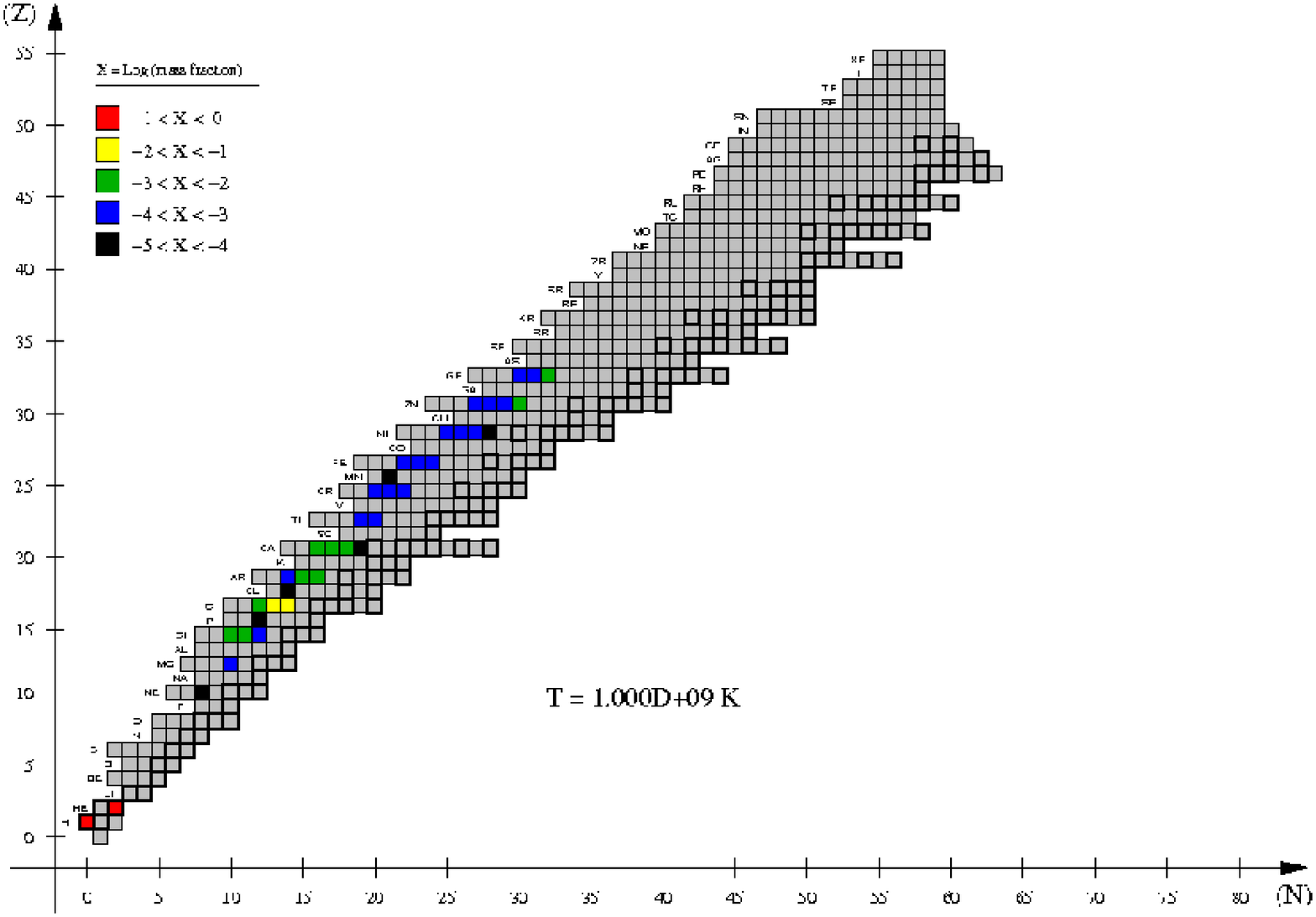}
   \includegraphics[width=0.90\textwidth]{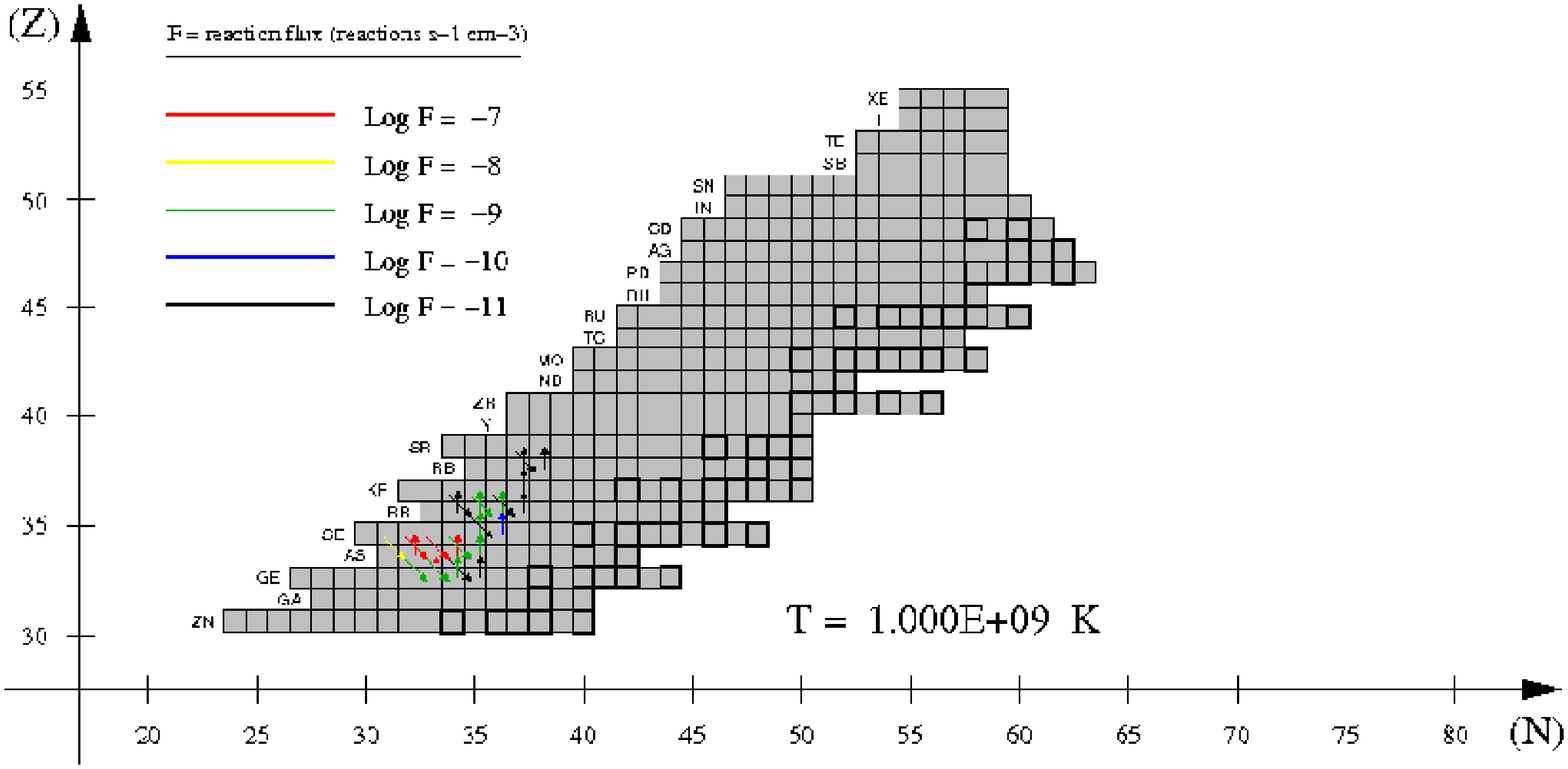}
\caption{Main nuclear activity at the innermost envelope shell for model 3
($\rm{M}_{NS}$ = 1.4 M$_\odot$, $\rm{\dot M}$$_{acc}$ = $1.75 \times 10^{-9}$ M$_\odot$ yr$^{-1}$, 
 Z = $10^{-3}$), at the time when temperature at the envelope base reaches T$_{base} = 10^{9}$ K. 
 Upper panel: mass fractions of the most abundant species (X $> 10^{-5}$); 
 Lower panel: main reaction fluxes (F $\geq 10^{-11}$ reactions s$^{-1}$ cm$^{-3}$) responsible for the nuclear activity in the A=65-100 mass region 
 --except for equilibrium (p, $\gamma$)-($\gamma$, p) pairs.}
\label{fig:3M222998}
\end{figure}

 One second later (t = 65,091 s), T$_{base}$ achieves $10^9$ K. The nuclear
 activity (with $\epsilon_{nuc} \sim 1.5 \times 10^{17}$ erg g$^{-1}$ s$^{-1}$; see Fig. \ref{fig:3M222998}) 
 continues to reduce  the H and $^4$He abundances down to 0.680 and 0.224, respectively. 
 The next most abundant species are now $^{28,29,30}$S, $^{33,34}$Ar, 
 $^{25}$Si, $^{60}$Zn, and $^{38}$Ca (all with mass fractions $\sim 10^{-2}$),
  with the main nuclear path reaching $^{72}$Kr.  
  The largest absolute fluxes are achieved by nuclear interactions 
  between equilibrium (p, $\gamma$)-($\gamma$, p) pairs, which do not
  contribute to the net energy balance. 
  Instead, the most important contributors to the energy production at
  this stage are 
  $^{25}$Al(p, $\gamma$)$^{26}$Si, $^{27,29,30}$P(p, $\gamma$)$^{28,30,31}$S, 
  $^{28}$Si(p, $\gamma$)$^{29}$P, $^{32,33}$Cl(p, $\gamma$)$^{33,34}$Ar,
  $^{32,33}$Cl(p, $\gamma$)$^{33,34}$Ar, $^{35}$Ar(p, $\gamma$)$^{36}$K,
  $^{35,36,37}$K(p, $\gamma$)$^{36,37,38}$Ca, the chain
  3$\alpha \rightarrow$ $^{12}$C(p, $\gamma$)$^{13}$N(p, 
     $\gamma$)$^{14}$O($\alpha$, p)$^{17}$F(p, $\gamma$)$^{18}$Ne($\alpha$, 
  p)$^{21}$Na(p, $\gamma$)$^{22}$Mg, and a suite of $\beta^+$-decays, such as 
  $^{25}$Si($\beta^+$)$^{25}$Al, $^{28,29,30}$S($\beta^+$)$^{28,29,30}$P,
  and $^{33}$Ar($\beta^+$)$^{33}$Cl.
  The activity in the A=65-100 mass region is dominated by
  the suite of reactions depicted in Fig. \ref{fig:3M222998}, lower panel,
  mainly $^{65}$Ge(p, $\gamma$)$^{66}$As, 
  $^{67}$Se($\beta^+$)$^{67}$As(p, $\gamma$)$^{68}$Se, 
  $^{65}$As(p, $\gamma$)$^{66}$Se($\beta^+$)$^{66}$As (with log F $\sim$ -7),
  and $^{65}$Se($\beta^+$)$^{65}$As (log F $\sim$ -8).

  In contrast to model 1, which achieved a peak temperature of $1.06 \times 10^9$ K, model 3 reaches relatively higher
  values. Hence, at t = 65,093 s, T$_{base}$ achieves $1.2 \times 10^9$ K.   
  The H and $^4$He abundances have been reduced to 0.648 and 0.205, respectively. 
  The next most abundant nucleus is still $^{30}$S ($3.1 \times 10^{-2}$), followed by $^{38}$Ca ($2 \times 10^{-2}$), 
  and by a large number of species with abundances of the order of $10^{-3}$, such as 
  $^{36,37}$Ca, $^{28,29}$S, $^{32,33,34}$Ar, $^{58,59,60}$Zn, $^{62,63,64}$Ge, $^{48,49,50}$Fe, $^{53,54,55}$Ni, $^{41}$Ti, and
  $^{44,45,46}$Cr. At this stage, the main nuclear path reaches $^{76}$Sr. 

  One second later, at t = 65,094 s, the rate of nuclear energy generation achieves a maximum value of
  $\epsilon_{nuc,max} \sim 2.1 \times 10^{17}$ erg g$^{-1}$ s$^{-1}$.

\begin{figure}[htbp]
 \centering
   \includegraphics[width=1.00\textwidth]{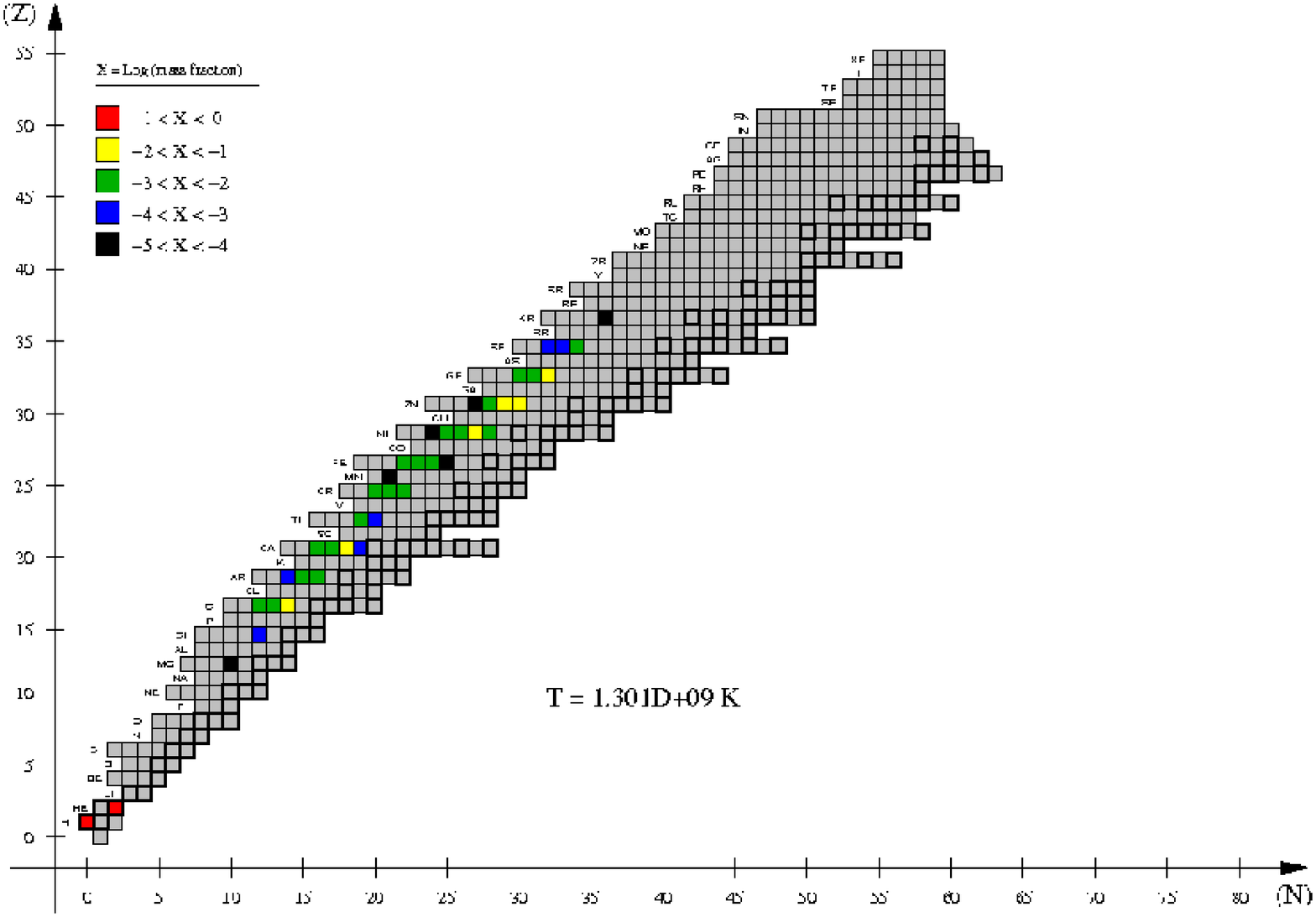}
   \includegraphics[width=1.00\textwidth]{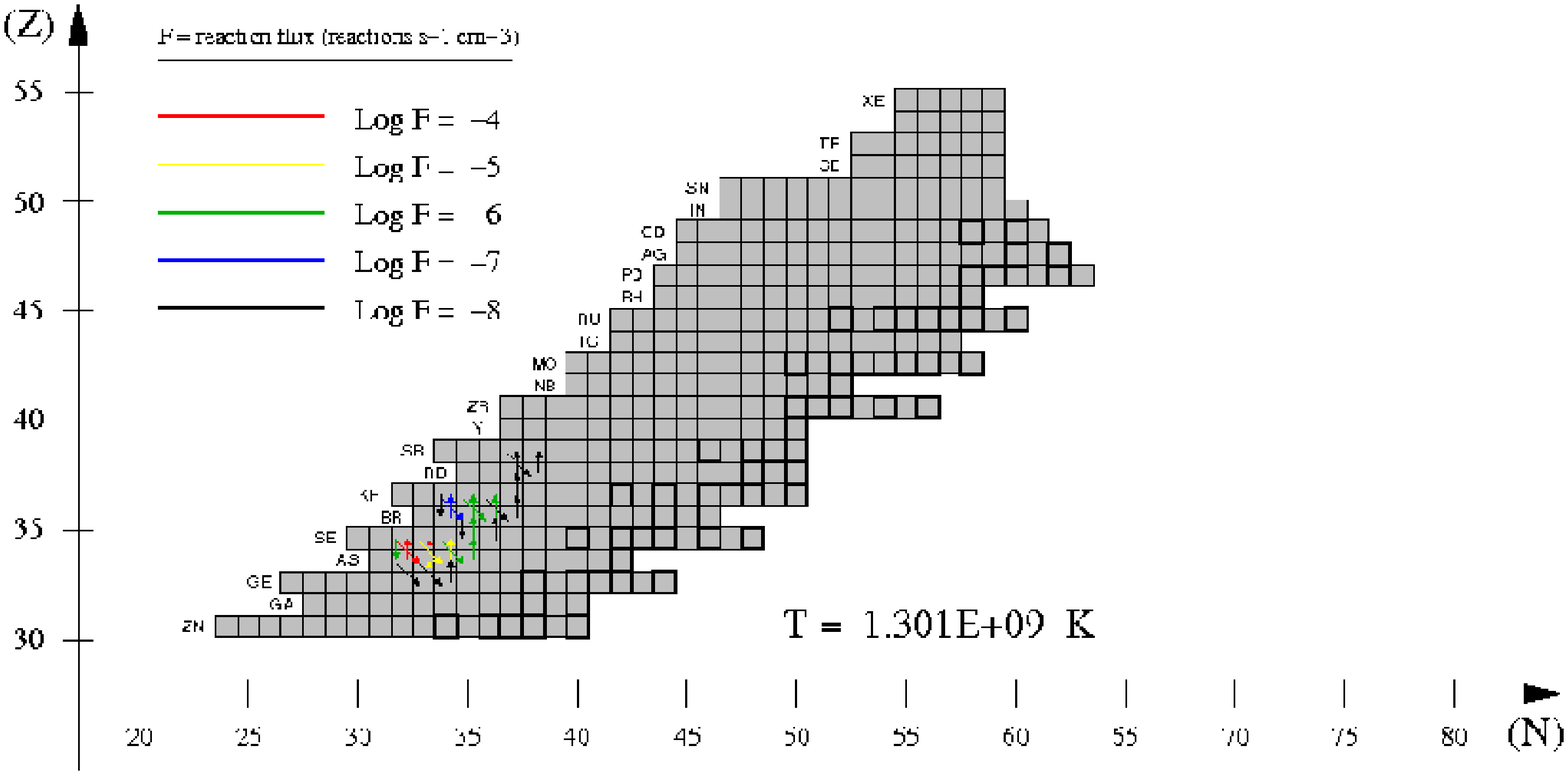}
\caption{Same as Fig. \ref{fig:3M222998}, but for $T_{base} = 1.3 \times 10^{9}$ K.}
\label{fig:3M889998}
\end{figure}

  At t = 65,095 s, while T$_{base}$ = $1.3 \times 10^9$ K, 
  the main nuclear path reaches $^{80}$Zr.
  Because of the large temperature achieved, the number of proton- and $\alpha$-captures
  increases, which in turn efficiently reduces the H (0.621) and $^4$He (0.197) abundances.   
  The next most abundant nucleus is $^{60}$Zn ($2.8 \times 10^{-2}$), followed by $^{64}$Ge,
   $^{38}$Ca, $^{30}$S, $^{55}$Ni, 
   and $^{59}$Zn (with X$_i \sim 10^{-2}$, see Fig. \ref{fig:3M889998}).
   The nuclear activity in the A=65-100 mass region is now dominated by 
   $^{65}$As(p, $\gamma$)$^{66}$Se($\beta^+$)$^{66}$As(p, $\gamma$)$^{67}$Se 
   (log F $\sim$ -4),  $^{65}$Ge(p, $\gamma$)$^{66}$As, and
     $^{67}$Se($\beta^+$)$^{67}$As(p, $\gamma$)$^{68}$Se
   (log F $\sim$ -5; see Fig. \ref{fig:3M889998}). 
   Energy production is mainly due to suite of
   (p, $\gamma$) reactions and $\beta^+$-decays 
   involving nuclear species in the mass range A=30-62.

\begin{figure}[htbp]
 \centering
   \includegraphics[width=1.00\textwidth]{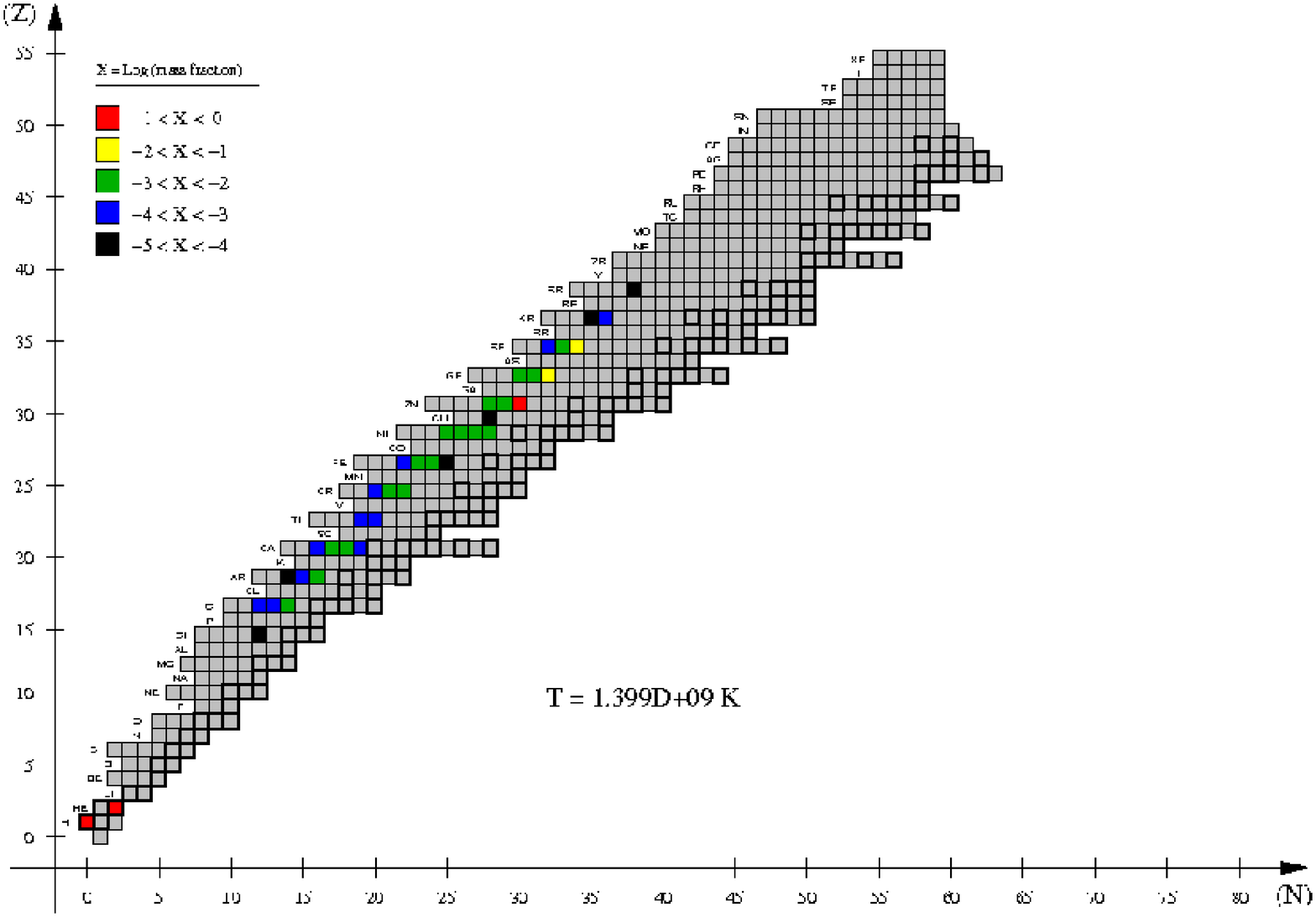}
   \includegraphics[width=1.00\textwidth]{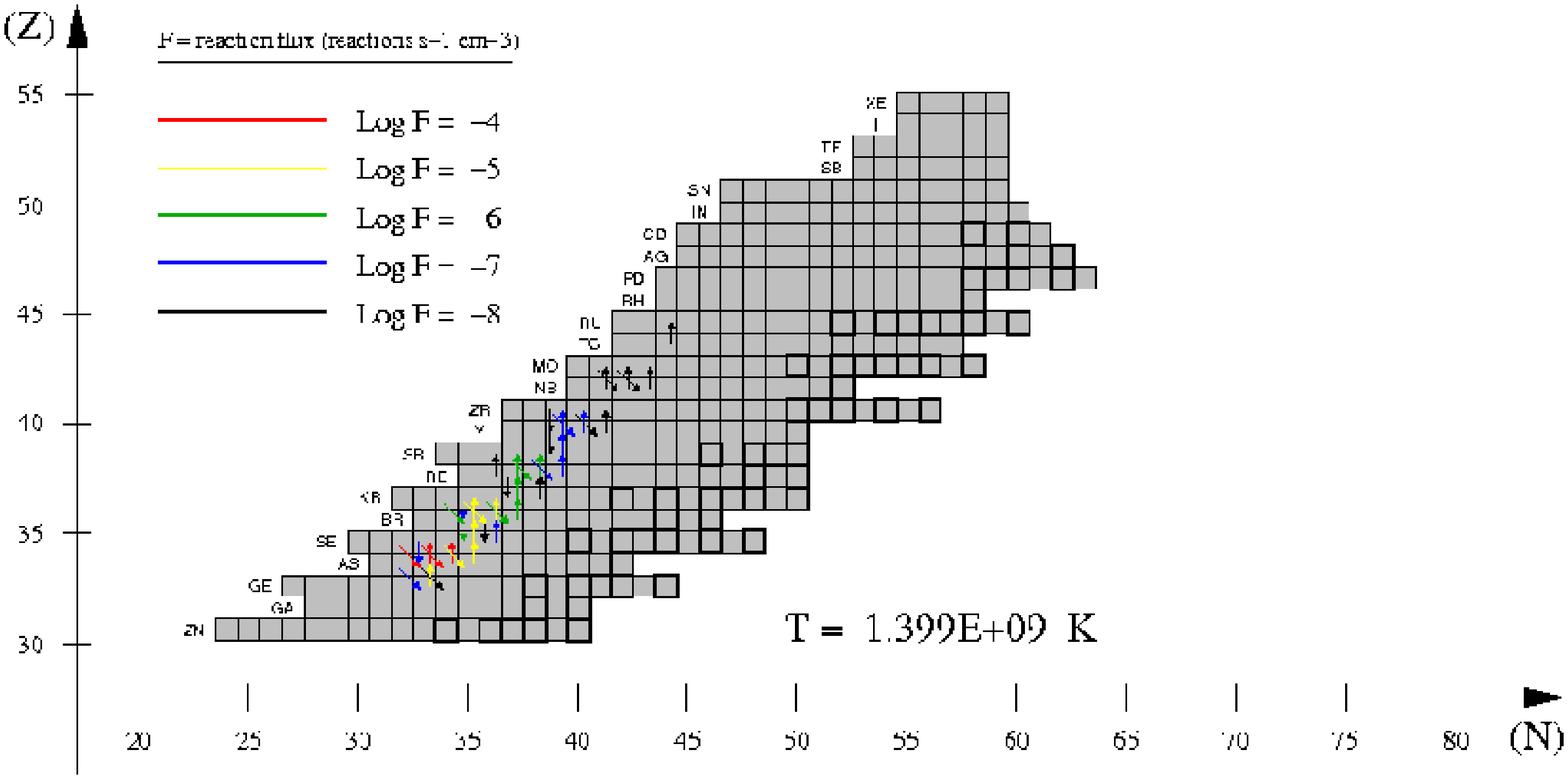}
\caption{Same as Fig. \ref{fig:3M222998}, but for the time when temperature at the envelope base reaches a peak value of
          $T_{peak} = 1.4 \times 10^{9}$ K.  }
\label{fig:3M1091998}
\end{figure}

   Shortly after, at t = 65,098 s, a peak temperature of T$_{peak}$ = $1.4 \times 10^9$ K is achieved at the envelope base. 
   This is followed, less than a second later, by a maximum expansion of the envelope, $\Delta z_{max} \sim$ 73.9 m,
   and by a maximum luminosity, L$_{max}$ = $4.0 \times 10^{38}$ erg s$^{-1}$ ($10^5$ L$_\odot$).
   The main nuclear path reaches $^{93}$Pd (already beyond the nucleosynthesis endpoint achieved in model 1). 
    With respect to the  chemical abundances, the envelope base is still
    dominated by H (0.560) and $^4$He (0.175), with  $^{60}$Zn reaching a mass fraction of 0.111. The next
    most abundant species are $^{64}$Ge ($6.3 \times 10^{-2}$),
    and $^{68}$Se ($2 \times 10^{-2}$). 
    Regarding the activity in the A$>$65 mass region, at this stage
    is dominated by $^{66}$Se($\beta^+$)$^{66}$As(p, 
    $\gamma$)$^{67}$Se($\beta^+$)$^{67}$As(p, $\gamma$)$^{68}$Se
    (log F $\sim$ -4), 
    $^{65}$Ge(p, $\gamma$)$^{66}$As, and
    $^{68}$Se($\beta^+$)$^{68}$As(p, $\gamma$)$^{69}$Se(p, $\gamma$)$^{70}$Br(p, $\gamma$)$^{71}$Kr($\beta^+$)$^{71}$Br(p, $\gamma$)$^{72}$Kr
    (log F $\sim$ -5; see Fig. \ref{fig:3M1091998}, for additional processes
    down to log F $\sim$ -8).
    The most important contributors to the energy production are at this stage
    $^{44,45}$V(p, $\gamma$)$^{45,46}$Cr, $^{49}$Mn(p, $\gamma$)$^{50}$Fe($\beta^+$)$^{50}$Mn(p, $\gamma$)$^{51}$Fe, 
    $^{52}$Co(p, $\gamma$)$^{53}$Ni,
    $^{53}$Co(p, $\gamma$)$^{54}$Ni($\beta^+$)$^{54}$Co(p, $\gamma$)$^{55}$Ni($\beta^+$)$^{55}$Co(p, $\gamma$)$^{56}$Ni,
    and 
    $^{58}$Zn($\beta^+$)$^{58}$Cu(p, $\gamma$)$^{59}$Zn($\beta^+$)$^{59}$Cu(p, $\gamma$)$^{60}$Zn (log F $\sim$ -3).

\begin{figure}[htbp]
 \centering
   \includegraphics[width=1.00\textwidth]{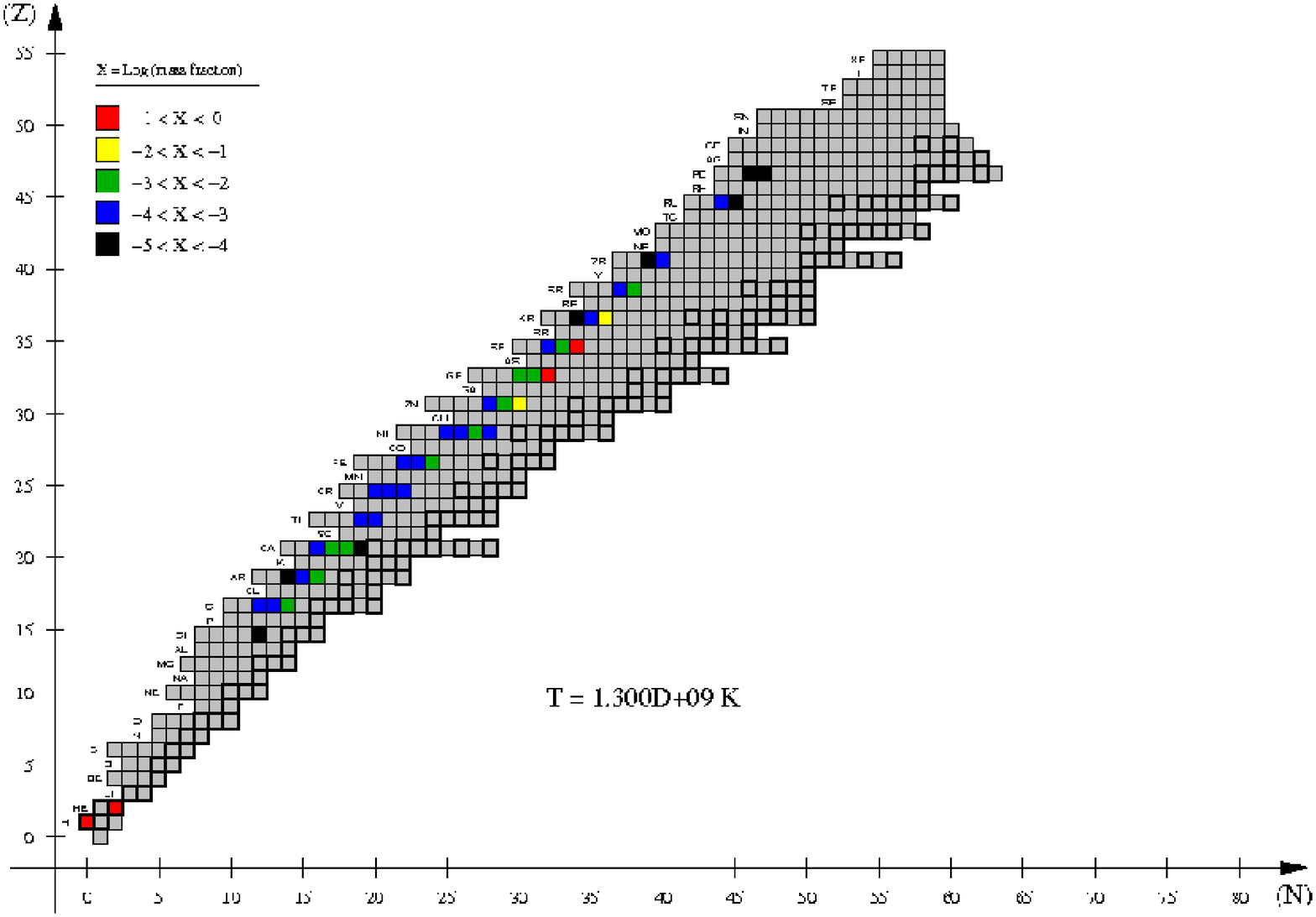}
   \includegraphics[width=1.00\textwidth]{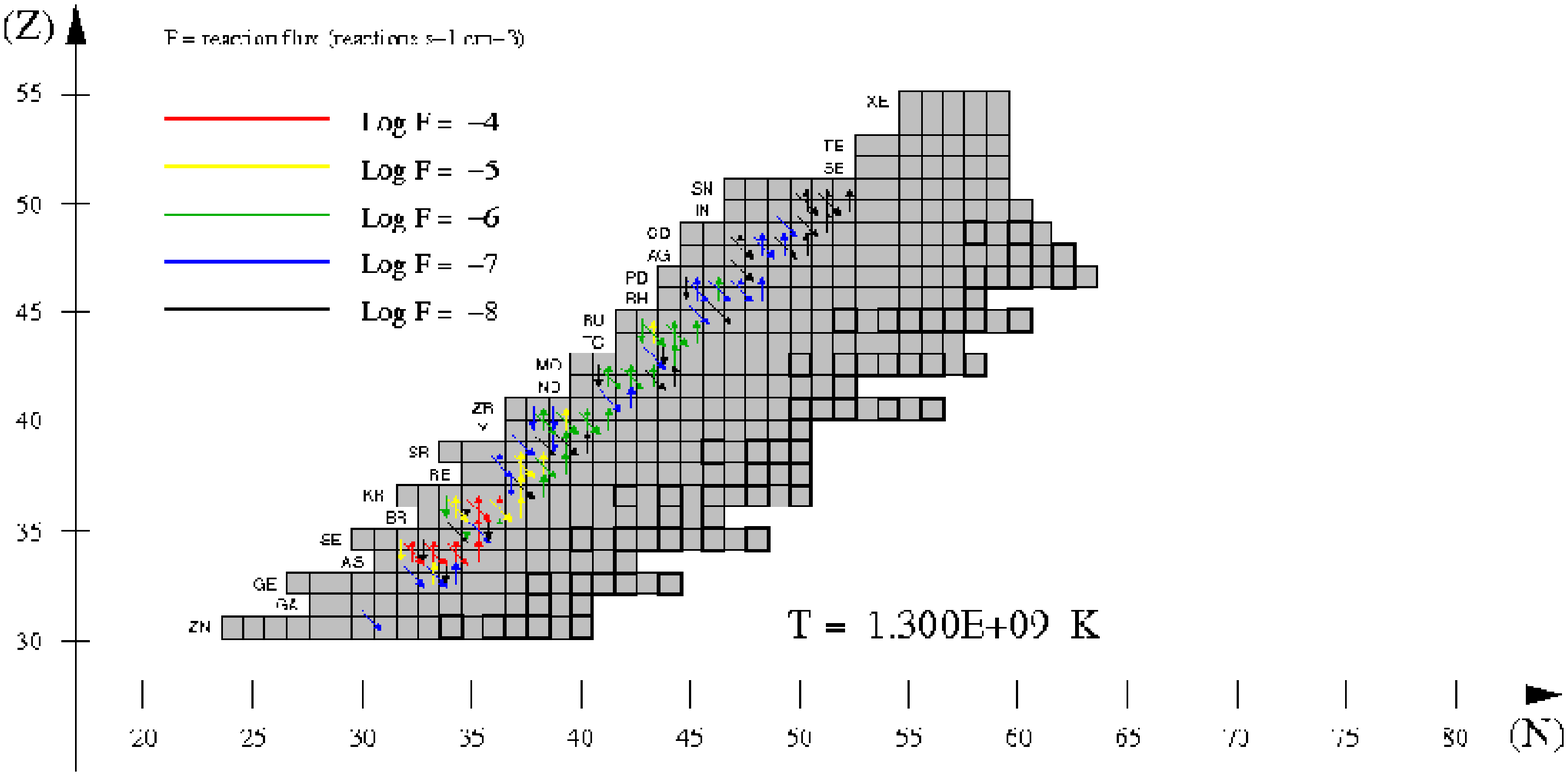}
\caption{Same as Fig. \ref{fig:3M222998}, but for $T_{base} = 1.3 \times 10^{9}$ K.}
\label{fig:3M1749998}
\end{figure}

    At t=65,110 s, following the decline from peak temperature, 
    the envelope base achieves $T_{base} = 1.3 \times 10^{9}$ K 
    (Fig. \ref{fig:3M1749998}). At this
    stage, the main nuclear activity has already reached the SnSbTe-mass region
    ($^{104}$Sn, in particular).
    The chemical abundances at the envelope base are still dominated by
    H (0.471), now followed by $^{64}$Ge (0.162), and $^{68}$Se (0.161),
    while $^4$He has dropped to 0.131. 
    The next most abundant species shift to $^{60}$Zn ($2.1 \times 10^{-2}$),
    and $^{72}$Kr ($1.9 \times 10^{-2}$), with a suite of nuclei reaching $\sim 10^{-3}$ 
    ($^{30}$S, $^{67}$Se, $^{37,38}$Ca, $^{76}$Sr, $^{62,63}$Ge, $^{59}$Zn, $^{55}$Ni, 
    $^{34}$Ar, or $^{50}$Fe).
    The nuclear activity in the A$>$65 mass region is now powered by 
    $^{65}$As(p, $\gamma$)$^{66}$Se($\beta^+$)$^{66}$As(p,
        $\gamma$)$^{67}$Se($\beta^+$)$^{67}$As(p, 
	$\gamma$)$^{68}$Se($\beta^+$)$^{68}$As(p, $\gamma$)$^{69}$Se(p, 
	$\gamma$)$^{70}$Br(p, $\gamma$)$^{71}$Kr($\beta^+$)$^{71}$Br(p, 
	$\gamma$)$^{72}$Kr (log F $\sim$ -4),
    $^{65}$Ge(p, $\gamma$)$^{66}$As, 
    $^{69}$Br(p, $\gamma$)$^{70}$Kr($\beta^+$)$^{70}$Br, $^{72}$Kr($\beta^+$)$^{72}$Br(p, $\gamma$)$^{73}$Kr(p, $\gamma$)$^{74}$Rb(p, 
    $\gamma$)$^{75}$Sr($\beta^+$)$^{75}$Rb(p, $\gamma$)$^{76}$Sr, $^{78}$Y(p, $\gamma$)$^{79}$Zr, and $^{86}$Tc(p, $\gamma$)$^{87}$Ru (log F $\sim$ -5; see Fig. \ref{fig:3M1749998}). 
      Energy production is not due to a handful of nuclear processes but
      to dozens of different reactions (from 3$\alpha \rightarrow$ $^{12}$C
      all the way to $^{71}$Br(p, $\gamma$)$^{72}$Kr).

\begin{figure}[htbp]
 \centering
   \includegraphics[width=1.00\textwidth]{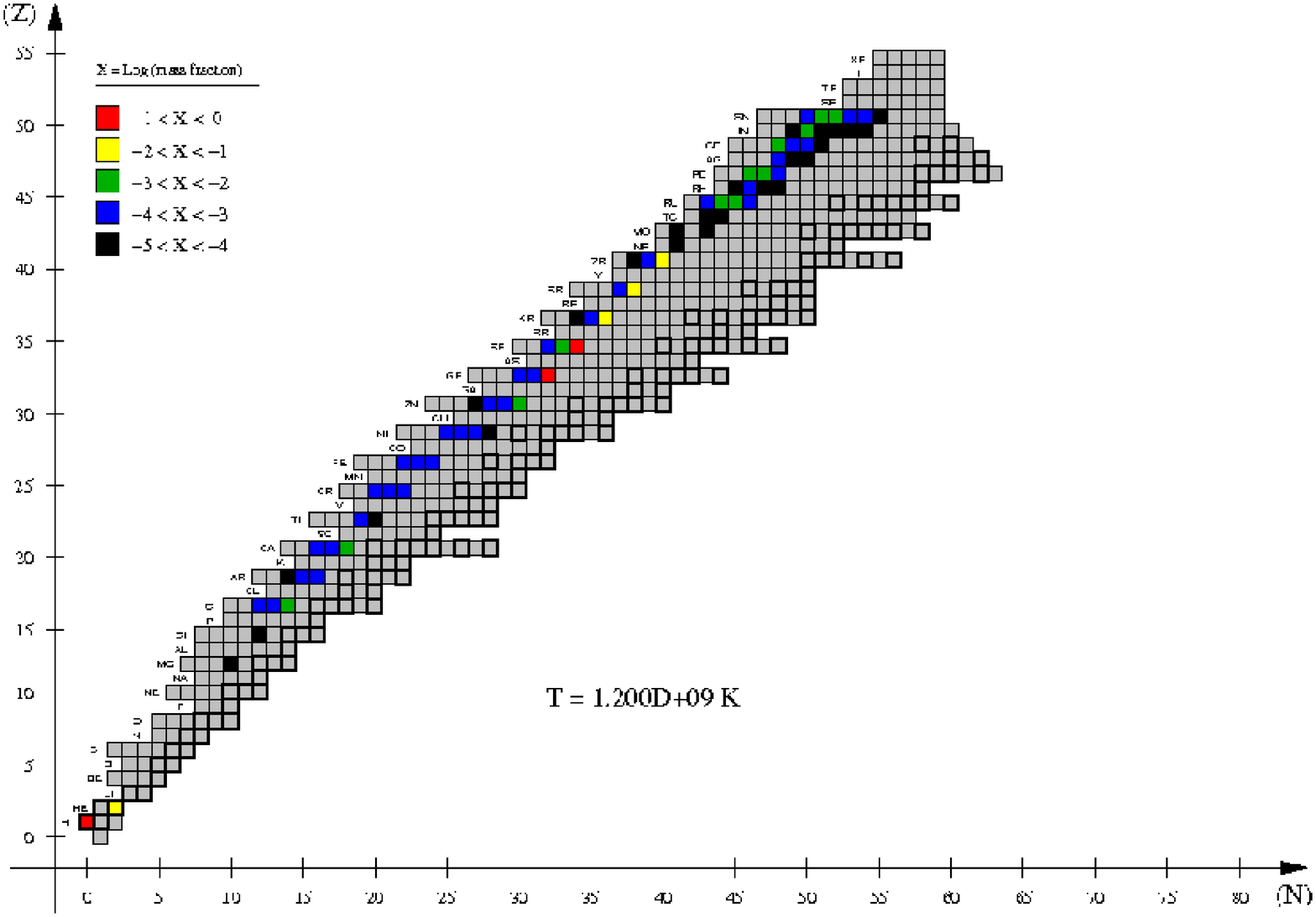}
    \includegraphics[width=1.00\textwidth]{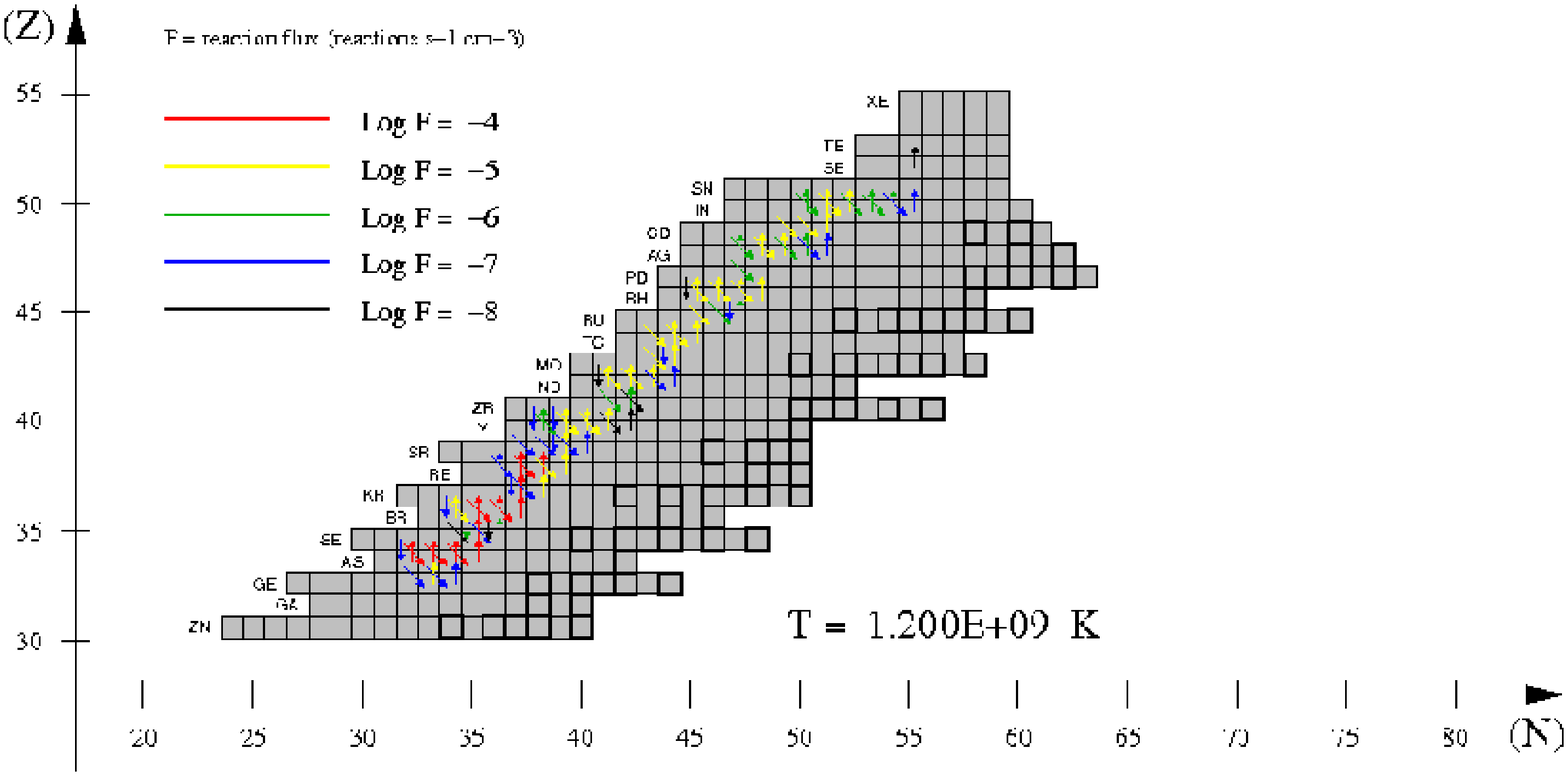}
\caption{Same as Fig. \ref{fig:3M222998}, but for $T_{base} = 1.2 \times 10^{9}$ K. }
\label{fig:3M2958998}
\end{figure}

      Twenty-two seconds later (t=65,132 s), the temperature at the envelope base 
      has decreased to $T_{base} = 1.2 \times 10^{9}$ K (Fig. \ref{fig:3M2958998}).
      The H content has been slightly reduced to 0.370, whereas $^4$He reaches
      $8.74 \times 10^{-2}$. Indeed, after H, the most abundant species at the envelope
      base are now $^{68}$Se (0.270), and $^{64}$Ge (0.104), followed by
      $^{72}$Kr ($8.7 \times 10^{-2}$), $^{76}$Sr ($2.8 \times 10^{-2}$), and
      $^{80}$Zr ($1.2 \times 10^{-2}$). At this stage, some of the heaviest species
      of the network have already achieved an abundance of $\sim 10^{-3}$
      (such as $^{88,89}$Ru, $^{92,93}$Pd, $^{96}$Cd, $^{99}$In, or $^{101,102}$Sn, 
       together with the lighter nuclei $^{30}$S, $^{38}$Ca, $^{59,60}$Zn, and
        $^{67}$Se), with the nuclear activity extending all the way up to $^{107}$Te.
 The nuclear activity in the A=65-100 mass region is similar to that
described above for  $T_{base} = 1.3 \times 10^{9}$ K, and is depicted
in Fig. \ref{fig:3M2958998}, lower panel.

\begin{figure}[htbp]
 \centering
   \includegraphics[width=1.00\textwidth]{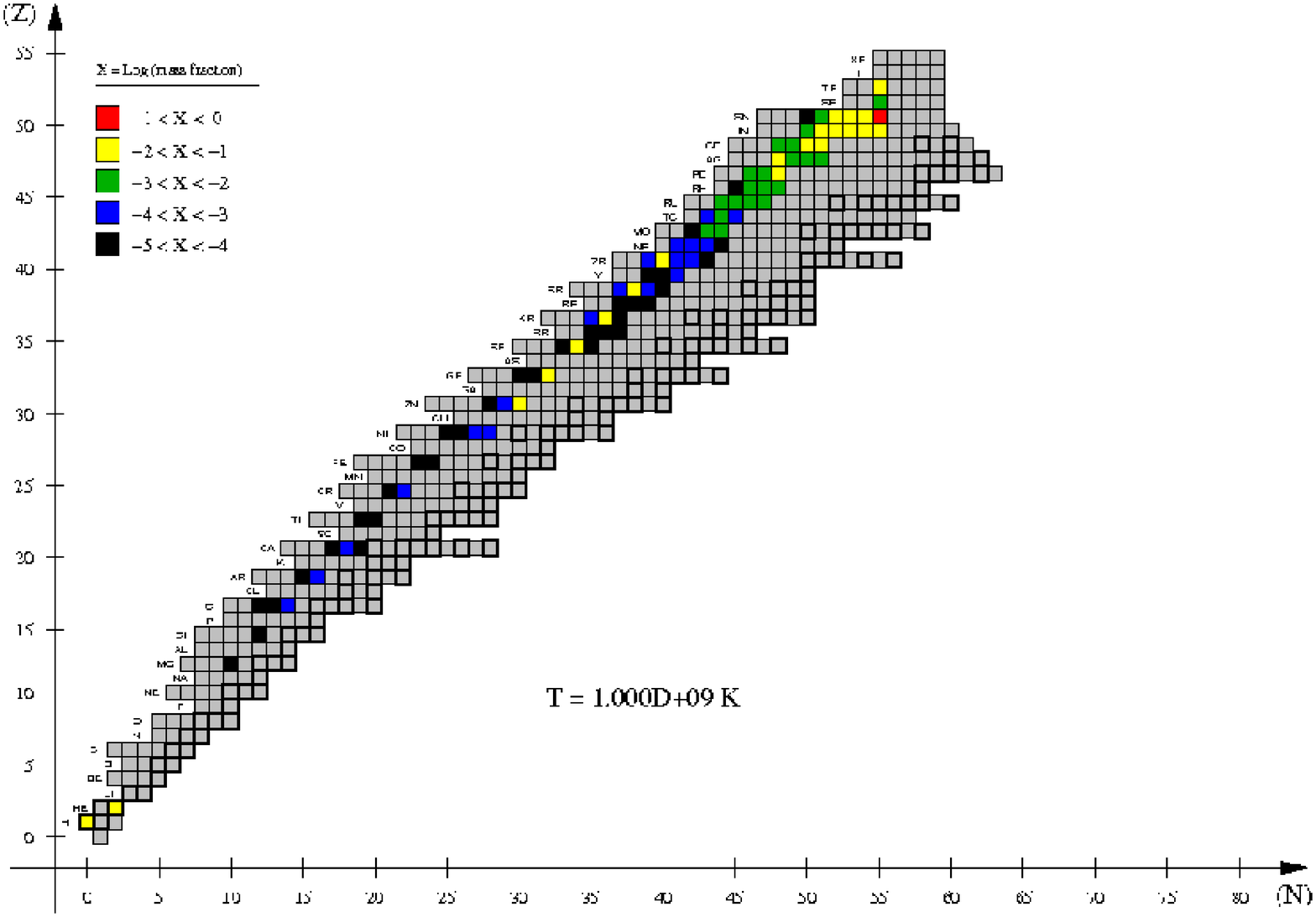}
   \includegraphics[width=1.00\textwidth]{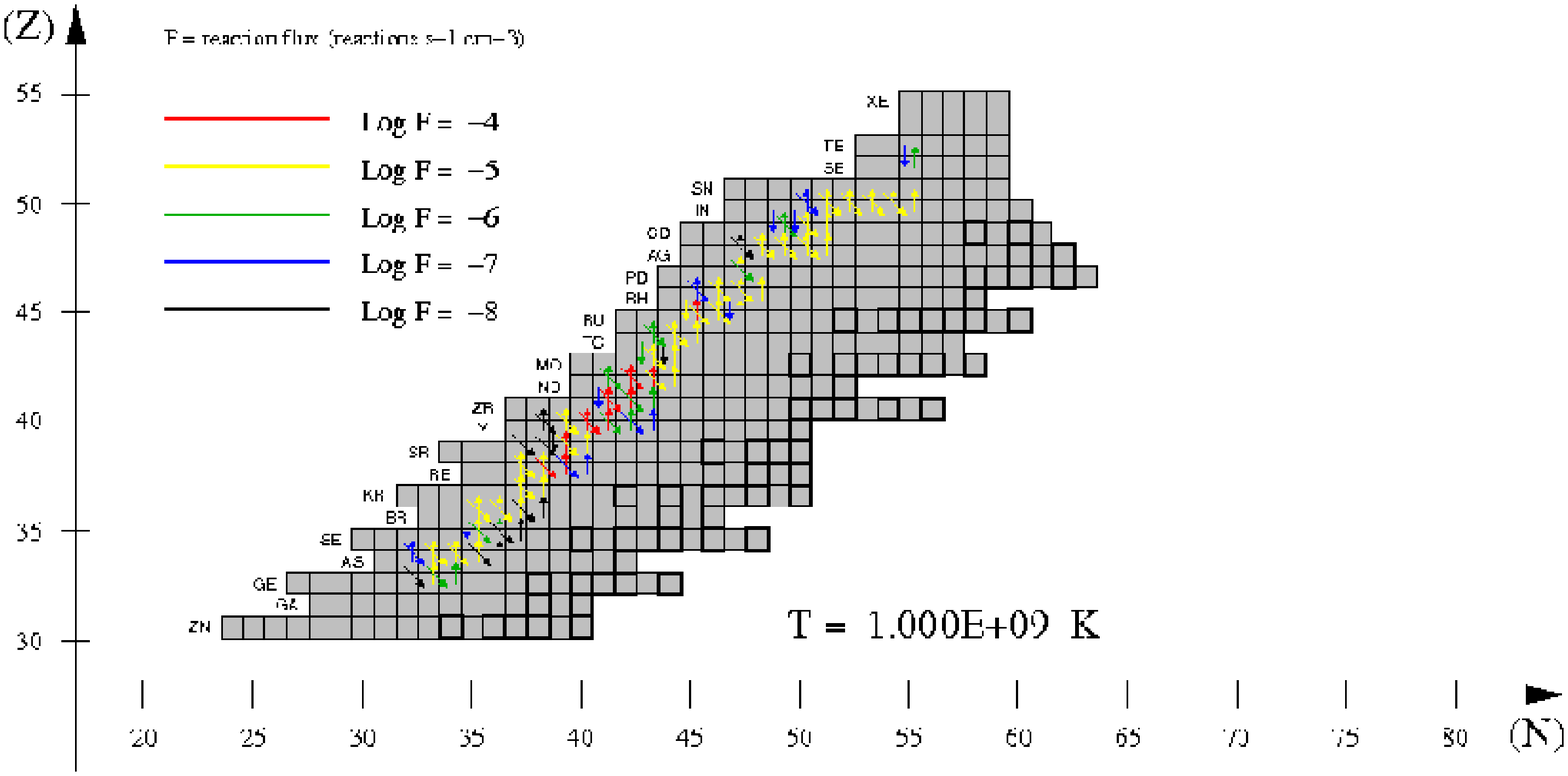}
\caption{Same as Fig. \ref{fig:3M222998}, but for $T_{base} = 10^{9}$ K.}
\label{fig:3M8950998}
\end{figure}

	At t=65,264 s, the envelope base 
	achieves $T_{base} =  10^{9}$ K (Fig. \ref{fig:3M8950998}).
	Now, the most abundant element at the envelope base is $^{105}$Sn (0.228).
This is
followed by a large number of species with abundances $\sim 10^{-2}$, such
as $^{104}$Sn, $^{68}$Se, $^{72}$Kr, $^{104}$In, $^{94}$Pd, $^{64}$Ge, 
	$^{103}$In, $^{76}$Sr --all more
	abundant than H ($1.8 \times 10^{-2}$) and $^4$He ($2.6 \times 10^{-2}$), at this
	stage--, together with $^{102,103}$Sn, $^{95}$Ag, 
	$^{107}$Te, $^{100,101,102}$In, 
	$^{80}$Zr, $^{60}$Zn, and $^{98,99}$Cd.
	Notice that, since the heaviest element included in our network, $^{107}$Te, achieved
	already an abundance of $2.2 \times 10^{-2}$, leakage from the SnSbTe-mass region cannot
	be discarded. Notice, however, that $\alpha$-emission for $^{107}$Te was not included, which may account for the high abundances reported here. 
Further studies to explore possible nucleosynthesis
beyond the SnSbTe-mass region are underway with a larger network, so detailed
abundances in this region should be taken with caution.
	The set of equilibrium (p, $\gamma$)-($\gamma$, p) pairs is now
	accompanied by a handful of $\beta^+$-decays
	(such as $^{80}$Zr($\beta^+$)$^{80}$Y,  $^{76}$Sr($\beta^+$)$^{76}$Rb,  
	$^{84}$Mo($\beta^+$)$^{84}$Nb, and   $^{82}$Nb($\beta^+$)$^{82}$Zr) as
	the nuclear processes with largest absolute fluxes, 
	since H depletion and the low temperature limit the extent of charged-particle reactions. 
	Indeed, these weak interactions will become 
	progressively more important during the last 
	stages of the burst.  
	At this stage, the activity in the A=65-100 mass region is 
	dominated by $^{76}$Sr($\beta^+$)$^{76}$Rb(p, 
	$\gamma$)$^{77}$Sr(p, $\gamma$)$^{78}$Y, 
        $^{79}$Y(p, $\gamma$)$^{80}$Zr($\beta^+$)$^{80}$Y(p, 
	$\gamma$)$^{81}$Zr(p, $\gamma$)$^{82}$Nb($\beta^+$)$^{82}$Zr(p,
	$\gamma$)$^{83}$Nb(p, $\gamma$)$^{84}$Mo($\beta^+$)$^{84}$Nb(p,
	$\gamma$)$^{85}$Mo, $^{89}$Ru(p, $\gamma$)$^{90}$Rh
	(log F $\sim$ -4), and more than 60 different nuclear processes
	with log F $\sim$ -5 (see Fig. \ref{fig:3M8950998}, lower panel),
	involving nuclei in the mass range A= 65-104. Indeed, 
	energy production is driven by (p, $\gamma$) and $\beta^+$ processes
	involving species in this mass range.

\begin{figure}[htbp]
 \centering
   \includegraphics[width=1.00\textwidth]{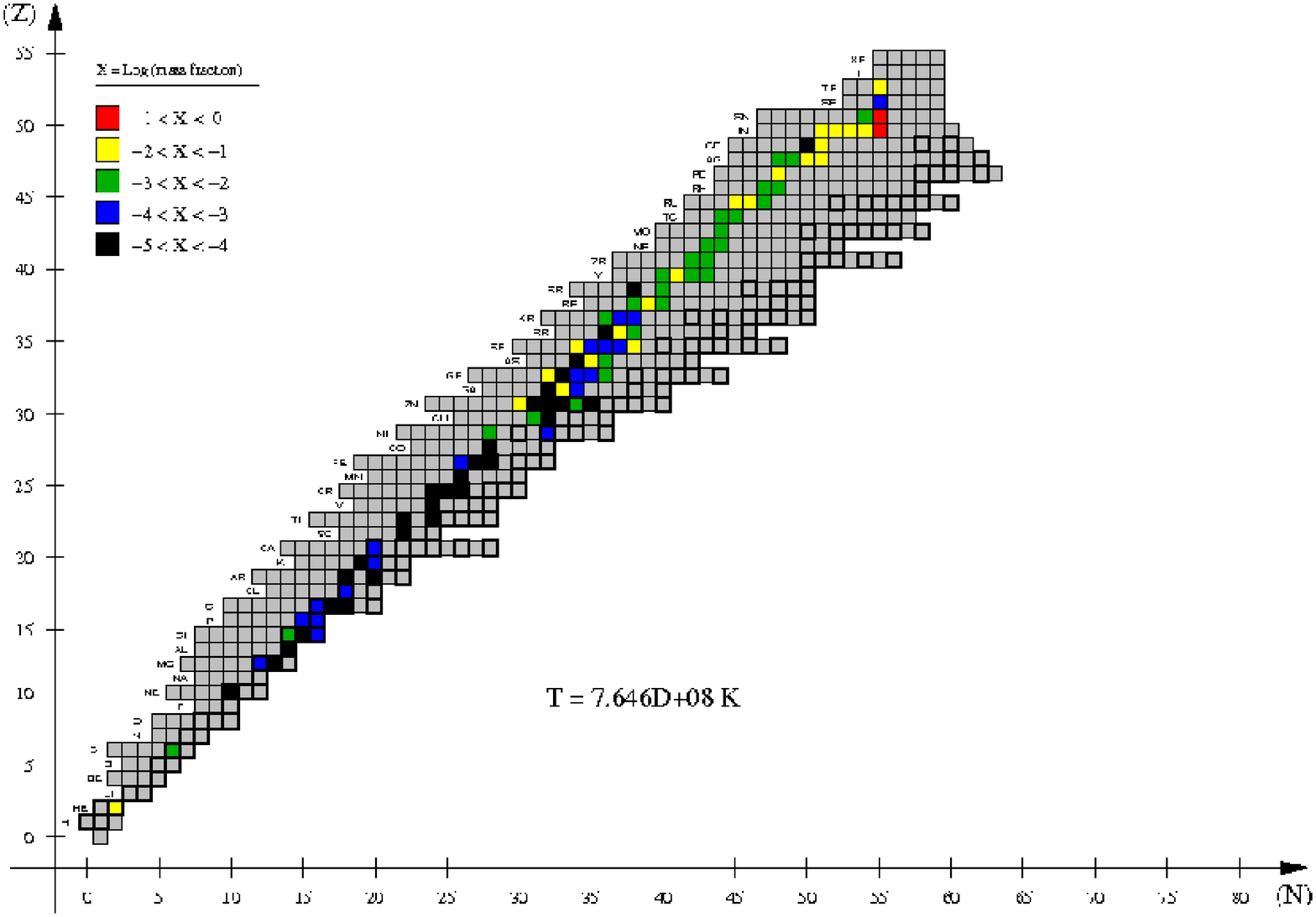}
   \includegraphics[width=1.00\textwidth]{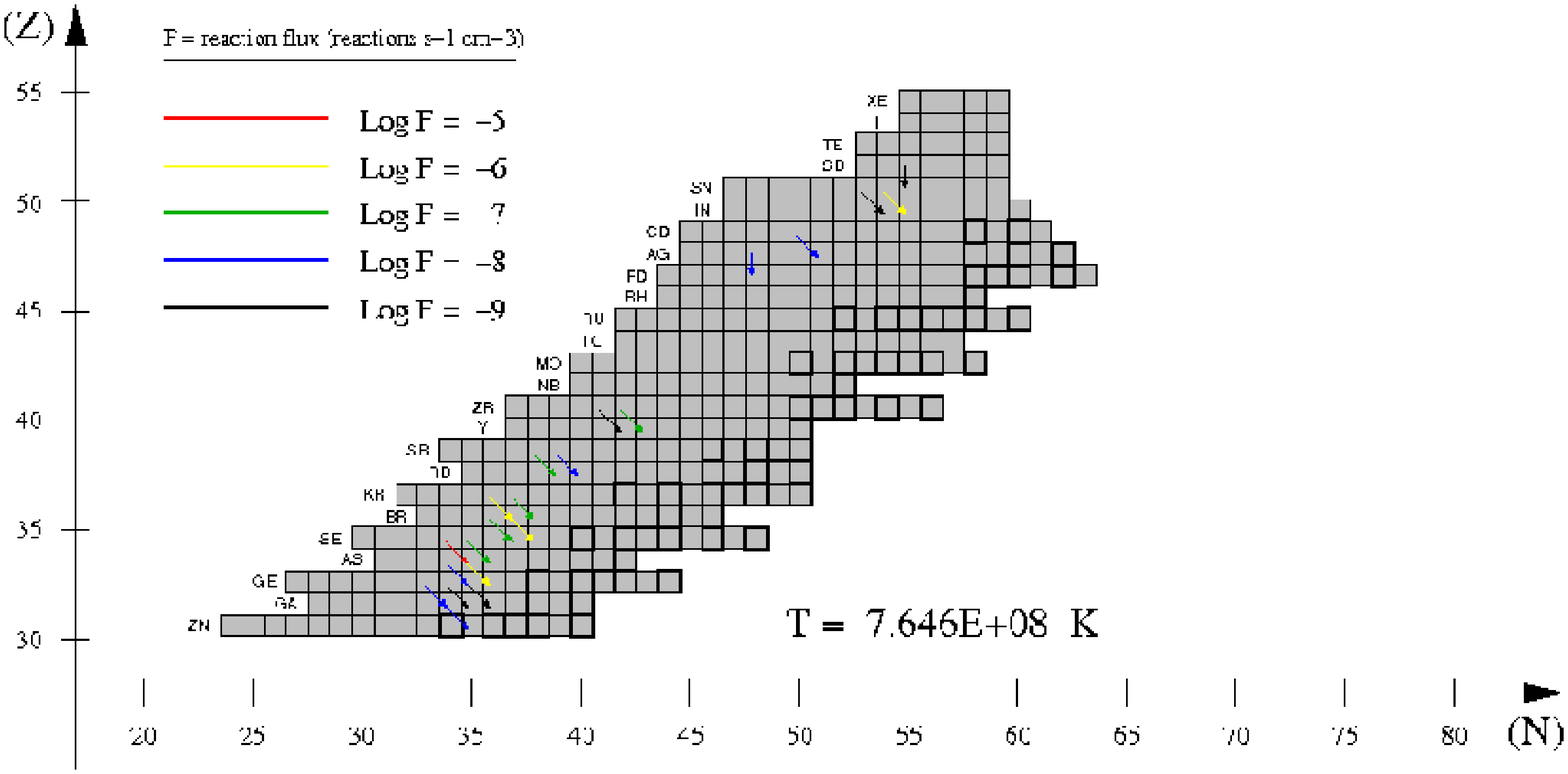}
\caption{Same as Fig. \ref{fig:3M222998}, but for 
         $T_{base} = 7.6 \times 10^{8}$ K.}
\label{fig:3M9755998}
\end{figure}

At t=65,362 s,  the temperature at the envelope base has already declined to 
$T_{base} = 7.6 \times 10^{8}$ K (Fig. \ref{fig:3M9755998}).
H is now fully depleted ($7.6 \times 10^{-12}$), while $^4$He barely reaches $1.8 \times 10^{-2}$.
As before, the most abundant element at the envelope base is $^{105}$Sn (0.251),
followed by $^{104}$In (0.142), 
and by a large number of species with abundances $\sim 10^{-2}$.
The depletion of H dramatically reduces the fluxes of most of the (p, $\gamma$) reactions, which are now overcome
by many $\beta^+$-decays (such as those affecting
$^{68,69}$Se, $^{68}$As, $^{64}$Ge,  $^{71,72}$Br, $^{60}$Zn, $^{60}$Cu,
$^{72,73}$Kr,  $^{104}$Sn,  $^{64}$Ga, $^{82}$Zr, or $^{76}$Sr), 
and by a suite  of $\alpha$-capture reactions, such as 3$\alpha \rightarrow 
^{12}$C($\alpha$, $\gamma$)$^{16}$O($\alpha$, $\gamma$)$^{20}$Ne($\alpha$, $\gamma$)$^{24}$Mg($\alpha$, 
$\gamma$)$^{28}$Si($\alpha$, $\gamma$)$^{32}$S, or $^{13}$N($\alpha$, p)$^{16}$O.

\begin{figure}[htbp]
 \centering
   \includegraphics[width=1.00\textwidth]{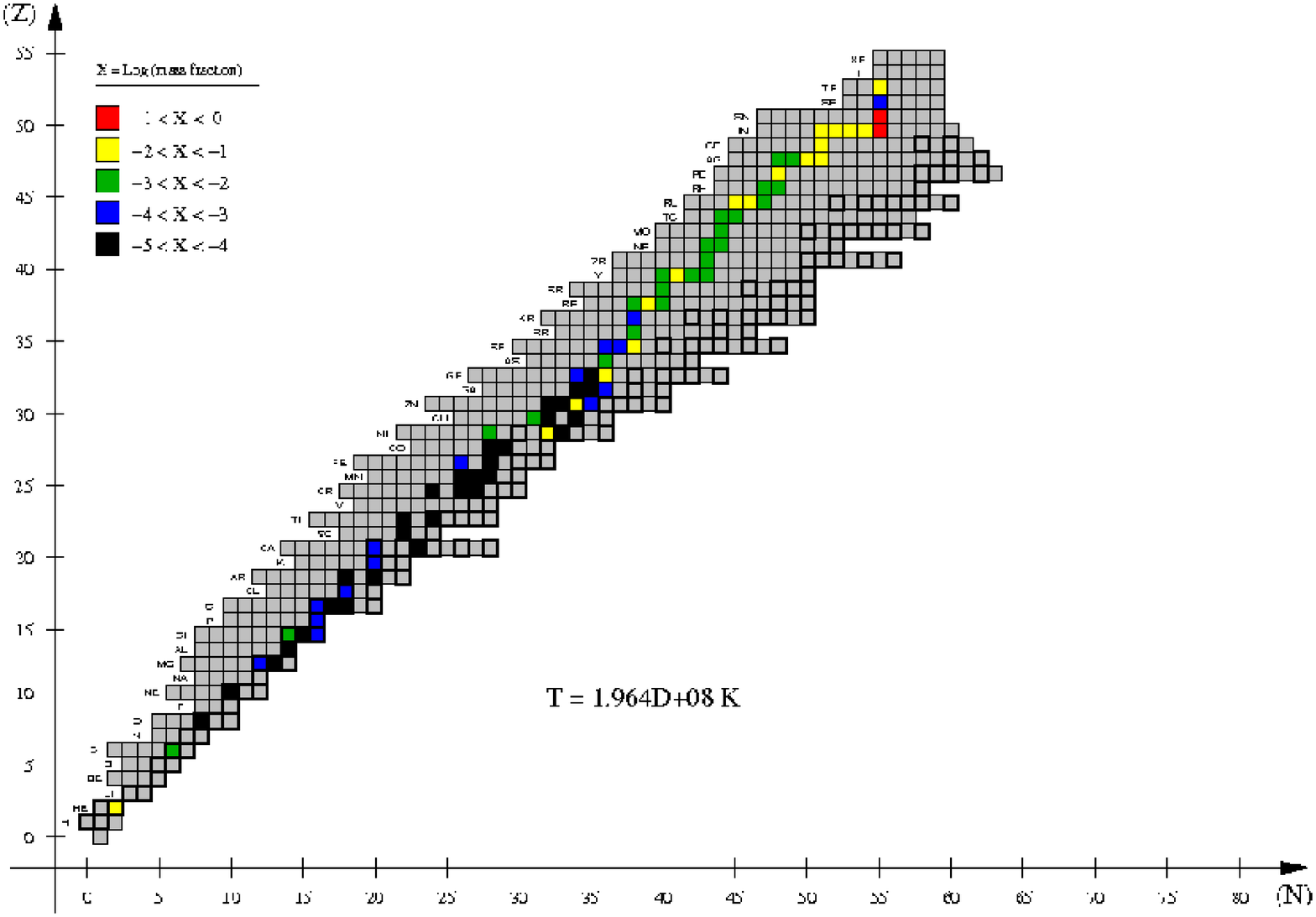}
   \includegraphics[width=1.00\textwidth]{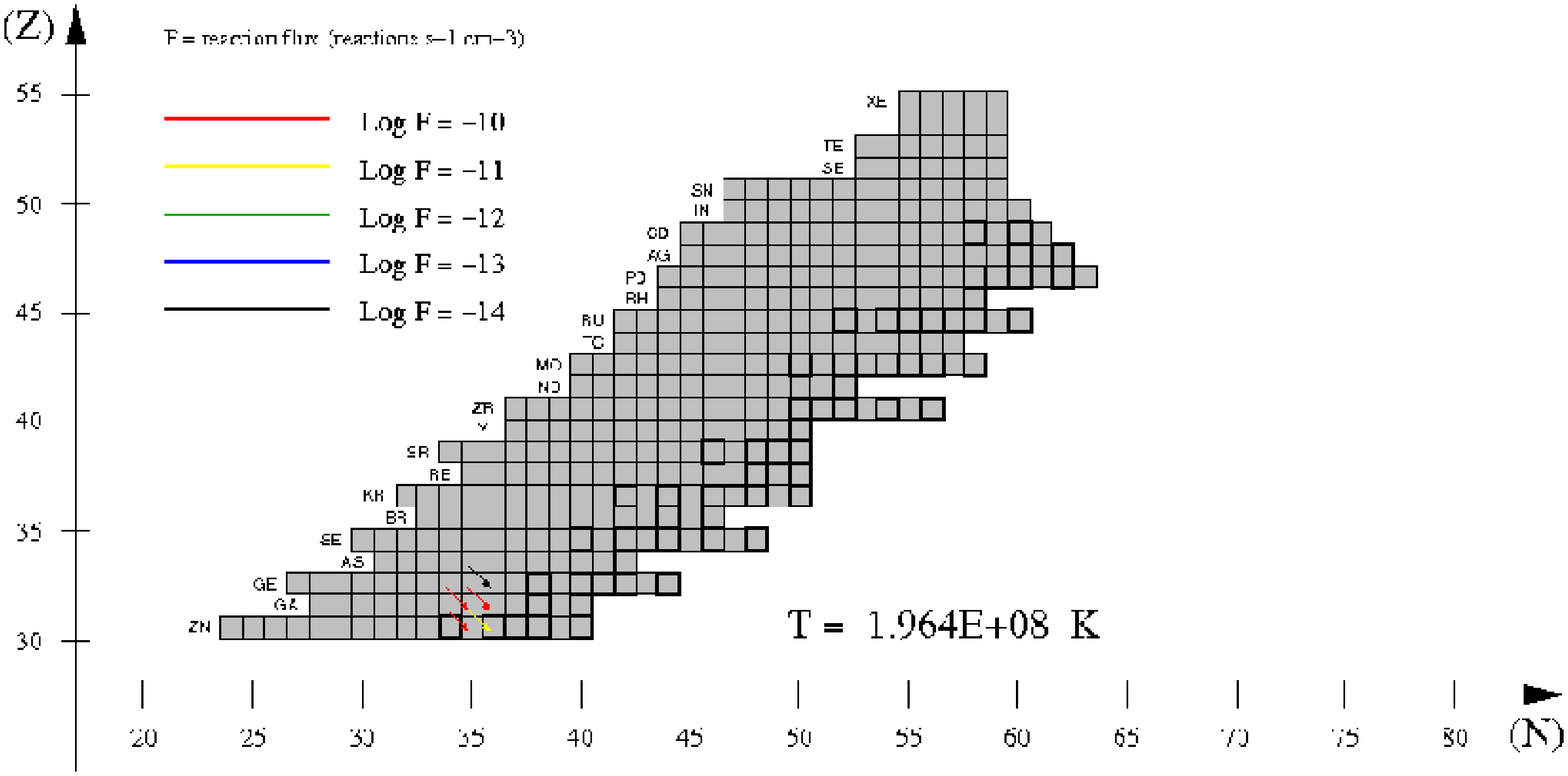}
\caption{Same as Fig. \ref{fig:3M222998}, but for the time when temperature 
at the envelope base achieves a minimum value of $T_{min} = 2 \times 10^{8}$ K.}
\label{fig:3M9774998}
\end{figure}

At t=69,715 s, after a long decline, a minimum temperature is achieved at the envelope base, 
$T_{base} = 2 \times 10^{8}$ K (Fig. \ref{fig:3M9774998}), which we consider
 to mark the end of the first
bursting episode for this model. At this stage, H is fully depleted ($3.2 \times 10^{-23}$) at the envelope base, while $^4$He has a mass fraction of
$1.5 \times 10^{-2}$ only.
The distribution of the most abundant elements almost follows the one described for
$T_{base} = 7.6 \times 10^{8}$ K (Fig. \ref{fig:3M9755998}), 
and is dominated by $^{105}$Sn (0.251),
followed by $^{104}$In (0.147), 
and by a large number of species with abundances $\sim 10^{-2}$, 
such as $^{94}$Pd,  $^{100,101,102,103}$In, 
$^{68}$Ge, $^{64}$Zn, $^{72}$Se, 
$^{76}$Rb, $^{107}$Te, $^{60}$Ni, $^{99}$Cd, $^{97,98}$Ag,
$^{89,90}$Ru, and $^{80}$Y. 
At this stage, the dominant interactions are all $\beta^+$-decays  
($^{60,61}$Cu($\beta^+$)$^{60,61}$Ni, $^{66,67}$Ge($\beta^+$)$^{66,67}$Ga, $^{65}$Ga($\beta^+$)$^{65}$Zn,  
$^{51}$Mn($\beta^+$)$^{51}$Cr, $^{52}$Fe($\beta^+$)$^{52}$Mn,
$^{63}$Zn($\beta^+$)$^{63}$Cu, $^{56}$Ni($\beta^+$)$^{56}$Co,
or $^{43}$Sc($\beta^+$)$^{43}$Ca),  
except for the triple-$\alpha$ reaction.  

 Depth also influences the
extent of the nuclear activity throughout the envelope, but in contrast to
Model 1, the nuclear activity in all shells of our computational domain
essentially reach the SnSbTe-mass region. Indeed, the inner part
of the envelope (encompassing
$3.4 \times 10^{21}$ g) is, at the end of the burst, 
dominated by large amounts of $^{105}$Sn and $^{104}$In, the most abundant
nuclei with mass fractions $\sim$ 0.1 - 0.2;   
at $5.6 \times 10^{21}$ g above the core-envelope interface, the 
most abundant isotopes are H (0.26) and $^4$He (0.12), while the
most abundant species in the SnSbTe-mass region achieve a mass fraction
$\sim 10^{-3}$; and  
close to the surface ($7.7 \times 10^{21}$ g), shells are largely dominated
by the presence of unburned H (0.75) and $^4$He (0.24), 
with X(SnSbTe) $\sim 10^{-7}$.

 The mean, mass-averaged chemical composition of the whole envelope, at the end of the first bursting episode, is dominated
 by the presence of unburned H (0.18) and $^4$He (0.084), followed by $^{105}$Ag (0.075), $^{104}$Pd (0.053), 
 $^{64}$Zn (0.042), $^{95}$Ru (0.031), $^{68}$Ge (0.028),  $^{94}$Tc (0.026), and  $^{103}$Ag (0.026),
 (see Table 8, for the mean composition of species -stable or with a half-life $> 1$ hr- which achieve X$_i > 10^{-9}$),
 with a nucleosynthesis endpoint around $^{107}$Cd.
 In contrast, the first burst computed in model 1 yielded, in general, lighter nuclei,
 $^{60}$Ni, $^4$He, $^1$H, $^{64}$Zn, $^{12}$C, and $^{52,56}$Fe, with a more modest nucleosynthesis endpoint around 
 $^{89}$Nb (Table 2). 
 
 In terms of overproduction factors, $f$ (Fig. \ref{fig:abun_b3}), while model 1 showed moderate
 values ($f \sim 10^4$) for a handful of intermediate-mass elements, such as  
 $^{43}$Ca, $^{45}$Sc, $^{49}$Ti, $^{51}$V, $^{60,61}$Ni,
 $^{63,65}$Cu, $^{64,67,68}$Zn, $^{69}$Ga, $^{74}$Se, or $^{78}$Kr, model 3 achieves moderate overproduction
 factors ($\geq 10^4$), for all stable species heavier than $^{64}$Zn, and as high as $\sim 10^8$ for 
 $^{98}$Ru, $^{102,104}$Pd, and $^{106}$Cd.

\subsection{Second, third, fourth, and fifth bursts}

Table 7 summarizes the most relevant properties that characterize the five bursting episodes computed for model 3.
Recurrence times between bursts of $\tau_{rec} \sim 9$ hr (except for the first one, for which 
$\tau_{rec} \sim 18$ hr), ratios between persistent 
and burst luminosities of $\alpha \sim 20 - 30$, and peak luminosities around $L_{peak} \sim 10^5 L_\odot$ 
represent the basic observables associated with this model. Indeed, the recurrence times obtained are
in agreement with the values reported for the XRB sources (see Galloway et al. 2008) 1A 1905+00 [$\tau_{rec} = 8.9$ hr],
4U 1254-69 [$\tau_{rec} = 9.2$ hr], or XTE J1710-281 [$\tau_{rec} = 8.9$ hr, $\alpha = 22-100$],
A striking result is the quick stabilization of the recurrence times, that show a regular periodicity after the
second burst. 

It is worth noting that both the recurrence periods and the ratios between 
persistent and burst luminosities
are larger than those reported for model 1 (see Table 6, for comparison), showing a clear dependence on the metallicity
of the accreted material: the smaller the metal content, the larger the recurrence time (and the smaller the value of $\alpha$). 

The corresponding light curves (see Fig. \ref{fig:lum_m3_5b}) exhibit, in turn, a clear pattern: 
as shown in Fig. \ref{fig:LUMI_OVER} (left panel), where light curves of the 
third bursting episode computed in models 1 \& 3 are compared, explosions in metal-deficient envelopes (such as model 3) 
are characterized by lower peak luminosities and longer decline times.  
A similar pattern has been reported by Heger et al. (2007), in the framework of 1-D, hydrodynamic models 
of XRBs performed with the KEPLER code.  
It is worth noting that no double-peaked bursts have been obtained in model 3.

 Larger peak temperatures, around $T_{peak} \sim (1.3 - 1.4) \times 10^9$ K, have also been obtained in model 3. 
 This, together with the longer exposure times to high temperatures (driven by the slower decline phase) cause 
 a dramatic extension of the main nuclear path towards the SnSbTe-mass region or beyond.

\begin{figure}[htbp]
 \centering
   \includegraphics[width=0.90\textwidth]{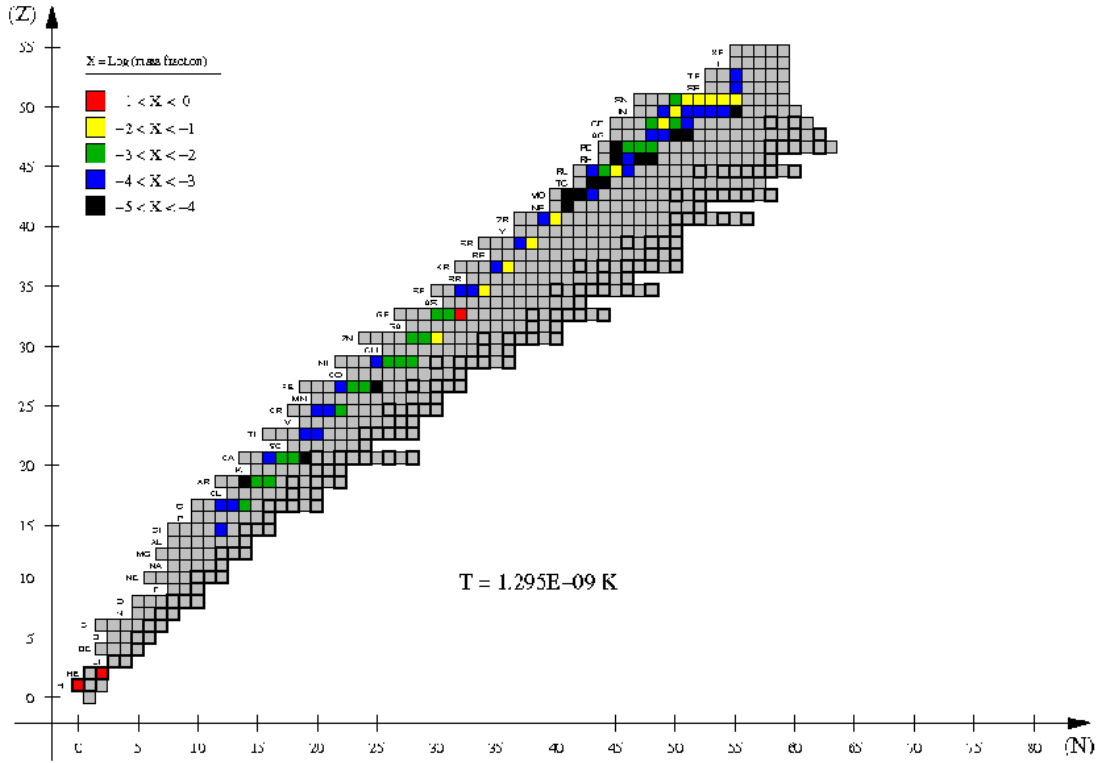}
   \includegraphics[width=0.90\textwidth]{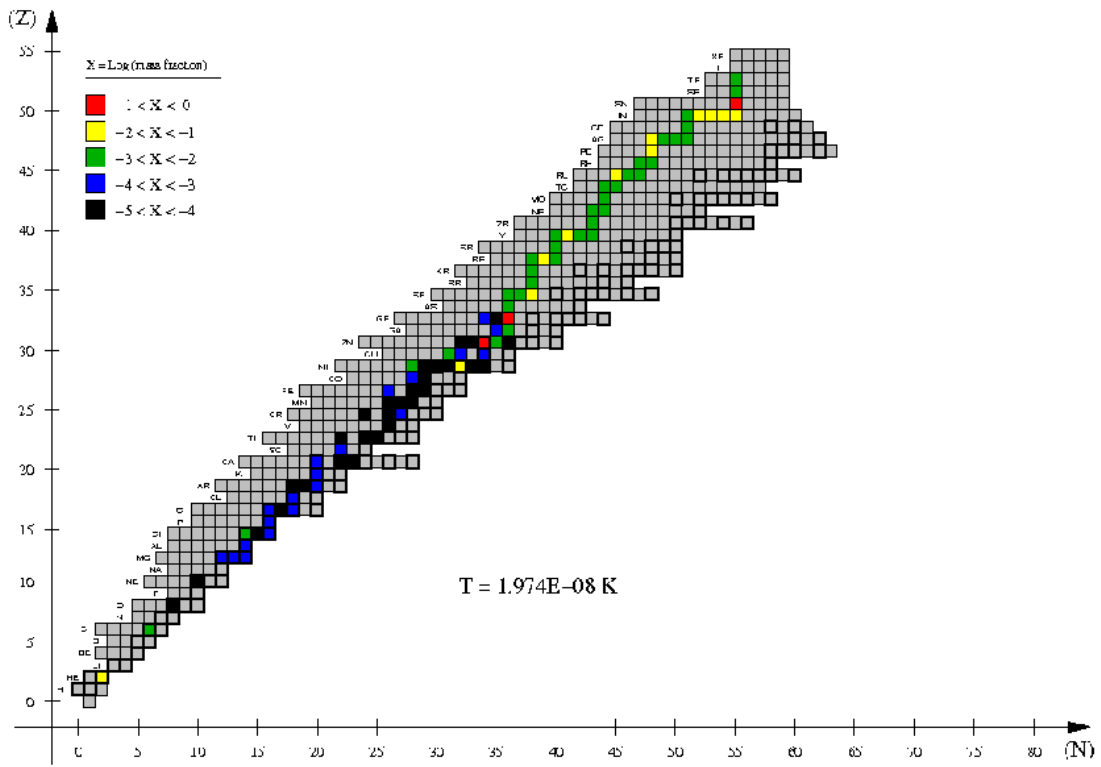}
\caption{Upper panel: main nuclear activity at the ignition shell 
 when temperature reaches a peak value of $T_{peak} = 1.23 \times 10^{9}$ K, during the 4$^{th}$ burst computed
 for model 3. 
 Lower panel: same as in the upper panel, but at the end of the burst.  }
\label{fig:M3B4TPK}
\end{figure}

 From the nucleosynthesis viewpoint, and as shown in Table 8 and 
 Figs. \ref{fig:lum_m3_5b} \& \ref{fig:LUMI_OVER},
 the nuclear activity already reaches the end of the network ($^{107}$Te) at the late stages of the first bursting episode.
 The overall mean metallicity of the envelope at the end of each burst is now 0.74 (1$^{st}$ burst), 0.85 (2$^{nd}$ burst), 0.90 (3$^{rd}$ burst), 
 0.92 (4$^{th}$ burst), and 0.93 (5$^{th}$ burst). 
   Notice that, although the accreted material is more metal-deficient in model 3 than in model 1, the post-burst 
 mean metallicity of the envelope is larger in model 3. This results from the combination of higher temperatures and longer 
 burst durations, which favors the extension of the nuclear activity: for short bursts (like those obtained in metal-rich envelopes), only the
fastest p- and/or $\alpha$-capture reactions can naturally occur (those that proceed with a characteristic time shorter than the overall exposure time 
to high temperatures); in contrast, for long-duration bursts, the overall  number of p- and/or $\alpha$-capture reactions increases
dramatically. This, in particular, affects CNO-breakout through $^{15}$O($\alpha$, $\gamma$)$^{19}$Ne and $^{14}$O($\alpha$, p)$^{17}$F, 
which are favored in the longer bursts obtained for model 3. It is
also worth noting that the abundance pattern obtained after the different
bursts is very similar. This is clearly shown in Fig. \ref{fig:M3B4TPK},
that depicts the main nuclear activity at peak and at the end of the fourth 
burst computed for Model 3, when compared with that
 corresponding to the first burst -- Figs. \ref{fig:3M1091998} \&
\ref{fig:3M9774998}. This fact justifies our emphasis on the reaction
sequences that characterize the first burst (see discussion in Sect. 5.2).  

 A final $^{12}$C yield of $\sim 2 \times 10^{-3}$ is obtained at the end
 of each bursting episode (except in the first one, for which X($^{12}$C)=$8 \times 10^{-4}$).  As reported for model 1, the amount of unburned $^{12}$C left over turns out to be too small to power a superburst.
 Finally, huge overproduction factors (see 
 Figs. \ref{fig:lum_m3_5b} \& \ref{fig:LUMI_OVER}), 
 involving heavy species such as 
 $^{102,104,105}$Pd, $^{98}$Ru, or $^{94}$Mo (with $f \sim 10^8$) 
 have been obtained in model 3, in contrast
 with the somewhat more modest values achieved in model 1, where maximum overproduction factors are about $f \sim 10^6$, and 
 involving lighter species, such as $^{76}$Se, $^{78,80}$Kr, or $^{84}$Sr.

\section{Discussion}

\subsection{General relativity corrections}
The calculations reported here have been performed assuming
Newtonian gravity. Since the envelope layers are very thin, it is 
easy to introduce general relativity corrections to this Newtonian framework 
(see Ayasli \& Joss 1982, Lewin et al. 1993, Taam et al. 1993, Cumming et al. 2002, and Woosley et al. 2004).
To this end, the surface gravity is rewritten as $g = G M_*/R_*^2 (1+z)$, where 
M$_*$ is the mass, R$_*$ is the stellar radius (defined in such a way 
that the surface area is 4$\pi R_*^2$), and $z$ is the gravitational
redshift given by $1 + z = (1 - 2G M_*/R_*c^2)^{-1/2}$.
Our models of M$_* = 1.4$ M$_\odot$ require R$_* = 14.3$ km, and a gravitational redshift of $z = 0.19$.

Following Woosley et al. (2004), once the redshift and radius are determined, it is straightforward to derive the set
of correcting factors to the physical magnitudes described above for a 
suitable observer at infinity. Hence,
recurrence times and burst durations should be increased by a factor $1 + z$. 
The mass-accretion rate as well as the burst luminosity have to take into account both 
the difference in surface area (compared to the Newtonian
framework) and the gravitational redshift term. The energy and rest mass-accretion rate scale as R$_*^2$/$(1+z)$, while the
luminosity $\propto$ R$_*^2$/$(1+z)^2$.
However, when M$_*$ is taken exactly as $M_{NS}$ (Newtonian framework), 
the surface area and redshift corrections for energy and mass
accretion rate cancel out, since g $\propto$ $(1+z)/$R$_*^2 = const$, and hence, no correction to the observed burst energy or
mass-accretion rate is necessary, while the luminosity correction is simply given by $1/(1+z) = 0.84$.
In addition, the accretion luminosity
for an observer at infinity changes only by a factor
1.012, that is, the ratio between gravitational energy released per unit mass in
general relativity, $c^2.z/(1+z)$, and the Newtonian value, $GM_{NS}/R_{NS}$.
Finally, the luminosity measured at infinity will be smaller by a factor of $(1+z)=1.19$. 

\subsection{Comparison with previous work}

 For consistency, the results discussed in this paper have been compared 
with those reported in previous work (obtained with similar hydrodynamic 
codes or in the framework of one-zone models). 

As emphasized in Section 2, model 1 is qualitatively similar to model ZM 
of Woosley et al. (2004).  The twelve bursts computed by Woosley et al. (2004) 
in a Newtonian frame  were characterized by recurrence times of about 
$\sim 2.7$ hr, peak luminosities of L$_{peak} \sim (1.5 - 2) \times 10^{38}$ 
erg s$^{-1}$, and ratios between persistent and burst luminosities of  
$\alpha \sim 60-65$. 
Our calculations (model 1, Newtonian frame) yield $\tau_{rec} \sim 5 - 6.5$ hr, 
L$_{peak} \sim (3 - 7) \times 10^{38}$ erg s$^{-1}$, and $\alpha \sim 35-40$. 

The role played by the metallicity of the accreted material (model 3, with 
Z = Z$_\odot$/20 = 0.001) qualitatively agrees with the pattern reported 
by Woosley et al. (2004) (see also, Heger et al. 2007).
Longer recurrence times of $\sim 9$ hours, peak temperatures of about 
$(1.3-1.4) \times 10^{9}$ K, and ratios between persistent and burst 
luminosities of $\alpha \sim 20-30$ (with L$_{peak}$ $\sim 10^{38}$ 
erg s$^{-1}$) have been obtained in the 5 bursts computed in model 3. 
In turn, the fifteen bursts computed by Woosley et al. (2004) for model 
zM are characterized by recurrence times of about $3 - 3.5$ hr,
peak luminosities of L$_{peak} \sim 10^{38}$ erg s$^{-1}$, and 
ratios between persistent and burst luminosities of  $\alpha \sim 50-60$. 
Results reveal a dependence of burst properties on the metallicity of the 
accreted material: the smaller the metal
content, the larger the recurrence time (and the smaller the $\alpha$). 
In turn, explosions in metal-deficient envelopes (i.e., model 3) are
characterized by lower peak luminosities and longer decline times, in 
agreement with the pattern described in Woosley et al. (2004) 
and Heger et al. (2007).
Model 3 bears as well a clear resemblance with the model computed by 
Fisker et al. (2008). In that work, five representative bursting sequences 
were analyzed, with $\tau_{rec} \sim 3.5 - 4$ hr, L$_{peak} \sim (7 - 8) 
\times 10^{37}$ erg s$^{-1}$, and $\alpha \sim 65-70$, as measured at infinity. 

\begin{figure}[htbp]
 \centering
   \includegraphics[width=0.45\textwidth]{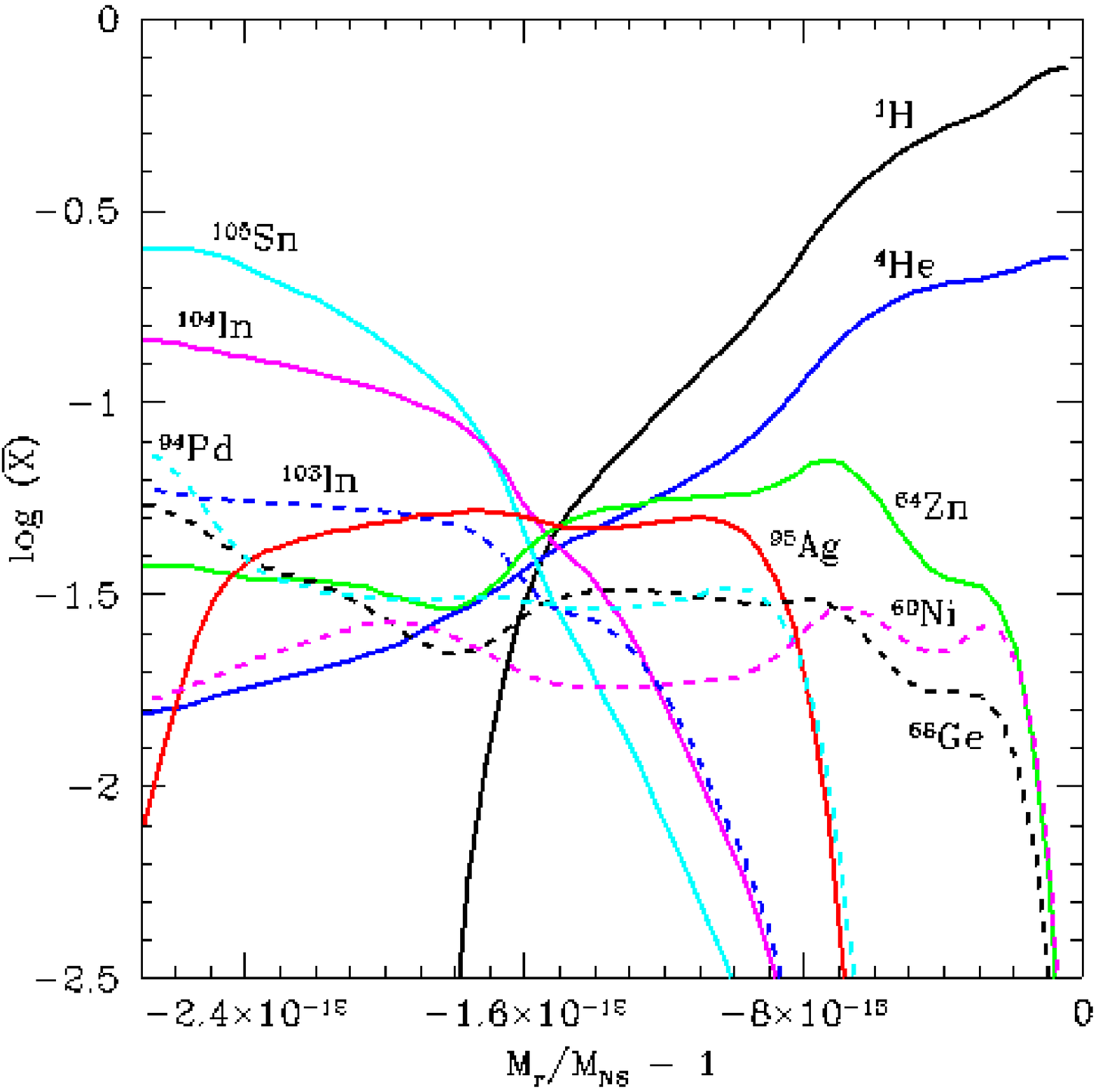}
   \includegraphics[width=0.45\textwidth]{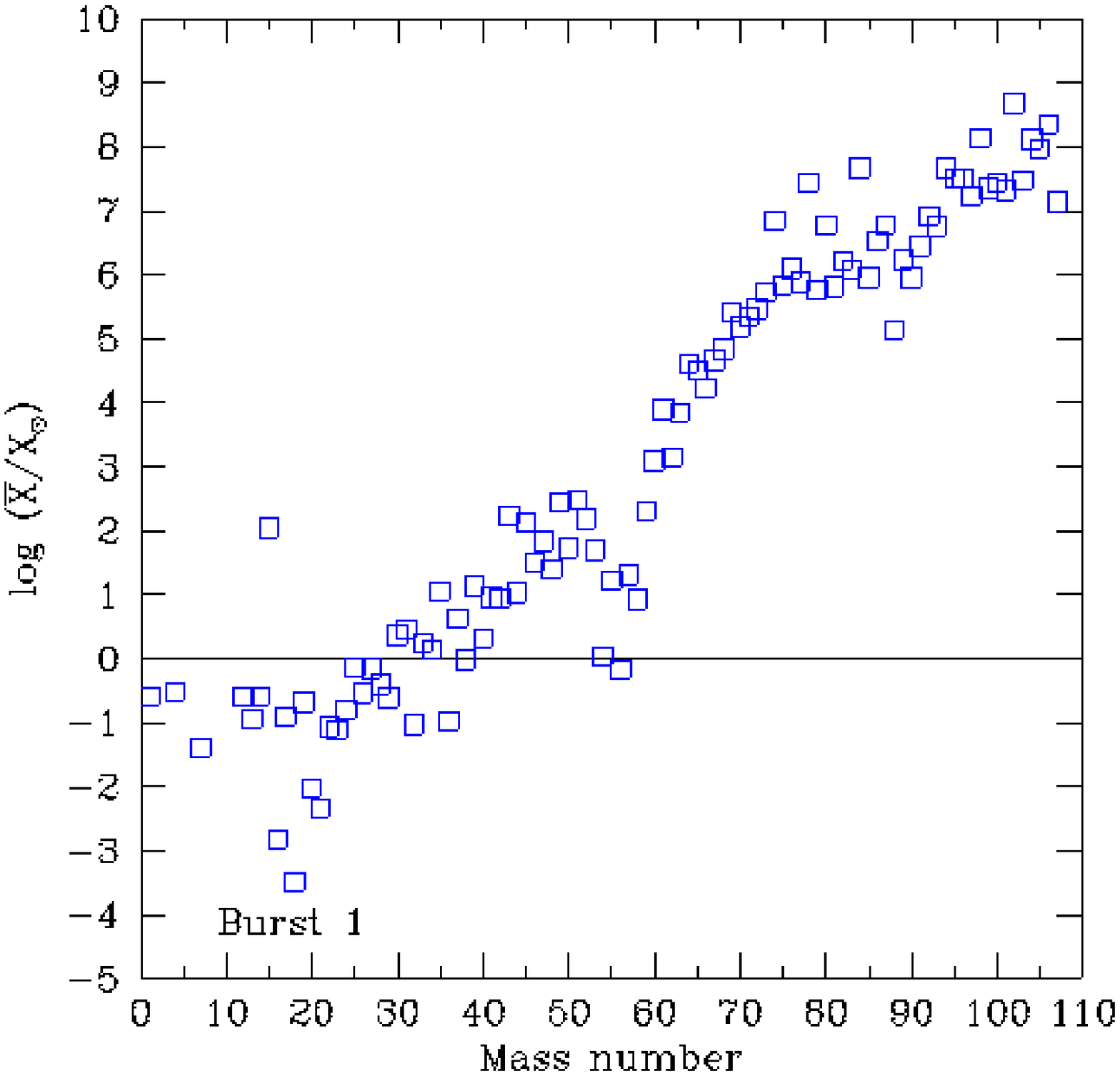}
  \includegraphics[width=0.45\textwidth]{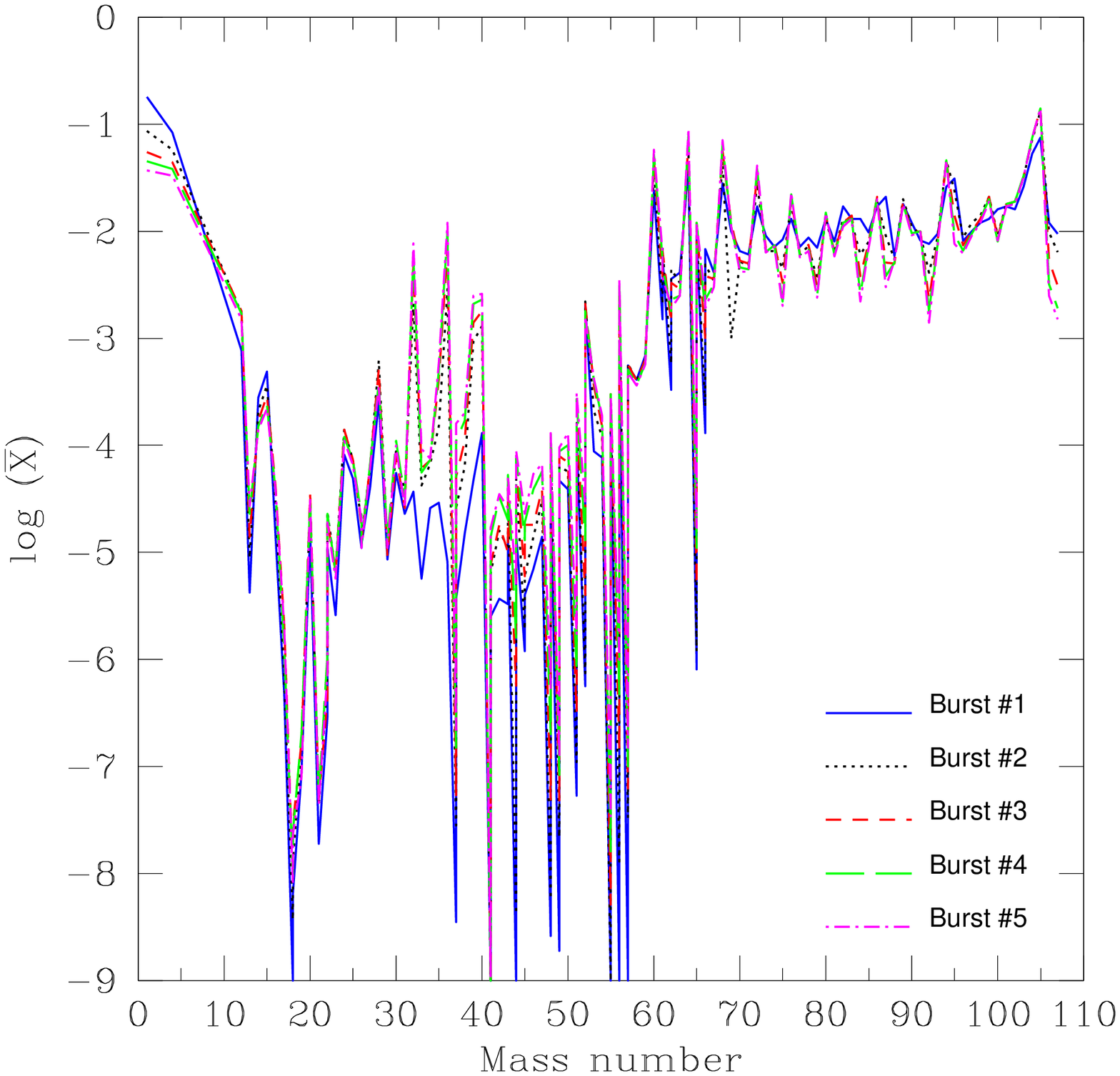}
  \includegraphics[width=0.45\textwidth]{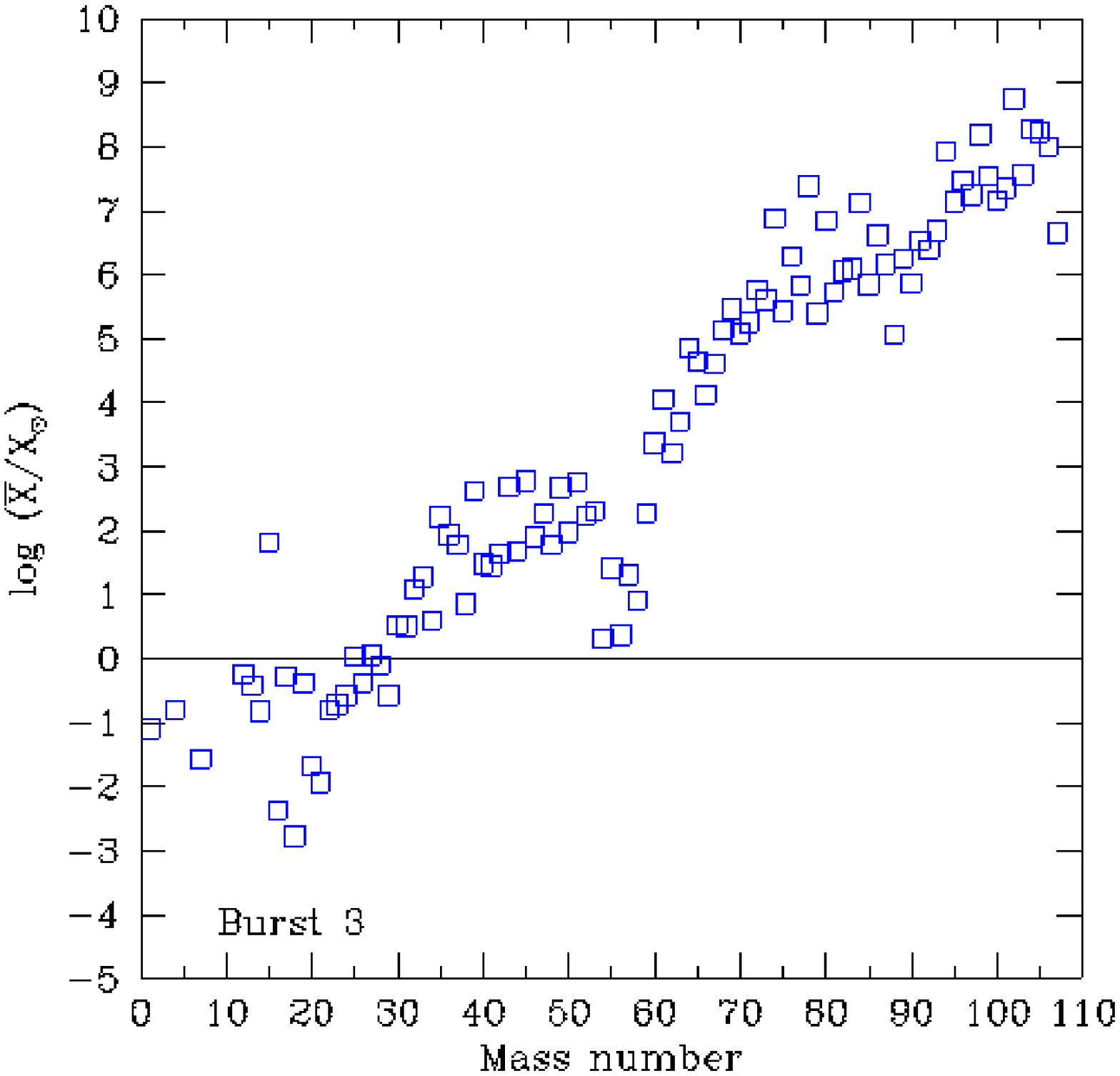}
\caption{Upper left panel: mass fractions of the ten most abundant stable (or $\tau > 1$ hr) isotopes, 
at the end of the first burst, for model 3. 
         Lower left panel: mean post-burst composition in the envelope for each of the bursting episodes computed for model 3. 
Right panels: overproduction factors relative to solar ($f > 10^{-5}$), at the end of the first (upper panel) and third bursts (lower panel),
   for model 3.}  
\label{fig:abun_b3}
\end{figure}

\begin{figure}[htbp]
 \centering
 \includegraphics[width=0.45\textwidth]{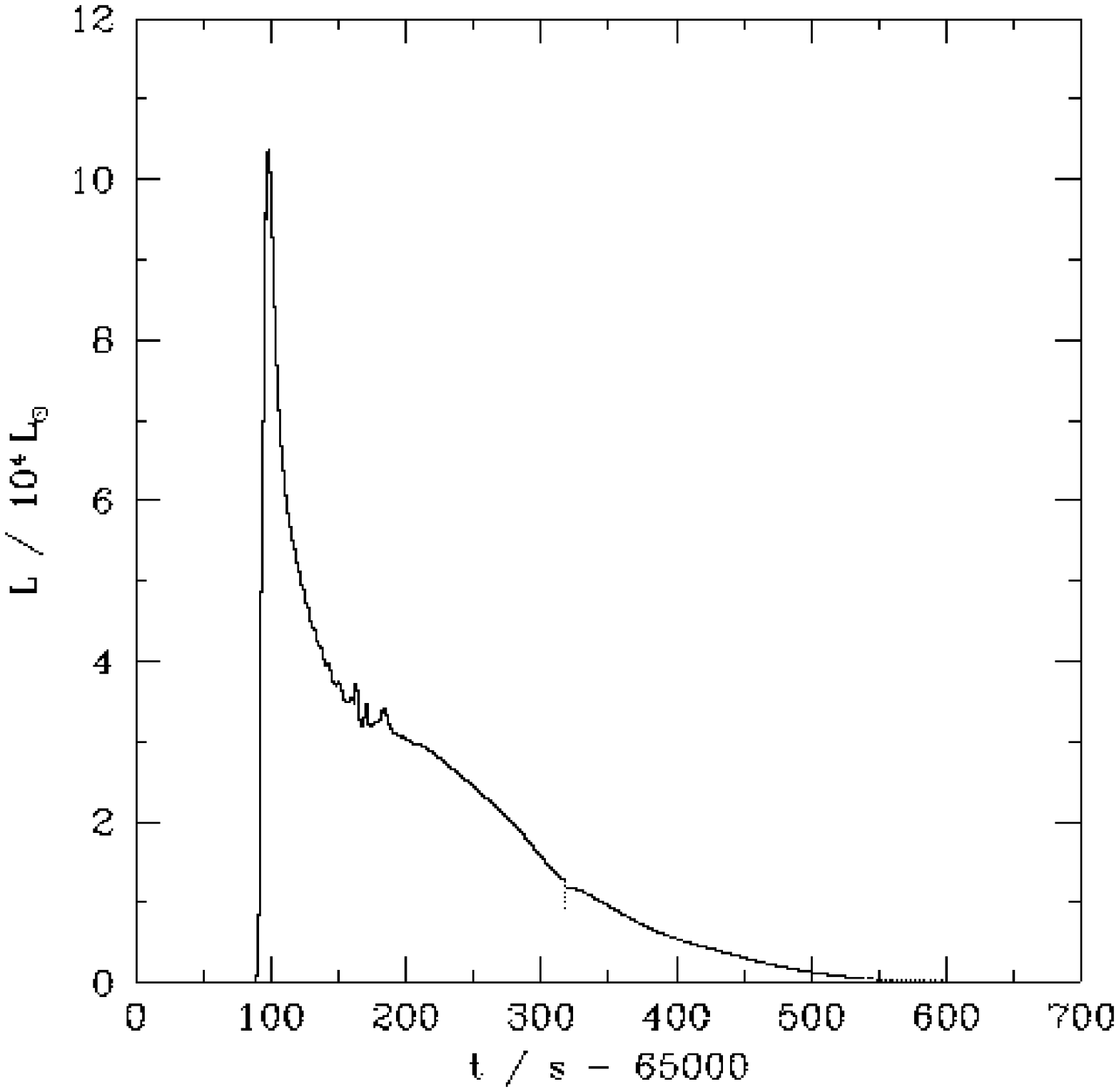}
 \includegraphics[width=0.45\textwidth]{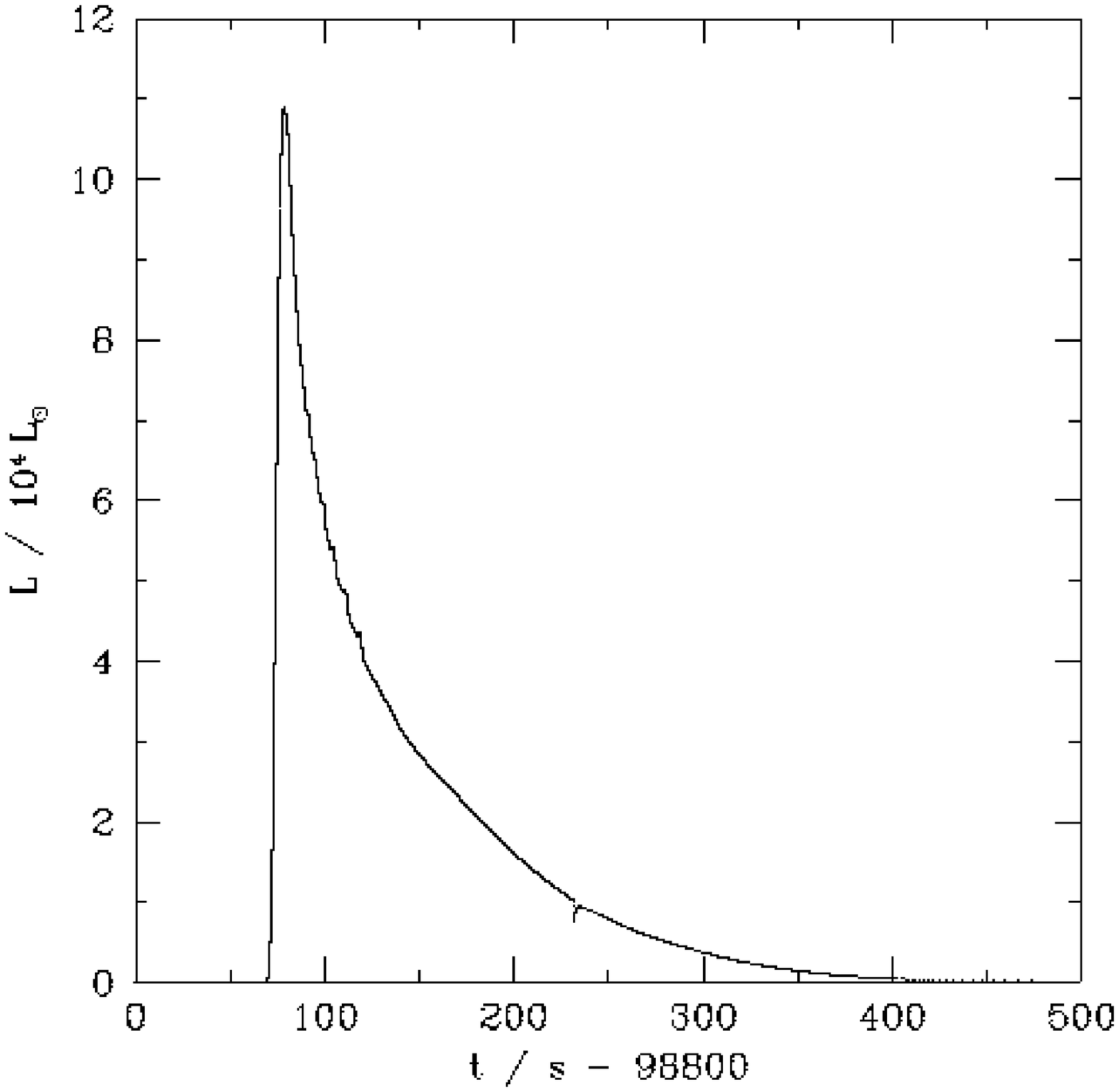}
 \includegraphics[width=0.45\textwidth]{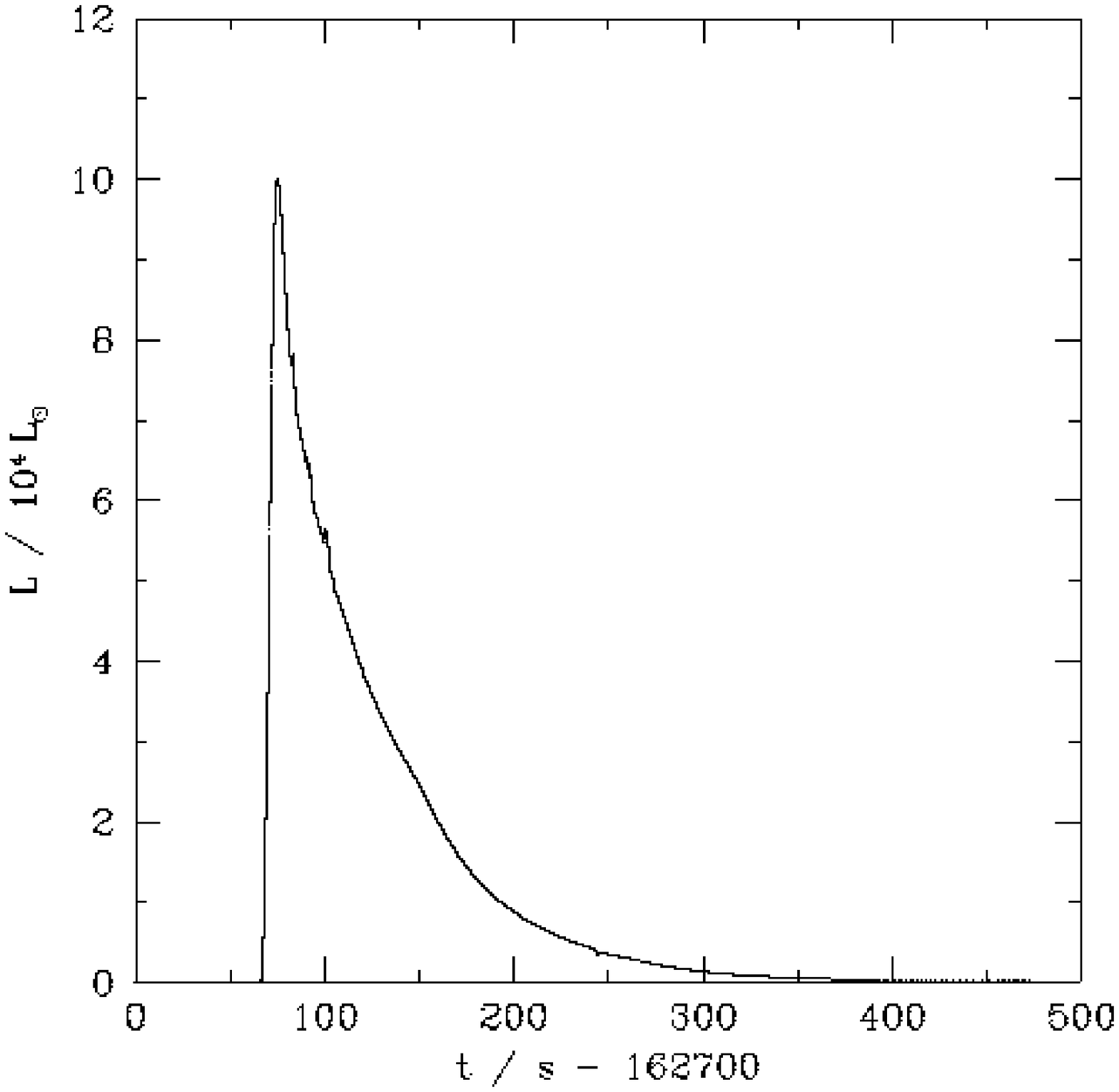}
 \includegraphics[width=0.45\textwidth]{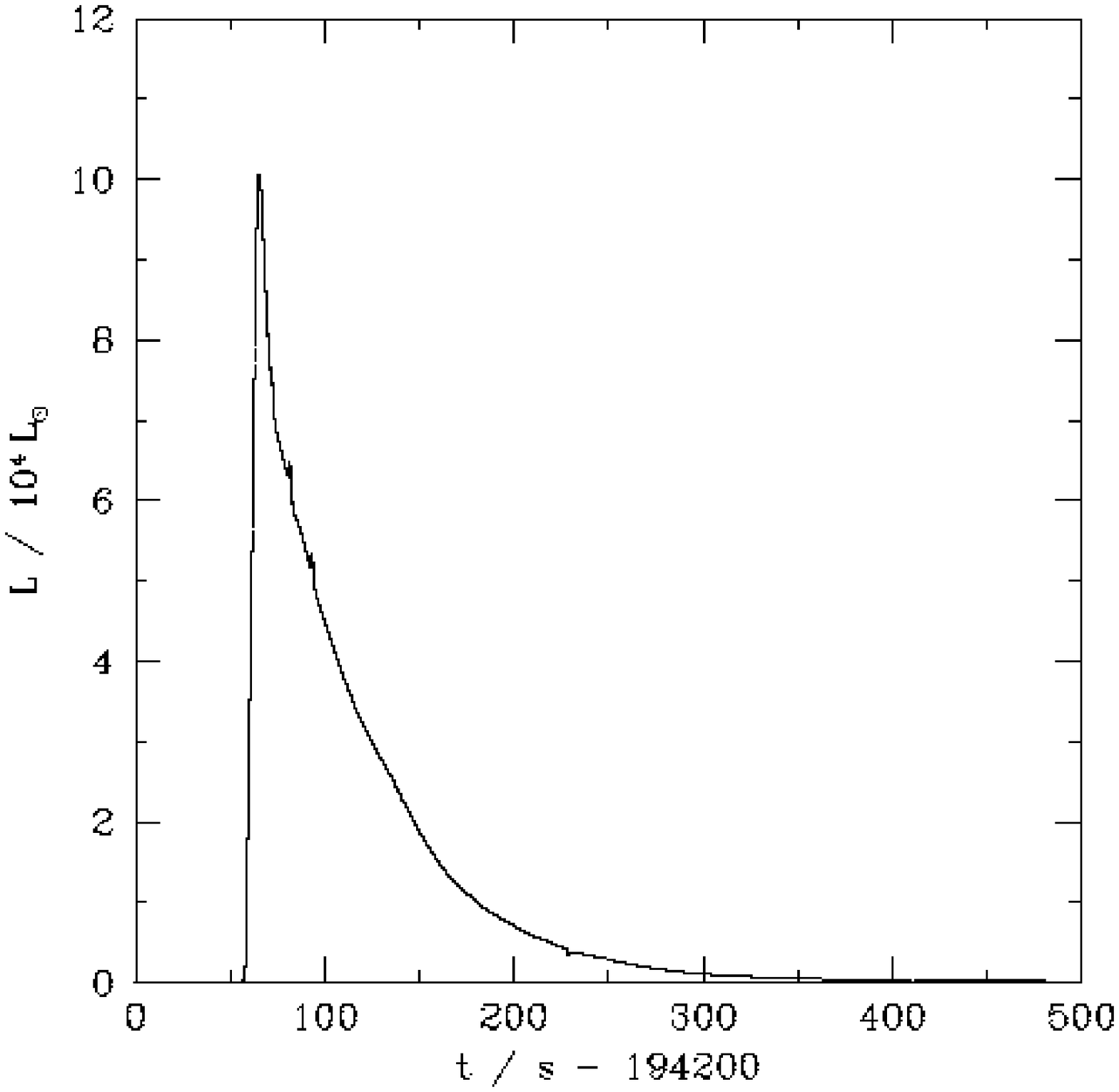}
\caption{Light curves corresponding to the first (upper left panel), second (upper right), fourth (lower left), and fifth bursts (lower right), 
  computed for model 3. }
\label{fig:lum_m3_5b}
\end{figure}

\begin{figure}[htbp]
    \centering
   \includegraphics[width=0.45\textwidth]{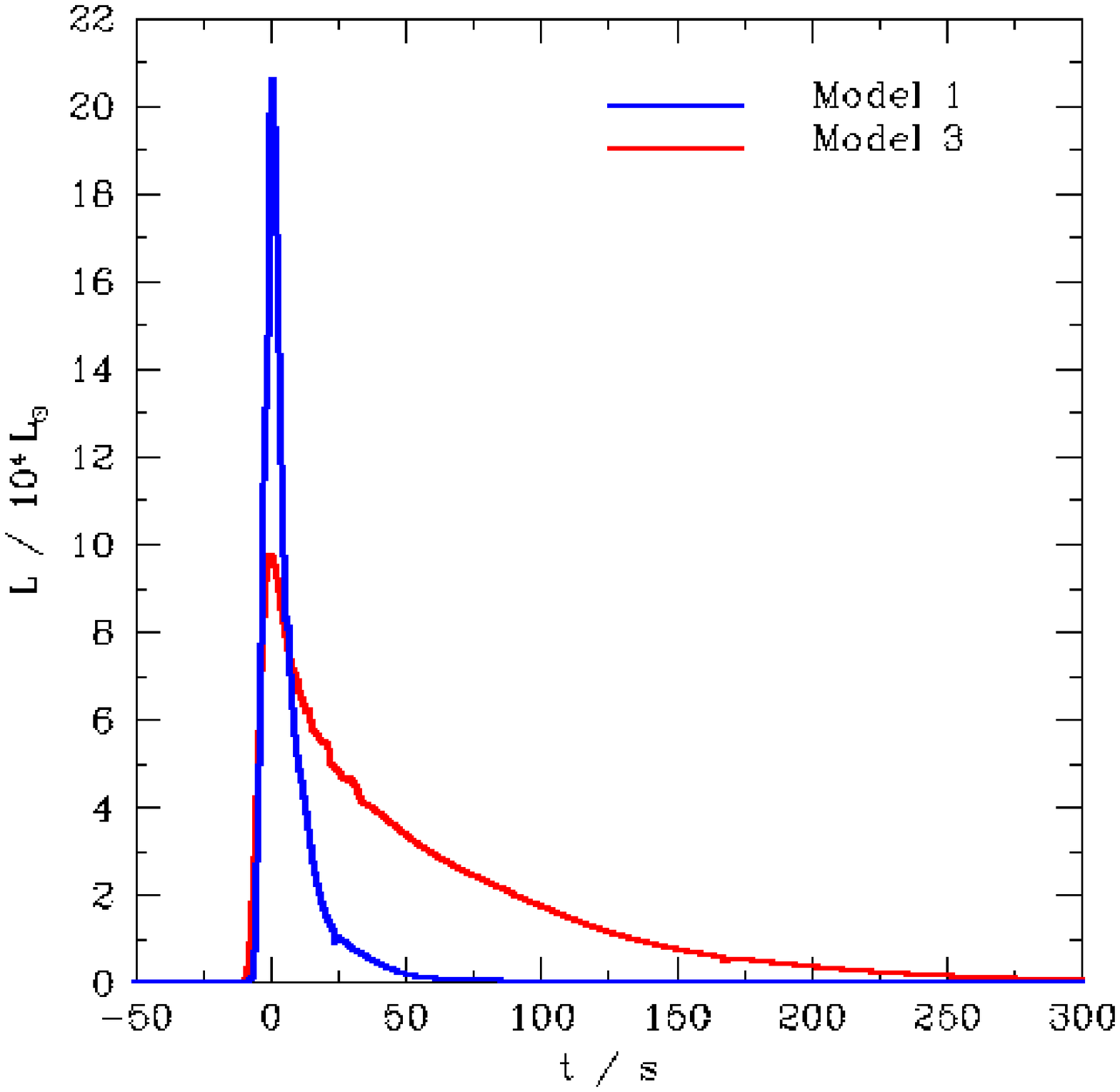}
   \includegraphics[width=0.45\textwidth]{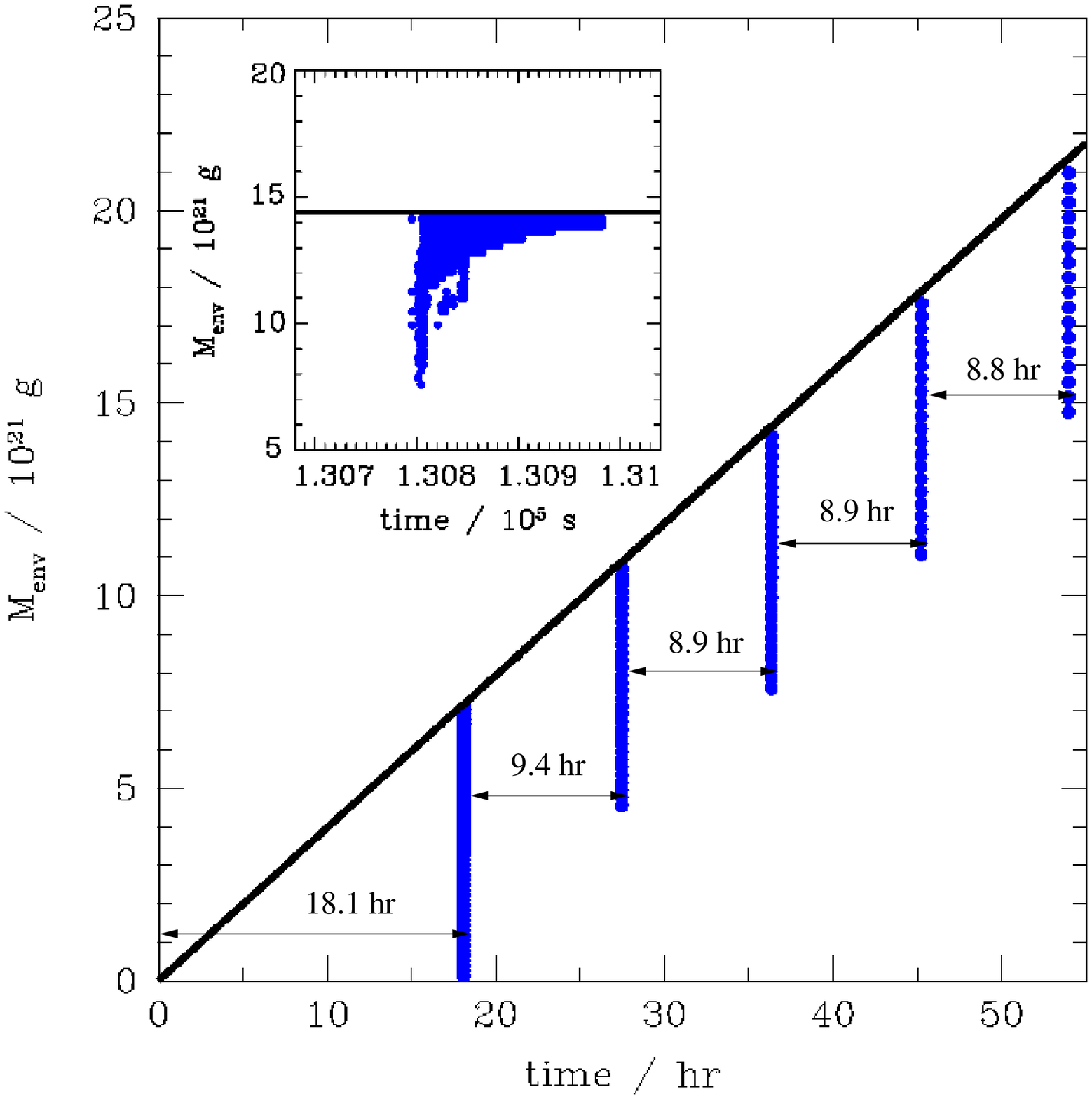}
   \caption{Left panel: light curve comparison for the third burst computed in 
           models 1 \& 3.
	   Right panel: Development of convective regions in model 3, as a 
	   function of time. For illustrative purposes, the convective pattern 
	   obtained during the third burst is shown in the accompanying inset.}
   \label{fig:LUMI_OVER}
   \end{figure}

Despite the qualitative similaries in the gross properties of the bursts 
presented in this paper (as well as in the role played by the metallicity of 
the accreted material) and  those reported in previous work, a quantitative 
comparison reveals some discrepancies that are worth analyzing. 
In model 1 (with Z = Z$_\odot$), our computations yield systematically larger 
(by a factor of $\sim$ 2) recurrence times and peak luminosities (and hence, 
lower $\alpha$) than model ZM of Woosley et al. (2004). 
Similar results are found in the low-metallicity case (model 3, with Z = 
Z$_\odot$/20) when compared with model zM of Woosley et al. (2004),
except for the peak luminosities that turn out to be very similar. 
It is also worth noting that the values reported by Fisker et al. (2008) 
show discrepancies with respect to Woosley et al. (2004), in particular, 
lower peak luminosities (and larger $\alpha$). 
A major difference concerns the much larger effect played by the metallicity 
of the accreted material in this work as compared with Woosley et al. (2004), 
who explained the moderate effect found as due to compositional inertia 
washing out the influence of the initial metallicity.  
Another striking issue concerns the extremely large differences in the gross 
physical characteristics -nucleosynthesis, energies or recurrence times- 
between the first and subsequent bursts, as reported by Woosley et al. (2004).
In terms of nucleoynthesis or nuclear activity,
Figs. \ref{fig:abun_b3}, \ref{fig:lum_m3_5b}, 
\& \ref{fig:LUMI_OVER}, reveal a similar behavior for the different
bursts (although a somewhat lower production of intermediate-mass elements 
as well as of the heaviest elements is reported for the first burst computed
in model 3). 

Very limited information on the nucleosynthetic yields obtained in model ZM is
given in Woosley et al. (2004). Thus, we will restrict the discussion on the
extent of the nuclear activity and on the resulting chemical abundance pattern
to model 3, through a brief comparison with the work reported by
Schatz et al. (2001), Fisker et al. (2008) and Woosley et al. (2004) (for model
zM).
It is worth noting that both the nucleosynthetic end-point (located in the SnSbTe-mass region) 
and the main nuclear path in the A $\sim 50 - 100$ mass region obtained
in this work (passing through a suite of different nuclei such, as $^{55}$Co, $^{60}$Zn, $^{70}$Br, $^{75}$Rb,
$^{85}$Mo, $^{90}$Rh, or $^{100,105}$Sn) are similar to
those reported by Schatz et al. (2001) in the framework of one-zone calculations.
Whereas the main nuclear path in the first burst of model zM (Woosley et al. 2004)
is very similar to the one reported in this work, compositional inertia causes
a more limited extension of the nuclear activity in the successive bursts of
Woosley et al.: hence,
while the three most abundant nuclei at the bottom of the envelope are $^{106}$Sn and $^{104,106}$In
at the end of the first burst, this switches to $^{64}$Zn, $^{68}$Se, and $^{32}$S 
(a similar trend is also reported by Fisker et al. 2008). In this work,
 the mass-averaged composition at the end of the first burst computed for
 model 3 (see Table 8)
is dominated (aside from some residual H and $^{4}$He) by the presence of 
$^{104}$Pd (0.05, by mass) and $^{105}$Ag (0.08), 
while X($^{64}$Zn) $\sim$ 0.04. 
But the peak at the end of the abundance distribution (see
Fig. \ref{fig:abun_b3})
increases with subsequent bursts up to a plateau value, which indicates
that these heavy nuclei are still produced in similar quantities. This is 
very different to the results reported by Woosley et al. (2004).
Indeed, at the end of the fifth burst, the abundance pattern, 
shows still a significant presence of 
heavy species (i.e., X($^{105}$Ag) $\sim$ 0.1, X($^{104}$Pd) $\sim$ 0.08, and X($^{94}$Tc) $\sim$ 0.05), together with a simultaneous 
increase in the abundances of intermediate-mass elements, such as 
$^{60}$Ni (0.06), $^{64}$Zn (0.09), $^{68}$Ge (0.07), or $^{72}$Se (0.04).
It is finally worth noting that, in agreement with all previous hydrodynamic studies, both models 1 and
3 yield very small post-burst abundances of $^{12}$C, below 
 the threshold amount required to power superbursts.  Even though only a few bursts have been computed for these models, they already show a 
  trend on the amount of $^{12}$C that may be expected after many more bursts.

 Finally, it is also worth mentioning that large differences exist between the 
 hydrodynamic simulations reported here (see also Woosley et al. 2004, and 
 Fisker et al. 2008) and those based on one-zone models (i.e.,
  Schatz et al. 1999, 2001) as regards the shape of the light curve 
  accompanying the bursting episodes (the primary difference being the 
  presence of a long-lasting plateau in the latter).

\begin{figure}[thbp]
 \centering
\includegraphics[width=0.40\textwidth]{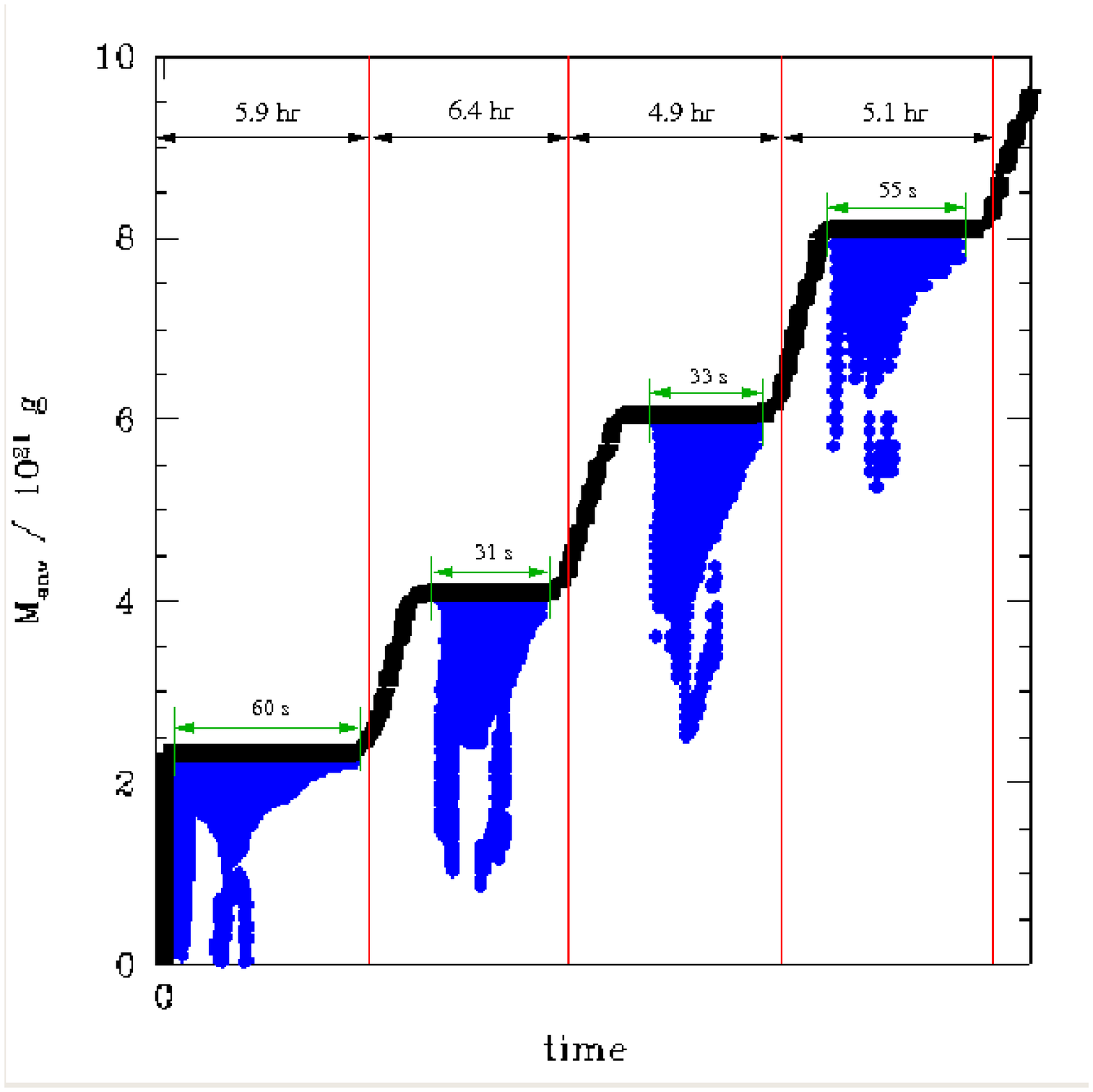}
\includegraphics[width=0.40\textwidth]{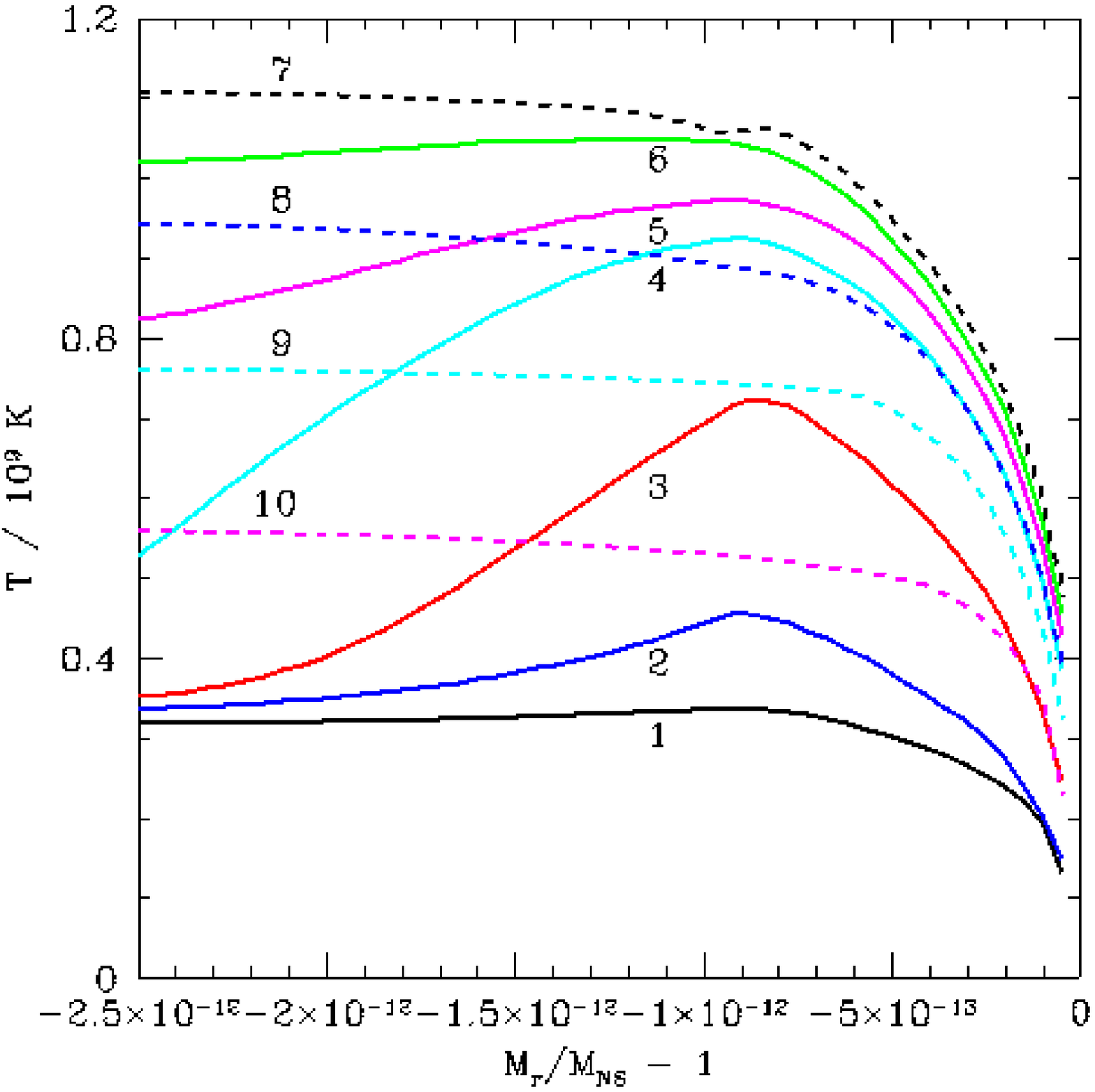}
\caption{Left panel: Schematic representation of the convective regions 
in the accreted envelope for the four bursts computed in model 1.
To simultaneously display the convective stages of the TNRs as compared with the
overall duration of the bursts, a discontinuous time axis is used.
  Right panel: Temperature profiles, showing the location of the ignition point ($\sim 5.6$ m above 
  the core-envelope interface), along the fourth bursting episode, for model 1
  ($\rm{M}_{NS}$ = 1.4 M$_\odot$, $\rm{\dot M}$$_{acc}$ = $1.75 \times 10^{-9}$ M$_\odot$ yr$^{-1}$,
  Z = 0.02).  Labels indicate different moments during the TNR,
  for which the temperature at the ignition shell reaches a value of:
 (1) $3.4 \times 10^8$ K, (2) $4.5 \times 10^8$ K,
 (3) $7 \times 10^8$ K, (4) $9.2 \times 10^8$ K,
 (5) $9.7 \times 10^8$ K, (6) $1.04 \times 10^9$ K,
 (7) $1.06 \times 10^9$ K, (8) $8.9 \times 10^8$ K,
 (9) $7.4 \times 10^8$ K, and (10) $5.3 \times 10^8$ K.  }
\label{fig:conv_tem_b4}
\end{figure}

  The origin of the discrepancies reported is not totally clear 
  and would require additional hydrodynamic studies. Notice, however, that 
  the local surface gravity of our model is somewhat smaller than that 
  adopted in the abovementioned works: whereas a 10 km radius is assigned 
  to the 1.4 M$_\odot$ neutron star in Woosley et al. (2004) 
  ($g = 1.86 \times 10^{14}$ cm s$^{-2}$),
  the integration of the neutron star structure from the core to its surface, 
  in hydrostatic equilibrium, yielded 13.1 km (14.3 km, after general 
  relativity corrections are introduced; see Subsection 5.1), 
  for our 1.4 M$_\odot$ neutron star (corresponding to a surface gravity 
  of $g = 1.08 \times 10^{14}$ cm s$^{-2}$); in turn,
  the calculations reported by Fisker et al. (2008), in a general 
  relativity framework, relied on a 11 km (1.4 M$_\odot$) neutron
  star, for which $g = 1.53 \times 10^{14}$ cm s$^{-2}$. Although XRB 
  properties depend weakly upon the neutron star mass (or surface gravity),
  part of the differences outlined between the three studies can be 
  attributed to the combined effect of the adopted neutron star
  size (surface gravity) and to differences in the input physics (i.e., 
  nuclear reaction network, opacities, treatment of convection). 
  In particular, the use of Iben's opacities may have some effect on 
  the peak luminosities achieved since the larger OPAL opacities
  will likely decrease the amount of energy radiated away from the star.  
  Moreover, the inclusion of semiconvection and thermohaline mixing 
  would have a minor effect in the properties of the explosions, likely 
  affecting the appearance of marginal convective transport between 
  bursts (see Woosley et al. 2004, 
  Fisker et al. 2008). It is however worth noting that the convective pattern
  shown in Figs. \ref{fig:conv_tem_b4} \& \ref{fig:LUMI_OVER}
  is similar to those reported in previous work:
  namely that convection sets in as soon as
  superadiabatic gradients are established in the envelope, following the early
  stages of the TNR and the corresponding rise in temperature; it reaches the
  surface and begins to recede before the observed burst properly commences, 
  shutting off thereafter (Woosley et al. 2004, Fisker et al. 2008).  

\begin{figure}[htbp]
 \centering
   \includegraphics[width=0.45\textwidth]{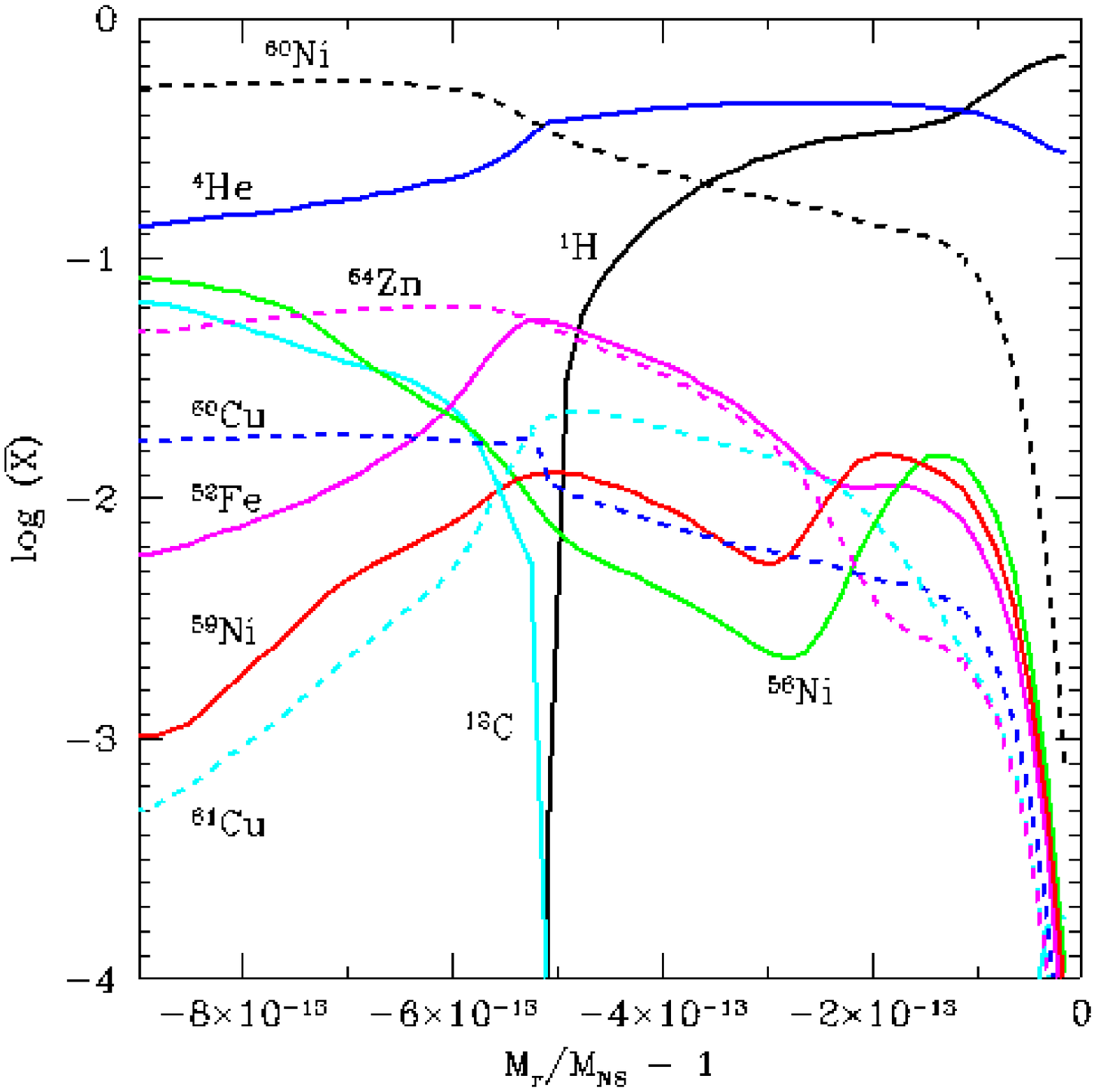}
   \includegraphics[width=0.45\textwidth]{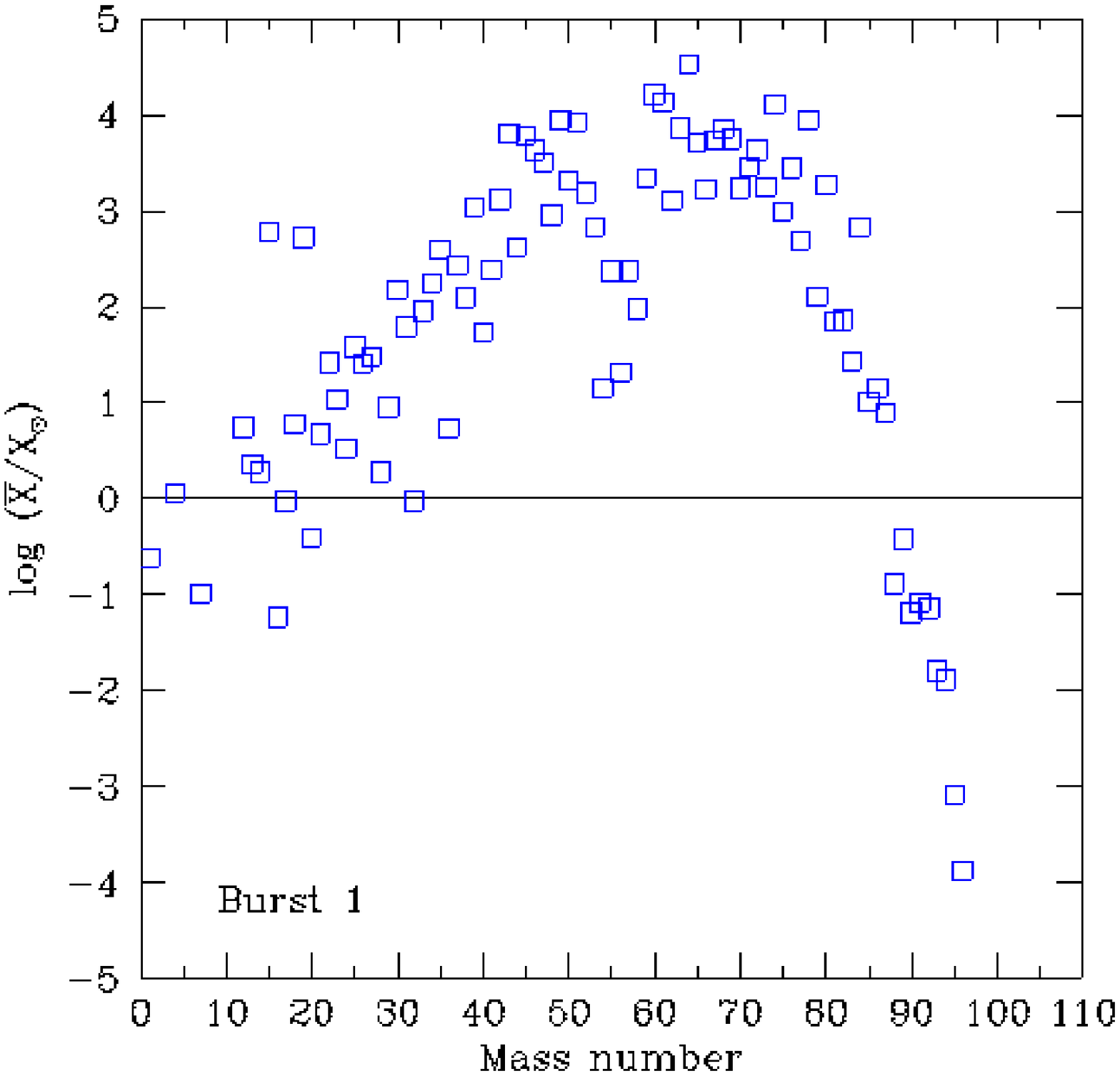}
   \includegraphics[width=0.45\textwidth]{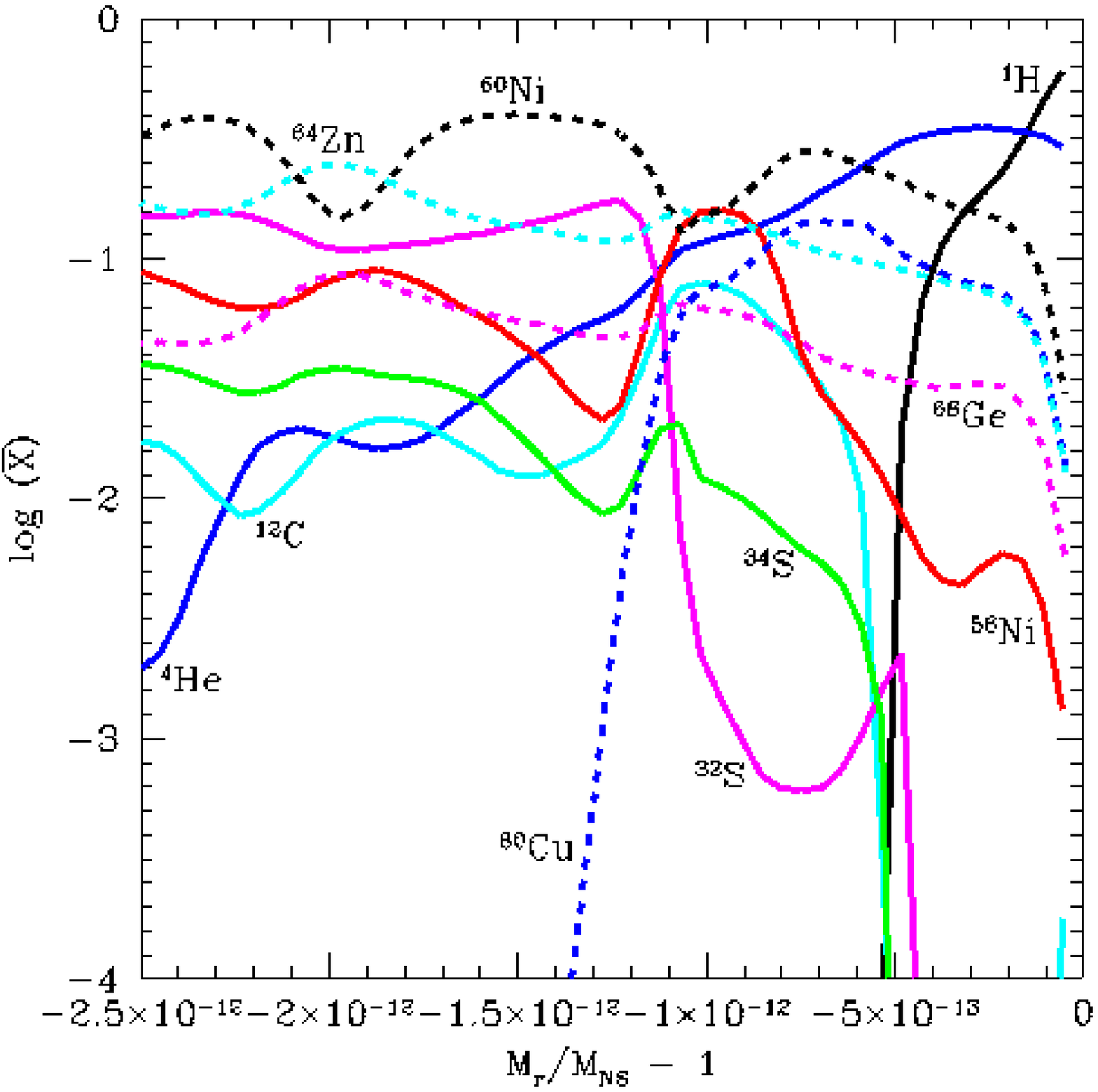}
   \includegraphics[width=0.45\textwidth]{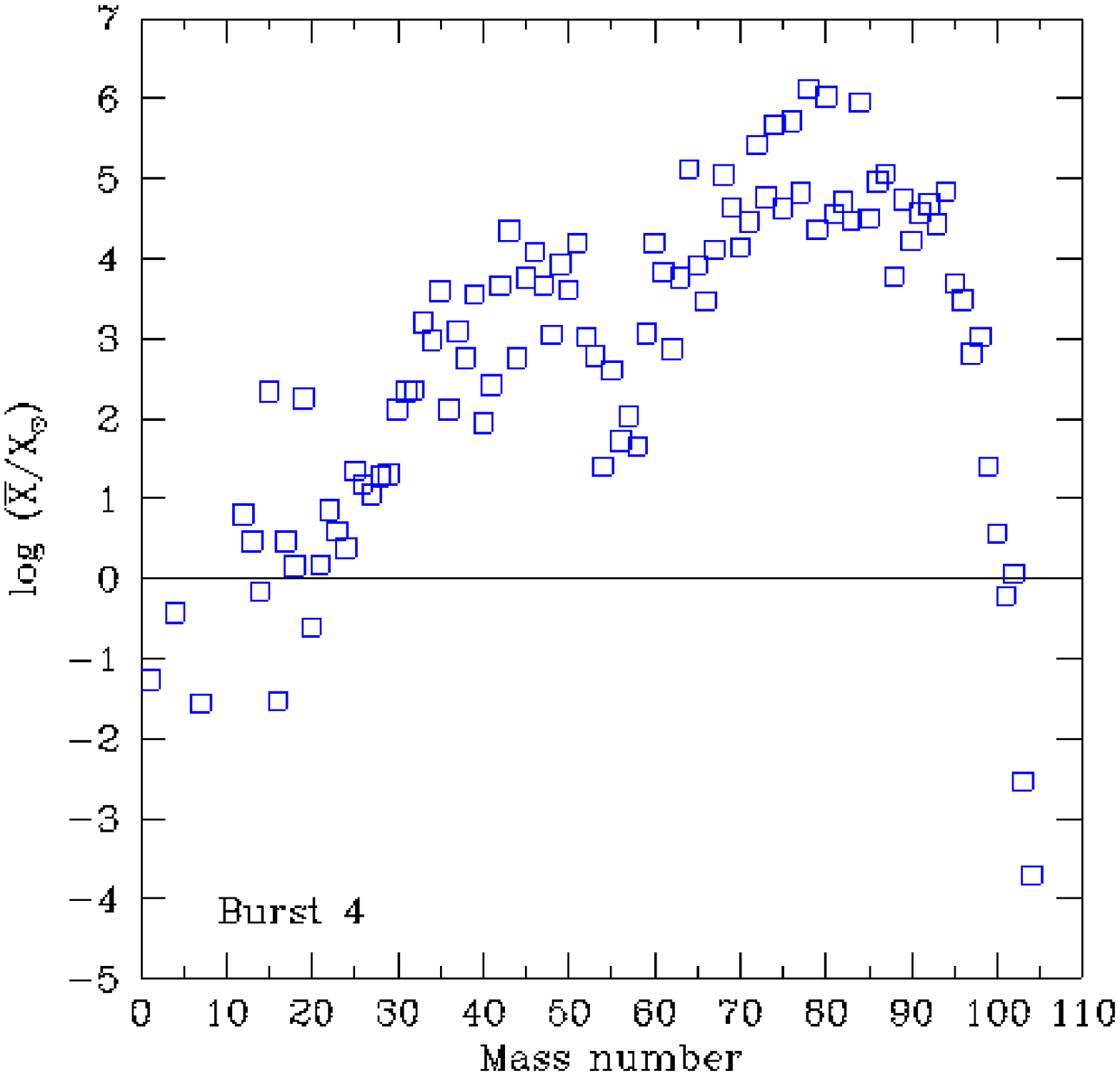}
\caption{Left  panels: mass fractions of the ten most abundant, stable (or $\tau > 1$ hr) isotopes for model 1, 
at the end of the first (upper panel) and fourth burst (lower panel), respectively. Right panels: same as left panels, but for overproduction factors relative to 
solar (for $f > 10^{-5}$).  }
\label{fig:abun_b1}
\end{figure}

\begin{figure}[htbp]
   \centering
   \includegraphics[width=0.45\textwidth]{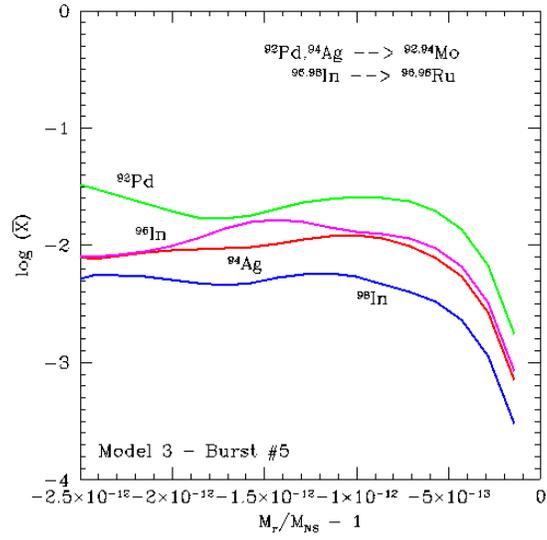}
 \caption{Distribution of the unstable species $^{92}$Pd, $^{94}$Ag, and 
   $^{96,98}$In (that power the light p-nuclei $^{92,94}$Mo and $^{96,98}$Ru)
          throughout the accreted envelope, at the end of the fifth bursting
	  episode of model 3. }
 \label{fig:MORU}
 \end{figure}

  The potential impact of XRB nucleosynthesis on Galactic abundances 
  is still a matter of debate. 
  Matter accreted onto a neutron star of mass $M$ and radius $R$ releases 
  $G M m_p/R \sim 200$ MeV nucleon$^{-1}$, whereas only a 
  few MeV nucleon$^{-1}$ are released from thermonuclear fusion. Thus
  ejection from a neutron star is unlikely.
  However, it has been suggested that radiation-driven winds during 
  photospheric radius expansion may lead to ejection of a tiny fraction of 
  the envelope (containing nuclear processed material; see Weinberg et al.
  2006; MacAlpine et al. 2007). Indeed, XRBs have been proposed as a 
  possible source of the light p-nuclei 
  $^{92,94}$Mo and $^{96,98}$Ru (Schatz et al. 1998, 2001). 
  No matter is ejected in any of the models reported in this work, 
  a result fully independent of the adopted resolution and in agreement 
  with all previous hydrodynamic simulations 
  (Woosley et al. 2004, Fisker et al. 2008).
  Moreover, it is worth noting that, as shown in 
  Figs. \ref{fig:abun_b1} \& \ref{fig:abun_b3}, the abundances of many 
 species synthesized during the bursts decrease remarkably towards the outer
 envelope layers, because of inefficient convective transport (see Figs.
 \ref{fig:conv_tem_b4} \& \ref{fig:LUMI_OVER}).
 To assess the possible contribution to the Galactic abundances, one 
 has to rely on the abundances of the outer envelope layers (the only ones that
 have a chance to be ejected by radiation-driven winds). 
 This shows the limitations posed by one-zone nucleosynthesis calculations, 
 in which the chemical species synthesized in the innermost layers are, 
 by construction, assumed to represent the whole (chemically homogeneous) 
 envelope. The mass fractions of these p-nuclei, obtained in
 model 3, drop by more than an order of magnitude in the outer envelope 
 layers (as compared with the values achieved at the innermost
 envelope; see Fig. \ref{fig:MORU}); 
 the resulting overproduction factors, $f \sim 10^6$, are several orders 
 of magnitude smaller than those required to account
 for the origin of these problematic nuclei 
 (see Weinberg et al. 2006, Bazin et al. 2008), in sharp contrast with the 
 results obtained on the basis of one-zone calculations 
 (Schatz et al. 1998, 2001).

\acknowledgments

The authors are grateful to the anonymous referee
for a critical reading of the manuscript and for valuable comments
that have improved the presentation of this paper.
This work has been partially supported by the Spanish MICINN grants
AYA2007-66256 and EUI2009-04167, by the E.U. FEDER funds, by the U.S. Department of Energy under Contract No. DE-FG02-97ER41041, by the DFG
cluster of excellence 'Origin and Structure of the Universe', and by
the ESF EUROCORES Program EuroGENESIS.

\clearpage

\input{table_mods.tex}
\input{table_nucl1.tex}
\input{table_prop_m1.tex}
\input{table_prop_m2.tex}
\input{table_nucl2.tex}
\input{table6.tex}
\input{table_prop_m3.tex}
\input{table_nucl3.tex}

\clearpage

\clearpage

\clearpage

\clearpage

\clearpage

\clearpage

\clearpage

\clearpage

\clearpage

\clearpage

\clearpage

\clearpage

\clearpage

\clearpage

\clearpage

\clearpage

\clearpage

\clearpage

\clearpage

\clearpage

\clearpage

\clearpage

\clearpage

\clearpage

\clearpage

\clearpage

\clearpage

\clearpage

\clearpage

\clearpage

\clearpage

\clearpage

\clearpage

\clearpage
\end{document}

%% file: table_mods.tex
\begin{deluxetable}{ccccc}
\small
\tablecaption{Summary of the models computed in this work. \label{tab:Models}}
		  \tablehead{
		  \colhead{Model}   &
		  \colhead{M$_{NS}$ (M$_\odot$)}   &
		  \colhead{Metallicity} &
		  \colhead{Envelope shells} &
		  \colhead{Bursts computed}
		  }
\startdata
 1    & 1.4                  &  0.02       &  60             &  4            \\
 2    & 1.4                  &  0.02       &  200            &  2            \\
 3    & 1.4           & $1 \times 10^{-3}$ &  60             &  5            \\
\enddata
\tablenotetext{a}{A mass-accretion rate of $1.75 \times 10^{-9}$ $\rm{M_\odot yr^{-1}}$ has been adopted for all models.}
\end{deluxetable}

%% file: table_nucl1.tex
\begin{deluxetable}{ccccc}
\small
\tablecaption{Mean composition of the envelope (X$_i > 10^{-9}$) at the end of each burst, for Model 1.}
                 \tablehead{
		 \colhead{Nucleus}   &
		 \colhead{Burst 1}   &
		 \colhead{Burst 2} &
		 \colhead{Burst 3} &
		 \colhead{Burst 4}
		          }
\startdata
   $^1$H     & $1.7 \times 10^{-1}$   &  $8.8 \times 10^{-2}$      &  $6.3 \times 10^{-2}$     &   $3.9 \times 10^{-2}$      \\
   $^4$He    & $3.1 \times 10^{-1}$   &  $2.0 \times 10^{-1}$      &  $1.4 \times 10^{-1}$     &   $1.0 \times 10^{-1}$      \\ 
   $^{12}$C  & $1.7 \times 10^{-2}$   &  $2.5 \times 10^{-2}$      &  $2.0 \times 10^{-2}$     &   $1.9 \times 10^{-2}$      \\
   $^{13}$C  & $8.4 \times 10^{-5}$   &  $1.4 \times 10^{-4}$      &  $1.3 \times 10^{-4}$     &   $1.1 \times 10^{-4}$      \\
   $^{14}$N  & $2.1 \times 10^{-3}$   &  $1.5 \times 10^{-3}$      &  $1.1 \times 10^{-3}$     &   $7.6 \times 10^{-4}$      \\
   $^{15}$N  & $2.6 \times 10^{-3}$   &  $2.0 \times 10^{-3}$      &  $1.4 \times 10^{-3}$     &   $9.6 \times 10^{-4}$      \\
   $^{16}$O  & $5.7 \times 10^{-4}$   &  $5.2 \times 10^{-4}$      &  $3.3 \times 10^{-4}$     &   $2.9 \times 10^{-4}$      \\
   $^{17}$O  & $3.7 \times 10^{-6}$   &  $6.3 \times 10^{-6}$      &  $6.8 \times 10^{-6}$     &   $1.2 \times 10^{-5}$      \\
   $^{18}$O  & $6.7 \times 10^{-5}$   &  $2.3 \times 10^{-5}$      &  $1.7 \times 10^{-5}$     &   $6.1 \times 10^{-6}$      \\
   $^{18}$F  & $6.2 \times 10^{-5}$   &  $5.4 \times 10^{-5}$      &  $3.1 \times 10^{-5}$     &   $2.5 \times 10^{-5}$      \\
   $^{19}$F  & $2.1 \times 10^{-4}$   &  $1.8 \times 10^{-4}$      &  $1.1 \times 10^{-4}$     &   $7.3 \times 10^{-5}$      \\
   $^{20}$Ne & $6.3 \times 10^{-4}$   &  $5.4 \times 10^{-4}$      &  $3.5 \times 10^{-4}$     &   $4.0 \times 10^{-4}$      \\
   $^{21}$Ne & $2.0 \times 10^{-5}$   &  $9.5 \times 10^{-6}$      &  $6.7 \times 10^{-6}$     &   $6.2 \times 10^{-6}$      \\
   $^{22}$Ne & $9.9 \times 10^{-5}$   &  $4.2 \times 10^{-5}$      &  $3.2 \times 10^{-5}$     &   $2.9 \times 10^{-5}$      \\
   $^{22}$Na & $3.3 \times 10^{-3}$   &  $1.9 \times 10^{-3}$      &  $1.2 \times 10^{-3}$     &   $9.2 \times 10^{-4}$      \\
   $^{23}$Na & $3.7 \times 10^{-4}$   &  $2.1 \times 10^{-4}$      &  $1.2 \times 10^{-4}$     &   $1.3 \times 10^{-4}$      \\
   $^{24}$Mg & $1.7 \times 10^{-3}$   &  $1.4 \times 10^{-3}$      &  $9.1 \times 10^{-4}$     &   $1.3 \times 10^{-3}$      \\
   $^{25}$Mg & $2.6 \times 10^{-3}$   &  $1.7 \times 10^{-3}$      &  $1.2 \times 10^{-3}$     &   $1.5 \times 10^{-3}$      \\
   $^{26}$Mg & $1.8 \times 10^{-3}$   &  $2.5 \times 10^{-3}$      &  $1.7 \times 10^{-3}$     &   $1.1 \times 10^{-3}$      \\
$^{26}$Al$^g$& $2.6 \times 10^{-4}$   &  $9.8 \times 10^{-5}$      &  $5.7 \times 10^{-5}$     &   $1.1 \times 10^{-4}$      \\
   $^{27}$Al & $1.8 \times 10^{-3}$   &  $1.5 \times 10^{-3}$      &  $8.8 \times 10^{-4}$     &   $6.7 \times 10^{-4}$      \\
   $^{28}$Si & $1.2 \times 10^{-3}$   &  $5.3 \times 10^{-3}$      &  $7.0 \times 10^{-3}$     &   $1.3 \times 10^{-2}$      \\
   $^{29}$Si & $3.1 \times 10^{-4}$   &  $8.6 \times 10^{-4}$      &  $4.6 \times 10^{-4}$     &   $6.9 \times 10^{-4}$      \\
   $^{30}$Si & $3.5 \times 10^{-3}$   &  $5.6 \times 10^{-3}$      &  $3.8 \times 10^{-3}$     &   $3.0 \times 10^{-3}$      \\
   $^{31}$P  & $5.1 \times 10^{-4}$   &  $1.4 \times 10^{-3}$      &  $1.7 \times 10^{-3}$     &   $1.8 \times 10^{-3}$      \\
   $^{32}$S  & $3.8 \times 10^{-4}$   &  $2.9 \times 10^{-2}$      &  $5.8 \times 10^{-2}$     &   $9.0 \times 10^{-2}$      \\
   $^{33}$S  & $2.9 \times 10^{-4}$   &  $3.1 \times 10^{-3}$      &  $4.3 \times 10^{-3}$     &   $5.1 \times 10^{-3}$      \\
   $^{34}$S  & $3.2 \times 10^{-3}$   &  $1.1 \times 10^{-2}$      &  $1.6 \times 10^{-2}$     &   $1.8 \times 10^{-2}$      \\
   $^{35}$Cl & $1.0 \times 10^{-3}$   &  $4.8 \times 10^{-3}$      &  $1.0 \times 10^{-2}$     &   $1.0 \times 10^{-2}$      \\
   $^{36}$Ar & $4.2 \times 10^{-4}$   &  $4.0 \times 10^{-3}$      &  $9.2 \times 10^{-3}$     &   $1.0 \times 10^{-2}$      \\
   $^{37}$Cl & $3.6 \times 10^{-7}$   &  $6.4 \times 10^{-7}$      &  $1.8 \times 10^{-6}$     &   $4.1 \times 10^{-6}$      \\ 
   $^{37}$Ar & $2.3 \times 10^{-4}$   &  $5.7 \times 10^{-4}$      &  $1.0 \times 10^{-3}$     &   $1.1 \times 10^{-3}$      \\
   $^{38}$Ar & $1.9 \times 10^{-3}$   &  $4.9 \times 10^{-3}$      &  $8.5 \times 10^{-3}$     &   $8.9 \times 10^{-3}$      \\
   $^{39}$K  & $3.8 \times 10^{-3}$   &  $7.8 \times 10^{-3}$      &  $1.2 \times 10^{-2}$     &   $1.3 \times 10^{-2}$      \\
   $^{40}$Ca & $3.2 \times 10^{-3}$   &  $4.3 \times 10^{-3}$      &  $5.6 \times 10^{-3}$     &   $5.3 \times 10^{-3}$      \\
   $^{41}$K  &  -                     &  $9.3 \times 10^{-9}$      &  $4.8 \times 10^{-8}$     &   $7.6 \times 10^{-8}$      \\
   $^{41}$Ca & $6.5 \times 10^{-5}$   &  $6.9 \times 10^{-5}$      &  $6.0 \times 10^{-5}$     &   $7.2 \times 10^{-5}$      \\
   $^{42}$Ca & $5.5 \times 10^{-4}$   &  $1.2 \times 10^{-3}$      &  $1.7 \times 10^{-3}$     &   $1.9 \times 10^{-3}$      \\
   $^{43}$Ca & $1.8 \times 10^{-4}$   &  $3.6 \times 10^{-4}$      &  $7.7 \times 10^{-4}$     &   $1.2 \times 10^{-3}$      \\
   $^{43}$Sc & $4.3 \times 10^{-4}$   &  $9.5 \times 10^{-4}$      &  $1.2 \times 10^{-3}$     &   $9.7 \times 10^{-4}$      \\
   $^{44}$Ca & $4.3 \times 10^{-8}$   &  $3.7 \times 10^{-6}$      &  $1.8 \times 10^{-5}$     &   $3.2 \times 10^{-5}$      \\
   $^{44}$Sc & $2.5 \times 10^{-7}$   &  $1.9 \times 10^{-5}$      &  $3.6 \times 10^{-5}$     &   $2.9 \times 10^{-5}$      \\
   $^{44}$Ti & $5.8 \times 10^{-4}$   &  $7.2 \times 10^{-4}$      &  $7.3 \times 10^{-4}$     &   $7.5 \times 10^{-4}$      \\
   $^{45}$Sc & $8.2 \times 10^{-5}$   &  $7.9 \times 10^{-5}$      &  $9.3 \times 10^{-5}$     &   $1.2 \times 10^{-4}$      \\
   $^{45}$Ti & $1.5 \times 10^{-4}$   &  $1.7 \times 10^{-4}$      &  $1.1 \times 10^{-4}$     &   $1.0 \times 10^{-4}$      \\
   $^{46}$Ti & $9.5 \times 10^{-4}$   &  $1.9 \times 10^{-3}$      &  $2.4 \times 10^{-3}$     &   $2.7 \times 10^{-3}$      \\
   $^{47}$Ti & $6.6 \times 10^{-4}$   &  $8.7 \times 10^{-4}$      &  $9.2 \times 10^{-4}$     &   $9.5 \times 10^{-4}$      \\
   $^{48}$Ti & $2.1 \times 10^{-7}$   &  $5.6 \times 10^{-7}$      &  $1.6 \times 10^{-6}$     &   $3.4 \times 10^{-6}$      \\
   $^{48}$V  & $1.2 \times 10^{-4}$   &  $1.4 \times 10^{-4}$      &  $2.4 \times 10^{-4}$     &   $3.7 \times 10^{-4}$      \\
   $^{48}$Cr & $1.8 \times 10^{-3}$   &  $2.1 \times 10^{-3}$      &  $2.0 \times 10^{-3}$     &   $2.0 \times 10^{-3}$      \\
   $^{49}$Ti & $1.4 \times 10^{-7}$   &  $4.4 \times 10^{-7}$      &  $6.9 \times 10^{-7}$     &   $8.9 \times 10^{-7}$      \\
   $^{49}$V  & $1.5 \times 10^{-3}$   &  $1.5 \times 10^{-3}$      &  $1.2 \times 10^{-3}$     &   $1.4 \times 10^{-3}$      \\
   $^{50}$Cr & $1.5 \times 10^{-3}$   &  $2.3 \times 10^{-3}$      &  $2.7 \times 10^{-3}$     &   $3.0 \times 10^{-3}$      \\
   $^{51}$V  & $3.4 \times 10^{-6}$   &  $6.4 \times 10^{-6}$      &  $1.6 \times 10^{-5}$     &   $3.0 \times 10^{-5}$      \\
   $^{51}$Cr & $3.2 \times 10^{-3}$   &  $4.5 \times 10^{-3}$      &  $5.1 \times 10^{-3}$     &   $5.8 \times 10^{-3}$      \\
   $^{52}$Cr & $1.5 \times 10^{-5}$   &  $2.2 \times 10^{-5}$      &  $4.9 \times 10^{-5}$     &   $9.4 \times 10^{-5}$      \\
   $^{52}$Mn & $3.4 \times 10^{-3}$   &  $2.3 \times 10^{-3}$      &  $2.7 \times 10^{-3}$     &   $4.0 \times 10^{-3}$      \\
   $^{52}$Fe & $2.0 \times 10^{-2}$   &  $1.5 \times 10^{-2}$      &  $9.9 \times 10^{-3}$     &   $1.2 \times 10^{-2}$      \\
   $^{53}$Mn & $1.2 \times 10^{-3}$   &  $1.1 \times 10^{-3}$      &  $8.6 \times 10^{-4}$     &   $1.1 \times 10^{-3}$      \\
   $^{54}$Fe & $1.0 \times 10^{-3}$   &  $1.4 \times 10^{-3}$      &  $1.6 \times 10^{-3}$     &   $1.8 \times 10^{-3}$      \\
   $^{55}$Mn & $6.6 \times 10^{-9}$   &  $2.3 \times 10^{-8}$      &  $7.1 \times 10^{-8}$     &   $1.6 \times 10^{-7}$      \\
   $^{55}$Fe & $2.3 \times 10^{-4}$   &  $3.1 \times 10^{-4}$      &  $6.4 \times 10^{-4}$     &   $1.0 \times 10^{-3}$      \\
   $^{55}$Co & $2.9 \times 10^{-3}$   &  $3.8 \times 10^{-3}$      &  $4.0 \times 10^{-3}$     &   $4.3 \times 10^{-3}$      \\
   $^{56}$Fe & $8.0 \times 10^{-8}$   &  $5.4 \times 10^{-7}$      &  $1.7 \times 10^{-6}$     &   $3.5 \times 10^{-6}$      \\
   $^{56}$Co & $2.2 \times 10^{-4}$   &  $5.3 \times 10^{-4}$      &  $1.1 \times 10^{-3}$     &   $1.8 \times 10^{-3}$      \\
   $^{56}$Ni & $2.4 \times 10^{-2}$   &  $4.5 \times 10^{-2}$      &  $5.2 \times 10^{-2}$     &   $5.9 \times 10^{-2}$      \\
   $^{57}$Fe & $2.6 \times 10^{-8}$   &  $1.6 \times 10^{-7}$      &  $2.0 \times 10^{-7}$     &   $2.5 \times 10^{-7}$      \\
   $^{57}$Co & $2.5 \times 10^{-4}$   &  $1.2 \times 10^{-4}$      &  $1.4 \times 10^{-4}$     &   $2.0 \times 10^{-4}$      \\
   $^{57}$Ni & $6.5 \times 10^{-3}$   &  $3.9 \times 10^{-3}$      &  $2.7 \times 10^{-3}$     &   $2.9 \times 10^{-3}$      \\
   $^{58}$Ni & $4.7 \times 10^{-3}$   &  $2.7 \times 10^{-3}$      &  $1.9 \times 10^{-3}$     &   $2.2 \times 10^{-3}$      \\
   $^{59}$Ni & $7.3 \times 10^{-3}$   &  $4.3 \times 10^{-3}$      &  $3.3 \times 10^{-3}$     &   $3.8 \times 10^{-3}$      \\
   $^{60}$Ni & $3.2 \times 10^{-1}$   &  $3.3 \times 10^{-1}$      &  $3.1 \times 10^{-1}$     &   $3.1 \times 10^{-1}$      \\
   $^{61}$Ni & $3.9 \times 10^{-3}$   &  $1.7 \times 10^{-3}$      &  $1.9 \times 10^{-3}$     &   $2.4 \times 10^{-3}$      \\
   $^{61}$Cu & $8.2 \times 10^{-3}$   &  $6.0 \times 10^{-3}$      &  $3.5 \times 10^{-3}$     &   $3.8 \times 10^{-3}$      \\
   $^{62}$Ni & $5.0 \times 10^{-4}$   &  $3.1 \times 10^{-4}$      &  $4.4 \times 10^{-4}$     &   $5.8 \times 10^{-4}$      \\
   $^{62}$Zn & $3.2 \times 10^{-3}$   &  $2.5 \times 10^{-3}$      &  $1.4 \times 10^{-3}$     &   $1.5 \times 10^{-3}$      \\
   $^{63}$Cu & $4.1 \times 10^{-3}$   &  $4.4 \times 10^{-3}$      &  $3.1 \times 10^{-3}$     &   $3.3 \times 10^{-3}$      \\
   $^{64}$Zn & $3.4 \times 10^{-2}$   &  $1.0 \times 10^{-1}$      &  $1.3 \times 10^{-1}$     &   $1.3 \times 10^{-1}$      \\
   $^{65}$Cu & $2.5 \times 10^{-7}$   &  $2.9 \times 10^{-7}$      &  $5.1 \times 10^{-7}$     &   $9.2 \times 10^{-7}$      \\
   $^{65}$Zn & $1.4 \times 10^{-3}$   &  $2.5 \times 10^{-3}$      &  $1.9 \times 10^{-3}$     &   $2.2 \times 10^{-3}$      \\
   $^{66}$Zn & $3.2 \times 10^{-5}$   &  $5.0 \times 10^{-5}$      &  $1.3 \times 10^{-4}$     &   $2.4 \times 10^{-4}$      \\
   $^{66}$Ga & $4.0 \times 10^{-4}$   &  $5.1 \times 10^{-4}$      &  $5.5 \times 10^{-4}$     &   $6.3 \times 10^{-4}$      \\
   $^{66}$Ge & $5.4 \times 10^{-4}$   &  $1.3 \times 10^{-3}$      &  $7.5 \times 10^{-4}$     &   $8.7 \times 10^{-4}$      \\
   $^{67}$Ga & $4.7 \times 10^{-4}$   &  $1.0 \times 10^{-3}$      &  $9.7 \times 10^{-4}$     &   $1.1 \times 10^{-3}$      \\
   $^{68}$Ge & $2.9 \times 10^{-3}$   &  $2.6 \times 10^{-2}$      &  $4.4 \times 10^{-2}$     &   $4.5 \times 10^{-2}$      \\
   $^{69}$Ge & $2.2 \times 10^{-4}$   &  $1.3 \times 10^{-3}$      &  $1.5 \times 10^{-3}$     &   $1.8 \times 10^{-3}$      \\
   $^{70}$Ge & $7.3 \times 10^{-5}$   &  $4.1 \times 10^{-4}$      &  $4.6 \times 10^{-4}$     &   $6.0 \times 10^{-4}$      \\
   $^{71}$As & $7.7 \times 10^{-5}$   &  $4.7 \times 10^{-4}$      &  $6.0 \times 10^{-4}$     &   $7.7 \times 10^{-4}$      \\
   $^{72}$Se & $2.5 \times 10^{-4}$   &  $5.7 \times 10^{-3}$      &  $1.4 \times 10^{-2}$     &   $1.6 \times 10^{-2}$      \\
   $^{73}$Se & $3.0 \times 10^{-5}$   &  $4.2 \times 10^{-4}$      &  $7.6 \times 10^{-4}$     &   $1.0 \times 10^{-3}$      \\
   $^{74}$Se & $1.3 \times 10^{-5}$   &  $1.8 \times 10^{-4}$      &  $3.3 \times 10^{-4}$     &   $4.8 \times 10^{-4}$      \\
   $^{75}$Br & $1.2 \times 10^{-5}$   &  $1.8 \times 10^{-4}$      &  $3.7 \times 10^{-4}$     &   $5.3 \times 10^{-4}$      \\
   $^{76}$Kr & $3.0 \times 10^{-5}$   &  $1.4 \times 10^{-3}$      &  $4.5 \times 10^{-3}$     &   $5.6 \times 10^{-3}$      \\
   $^{77}$Kr & $4.5 \times 10^{-6}$   &  $1.5 \times 10^{-4}$      &  $4.1 \times 10^{-4}$     &   $6.1 \times 10^{-4}$      \\
   $^{78}$Kr & $2.7 \times 10^{-6}$   &  $9.2 \times 10^{-5}$      &  $2.5 \times 10^{-4}$     &   $4.1 \times 10^{-4}$      \\
   $^{79}$Kr & $1.6 \times 10^{-6}$   &  $5.9 \times 10^{-5}$      &  $1.7 \times 10^{-4}$     &   $2.7 \times 10^{-4}$      \\
   $^{80}$Sr & $3.9 \times 10^{-6}$   &  $4.0 \times 10^{-4}$      &  $1.6 \times 10^{-3}$     &   $2.2 \times 10^{-3}$      \\
   $^{81}$Rb & $8.4 \times 10^{-7}$   &  $6.8 \times 10^{-5}$      &  $2.7 \times 10^{-4}$     &   $4.3 \times 10^{-4}$      \\
   $^{82}$Sr & $7.9 \times 10^{-7}$   &  $8.2 \times 10^{-5}$      &  $3.2 \times 10^{-4}$     &   $5.6 \times 10^{-4}$      \\
   $^{83}$Sr & $3.0 \times 10^{-7}$   &  $4.3 \times 10^{-5}$      &  $1.9 \times 10^{-4}$     &   $3.2 \times 10^{-4}$      \\
   $^{84}$Sr & $1.9 \times 10^{-7}$   &  $3.8 \times 10^{-5}$      &  $1.4 \times 10^{-4}$     &   $2.6 \times 10^{-4}$      \\
   $^{85}$Y  & $1.1 \times 10^{-7}$   &  $4.0 \times 10^{-5}$      &  $2.1 \times 10^{-4}$     &   $3.5 \times 10^{-4}$      \\
   $^{86}$Zr & $7.1 \times 10^{-8}$   &  $4.4 \times 10^{-5}$      &  $2.3 \times 10^{-4}$     &   $4.6 \times 10^{-4}$      \\
   $^{87}$Zr & $2.7 \times 10^{-8}$   &  $3.6 \times 10^{-5}$      &  $1.8 \times 10^{-4}$     &   $3.9 \times 10^{-4}$      \\
   $^{88}$Zr & $5.6 \times 10^{-9}$   &  $1.8 \times 10^{-5}$      &  $1.4 \times 10^{-4}$     &   $2.6 \times 10^{-4}$      \\
   $^{89}$Nb & $4.0 \times 10^{-9}$   &  $3.1 \times 10^{-5}$      &  $2.9 \times 10^{-4}$     &   $5.8 \times 10^{-4}$      \\
   $^{90}$Mo & -                      &  $1.1 \times 10^{-5}$      &  $1.1 \times 10^{-4}$     &   $2.2 \times 10^{-4}$      \\
   $^{91}$Nb & -                      &  $5.4 \times 10^{-6}$      &  $5.3 \times 10^{-5}$     &   $1.1 \times 10^{-4}$      \\
   $^{92}$Mo & -                      &  $2.1 \times 10^{-6}$      &  $1.9 \times 10^{-5}$     &   $4.5 \times 10^{-5}$      \\
   $^{93}$Tc & -                      &  $1.3 \times 10^{-6}$      &  $1.3 \times 10^{-5}$     &   $4.5 \times 10^{-5}$      \\
   $^{94}$Tc & -                      &  $6.9 \times 10^{-7}$      &  $7.6 \times 10^{-6}$     &   $3.8 \times 10^{-5}$      \\
   $^{95}$Ru & -                      &  $1.1 \times 10^{-7}$      &  $1.1 \times 10^{-6}$     &   $4.7 \times 10^{-6}$      \\
   $^{96}$Ru & -                      &  $9.4 \times 10^{-9}$      &  $1.1 \times 10^{-7}$     &   $7.6 \times 10^{-7}$      \\
   $^{97}$Ru & -                      &  $3.9 \times 10^{-9}$      &  $5.1 \times 10^{-8}$     &   $3.9 \times 10^{-7}$      \\
   $^{98}$Ru & -                      &  -                         &  $1.3 \times 10^{-8}$     &   $9.2 \times 10^{-8}$      \\
   $^{99}$Rh & -                      &  -                         &  $2.4 \times 10^{-9}$     &   $1.5 \times 10^{-8}$      \\
  $^{100}$Pd & -                      &  -                         &  -                        &   $2.2 \times 10^{-9}$      \\[0.5ex]
\enddata
\end{deluxetable}

%% file: table_prop_m1.tex
\begin{deluxetable}{ccccccc}
\small
\tablecaption{ Summary of burst properties for Model 1. }
                  \tablehead{
		  \colhead{Burst}   &
		  \colhead{T$_{peak}$ (K)}   &
		  \colhead{t(T$_{peak}$) (s)} &
		  \colhead{$\tau_{rec}$ (hr)} &
		  \colhead{L$_{peak}$ (L$_\odot$)} &
		  \colhead{$\tau_{0.01}$ (s)} &
		  \colhead{$\alpha$} 
		                    }
\startdata
 1    & $1.06 \times 10^9$& 21192    &  5.9        & $9.7 \times 10^4$&  75.8      &  60 \\
 2    & $1.15 \times 10^9$& 44342    &  6.4        & $1.7 \times 10^5$&  62.3      &  40 \\
 3    & $1.26 \times 10^9$& 62137    &  4.9        & $2.1 \times 10^5$&  55.4      &  34 \\
 4    & $1.12 \times 10^9$& 80568    &  5.1        & $1.2 \times 10^5$&  75.7      &  36 \\ [0.5ex]
\enddata
\end{deluxetable}

%% file: table_prop_m2.tex
\begin{deluxetable}{ccccccc}
\small
\tablecaption{ Summary of burst properties for Model 2. }
                  \tablehead{
		  \colhead{Burst}   &
		  \colhead{T$_{peak}$ (K)}   &
		  \colhead{t(T$_{peak}$) (s)} &
		  \colhead{$\tau_{rec}$ (hr)} &
		  \colhead{L$_{peak}$ (L$_\odot$)} &
		  \colhead{$\tau_{0.01}$ (s)} &
		  \colhead{$\alpha$} 
		                    }
\startdata
 1    & $1.05 \times 10^9$& 21189    &  5.9        & $9.0 \times 10^4$&  59.2      &  62 \\
 2    & $1.20 \times 10^9$& 37783    &  4.6        & $1.5 \times 10^5$&  73.9      &  31 \\ [0.5ex]
\enddata
\end{deluxetable}

%% file: table_nucl2.tex
\begin{deluxetable}{ccc}
\small
\tablecaption{Mean composition of the envelope (X$_i > 10^{-9}$) 
at the end of each burst, for Model 2.}
                 \tablehead{
		 \colhead{Nucleus}   &
		 \colhead{Burst 1}   &
		 \colhead{Burst 2} 
		           }
\startdata
   $^1$H     & $2.3 \times 10^{-1}$   &  $1.0 \times 10^{-1}$   \\   
   $^4$He    & $3.0 \times 10^{-1}$   &  $1.9 \times 10^{-1}$     \\
   $^{12}$C  & $1.8 \times 10^{-2}$   &  $2.2 \times 10^{-2}$     \\
   $^{13}$C  & $6.8 \times 10^{-5}$   &  $6.2 \times 10^{-5}$     \\
   $^{14}$N  & $2.0 \times 10^{-3}$   &  $1.4 \times 10^{-3}$     \\
   $^{15}$N  & $2.6 \times 10^{-3}$   &  $2.0 \times 10^{-3}$     \\
   $^{16}$O  & $5.3 \times 10^{-4}$   &  $3.9 \times 10^{-4}$     \\
   $^{17}$O  & $4.8 \times 10^{-6}$   &  $3.0 \times 10^{-6}$     \\
   $^{18}$O  & $6.3 \times 10^{-5}$   &  $2.8 \times 10^{-5}$     \\
   $^{18}$F  & $2.1 \times 10^{-4}$   &  $5.7 \times 10^{-5}$     \\
   $^{19}$F  & $3.6 \times 10^{-4}$   &  $1.7 \times 10^{-4}$     \\
   $^{20}$Ne & $5.4 \times 10^{-4}$   &  $4.1 \times 10^{-4}$     \\
   $^{21}$Ne & $2.2 \times 10^{-5}$   &  $7.3 \times 10^{-6}$     \\
   $^{22}$Ne & $7.7 \times 10^{-5}$   &  $4.4 \times 10^{-5}$     \\
   $^{22}$Na & $3.6 \times 10^{-3}$   &  $1.6 \times 10^{-3}$     \\
   $^{23}$Na & $2.3 \times 10^{-4}$   &  $1.7 \times 10^{-4}$    \\
   $^{24}$Mg & $1.2 \times 10^{-3}$   &  $1.3 \times 10^{-3}$    \\
   $^{25}$Mg & $1.8 \times 10^{-3}$   &  $1.4 \times 10^{-3}$    \\
   $^{26}$Mg & $3.1 \times 10^{-3}$   &  $1.6 \times 10^{-3}$    \\
$^{26}$Al$^g$& $1.0 \times 10^{-4}$   &  $1.1 \times 10^{-4}$    \\
   $^{27}$Al & $1.6 \times 10^{-3}$   &  $1.1 \times 10^{-3}$    \\
   $^{28}$Si & $1.2 \times 10^{-3}$   &  $7.5 \times 10^{-3}$    \\
   $^{29}$Si & $4.6 \times 10^{-4}$   &  $6.7 \times 10^{-4}$    \\
   $^{30}$Si & $4.7 \times 10^{-3}$   &  $4.6 \times 10^{-3}$    \\
   $^{31}$P  & $6.1 \times 10^{-4}$   &  $1.4 \times 10^{-3}$    \\
   $^{32}$S  & $4.7 \times 10^{-4}$   &  $2.7 \times 10^{-2}$    \\
   $^{33}$S  & $4.0 \times 10^{-4}$   &  $2.4 \times 10^{-3}$    \\
   $^{34}$S  & $4.3 \times 10^{-3}$   &  $9.6 \times 10^{-3}$    \\
   $^{35}$Cl & $1.1 \times 10^{-3}$   &  $3.8 \times 10^{-3}$    \\
   $^{36}$Ar & $5.8 \times 10^{-4}$   &  $2.9 \times 10^{-3}$    \\
   $^{37}$Cl & $1.6 \times 10^{-7}$   &  $5.3 \times 10^{-7}$    \\
   $^{37}$Ar & $2.9 \times 10^{-4}$   &  $4.0 \times 10^{-4}$    \\
   $^{38}$Ar & $2.6 \times 10^{-3}$   &  $5.2 \times 10^{-3}$    \\
   $^{39}$K  & $4.9 \times 10^{-3}$   &  $8.7 \times 10^{-3}$    \\
   $^{40}$Ca & $3.1 \times 10^{-3}$   &  $3.4 \times 10^{-3}$   \\
   $^{41}$K  &  -                     &  $8.7 \times 10^{-9}$  \\
   $^{41}$Ca & $6.8 \times 10^{-5}$   &  $4.5 \times 10^{-5}$  \\
   $^{42}$Ca & $7.3 \times 10^{-4}$   &  $1.2 \times 10^{-3}$  \\
   $^{43}$Ca & $9.6 \times 10^{-5}$   &  $4.3 \times 10^{-4}$  \\
   $^{43}$Sc & $7.2 \times 10^{-4}$   &  $1.2 \times 10^{-3}$  \\
   $^{44}$Ca & $5.8 \times 10^{-9}$   &  $2.3 \times 10^{-6}$  \\
   $^{44}$Sc & $9.6 \times 10^{-8}$   &  $1.1 \times 10^{-5}$  \\
   $^{44}$Ti & $5.6 \times 10^{-4}$   &  $4.8 \times 10^{-4}$  \\
   $^{45}$Sc & $3.5 \times 10^{-5}$   &  $4.9 \times 10^{-5}$  \\
   $^{45}$Ti & $2.1 \times 10^{-4}$   &  $1.2 \times 10^{-4}$  \\
   $^{46}$Ti & $1.2 \times 10^{-3}$   &  $1.9 \times 10^{-3}$ \\
   $^{47}$Ti & $6.8 \times 10^{-4}$   &  $6.3 \times 10^{-4}$ \\
   $^{48}$Ti & $2.8 \times 10^{-8}$   &  $2.5 \times 10^{-7}$ \\
   $^{48}$V  & $4.4 \times 10^{-5}$   &  $9.4 \times 10^{-5}$ \\
   $^{48}$Cr & $1.9 \times 10^{-3}$   &  $1.7 \times 10^{-3}$ \\
   $^{49}$Ti & $2.4 \times 10^{-8}$   &  $2.9 \times 10^{-7}$ \\
   $^{49}$V  & $1.4 \times 10^{-3}$   &  $1.2 \times 10^{-3}$ \\
   $^{50}$Cr & $1.6 \times 10^{-3}$   &  $2.2 \times 10^{-3}$ \\
   $^{51}$V  & $6.4 \times 10^{-7}$   &  $5.5 \times 10^{-6}$ \\
   $^{51}$Cr & $3.4 \times 10^{-3}$   &  $4.9 \times 10^{-3}$ \\
   $^{52}$Cr & $1.6 \times 10^{-6}$   &  $8.1 \times 10^{-6}$ \\
   $^{52}$Mn & $1.0 \times 10^{-3}$   &  $1.5 \times 10^{-3}$ \\
   $^{52}$Fe & $1.7 \times 10^{-2}$   &  $1.3 \times 10^{-2}$ \\
   $^{53}$Mn & $9.9 \times 10^{-4}$   &  $8.4 \times 10^{-4}$ \\
   $^{54}$Fe & $9.2 \times 10^{-4}$   &  $1.2 \times 10^{-3}$ \\
   $^{55}$Mn & -                      &  $1.3 \times 10^{-8}$ \\
   $^{55}$Fe & $7.9 \times 10^{-5}$   &  $2.6 \times 10^{-4}$ \\
   $^{55}$Co & $2.8 \times 10^{-3}$   &  $3.7 \times 10^{-3}$ \\
   $^{56}$Fe & $1.4 \times 10^{-8}$   &  $3.6 \times 10^{-7}$ \\
   $^{56}$Co & $1.1 \times 10^{-4}$   &  $5.0 \times 10^{-4}$ \\
   $^{56}$Ni & $3.2 \times 10^{-2}$   &  $5.0 \times 10^{-2}$ \\
   $^{57}$Fe & $2.6 \times 10^{-9}$   &  $6.6 \times 10^{-8}$   \\  
   $^{57}$Co & $6.9 \times 10^{-5}$   &  $7.4 \times 10^{-5}$     \\
   $^{57}$Ni & $5.1 \times 10^{-3}$   &  $3.4 \times 10^{-3}$     \\
   $^{58}$Ni & $3.4 \times 10^{-3}$   &  $2.3 \times 10^{-3}$     \\
   $^{59}$Ni & $4.8 \times 10^{-3}$   &  $3.7 \times 10^{-3}$     \\
   $^{60}$Ni & $2.9 \times 10^{-1}$   &  $3.2 \times 10^{-1}$     \\
   $^{61}$Ni & $9.1 \times 10^{-4}$   &  $1.4 \times 10^{-3}$     \\
   $^{61}$Cu & $6.3 \times 10^{-3}$   &  $5.8 \times 10^{-3}$     \\
   $^{62}$Ni & $9.9 \times 10^{-5}$   &  $1.9 \times 10^{-4}$     \\
   $^{62}$Zn & $1.8 \times 10^{-3}$   &  $2.1 \times 10^{-3}$     \\
   $^{63}$Cu & $2.2 \times 10^{-3}$   &  $3.2 \times 10^{-3}$     \\
   $^{64}$Zn & $2.3 \times 10^{-2}$   &  $1.2 \times 10^{-1}$     \\
   $^{65}$Cu & $2.4 \times 10^{-8}$   &  $2.7 \times 10^{-7}$     \\
   $^{65}$Zn & $5.4 \times 10^{-4}$   &  $3.1 \times 10^{-3}$     \\
   $^{66}$Zn & $1.7 \times 10^{-6}$   &  $2.9 \times 10^{-5}$     \\
   $^{66}$Ga & $6.4 \times 10^{-5}$   &  $6.2 \times 10^{-4}$    \\ 
   $^{66}$Ge & $2.8 \times 10^{-4}$   &  $1.7 \times 10^{-3}$    \\
   $^{67}$Ga & $1.8 \times 10^{-4}$   &  $1.2 \times 10^{-3}$    \\
   $^{68}$Ge & $1.6 \times 10^{-3}$   &  $2.8 \times 10^{-2}$    \\
   $^{69}$Ge & $7.3 \times 10^{-5}$   &  $1.8 \times 10^{-3}$    \\
   $^{70}$Ge & $2.0 \times 10^{-5}$   &  $5.9 \times 10^{-4}$    \\
   $^{71}$As & $2.4 \times 10^{-5}$   &  $6.2 \times 10^{-4}$    \\
   $^{72}$Se & $1.2 \times 10^{-4}$   &  $5.8 \times 10^{-3}$    \\
   $^{73}$Se & $8.0 \times 10^{-6}$   &  $6.1 \times 10^{-4}$    \\
   $^{74}$Se & $3.2 \times 10^{-6}$   &  $2.7 \times 10^{-4}$    \\
   $^{75}$Br & $3.3 \times 10^{-6}$   &  $2.7 \times 10^{-4}$    \\
   $^{76}$Kr & $1.2 \times 10^{-5}$   &  $1.6 \times 10^{-3}$    \\
   $^{77}$Kr & $1.2 \times 10^{-6}$   &  $2.3 \times 10^{-4}$    \\
   $^{78}$Kr & $6.7 \times 10^{-7}$   &  $1.5 \times 10^{-4}$    \\
   $^{79}$Kr & $4.0 \times 10^{-7}$   &  $9.2 \times 10^{-5}$    \\
   $^{80}$Sr & $1.5 \times 10^{-6}$   &  $4.9 \times 10^{-4}$    \\
   $^{81}$Rb & $2.3 \times 10^{-7}$   &  $1.1 \times 10^{-4}$    \\
   $^{82}$Sr & $2.2 \times 10^{-7}$   &  $1.4 \times 10^{-4}$    \\
   $^{83}$Sr & $9.4 \times 10^{-8}$   &  $7.0 \times 10^{-5}$    \\
   $^{84}$Sr & $6.4 \times 10^{-8}$   &  $6.3 \times 10^{-5}$    \\
   $^{85}$Y  & $4.5 \times 10^{-8}$   &  $6.4 \times 10^{-5}$    \\
   $^{86}$Zr & $3.1 \times 10^{-8}$   &  $7.8 \times 10^{-5}$    \\
   $^{87}$Zr & $1.4 \times 10^{-8}$   &  $6.9 \times 10^{-5}$   \\
   $^{88}$Zr & $3.5 \times 10^{-9}$   &  $3.3 \times 10^{-5}$   \\
   $^{89}$Nb & $2.5 \times 10^{-9}$   &  $5.7 \times 10^{-5}$   \\
   $^{90}$Mo & -                      &  $2.4 \times 10^{-5}$  \\
   $^{91}$Nb & -                      &  $1.3 \times 10^{-5}$   \\
   $^{92}$Mo & -                      &  $7.3 \times 10^{-6}$  \\
   $^{93}$Tc & -                      &  $6.9 \times 10^{-6}$  \\
   $^{94}$Tc & -                      &  $5.0 \times 10^{-6}$  \\
   $^{95}$Ru & -                      &  $9.2 \times 10^{-7}$  \\
   $^{96}$Ru & -                      &  $9.1 \times 10^{-8}$  \\
   $^{97}$Ru & -                      &  $3.7 \times 10^{-8}$  \\ 
   $^{98}$Ru & -                      &  $9.1 \times 10^{-9}$  \\ 
   $^{99}$Rh & -                      &  $1.8 \times 10^{-9}$  \\ [0.5ex]
\enddata
\end{deluxetable}

%% file: table6.tex
 \begin{deluxetable}{ccc}
 \small
 \tablecaption{Properties of the last burst computed in Models 1 \& 3.}
                   \tablehead{
                     \colhead{}   &
                     \colhead{Model 1}   &
                     \colhead{Model 3}
	                       }
\startdata
 $\rho_{max,base}$ (g.cm$^{-3}$)  & $1.3 \times 10^6$     & $2.6 \times 10^6$  \\
 P$_{max,base}$ (dyn.cm$^{-2}$)   & $4.2 \times 10^{22}$  & $1.1 \times 10^{23}$  \\
 $\rho_{max,ign}$ (g.cm$^{-3}$)  & $5.4 \times 10^5$     & $1.0 \times 10^6$  \\
 P$_{max,ign}$ (dyn.cm$^{-2}$)   & $1.3 \times 10^{22}$  & $2.9 \times 10^{22}$  \\
 $\tau_{acc}$ (hr)           & 5.1                   & 8.8  \\
 $\Delta m_{acc}$ (M$_\odot$)& $1.0 \times 10^{-12}$ &  $1.8 \times 10^{-12}$ \\
 T$_{peak}$ (K)              & $1.1 \times 10^9$     &  $1.3 \times 10^9$  \\ 
 L$_{peak}$ (L$_\odot$)      & $1.2 \times 10^5$     &  $1.0 \times 10^5$  \\
 $\Delta z_{max}$ (m)        & 40                    &  44  \\
 $\alpha$                    & 36                    &  30  \\[0.5ex]
 \enddata
 \tablenotetext{a}{$\rho_{max,base}$ and  P$_{max,base}$ are the maximum density and pressure achieved at the base of the envelope,
 whereas  $\rho_{max,ign}$ and  P$_{max,ign}$ correspond to the maximum values attained at the ignition shell (defined as the
 first shell that reaches $T > 4 \times 10^8$ K, for this burst).
 }
\end{deluxetable}

%% file: table_prop_m3.tex
\begin{deluxetable}{ccccccc}
\small
\tablecaption{ Summary of burst properties for Model 3. }
                  \tablehead{
		  \colhead{Burst}   &
		  \colhead{T$_{peak}$ (K)}   &
		  \colhead{t(T$_{peak}$) (s)} &
		  \colhead{$\tau_{rec}$ (hr)} &
		  \colhead{L$_{peak}$ (L$_\odot$)} &
		  \colhead{$\tau_{0.01}$ (s)} &
		  \colhead{$\alpha$} 
		                    }
\startdata
 1    & $1.40 \times 10^9$& 65110    &  18.1       & $1.0 \times 10^5$&  423      &  34 \\
 2    & $1.39 \times 10^9$& 98879    &  9.4        & $1.1 \times 10^5$&  296      &  24 \\
 3    & $1.32 \times 10^9$& 130816   &  8.9        & $9.8 \times 10^4$&  281      &  24 \\
 4    & $1.30 \times 10^9$& 162777   &  8.9        & $1.0 \times 10^5$&  252      &  27 \\ 
 5    & $1.26 \times 10^9$& 194266   &  8.8        & $1.0 \times 10^5$&  250      &  30 \\ [0.5ex]
\enddata
\end{deluxetable}

%% file: table_nucl3.tex
\begin{deluxetable}{cccccc}
\small
\tablecaption{Mean composition of the envelope (X$_i > 10^{-9}$) at the 
end of each burst, for Model 3.}
                 \tablehead{
		 \colhead{Nucleus}   &
		 \colhead{Burst 1}   &
		 \colhead{Burst 2} &
		 \colhead{Burst 3} &
		 \colhead{Burst 4} &
		 \colhead{Burst 5}
				   }
\startdata
   $^1$H     & $1.8 \times 10^{-1}$   &  $8.6 \times 10^{-2}$      &  $5.5 \times 10^{-2}$     &   $4.5 \times 10^{-2}$   &  $3.7 \times 10^{-2}$    \\
   $^4$He    & $8.4 \times 10^{-2}$   &  $5.8 \times 10^{-2}$      &  $4.4 \times 10^{-2}$     &   $3.8 \times 10^{-2}$   &  $3.3 \times 10^{-2}$ \\ 
   $^{12}$C  & $7.7 \times 10^{-4}$   &  $1.7 \times 10^{-3}$      &  $1.8 \times 10^{-3}$     &   $1.7 \times 10^{-3}$   &  $1.5 \times 10^{-3}$ \\
   $^{13}$C  & $4.2 \times 10^{-6}$   &  $8.9 \times 10^{-6}$      &  $1.4 \times 10^{-5}$     &   $2.3 \times 10^{-5}$   &  $2.7 \times 10^{-5}$ \\
   $^{14}$N  & $2.8 \times 10^{-4}$   &  $2.1 \times 10^{-4}$      &  $1.7 \times 10^{-4}$     &   $1.5 \times 10^{-4}$   &   $1.4 \times 10^{-4}$\\
   $^{15}$N  & $4.9 \times 10^{-4}$   &  $3.6 \times 10^{-4}$      &  $2.9 \times 10^{-4}$     &   $2.4 \times 10^{-4}$   &  $2.2 \times 10^{-4}$ \\
   $^{16}$O  & $1.4 \times 10^{-5}$   &  $2.9 \times 10^{-5}$      &  $4.1 \times 10^{-5}$     &   $3.7 \times 10^{-5}$   &  $3.8 \times 10^{-5}$ \\
   $^{17}$O  & $4.8 \times 10^{-7}$   &  $6.8 \times 10^{-7}$      &  $2.1 \times 10^{-6}$     &   $1.3 \times 10^{-6}$   &  $2.0 \times 10^{-6}$ \\
   $^{18}$O  & -                      &  $3.7 \times 10^{-9}$      &  $9.4 \times 10^{-9}$     &   $9.5 \times 10^{-9}$   &  $8.1 \times 10^{-9}$ \\
   $^{18}$F  & $6.4 \times 10^{-9}$   &  $1.3 \times 10^{-8}$      &  $2.8 \times 10^{-8}$     &   $2.4 \times 10^{-8}$   &  $2.3 \times 10^{-8}$ \\
   $^{19}$F  & $8.7 \times 10^{-8}$   &  $1.6 \times 10^{-7}$      &  $1.7 \times 10^{-7}$     &   $2.0 \times 10^{-7}$   &   $8.0 \times 10^{-8}$\\
   $^{20}$Ne & $1.5 \times 10^{-5}$   &  $2.9 \times 10^{-5}$      &  $3.5 \times 10^{-5}$     &   $3.0 \times 10^{-5}$   &  $3.1 \times 10^{-5}$ \\
   $^{21}$Ne & $1.9 \times 10^{-8}$   &  $4.4 \times 10^{-8}$      &  $4.9 \times 10^{-8}$     &   $5.2 \times 10^{-8}$   &  $4.4 \times 10^{-8}$ \\
   $^{22}$Ne & $2.9 \times 10^{-7}$   &  $6.6 \times 10^{-7}$      &  $4.3 \times 10^{-7}$     &   $1.1 \times 10^{-6}$   &   $7.5 \times 10^{-7}$\\
   $^{22}$Na & $1.1 \times 10^{-5}$   &  $2.4 \times 10^{-5}$      &  $2.0 \times 10^{-5}$     &   $2.3 \times 10^{-5}$   &  $1.9 \times 10^{-5}$ \\
   $^{23}$Na & $2.6 \times 10^{-6}$   &  $6.5 \times 10^{-6}$      &  $6.5 \times 10^{-6}$     &   $6.5 \times 10^{-6}$   &   $5.7 \times 10^{-6}$\\
   $^{24}$Mg & $8.2 \times 10^{-5}$   &  $1.4 \times 10^{-4}$      &  $1.4 \times 10^{-4}$     &   $1.2 \times 10^{-4}$   &   $1.1 \times 10^{-4}$\\
   $^{25}$Mg & $4.9 \times 10^{-5}$   &  $7.3 \times 10^{-5}$      &  $7.4 \times 10^{-5}$     &   $7.2 \times 10^{-5}$   &  $6.6 \times 10^{-5}$ \\
   $^{26}$Mg & $1.1 \times 10^{-5}$   &  $1.9 \times 10^{-5}$      &  $1.8 \times 10^{-5}$     &   $2.2 \times 10^{-5}$   &   $2.0 \times 10^{-5}$\\
$^{26}$Al$^g$& $1.2 \times 10^{-5}$   &  $1.4 \times 10^{-5}$      &  $1.3 \times 10^{-5}$     &   $1.2 \times 10^{-5}$   &   $1.1 \times 10^{-5}$\\
   $^{27}$Al & $4.1 \times 10^{-5}$   &  $6.3 \times 10^{-5}$      &  $6.5 \times 10^{-5}$     &   $6.8 \times 10^{-5}$   &  $6.2 \times 10^{-5}$ \\
   $^{28}$Si & $2.6 \times 10^{-4}$   &  $6.2 \times 10^{-4}$      &  $5.2 \times 10^{-4}$     &   $3.9 \times 10^{-4}$   &  $3.6 \times 10^{-4}$ \\
   $^{29}$Si & $8.6 \times 10^{-6}$   &  $1.1 \times 10^{-5}$      &  $9.2 \times 10^{-6}$     &   $1.3 \times 10^{-5}$   &  $1.1 \times 10^{-5}$ \\
   $^{30}$Si & $5.5 \times 10^{-5}$   &  $8.8 \times 10^{-5}$      &  $8.0 \times 10^{-5}$     &   $1.1 \times 10^{-4}$   &  $9.9 \times 10^{-5}$ \\
   $^{31}$P  & $2.3 \times 10^{-5}$   &  $3.6 \times 10^{-5}$      &  $2.6 \times 10^{-5}$     &   $3.2 \times 10^{-5}$   &  $2.6 \times 10^{-5}$ \\
   $^{32}$S  & $3.7 \times 10^{-5}$   &  $2.1 \times 10^{-3}$      &  $4.9 \times 10^{-3}$     &   $5.8 \times 10^{-3}$   &  $7.7 \times 10^{-3}$ \\
   $^{33}$S  & $5.7 \times 10^{-6}$   &  $4.2 \times 10^{-5}$      &  $6.1 \times 10^{-5}$     &   $5.6 \times 10^{-5}$   &  $9.2 \times 10^{-5}$ \\
   $^{34}$S  & $2.6 \times 10^{-5}$   &  $6.7 \times 10^{-5}$      &  $7.2 \times 10^{-5}$     &   $7.3 \times 10^{-5}$   &  $7.5 \times 10^{-5}$ \\
   $^{35}$Cl & $2.9 \times 10^{-5}$   &  $1.5 \times 10^{-4}$      &  $4.2 \times 10^{-4}$     &   $5.0 \times 10^{-4}$   &  $5.9 \times 10^{-4}$ \\
   $^{36}$Ar & $8.2 \times 10^{-6}$   &  $2.9 \times 10^{-3}$      &  $6.8 \times 10^{-3}$     &   $9.9 \times 10^{-3}$   &  $1.2 \times 10^{-2}$ \\
   $^{37}$Cl & $3.5 \times 10^{-9}$   &  $2.8 \times 10^{-8}$      &  $5.4 \times 10^{-8}$     &   $1.5 \times 10^{-7}$   &  $2.2 \times 10^{-7}$ \\ 
   $^{37}$Ar & $3.7 \times 10^{-6}$   &  $2.6 \times 10^{-5}$      &  $5.1 \times 10^{-5}$     &   $1.1 \times 10^{-4}$   &  $1.6 \times 10^{-4}$ \\
   $^{38}$Ar & $1.5 \times 10^{-5}$   &  $6.4 \times 10^{-5}$      &  $1.1 \times 10^{-4}$     &   $1.7 \times 10^{-4}$   &  $2.0 \times 10^{-4}$ \\
   $^{39}$K  & $4.7 \times 10^{-5}$   &  $9.4 \times 10^{-4}$      &  $1.4 \times 10^{-3}$     &   $2.1 \times 10^{-3}$   &  $2.5 \times 10^{-3}$ \\
   $^{40}$Ca & $1.3 \times 10^{-4}$   &  $1.3 \times 10^{-3}$      &  $1.8 \times 10^{-3}$     &   $2.3 \times 10^{-3}$   &  $2.6 \times 10^{-3}$ \\
   $^{41}$K  &  -                     &  -                         &  -                        &   -                      &   $1.1 \times 10^{-9}$ \\
   $^{41}$Ca & $2.5 \times 10^{-6}$   &  $6.4 \times 10^{-6}$      &  $7.7 \times 10^{-6}$     &   $1.4 \times 10^{-5}$   &  $1.7 \times 10^{-5}$ \\
   $^{42}$Ca & $3.7 \times 10^{-6}$   &  $1.6 \times 10^{-5}$      &  $1.8 \times 10^{-5}$     &   $3.3 \times 10^{-5}$   &  $3.5 \times 10^{-5}$ \\
   $^{43}$Ca & $3.3 \times 10^{-6}$   &  $1.1 \times 10^{-5}$      &  $1.0 \times 10^{-5}$     &   $2.0 \times 10^{-5}$   &  $2.6 \times 10^{-5}$ \\
   $^{43}$Sc & $1.4 \times 10^{-5}$   &  $4.0 \times 10^{-5}$      &  $3.7 \times 10^{-5}$     &   $5.3 \times 10^{-5}$   &  $5.2 \times 10^{-5}$ \\
   $^{44}$Ca & -                      &  $4.2 \times 10^{-9}$      &  $5.7 \times 10^{-7}$     &   $1.6 \times 10^{-6}$   &  $5.1 \times 10^{-6}$ \\
   $^{44}$Sc & $6.0 \times 10^{-9}$   &  $2.8 \times 10^{-8}$      &  $2.2 \times 10^{-6}$     &   $3.3 \times 10^{-6}$   &  $9.2 \times 10^{-6}$ \\
   $^{44}$Ti & $1.5 \times 10^{-5}$   &  $4.9 \times 10^{-5}$      &  $6.3 \times 10^{-5}$     &   $8.0 \times 10^{-5}$   &  $9.0 \times 10^{-5}$ \\
   $^{45}$Sc & $1.2 \times 10^{-6}$   &  $1.9 \times 10^{-6}$      &  $6.0 \times 10^{-6}$     &   $1.3 \times 10^{-5}$   &  $2.3 \times 10^{-5}$ \\
   $^{45}$Ti & $4.0 \times 10^{-6}$   &  $6.0 \times 10^{-6}$      &  $1.8 \times 10^{-5}$     &   $1.9 \times 10^{-5}$   &  $2.3 \times 10^{-5}$ \\
   $^{46}$Ti & $7.0 \times 10^{-6}$   &  $1.4 \times 10^{-5}$      &  $1.8 \times 10^{-5}$     &   $3.9 \times 10^{-5}$   &  $5.4 \times 10^{-5}$ \\
   $^{47}$Ti & $1.4 \times 10^{-5}$   &  $3.0 \times 10^{-5}$      &  $3.9 \times 10^{-5}$     &   $5.7 \times 10^{-5}$   &  $6.7 \times 10^{-5}$ \\
   $^{48}$Ti & $2.6 \times 10^{-9}$   &  $5.8 \times 10^{-9}$      &  $4.9 \times 10^{-8}$     &   $2.0 \times 10^{-7}$   &  $5.4 \times 10^{-7}$ \\
   $^{48}$V  & $2.4 \times 10^{-6}$   &  $4.8 \times 10^{-6}$      &  $1.3 \times 10^{-5}$     &   $2.7 \times 10^{-5}$   &  $4.7 \times 10^{-5}$ \\
   $^{48}$Cr & $5.2 \times 10^{-5}$   &  $1.0 \times 10^{-4}$      &  $1.2 \times 10^{-4}$     &   $1.3 \times 10^{-4}$   &  $1.3 \times 10^{-4}$ \\
   $^{49}$Ti & $1.9 \times 10^{-9}$   &  $2.2 \times 10^{-8}$      &  $4.8 \times 10^{-8}$     &   $8.3 \times 10^{-8}$   &  $1.2 \times 10^{-7}$ \\
   $^{49}$V  & $4.7 \times 10^{-5}$   &  $6.3 \times 10^{-5}$      &  $7.9 \times 10^{-5}$     &   $8.8 \times 10^{-5}$   &  $9.5 \times 10^{-5}$ \\
   $^{50}$Cr & $3.9 \times 10^{-5}$   &  $5.7 \times 10^{-5}$      &  $6.9 \times 10^{-5}$     &   $1.0 \times 10^{-4}$   &  $1.3 \times 10^{-4}$ \\
   $^{51}$V  & $5.3 \times 10^{-8}$   &  $1.1 \times 10^{-7}$      &  $3.2 \times 10^{-7}$     &   $7.9 \times 10^{-7}$   &  $1.7 \times 10^{-6}$ \\
   $^{51}$Cr & $1.2 \times 10^{-4}$   &  $1.9 \times 10^{-4}$      &  $2.2 \times 10^{-4}$     &   $2.7 \times 10^{-4}$   &  $3.1 \times 10^{-4}$ \\
   $^{52}$Cr & $5.6 \times 10^{-7}$   &  $7.4 \times 10^{-7}$      &  $4.0 \times 10^{-6}$     &   $1.3 \times 10^{-5}$   &  $2.8 \times 10^{-5}$ \\
   $^{52}$Mn & $2.1 \times 10^{-4}$   &  $2.5 \times 10^{-4}$      &  $4.0 \times 10^{-4}$     &   $6.0 \times 10^{-4}$   &  $7.7 \times 10^{-4}$ \\
   $^{52}$Fe & $2.1 \times 10^{-3}$   &  $2.2 \times 10^{-3}$      &  $2.1 \times 10^{-3}$     &   $1.8 \times 10^{-3}$   &  $1.6 \times 10^{-3}$ \\
   $^{53}$Mn & $8.8 \times 10^{-5}$   &  $2.1 \times 10^{-4}$      &  $3.6 \times 10^{-4}$     &   $4.0 \times 10^{-4}$   &  $4.2 \times 10^{-4}$ \\
   $^{54}$Fe & $7.7 \times 10^{-5}$   &  $1.2 \times 10^{-4}$      &  $1.5 \times 10^{-4}$     &   $1.8 \times 10^{-4}$   &  $1.9 \times 10^{-4}$ \\
   $^{55}$Mn & -                      &  -                         &  $5.1 \times 10^{-9}$     &   $1.6 \times 10^{-8}$   &  $3.4 \times 10^{-8}$ \\
   $^{55}$Fe & $1.0 \times 10^{-5}$   &  $1.7 \times 10^{-5}$      &  $4.5 \times 10^{-5}$     &   $7.7 \times 10^{-5}$   &  $1.1 \times 10^{-4}$ \\
   $^{55}$Co & $2.2 \times 10^{-4}$   &  $3.1 \times 10^{-4}$      &  $3.0 \times 10^{-4}$     &   $3.0 \times 10^{-4}$   &  $2.9 \times 10^{-4}$ \\
   $^{56}$Fe & $1.0 \times 10^{-9}$   &  $1.3 \times 10^{-8}$      &  $1.4 \times 10^{-7}$     &   $4.7 \times 10^{-7}$   &  $9.9 \times 10^{-7}$ \\
   $^{56}$Co & $4.6 \times 10^{-6}$   &  $2.2 \times 10^{-5}$      &  $8.4 \times 10^{-5}$     &   $1.6 \times 10^{-4}$   &  $2.5 \times 10^{-4}$ \\
   $^{56}$Ni & $7.8 \times 10^{-4}$   &  $2.1 \times 10^{-3}$      &  $2.7 \times 10^{-3}$     &   $3.3 \times 10^{-3}$   &  $3.6 \times 10^{-3}$ \\
   $^{57}$Fe & -                      &  $3.3 \times 10^{-8}$      &  $6.2 \times 10^{-8}$     &   $1.0 \times 10^{-7}$   &  $1.5 \times 10^{-7}$ \\
   $^{57}$Co & $1.3 \times 10^{-5}$   &  $1.5 \times 10^{-5}$      &  $2.9 \times 10^{-5}$     &   $4.8 \times 10^{-5}$   &  $6.6 \times 10^{-5}$ \\
   $^{57}$Ni & $5.6 \times 10^{-4}$   &  $5.8 \times 10^{-4}$      &  $5.7 \times 10^{-4}$     &   $5.1 \times 10^{-4}$   &  $4.8 \times 10^{-4}$ \\
   $^{58}$Ni & $4.1 \times 10^{-4}$   &  $4.0 \times 10^{-4}$      &  $4.0 \times 10^{-4}$     &   $3.7 \times 10^{-4}$   &  $3.6 \times 10^{-4}$ \\
   $^{59}$Ni & $6.9 \times 10^{-4}$   &  $6.4 \times 10^{-4}$      &  $6.3 \times 10^{-4}$     &   $5.8 \times 10^{-4}$   &  $5.7 \times 10^{-4}$ \\
   $^{60}$Ni & $2.4 \times 10^{-2}$   &  $3.7 \times 10^{-2}$      &  $4.6 \times 10^{-2}$     &   $5.2 \times 10^{-2}$   &  $5.8 \times 10^{-2}$ \\
   $^{61}$Ni & $1.5 \times 10^{-3}$   &  $2.1 \times 10^{-3}$      &  $3.8 \times 10^{-3}$     &   $5.5 \times 10^{-3}$   &  $6.3 \times 10^{-3}$ \\
   $^{61}$Cu & $5.7 \times 10^{-3}$   &  $6.9 \times 10^{-3}$      &  $6.5 \times 10^{-3}$     &   $4.7 \times 10^{-3}$   &  $3.7 \times 10^{-3}$ \\
   $^{62}$Ni & $3.3 \times 10^{-4}$   &  $5.8 \times 10^{-4}$      &  $1.4 \times 10^{-3}$     &   $1.8 \times 10^{-3}$   &  $2.0 \times 10^{-3}$ \\
   $^{62}$Zn & $3.6 \times 10^{-3}$   &  $4.4 \times 10^{-3}$      &  $3.3 \times 10^{-3}$     &   $2.3 \times 10^{-3}$   &  $1.9 \times 10^{-3}$ \\
   $^{63}$Cu & $4.1 \times 10^{-3}$   &  $3.3 \times 10^{-3}$      &  $2.9 \times 10^{-3}$     &   $2.5 \times 10^{-3}$   &  $2.5 \times 10^{-3}$ \\
   $^{64}$Zn & $4.2 \times 10^{-2}$   &  $5.6 \times 10^{-2}$      &  $7.0 \times 10^{-2}$     &   $8.1 \times 10^{-2}$   &  $9.1 \times 10^{-2}$ \\
   $^{65}$Cu & $8.1 \times 10^{-7}$   &  $1.2 \times 10^{-6}$      &  $4.9 \times 10^{-6}$     &   $1.1 \times 10^{-5}$   &  $1.7 \times 10^{-5}$ \\
   $^{65}$Zn & $8.3 \times 10^{-3}$   &  $9.7 \times 10^{-3}$      &  $1.2 \times 10^{-2}$     &   $1.2 \times 10^{-2}$   &  $1.2 \times 10^{-2}$ \\
   $^{66}$Zn & $1.3 \times 10^{-4}$   &  $2.3 \times 10^{-4}$      &  $9.9 \times 10^{-4}$     &   $1.8 \times 10^{-3}$   &  $2.3 \times 10^{-3}$ \\
   $^{66}$Ga & $2.8 \times 10^{-3}$   &  $2.7 \times 10^{-3}$      &  $3.1 \times 10^{-3}$     &   $2.7 \times 10^{-3}$   &  $2.2 \times 10^{-3}$ \\
   $^{66}$Ge & $6.8 \times 10^{-3}$   &  $4.9 \times 10^{-3}$      &  $3.8 \times 10^{-3}$     &   $2.3 \times 10^{-3}$   &  $1.9 \times 10^{-3}$ \\
   $^{67}$Ga & $4.1 \times 10^{-3}$   &  $3.4 \times 10^{-3}$      &  $3.6 \times 10^{-3}$     &   $3.2 \times 10^{-3}$   &  $3.0 \times 10^{-3}$ \\
   $^{68}$Ge & $2.8 \times 10^{-2}$   &  $4.5 \times 10^{-2}$      &  $5.6 \times 10^{-2}$     &   $6.5 \times 10^{-2}$   &  $7.1 \times 10^{-2}$ \\
   $^{69}$Ge & $1.0 \times 10^{-2}$   &  $1.0 \times 10^{-3}$      &  $1.2 \times 10^{-2}$     &   $1.1 \times 10^{-2}$   &  $1.1 \times 10^{-2}$ \\
   $^{70}$Ge & $6.5 \times 10^{-3}$   &  $5.2 \times 10^{-3}$      &  $5.3 \times 10^{-3}$     &   $4.6 \times 10^{-3}$   &  $4.2 \times 10^{-3}$ \\
   $^{71}$As & $6.1 \times 10^{-3}$   &  $4.8 \times 10^{-3}$      &  $5.0 \times 10^{-3}$     &   $4.4 \times 10^{-3}$   &  $4.2 \times 10^{-3}$ \\
   $^{72}$Se & $1.7 \times 10^{-2}$   &  $2.8 \times 10^{-2}$      &  $3.4 \times 10^{-2}$     &   $3.8 \times 10^{-2}$   &  $4.1 \times 10^{-2}$ \\
   $^{73}$Se & $9.1 \times 10^{-3}$   &  $6.5 \times 10^{-3}$      &  $7.0 \times 10^{-3}$     &   $6.4 \times 10^{-3}$   &  $6.3 \times 10^{-3}$ \\
   $^{74}$Se & $7.2 \times 10^{-3}$   &  $6.5 \times 10^{-3}$      &  $7.7 \times 10^{-3}$     &   $7.4 \times 10^{-3}$   &  $7.2 \times 10^{-3}$ \\
   $^{75}$Br & $8.4 \times 10^{-3}$   &  $4.5 \times 10^{-3}$      &  $3.3 \times 10^{-3}$     &   $2.2 \times 10^{-3}$   &  $2.0 \times 10^{-3}$ \\
   $^{76}$Kr & $1.3 \times 10^{-2}$   &  $1.8 \times 10^{-2}$      &  $2.0 \times 10^{-2}$     &   $2.2 \times 10^{-2}$   &  $2.3 \times 10^{-2}$ \\
   $^{77}$Kr & $7.2 \times 10^{-3}$   &  $6.0 \times 10^{-3}$      &  $6.2 \times 10^{-3}$     &   $5.9 \times 10^{-3}$   &  $5.7 \times 10^{-3}$ \\
   $^{78}$Kr & $8.7 \times 10^{-3}$   &  $7.3 \times 10^{-3}$      &  $7.5 \times 10^{-3}$     &   $6.9 \times 10^{-3}$   &  $6.5 \times 10^{-3}$ \\
   $^{79}$Kr & $7.0 \times 10^{-3}$   &  $3.5 \times 10^{-3}$      &  $3.0 \times 10^{-3}$     &   $2.5 \times 10^{-3}$   &  $2.4 \times 10^{-3}$ \\
   $^{80}$Sr & $1.3 \times 10^{-2}$   &  $1.3 \times 10^{-2}$      &  $1.4 \times 10^{-2}$     &   $1.5 \times 10^{-2}$   &  $1.5 \times 10^{-2}$ \\
   $^{81}$Rb & $8.0 \times 10^{-3}$   &  $6.4 \times 10^{-3}$      &  $6.4 \times 10^{-3}$     &   $6.0 \times 10^{-3}$   &  $5.8 \times 10^{-3}$ \\
   $^{82}$Sr & $1.7 \times 10^{-2}$   &  $1.3 \times 10^{-2}$      &  $1.2 \times 10^{-2}$     &   $1.2 \times 10^{-2}$   &  $1.1 \times 10^{-2}$ \\
   $^{83}$Sr & $1.3 \times 10^{-2}$   &  $1.2 \times 10^{-2}$      &  $1.4 \times 10^{-2}$     &   $1.3 \times 10^{-2}$   &  $1.3 \times 10^{-2}$ \\
   $^{84}$Sr & $1.3 \times 10^{-2}$   &  $6.1 \times 10^{-3}$      &  $3.8 \times 10^{-3}$     &   $2.6 \times 10^{-3}$   &  $2.2 \times 10^{-3}$ \\
   $^{85}$Y  & $9.8 \times 10^{-3}$   &  $7.9 \times 10^{-3}$      &  $7.7 \times 10^{-3}$     &   $7.2 \times 10^{-3}$   &  $6.9 \times 10^{-3}$ \\
   $^{86}$Zr & $1.7 \times 10^{-2}$   &  $1.9 \times 10^{-2}$      &  $2.1 \times 10^{-2}$     &   $2.1 \times 10^{-2}$   &  $2.0 \times 10^{-2}$ \\
   $^{87}$Zr & $2.1 \times 10^{-2}$   &  $9.1 \times 10^{-3}$      &  $5.1 \times 10^{-3}$     &   $3.8 \times 10^{-3}$   &  $3.0 \times 10^{-3}$ \\
   $^{88}$Zr & $5.9 \times 10^{-3}$   &  $5.5 \times 10^{-3}$      &  $5.0 \times 10^{-3}$     &   $5.1 \times 10^{-3}$   &  $5.3 \times 10^{-3}$ \\
   $^{89}$Nb & $1.8 \times 10^{-2}$   &  $2.0 \times 10^{-2}$      &  $1.9 \times 10^{-2}$     &   $1.8 \times 10^{-2}$   &  $1.8 \times 10^{-2}$ \\
   $^{90}$Mo & $1.2 \times 10^{-2}$   &  $1.1 \times 10^{-2}$      &  $9.9 \times 10^{-3}$     &   $9.6 \times 10^{-3}$   &  $9.2 \times 10^{-3}$ \\
   $^{91}$Nb & $8.2 \times 10^{-3}$   &  $9.4 \times 10^{-3}$      &  $1.0 \times 10^{-2}$     &   $1.0 \times 10^{-2}$   &  $9.9 \times 10^{-3}$ \\
   $^{92}$Mo & $7.6 \times 10^{-3}$   &  $4.1 \times 10^{-3}$      &  $2.3 \times 10^{-3}$     &   $1.8 \times 10^{-3}$   &  $1.4 \times 10^{-3}$ \\
   $^{93}$Tc & $9.5 \times 10^{-3}$   &  $8.8 \times 10^{-3}$      &  $8.1 \times 10^{-3}$     &   $7.6 \times 10^{-3}$   &  $7.4 \times 10^{-3}$ \\
   $^{94}$Tc & $2.6 \times 10^{-2}$   &  $4.6 \times 10^{-2}$      &  $4.6 \times 10^{-2}$     &   $4.6 \times 10^{-2}$   &  $4.6 \times 10^{-2}$ \\
   $^{95}$Ru & $3.1 \times 10^{-2}$   &  $2.4 \times 10^{-2}$      &  $1.4 \times 10^{-2}$     &   $9.3 \times 10^{-3}$   &  $7.6 \times 10^{-3}$ \\
   $^{96}$Ru & $8.1 \times 10^{-3}$   &  $8.7 \times 10^{-3}$      &  $7.4 \times 10^{-3}$     &   $6.6 \times 10^{-3}$   &  $6.2 \times 10^{-3}$ \\
   $^{97}$Ru & $1.0 \times 10^{-2}$   &  $1.2 \times 10^{-2}$      &  $1.0 \times 10^{-2}$     &   $9.5 \times 10^{-3}$   &  $9.1 \times 10^{-3}$ \\
   $^{98}$Ru & $1.2 \times 10^{-2}$   &  $1.4 \times 10^{-2}$      &  $1.3 \times 10^{-2}$     &   $1.3 \times 10^{-2}$   &  $1.2 \times 10^{-2}$ \\
   $^{99}$Rh & $1.3 \times 10^{-2}$   &  $2.1 \times 10^{-2}$      &  $2.1 \times 10^{-2}$     &   $2.0 \times 10^{-2}$   &  $1.9 \times 10^{-2}$ \\
  $^{100}$Pd & $1.6 \times 10^{-2}$   &  $9.2 \times 10^{-3}$      &  $8.7 \times 10^{-3}$     &   $8.0 \times 10^{-3}$   &  $8.2 \times 10^{-3}$ \\
  $^{101}$Pd & $1.7 \times 10^{-2}$   &  $1.8 \times 10^{-2}$      &  $1.8 \times 10^{-2}$     &   $1.8 \times 10^{-2}$   &  $1.7 \times 10^{-2}$ \\
  $^{102}$Pd & $1.6 \times 10^{-2}$   &  $1.8 \times 10^{-2}$      &  $1.9 \times 10^{-2}$     &   $1.9 \times 10^{-2}$   &  $1.8 \times 10^{-2}$ \\
  $^{103}$Ag & $2.6 \times 10^{-2}$   &  $3.2 \times 10^{-2}$      &  $3.2 \times 10^{-2}$     &   $3.3 \times 10^{-2}$   &  $3.2 \times 10^{-2}$ \\
  $^{104}$Pd & $5.3 \times 10^{-2}$   &  $7.2 \times 10^{-2}$      &  $7.4 \times 10^{-2}$     &   $7.5 \times 10^{-2}$   &  $7.5 \times 10^{-2}$ \\
  $^{105}$Ag & $7.5 \times 10^{-2}$   &  $1.3 \times 10^{-1}$      &  $1.4 \times 10^{-1}$     &   $1.4 \times 10^{-1}$   &  $1.4 \times 10^{-1}$ \\
  $^{106}$Cd & $1.2 \times 10^{-2}$   &  $9.8 \times 10^{-3}$      &  $5.3 \times 10^{-3}$     &   $3.2 \times 10^{-3}$   &  $2.5 \times 10^{-3}$ \\
  $^{107}$Cd & $9.5 \times 10^{-3}$   &  $6.4 \times 10^{-3}$      &  $3.1 \times 10^{-3}$     &   $1.9 \times 10^{-3}$   &  $1.5 \times 10^{-3}$ \\[0.5ex]
\enddata
\end{deluxetable}

%% file: xrb8v8.bbl
\begin{thebibliography}{}

\bibitem[]{Ang99} Angulo, C., et al., {\it Nucl. Phys. A}, {\bf 656}, 3 (1999)
\bibitem[]{Arn06} Arnould, M., \& Goriely, S., {\it Nucl. Phys. A}, {\bf 777}, 157 (2006)
\bibitem[]{Aud03a} Audi, G., Bersillon, O., Blachot, J., \& Wapstra, A. H., {\it Nucl. Phys. A}, {\bf 729}, 3 (2003a)
\bibitem[]{Aud03b} Audi, G., Wapstra, A. H., \& Thibault, C., {\it Nucl. Phys. A}, {\bf 729}, 337 (2003b)
\bibitem[]{Ayas82} Ayasli, S., and Joss, P.C., {\it ApJ}  {\bf 256}, 637 (1982)
\bibitem[]{Baz08} Bazin, D., et al., {\it Phys. Rev. Lett.}, {\bf 101}, 252501 (2008)
\bibitem[]{Bel76} Belian, R. D., Conner, J. P., \& Evans, W. D., {\it ApJ}, {\bf 206}, L135 (1976)
\bibitem[]{Bil98} Bildsten, L., in The Many Faces of Neutron Stars, ed. R. Buccheri, J. van
   Paradijs, \& M. A. Alpar (Dordrecht: Kluwer), 419 (1998)
\bibitem[]{Bli96} Blinnikov, S.I., Dunina-Barkovskaya, N.V., and 
       Nadyozhin, D.K., {\it ApJS} {\bf 106}, 171 (1996)
\bibitem[]{Bro04} Brown, E.F., {\it ApJ} {\bf 614}, L57 (2004)
\bibitem[]{Coo06} Cooper, R. L., Mukhopadhyay, B., Steeghs, D., and Narayan, R., {\it ApJ} {\bf 642}, 443 (2006)
\bibitem[]{Coo04} Cooper, R. L., and Narayan, R., {\it American Astronomical Society} {\bf 36}, 941 (2004)
\bibitem[]{Coo05} Cooper, R. L., and Narayan, R., {\it ApJ} {\bf }629, 422 (2005)
\bibitem[]{Cum05} Cumming, A., {\it Nuclear Physics A} {\bf 758}, 439c (2005)
\bibitem[]{Cum01} Cumming, A., and Bildsten, L., {\it ApJ} {\bf 559}, L127 (2001)
\bibitem[]{Cum02} Cumming, A., Morsink, S., Bildsten, L., Friedman, J.L., and Holz, D.E. {\it ApJ} {\bf 564}, 343 (2002)
\bibitem[]{DGC73} DeWitt,~H.E., Graboske,~H.C., and Cooper,~M.S.,
 {\it ApJ} {\bf 181}, 439 (1973)
\bibitem[]{Elo09} Elomaa, V.-V., et al., {\it Phys. Rev. Lett.}, {\bf 102}, 252501 (2009)
\bibitem[]{Fis04} Fisker,~J.L., Thielemann, F.-K., and Wiescher, M.C., {\it ApJ} {\bf 608}, L61 (2004)
\bibitem[]{Fis06} Fisker,~J.L., G\"orres, J., Wiescher, M.C., and Davids, B., {\it ApJ} {\bf 650}, 332 (2006)
\bibitem[]{Fis07} Fisker,~J.L., Tan, W.P., G\"orres, J., Wiescher, M.C., and Cooper, R.L. {\it ApJ} {\bf 665}, 637 (2007)
\bibitem[]{Fis08} Fisker,~J.L., Schatz, H., and Thielemann, F.-K., {\it ApJS} {\bf 174}, 261 (2008)
\bibitem[]{Fuj88} Fujimoto, M. Y., Sztajno, M., Lewin, W. H. G., and van Paradijs, J., {\it A\&A} {\bf 199}, L9 (1988)
\bibitem[]{Ful82a} Fuller, G.M., Fowler, W.A., and Newman, M.J., {\it ApJ} {\bf 25
2}, 715 (1982a)
\bibitem[]{Ful82b} Fuller, G.M., Fowler, W.A., and Newman, M.J., {\it ApJS} {\bf 4
8}, 279 (1982b)
\bibitem[]{Gal08} Galloway, D.K., Muno, M.P., Hartman, J.M., Savov, P., Psaltis, D., and Chakrabarty, D.,
                {\it ApJS} {\bf 179}, 360 (2008)
\bibitem[]{Gor98} Goriely, S., in {\it Nuclei in the Cosmos V}, N. Prantzos, S. Harissopulos (eds.),
		Paris: Ed. Frontieres, p. 314 (1998)
\bibitem[]{Gra73} Graboske,~H.C., DeWitt,~H.E., Grossman,~A.S., and Cooper,~M.S.,
 {\it ApJ} {\bf 181}, 457 (1973)
\bibitem[]{Gri76} Grindlay, J., Gursky, H., Schnopper, H., Parsignault, D. R., Heise, J., 
                Brinkman, A. C., and Schrijver, J.,    {\it ApJ} {\bf 205}, L127 (1976)
\bibitem[]{Han83} Hanawa, T., Sugimoto, D., and Hashimoto, M.-A.,  {\it PASJ} {\bf 35}, 491 (1983)
\bibitem[]{Heg07} Heger, A., Cumming, A., Galloway, D.K., and Woosley, S.E., {\it ApJ}  {\bf 671}, L141 (2007)
\bibitem[]{Ibe75} Iben,~I., {\it ApJ}  {\bf 196}, 525 (1975)
\bibitem[]{Ili01} Iliadis, C., D'Auria, J.M., Starrfield, S., Thompson, W.J., \& Wiescher, M.N., {\it ApJS} {\bf 134}, 151 (2001)
\bibitem[]{Jon04} Jonker, P. G., Galloway, D. K., McClintock, J. E., Buxton, M., Garcia, M., and
                 Murray, S., {\it MNRAS} {\bf 354}, 666 (2004)
\bibitem[]{JH98} Jos\'e, J., and Hernanz, M., {\it ApJ} {\bf 494}, 680 (1998)
\bibitem[]{JM06} Jos\'e, J., and Moreno, F., {\it Proc. Science}, PoS(NIC-IX) 123 (2006)
\bibitem[]{JM10} Jos\'e, J., and Moreno, F., {\it ApJ}, in preparation
\bibitem[]{Jos77} Joss,~P.C., {\it Nat} {\bf 270}, 310 (1977)
\bibitem[]{Koi99} Koike, O., Hashimoto, M., Arai, K., and Wanajo, S., {\it A\&A} {\bf 342}, 464 (1999)
\bibitem[]{Koi04} Koike, O., Hashimoto, M., Kuromizu, R., Fujimoto, S., {\it ApJ} {\bf 603}, 242 (2004)
\bibitem[]{Kuu02} Kuulkers, E., in't Zand, J. J. M., van Kerkwijk, M. H., Cornelisse, R., Smith, D. A.,
    Heise, J., Bazzano, A., Cocchi, M., Natalucci, L., and Ubertini, P., {\it A\&A} {\bf 382}, 503 (2002)
\bibitem[]{Kut80} Kutter, G. S., and Sparks, W. M., {\it ApJ} {\bf 239}, 988 (1980)
\bibitem[]{LM00} Langanke, K., and Martinez-Pinedo, G., {\it Nucl. Phys. A} {\bf 6
73}, 481 (2000)
\bibitem[]{Lat09} Lattimer, J., private com. (2009)
\bibitem[]{Lew93} Lewin, W.H.G., van Paradijs, J., and R.E. Taam, {\it Space Sci. Rev.} {\bf 62}, 233  (1993)
\bibitem[]{Lew95} Lewin, W.H.G., van Paradijs, J., and R.E. Taam, in {\it X-Ray Binaries}, W.H.G. Lewin, 
          J. van Paradijs, and E.P.J. van den Heuvel (eds.), Cambridge: Cambridge Univ. Press, p. 175 (1995)
\bibitem[]{Lie02} Liebend\"orfer, M., Rosswog, S., and Thielemann, F.-K., {\it ApJS} {\bf 141}, 229 (2002)
\bibitem[]{Mac07} MacAlpine, G.M., Ecklund, T.C., Lester, W.R., and Vanderveer, S.J., {\it AJ} {\bf 133}, 81 (2007)
\bibitem[]{Mar77} Maraschi, L., and Cavaliere, A., in {\it Highlights in Astronomy},
         E. Muller (ed.), Reidel:Dordrecht, {\bf vol. 4}, Part I, p. 127 (1977)
\bibitem[]{Par08} Parikh, A., Jos\'e, J., F. Moreno, and Iliadis, C., {\it ApJS} {\bf 178}, 110 (2008)
\bibitem[]{Pen89} Penninx, W., Damen, E., van Paradijs, J., Tan, J., and Lewin, W. H. G., {\it A\&A} {\bf 208}, 146 (1989)
\bibitem[]{PSS79} Prialnik,~D., Shara, ~M.M.,
        and Shaviv,~G., {\it A\&A} {\bf 72}, 192 (1979)
\bibitem[]{Psa06} Psaltis, D., in {\it Compact Stellar X-ray Sources}, 
    W. H. G. Lewin, and M. van der Klis (eds.), Cambridge: Cambridge Univ. Press, p. 1 (2006)
\bibitem[]{RT00} Rauscher, T., and Thielemann, F.-K., {\it At. Data Nucl. Data Tab.} {\bf 75}, 1 (2000)
\bibitem[]{Sch99} Schatz, H., Bildsten, L., Cumming, A., and Wiescher, M., {\it ApJ} {\bf 524}, 1014 (1999)
\bibitem[]{Sch01} Schatz, H., et al., {\it Phys. Rev. Lett.} {\bf 86}, 3471 (2001)
\bibitem[]{Sch98} Schatz, H., et al., {\it Phys. Rep.} {\bf 294}, 167 (1998)
\bibitem[]{SR06} Schatz, H., and Rehm, K.E., {\it Nucl. Phys. A} {\bf 777}, 601 (2006)
\bibitem[]{ST83} Shapiro, S.L., and Teukolsky, S.A, {\it Black holes, white
dwarfs, and neutron stars: the physics of compact objects}, New York, NY: Wiley, (1983)
\bibitem[]{Stro06} Strohmayer, T., \& Bildsten, L., in {\it Compact Stellar X-Ray Sources}, 
       W. H. G. Lewin, and M. van der Klis (eds.), Cambridge: Cambridge Univ. Press, p. 113 (2006)
\bibitem[]{Str02} Strohmayer, T., and Brown, E. F., {\it ApJ} {\bf 566}, 1045 (2002)
\bibitem[]{Taa80} Taam, R.E., {\it ApJ} {\bf 241}, 358 (1980) 
\bibitem[]{Taa93} Taam, R.E., Woosley, S.E., Weaver, T.A., and Lamb, D.Q., {\it ApJ} {\bf 413}, 324 (1993)
\bibitem[]{Taa96} Taam, R.E., Woosley, S.E., and Lamb, D.Q., {\it ApJ} {\bf 459}, 271 (1996)
\bibitem[]{Tan07} Tan, W.P., Fisker,~J.L., G\"orres, J., Couder, M., and Wiescher, M.C., {\it PRL} {\bf 98}, 242503 (2007)
\bibitem[]{Wal81} Wallace, R.K., and Woosley, S.E., {\it ApJS} {\bf 45}, 389 (1981)
\bibitem[]{Wal84} Wallace, R.K., and Woosley, S.E., in {\it High Energy Transients in
          Astrophysics},  S.E. Woosley (ed.), New York: AIP,  p. 319 (1984)
\bibitem[]{Wat07} Watts, A. L., and Maurer, I., {\it A\&A} {\bf 467}, L33 (2007)
\bibitem[]{Wei06} Weinberg, N., Bildsten, L., and Schatz, H., {\it ApJ} {\bf 639}, 1018 (2006)
\bibitem[]{Woo76} Woosley, S.E.,  and Taam, R.E., {\it Nat} {\bf 263}, 101 (1976)
\bibitem[]{Woo04} Woosley, S.E., et al., {\it ApJS} {\bf 151}, 75 (2004)
\bibitem[]{Woo84} Woosley, S.E., and Weaver, T.A., in {\it High Energy Transients in
          Astrophysics},  S.E. Woosley (ed.), New York: AIP,  p. 273 (1984)
\end{thebibliography}
